\documentclass{article}

\usepackage{amsmath,amssymb,mathtools}
\usepackage{jheppub}
\usepackage[utf8]{inputenc}
\usepackage[numbers]{natbib}
\usepackage{xcolor}
\usepackage{slashed,cancel}
\usepackage{physics}
\usepackage{listings}
\usepackage{grffile}
\usepackage[compat=1.1.0]{tikz-feynman}
\usepackage{placeins}
\usepackage{enumitem}
\usepackage{todonotes}
\usepackage{makecell,tablefootnote}

\newcommand{\standardwidth}{0.8}

\newcommand{\lambdaOLD}{\sqrt{\lambda}}
\newcommand{\pd}{\partial}
\newcommand{\alphaEM}{\alpha_{\rm EM}}
\newcommand{\alphaweak}{\alpha_{2}}
\newcommand{\mplanck}{m_{\rm Pl}}

\newcommand{\dphi}{\dot{\phi}}

\newcommand{\dtheta}{\dot{\theta}}
\newcommand{\dthetaap}{\dot{\theta}_{\rm ap}}

\newcommand{\mchitoday}{m_{\chi,\rm today}}
\newcommand{\meff}{m_{\phi,\rm eff}}
\newcommand{\mchieff}{m_{\chi,\rm eff}}
\newcommand{\mchiS}{m_{\chi,\phi}}
\newcommand{\mchiT}{m_{\chi,T}}
\newcommand{\dmeff}{\dot{m}_{\phi,\rm eff}}
\newcommand{\mth}{m_{\phi,\rm th}}
\newcommand{\mln}{m_{\phi,\rm ln}}
\newcommand{\kmin}{k_{\rm kin}}
\newcommand{\Nkin}{N_{\rm kin}}

\newcommand{\kick}{{\rm kick}}
\newcommand{\dom}{{\rm dom}}
\newcommand{\kin}{{\rm kin}}
\newcommand{\frag}{{\rm *}}
\newcommand{\damp}{{\rm damp}}
\newcommand{\eq}{{\rm eq}}
\newcommand{\pvt}{{\rm pivot}}

\newcommand{\Tdom}{T_{\rm dom}}
\newcommand{\Tkend}{T_{\rm kin,end}}
\newcommand{\Tfo}{T_{\rm fo}}
\newcommand{\Tfrag}{T_{*}}
\newcommand{\Tosc}{T_{\rm kick}}
\newcommand{\Tkick}{T_{\rm kick}}
\newcommand{\Tth}{T_{\rm damp}}
\newcommand{\TSth}{T_{\rm damp,heated}}
\newcommand{\Tmin}{T_{\rm kin}}
\newcommand{\Teq}{T_{\rm eq}}
\newcommand{\TBBN}{T_{\rm BBN}}
\newcommand{\Tdamp}{T_{\rm damp}}
\newcommand{\Tsd}{\Tdom}
\newcommand{\Tkin}{T_{\rm kin}}

\newcommand{\afrag}{a_*}

\newcommand{\akick}{a_{\rm kick}}

\newcommand{\amin}{a_{\rm kin}}

\newcommand{\Skick}{\phi_{\rm kick}}
\newcommand{\phikick}{\phi_{\rm kick}}

\newcommand{\mphiVprime}{m_{\phi,V'}}
\newcommand{\thetakick}{\theta_{\rm kick}}

\newcommand{\mphi}{m_{\phi}}
\newcommand{\mphiz}{m_{\phi,0}}
\newcommand{\nPQ}{n_{\rm PQ}}
\newcommand{\Ykick}{Y_{\rm kick}}
\newcommand{\Ydm}{Y_{\rm DM}}
\newcommand{\Ydiluted}{Y_{\rm diluted}}

\newcommand{\Asplanck}{A_s}

\newcommand{\Gammatot}{\Gamma_{\rm tot}}
\newcommand{\gchi}{g_s}
\newcommand{\alphachi}{\alpha_s}
\newcommand{\mHiggs}{m_{\rm H,0}}
\newcommand{\mHiggsth}{m_{\rm H,th}}

\definecolor{darkgreen}{HTML}{009B55}

\definecolor{tolblue}{HTML}{4477AA}
\definecolor{tolgreen}{HTML}{228833}
\definecolor{tolred}{HTML}{EE6677}

\begin{document}

\title{Model implementations of axion dark matter from kinetic misalignment}
\author[a,b]{Cem Eröncel}
\author[c]{Ryosuke Sato}
\author[d,e,f]{G\'eraldine Servant}
\author[g,h]{Philip Sørensen}

\emailAdd{cem.eroncel@itu.edu.tr}
\emailAdd{geraldine.servant@desy.de}
\emailAdd{rsato@het.phys.sci.osaka-u.ac.jp}
\emailAdd{philip.soerensen@pd.infn.it}

\affiliation[a]{Istanbul Technical University, Department of Physics, 34469 Maslak, Istanbul, Türkiye}
\affiliation[b]{MEF University, Faculty of Engineering, 34396 Maslak, Istanbul, Türkiye}
\affiliation[c]{Department of Physics, Osaka University, Toyonaka, Osaka 560-0043, Japan}
\affiliation[d]{Deutsches Elektronen-Synchrotron DESY, Notkestr. 85, 22607 Hamburg, Germany}
\affiliation[e]{II. Institute of Theoretical Physics, Universit\"{a}t  Hamburg, 22761 Hamburg, Germany}

\affiliation[f]{Theoretical Physics Department, CERN, 1211 Geneva 23, Switzerland}

\affiliation[g]{\small Dipartimento di Fisica e Astronomia ‘G. Galilei’, Università di Padova,\\ Via F. Marzolo 8, 35131 Padova, Italy}
\affiliation[h]{\small INFN Sezione di Padova, Via F. Marzolo 8, 35131 Padova, Italy}

\abstract{The axion kinetic misalignment mechanism (KMM) opens the possibility of explaining dark matter for almost any axion mass and decay constant that are not accessible by the standard misalignment mechanism, in particular at low values of the axion decay constant (i.e. large coupling).   This is a new opportunity for most axion experiments which could be sensitive to dark matter and probe new regimes of axion cosmology.
We scrutinise UV completions that lead to the KMM mechanism.
These mainly rely on the early dynamics of the axion partner, the radial mode of the complex scalar field, from which the axion inherits kinetic energy. The damping of the radial-mode energy density is then a necessary ingredient. We study in detail thermal damping from interactions in the plasma. 
A minimal and rather natural implementation consists of a KSVZ-type model with a nearly-quadratic potential for the radial mode extended by U(1)-breaking higher-dimensional operators. 
Furthermore, we study Higgs portal interactions as an alternative damping mechanism and improve upon previously proposed implementations based on quartic potentials.
These implementations
can lead to the QCD axion being dark matter and in the reach of IAXO, while  MADMAX, IAXO and ALPS II can be sensitive to a generic Axion-Like-Particle (ALP) as dark matter. Such models typically feature a kination era. We also show that ALP dark matter from KMM points to a particular realization of inflation.  } 

\begin{flushright}
\footnotesize
DESY 23-194\\
OU-HET-1152\\
CERN-TH-2024-093\\
\end{flushright}

\maketitle



\section{Introduction}
The axion kinetic misalignment mechanism (KMM) ~\cite{Co:2019jts,Chang:2019tvx} is an interesting possibility for axion-like-particles (ALPs) to constitute all of the observed dark matter (DM) in regions of the parameter space where ALPs are much too under-abundant in the common misalignment mechanism,  referred as the standard misalignment mechanism (SMM)~\cite{Preskill:1982cy,Abbott:1982af,Dine:1982ah}. The key feature of KMM is to consider scenarios in which the ALP field has acquired sufficient initial kinetic energy at early times to overcome the barriers of the axion potential when this potential develops. The rotation in field space persists until the field is eventually trapped when the kinetic energy becomes insufficient to overcome the potential barriers. This trapping can take place much later than the time when axion oscillations start in SMM. As a result, the energy stored in axion 
oscillations is less redshifted than in SMM, such that the correct dark matter abundance is reached in the experimentally accessible low-$ f_a $ region, opening the possibility for a large range of axion experiments to probe this new regime for axion DM. The KMM paradigm was further developed in~\cite{Co:2020dya,Co:2020jtv,Co:2020xlh,Barman:2021rdr,Co:2022aav,Barnes:2022ren,Harigaya:2023mhl,Co:2023mhe}
Our motivation is to investigate in detail which UV completions naturally lead to rotating axions in the first place.

It was shown in~\cite{Fonseca:2019ypl} that a rotating axion generally experiences parametric resonance leading to axion fragmentation. The impact of fragmentation on axion dark matter from KMM was studied in detail in a model-independent way in~\cite{Eroncel:2022vjg}. Fragmentation can take place either before or after trapping. The homogeneity of the ALP field is disrupted when fluctuations grow exponentially until the field fragments into axion particles.  The momentum spectrum was calculated in~\cite{Eroncel:2022vjg} in the regime of complete fragmentation and in the general case including incomplete fragmentation in \cite{miniclusterpaper}. While fragmentation has a rather limited impact on the final relic abundance, it can lead to important observable effects related to the formation of axion dark matter miniclusters \cite{miniclusterpaper}.

We will be considering the general class of models where the axion rotation is induced by the dynamics of the radial mode partner in models where the axion is the angular mode of the complex scalar field responsible for the spontaneous breaking of a global $U(1)$ symmetry. It would be interesting to investigate how KMM could be implemented for the alternative axion UV completions, namely for a    string theory axion which is not  accompanied by a radial mode partner, and whether the other moduli associated with the compactification of extra spacial dimensions could instead be responsible for transferring kinetic energy to the axion (like in \cite{Krippendorf:2018tei} where the effect of mixing kinetic terms between the volume modulus and its axionic partner is discussed). This would be a subject for a separate publication. In this paper, we restrict to axions identified as the angular mode of a complex (Peccei-Quinn) scalar field. Although minimal, these models already feature a very rich spectrum of possible phenomenologies. Rotating complex scalar fields have an interesting phenomenology beyond DM. In particular, they have been used to account for baryogenesis through the Affleck-Dine mechanism~\cite{Affleck:1984fy,Dine:1995kz,Dine:1995uk}. In~\cite{Co:2020jtv}, leptogenesis was implemented in KMM. Furthermore, rotating ALP scenarios can lead to a period in which the universe is dominated by the kinetic energy of the rotating ALP, known as kination. In this case,  the energy density of the universe redshifts as $ a^{-6} $, which leads to  a significant amplification of primordial gravitational-wave backgrounds~\cite{Gouttenoire:2021wzu,Gouttenoire:2021jhk,Co:2021lkc}.

In this work, we study UV completions that lead to such rotation. We refer to the event which starts the rotation as the \textit{kick}. So far, kick mechanisms rely on higher-dimensional operators involving the complex scalar field, as introduced in the Affleck-Dine mechanism\cite{Affleck:1984fy,Dine:1995kz,Dine:1995uk}. They were used in the context of rotating ALPs by Co et al.~\cite{Co:2019jts,Co:2020dya,Co:2020jtv,Co:2021lkc,Co:2020xlh} and Gouttenoire et al.~\cite{Gouttenoire:2021wzu,Gouttenoire:2021jhk}. An alternative to the Affleck-Dine-like kick mechanism is the trapped misalignment mechanism \cite{DiLuzio:2021pxd,DiLuzio:2021gos}, which exploits the fast change in the axion mass, without referring to radial-mode dynamics.

Our goal is to determine precisely for which axion mass $m_a$ and decay constant $f_a$ the KMM can be realistically implemented. A key ingredient of KMM is the early dynamics of the axion partner, the radial mode of the complex scalar field.
One major requirement for a successful implementation of KMM is an efficient damping of the radial-mode oscillations.
We perform a detailed treatment of this damping, supplementing earlier studies in the literature \cite{Co:2019jts,Co:2020dya,Co:2021lkc,Gouttenoire:2021jhk} with a systematic analysis over the full axion parameter space. For each point in the 
($m_a$, $f_a$) plane, we determine the damping temperature as well as the radial-mode mass required to account for the correct dark matter abundance and the size of the interactions between the radial mode and the plasma. We can therefore correlate the axion mass and the radial-mode mass. In a significant fraction of parameter space, the radial mode is rather light and in some cases could even be searched for experimentally. 

KMM takes place in the so-called {\it pre-inflationary scenario}, meaning that the global $U(1)$ is broken by the large vacuum expectation value of the Peccei-Quinn field during inflation. In this work, we consider the situation where the axion starts rotating after reheating, in the radiation era, shortly after the radial mode has started rolling down its potential.
This is a minimal assumption about the early cosmology, associated with only a weak lower bound on the reheat temperature of the universe.  In the literature on KMM, inflation is usually not discussed\footnote{see however \cite{Lee:2023dtw} which considers a kick happening before or during reheating.}. We will show however that our treatment, which assumes homogeneity of the field and does not rely on lattice simulations, imposes constraints on the size of the primordial fluctuations of the scalar field that points to a particular dynamics at the end of inflation. Lattice studies would be needed to determine whether such restriction is physical.

The paper is structured as follows: In Section \ref{sec:axion fragmentation review} we review the essentials of KMM and of axion fragmentation. In Section \ref{sec:overview} we present the general aspects of Affleck-Dine-like implementations of axion KMM. In Section \ref{sec:Constraints (general)} we go through the list of constraints from theory, cosmology  and observations. 
In Section~\ref{sec:potential}, we 
discuss the form of the Peccei-Quinn potential and justify why nearly-quadratic potentials are favored. Sections \ref{sec:initial conditions (NQ)}, \ref{sec:Nearly quadratic early damping results}, 
\ref{sec:nearly quadratic late damping}, 
\ref{sec:Nearly quadratic Yukawa section} and 
\ref{sec:NQHiggs} study in detail KMM implementations with a nearly-quadratic potential.
In Section \ref{sec:initial conditions (NQ)}, we discuss the initial dynamics of both the radial field and axion field after inflation. We then discuss the damping of radial-mode motion. This may happen early, through some unspecified dynamics, before the radial mode energy becomes dominant, as discussed in Section \ref{sec:Nearly quadratic early damping results}. Damping may instead happen after radial mode dominance as investigated in \ref{sec:nearly quadratic late damping}, through thermal effects. We investigate the thermal damping via Yukawa interactions in Section \ref{sec:Nearly quadratic Yukawa section} and via Higgs portal interactions in Section \ref{sec:NQHiggs}.
Section \ref{sec:quartic model} investigates the case of a viable implementation with a quartic potential. 
Section \ref{sec:power} discusses the need  for a suppressed primordial power spectrum of fluctuations at the time when the axion has a kination equation of state, which implies particular conditions for inflation.
In Section \ref{sec:summary}, we summarize our results.
The rest of the paper contains a large number of appendices.
Appendix \ref{app:notation} lists all subscripts, abbreviations and symbols that are used throughout the paper. Appendix \ref{app:Thermal relics} discusses the constraints from the thermalisation of the radial mode. Appendix \ref{app:damping rates} lists all the damping rates. Appendix \ref{app:TReheat} discusses the (weak) constraints on the reheating temperature. Appendix \ref{app:boltzmann equations} derives and solves the Boltzmann equations describing thermal damping. Finally, Appendix \ref{app:parameter values} collects a large number of complementary figures showing the range of model parameters and the viable parameter space in the $ [m_a,f_a] $ plane.


\section{Axion dark matter from kinetic misalignment}\label{sec:axion fragmentation review}
	
The dynamics required to start the rotation of the axion relies on radial dynamics of the full complex scalar field $ P $. 
We refer to the complex field $P$ as the Peccei-Quinn (PQ) \cite{Peccei:1977hh} field although our analysis is not restricted to the QCD axion but applies to all possible ALPs.
$ P $ is decomposed in terms of the angular mode $ \theta $ (the axion) and the radial mode $ \phi $ as
\begin{gather}
P = \frac{1}{\sqrt{2}}\phi e^{i \theta}.\label{eq:P definition}
\end{gather}
Rotation of the field then corresponds to $ \dtheta \neq 0 $, and it is conveniently measured by the PQ charge density
\begin{gather}
\nPQ = \dtheta \phi^2, \label{eq:PQ charge}
\end{gather}  
which is the conserved charge related to the ALP shift symmetry: $\theta \to \theta + \delta$ and which serves as the analogue of angular momentum of the field. 

We will consider potentials for $P$ which give the radial mode a ground state vacuum expectation value (VEV) of 
\begin{equation}
 \expval{\phi} =f_a. 
 \end{equation}
The dimensionful axion field is given by $\theta f_a$ and we will denote its mass by $m_a$. We will show how the rotation arises, but for now, let us consider the evolution after the onset of the rotation.

\subsection{Relic abundance}
As long as shift-symmetry-breaking effects are absent, $ \nPQ $ is conserved in comoving space, i.e. $ \nPQ \propto a^{-3} $ where $ a $ is the scale factor of the universe. It is therefore convenient to normalize the PQ charge to the entropy density, $ s $, and use the yield parameter
\begin{gather}
Y = \frac{\nPQ}{s},
\end{gather}
which is constant in time as long as PQ charge and entropy are conserved.

We consider the scenario of a QCD axion or an ALP with a similar temperature-dependent potential.  To fit the lattice result for QCD \cite{Borsanyi:2016ksw} we parametrize this temperature-dependent mass as
\begin{gather}
m_a^2(T) \approx m_a^2 \times \begin{cases}
(T/T_c)^{-\gamma} &\text{if } T>T_c \\
1 &\text{if } T<T_c
\end{cases}, \qq{where} 
\begin{gathered}
T_c = 2.12 \times \Lambda_{b,0},\\
\Lambda_{b,0} = \sqrt{f_a m_a}, \\
\gamma = 8.16,
\end{gathered}\label{eq:thermal potential}
\end{gather}
which sets the amplitude of a cosine potential for the axion:
\begin{gather}
V_a(T)\approx m_a(T)^2 f_a^2 \left[1-\cos(\theta)\right] .
\end{gather}
In the notation used here, $ m_a $ always refers to the zero-temperature value observed today and $ m_a(T) $ is used whenever the temperature-dependent value is referred to. As $ m_a(T) $ does not play a role in the initial kick dynamics and only enters near the time of fragmentation, modifying the fit of $ m_a(T) $ will at most affect constraints related to $ \Tfrag $ defined below.
In most of our analysis, $f_a$ and $m_a$ are free parameters and characterise both the QCD axion and a generic ALP. We apply the same formula Eq.~(\ref{eq:thermal potential}) to the generic ALP scenario for simplicity.

The relic axion number density is  $ n_a \approx 2 \nPQ $~\cite{Co:2019jts,Eroncel:2022vjg}, such that the present-day axion relic energy density is
\begin{gather}
\rho_a \approx 2 m_a \nPQ.
\end{gather}
In \cite{Eroncel:2022vjg} we show that the relation between $ n_a $ and $ \nPQ $ arises from the approximately conserved quantity known as the \textit{action variable}. The relevant quantity to calculate is therefore $ \nPQ $ as parametrized by the yield $ Y = \nPQ / s $. Demanding that the axion relic density matches the observed DM relic density is therefore equivalent to demanding $ 2Y = Y_{\rm DM} $, where the observed DM yield $ Y_{\rm DM}$ is
\begin{gather}
Y_{\rm DM} \approx \frac{3}{4}\frac{g_{*}(\Teq)}{g_{*s}(\Teq)}\frac{\Teq}{m_a}\approx 0.64\frac{\Teq}{m_a}.
\end{gather}
where $\Teq$ is the temperature of the universe at matter-radiation equality.

The axion field keeps overshooting the periodic barriers of the axion potential as long as its kinetic energy exceeds the barrier height. The rotation is slowed down by Hubble friction while the potential barriers grow larger according to Eq.~\eqref{eq:thermal potential}. Eventually, at temperature $ \Tfrag $ the field gets trapped. 
In \cite{Eroncel:2022vjg}, we find that $ \Tfrag $ can be written as
\begin{gather}
\Tfrag \approx  (2.12)^{\frac{\gamma}{2+\gamma}}\left(2\times 10^8\right)^{\frac{2}{6+\gamma}}\left(\frac{g_{*}}{72}\right)^{-\frac{2}{6+\gamma}}\left(m_a f_a\right)^{\frac{1}{2}+\frac{1}{6+\gamma}}\left(\frac{h^2\Omega_{\phi,0}}{h^2\Omega_{\rm DM}}\right)^{-\frac{2}{6+\gamma}},\label{eq:Tfrag}
\end{gather}
where \cite{Co:2019jts,Eroncel:2022vjg}
\begin{gather}
\left(\frac{h^2\Omega_{\phi,0}}{h^2\Omega_{\rm DM}}\right)\approx \qty(\frac{m_a}{5\times 10^{-3}\,\rm{eV}})\qty(\frac{Y}{40}),
\label{eq:relicabundance}
\end{gather}
One main point of this paper will be to extract predictions for $Y$ for any $(m_a, f_a)$ as a function of the UV parameters of the theory that control the size of the kick, such as the mass  of the radial mode $m_{\phi}$ and $n$ the dimension of the explicit PQ-breaking higher-dimensional operators, as we will introduce in Section \ref{sec:overview}. Before doing so, we discuss axion fragmentation that accompanies KMM in most of its parameter space.

\subsection{Axion fragmentation}
A main result of \cite{Eroncel:2022vjg} is that shortly before or shortly after trapping, the field may experience growth of fluctuations in a process known as axion fragmentation~\cite{Fonseca:2019ypl}. As such processes take place very close to $ \Tfrag $ we will here approximate the temperature of fragmentation to be the same as the trapping temperature $ \Tfrag $.
The efficiency of fragmentation depends crucially on the ratio of $ m_a(T) / H $ at $ \Tfrag $. Fragmentation can only take place before trapping if $ m_a(\Tfrag) / H(\Tfrag) $ is sufficiently large. In \cite{Eroncel:2022vjg} we found that even if $ m_a(T) / H  $ is not large enough to ensure efficient fragmentation before trapping, then fragmentation may still occur after trapping. For lower values of $ m_a(\Tfrag) / H(\Tfrag)  $ fragmentation becomes negligible and the phenomenology approaches that of pure kinetic misalignment. Ultimately, for $ m_a(T) / H < 3$ kinetic misalignment also cannot take place because the field becomes frozen upon trapping such that only the standard misalignment mechanism can take place. In \cite{Eroncel:2022vjg}, we categorized the parameter space according to the efficiency of fragmentation:
\begin{align*}
&\frac{m_a(\Tfrag)}{H(\Tfrag)} \lesssim 3 	&&\implies \text{Standard misalignment}		\\
3   \lesssim &\frac{m_a(\Tfrag)}{H(\Tfrag)} \lesssim 42	&&\implies \text{Kinetic misalignment with weak fragmentation}		\\
42  \lesssim &\frac{m_a(\Tfrag)}{H(\Tfrag)} \lesssim 900	&&\implies \text{Fragmentation after trapping} \\
900 \lesssim &\frac{m_a(\Tfrag)}{H(\Tfrag)}  		&&\implies \text{Fragmentation before trapping}
\end{align*}

\begin{figure}
	\centering
	\includegraphics[width=0.9\textwidth]{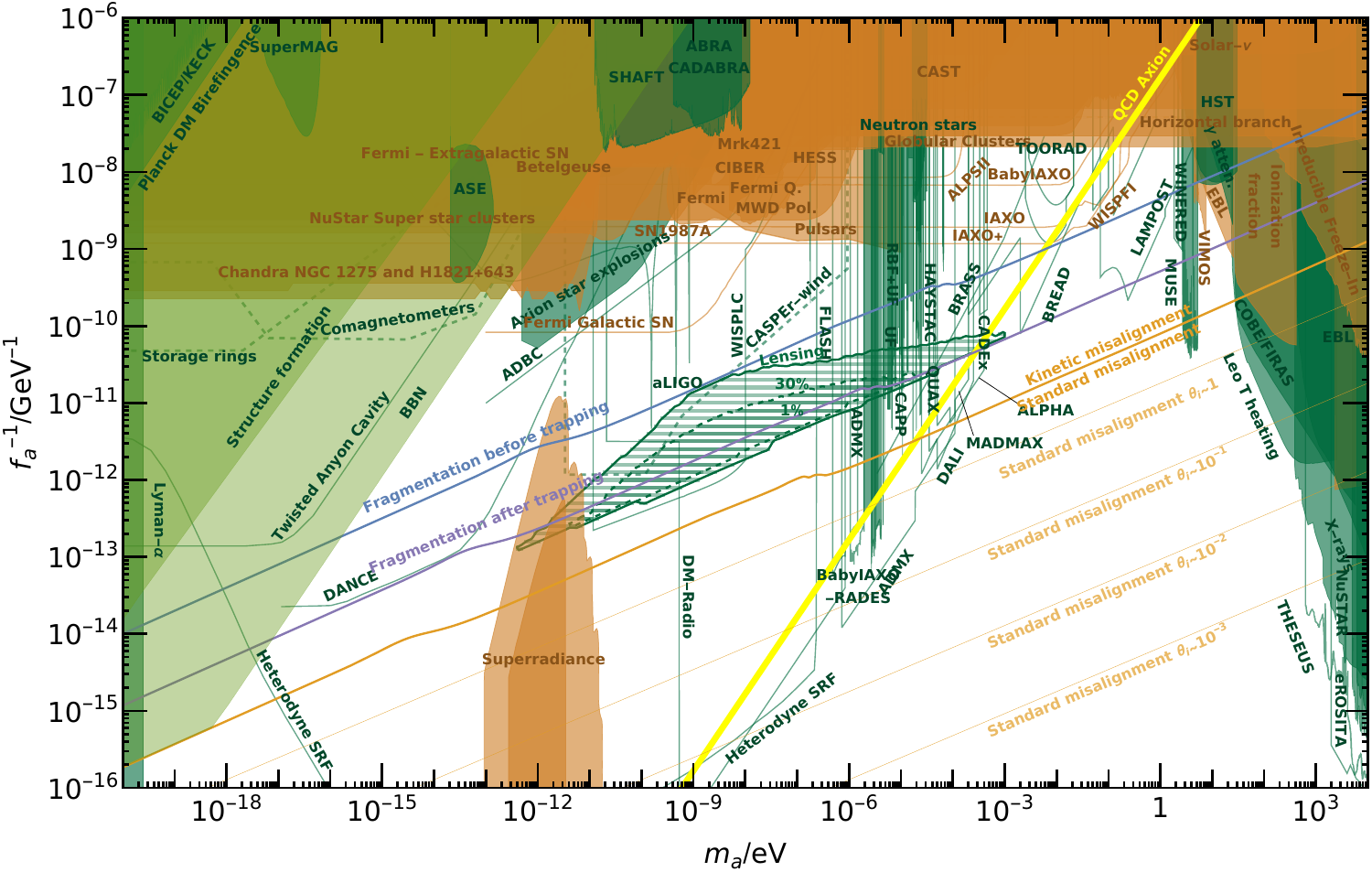}
	\caption{\small\it Potentially available  ALP DM parameter space (in white) together with experimental bounds~\cite{Salemi:2021gck,ADMX:2009iij,ADMX:2018gho,ADMX:2019uok,ADMX:2018ogs,Crisosto:2019fcj,ADMX:2021nhd,Bartram:2021ysp,Devlin:2021fpq,Lee:2020cfj,Jeong:2020cwz,CAPP:2020utb,CAST:2020rlf,Oshima:2021irp,Grenet:2021vbb,HAYSTAC:2018rwy,HAYSTAC:2020kwv,McAllister:2017lkb,Alesini:2019ajt,Alesini:2020vny,PhysRevLett.59.839,Gramolin:2020ict,Arza:2021ekq,PhysRevD.42.1297,Thomson:2019aht,Liu:2018icu,Stern:2016bbw,Nagano:2019rbw,Lawson:2019brd,BRASSwebpage,BREAD:2021tpx,Michimura:2019qxr,DMRadio:2022pkf,Brouwer:2022bwo,Alesini:2019nzq,FLASHconference,Berlin:2020vrk,HeterodyneSRF,Baryakhtar:2018doz,Beurthey:2020yuq,QUAXprojection,Marsh:2018dlj,Schutte-Engel:2021bqm,Zhang:2021bpa,Ehret:2010mh,Ortiz:2020tgs,Betz:2013dza,OSQAR:2015qdv,DellaValle:2015xxa,CAST:2007jps,CAST:2017uph,Shilon:2012te,Armengaud:2014gea,Irastorza:2018dyq,Foster:2022fxn,Wouters:2013hua,Marsh:2017yvc,Reynolds:2019uqt,Reynes:2021bpe,Chan:2021gjl,Depta:2020wmr,Dessert:2021bkv,Bolliet:2020ofj,Buen-Abad:2020zbd,Calore:2020tjw,Calore:2021hhn,Meyer:2020vzy,Fermi-LAT:2016nkz,Blout:2000uc,Jacobsen:2022swa,HESS:2013udx,Ayala:2014pea,Wadekar:2021qae,Dessert:2022yqq,Regis:2020fhw,Li:2020pcn,Xiao:2020pra,Dessert:2020lil,Vinyoles:2015aba,Jaeckel:2017tud,Payez:2014xsa,Caputo:2021rux,Lee:2018lcj,Cadamuro:2011fd,Dolan:2021rya,Grin:2006aw,Foster:2021ngm,Dekker:2021bos,Meyer:2016wrm,Ge:2020zww,Thorpe-Morgan:2020rwc,JacksonKimball:2017elr,Centers:2019dyn,Wu:2019exd,Garcon:2019inh,Bloch:2021vnn,Abel:2017rtm,Vasilakis:2008yn,Bloch:2019lcy,Bhusal:2020bvx,Graham:2020kai,Buschmann:2021juv,Carenza:2019pxu,Arvanitaki:2014wva} and model-independent constraints on DM in the rotating axion scenario. The two lines ``fragmentation before trapping" and ``fragmentation after trapping" are derived in our companion paper \cite{Eroncel:2022vjg} while the hashed green region at the centre refers to the observable axion mini-cluster region derived in \cite{miniclusterpaper}. The goal of this work is to provide KMM model implementations in this plane above the thick orange line. 
 {Experimental sensitivity lines assume a KSVZ-like axion-photon coupling 
$g_{\theta\gamma\gamma}^{\rm KSVZ}=(\alphaEM/{2\pi})({1.92}/f_a)\approx 2.23\times 10^{-3}/{f_a}$.} For a list of experimental constraints and more details on the BBN and structure formation constraints see \cite{Eroncel:2022vjg}.}
	\label{fig:moneyPlotTwoCol}
\end{figure}

Why is fragmentation relevant? In fact, as discussed in \cite{Eroncel:2022vjg}, fragmentation does not significantly affect the prediction for the final axion relic energy density. On the other hand, it has observable consequences.
The growth of fluctuations caused by the delay in the onset of oscillations significantly enhances the matter power spectrum at scales $k\sim m_a(\Tfrag)a(\Tfrag)$\,\cite{Zhang:2017dpp,Arvanitaki:2019rax,miniclusterpaper,Chatrchyan:2023cmz,Koschnitzke:2024qgm}. These large fluctuations experience gravitational collapse very early in the matter era, yielding very dense and compact ALP mini-clusters, similar to the axion mini-clusters predicted in the post-inflationary scenario~\cite{Hogan:1988mp,Kolb:1993zz,Kolb:1993hw,Kolb:1994fi,Eggemeier:2019khm,Ellis:2022grh}. These dark matter halos can be probed by future gravitational surveys via their direct gravitational interactions~\cite{Arvanitaki:2019rax}. A promising detection method for halos with scale mass $M_s\sim 10^{-6}\,M_{\odot}$ is the photometric lensing that looks for deviations in the brightness of a background light source due to the passage of a compact halo~\cite{1986ApJ3041P,Dai:2019lud}. 
By assuming a monochromatic mass distribution for the dark matter halos, and that 30\% of dark matter resides in these halos, Ref.~\cite{Arvanitaki:2019rax} derives a region $M_s-\rho_s$ plane\footnote{$M_s$ and $\rho_s$ refer to the scale mass and the scale density of the halo respectively. Both are used to parametrize dark matter halos that have a Navarro-Frenk-White (NFW) profile~\cite{Navarro:1996gj}.} which can be probed by future lensing observations. Based on this region, Ref.~\cite{miniclusterpaper} derives a prospective region in the $m_a-f_a$ plane which yields halos that are dense enough to be probed by future lensing surveys by assuming an ALP model with a temperature-independent potential. Although we here assume a temperature-dependent ALP potential, we provide the results of \cite{miniclusterpaper} in Fig.~\ref{fig:moneyPlotTwoCol}.

In summary, KMM is interesting because it opens vast new regions of $(m_a,f_a)$ parameter space where the axion can be naturally dark matter. The corresponding new regions of parameter space coincide with the sensitivity of a large range of upcoming experiments, as shown in Fig.~\ref{fig:moneyPlotTwoCol}. Everywhere in the white region, the axion can be dark matter. Above the orange thick line is the KMM domain.
In this work, we seek model implementations that provide the initial conditions for the above dynamics, producing a  homogeneous, rotating PQ field whose radial mode is relaxed to $ f_a $ and with a yield satisfying $ 2Y=Y_{\rm DM} $. In the following sections, we will explore two models (with nearly-quadratic or quartic potentials for the Peccei-Quinn field, respectively) which provide these initial conditions and discuss the challenges each model faces. We will determine the parameter space above the thick orange line in Fig.\ref{fig:moneyPlotTwoCol} where realistic UV completions can explain axion dark matter through the KMM.
Before discussing full implementations of KMM, we next provide an overview of its essential components.


\section{Ingredients for a KMM implementation}
\label{sec:overview}

\subsection{Starting the rotation}
\label{subsection:starting}
The Affleck-Dine-inspired mechanism which we explore in this work (and which is also reviewed in details in \cite{Gouttenoire:2021jhk}) relies on two key features:
\begin{description}
	\item[PQ-violating operators:] 
 These are needed to give a kick to the axion from the motion of the radial mode.
 They arise as higher-dimensional operators and can be motivated in SUSY models~\cite{Dine:1995uk}. They are of the form
	\begin{gather}
		\frac{A}{n}\frac{P^n}{M^{n-3}}+ \frac{A}{n}\frac{P^{*n}}{M^{n-3}} = \frac{2A}{n2^{n/2}} \cos(n\theta+\delta_\theta) \label{eq:higher dimensional PQ breaking potential} \frac{\phi^n}{M^{n-3}},
	\end{gather}
	where, $ M $ is a suppression scale which we typically set to the Planck scale $ \mplanck $, and $ \delta_\theta $ is a phase in the potential, which in general is not aligned with the minimum at $ \theta=0 $. As will be discussed further in Section \ref{sec:initial conditions (NQ)}, SUSY motivates $A\sim {\cal O}(\mphi)$. The key idea is that even though this potential plays a major role initially, it rapidly becomes negligible as $ \phi $ decreases from a large VEV to $f_a$.

	\item[Large VEV driver:] To probe such higher-dimensional operators, the VEV of the PQ field has to be driven to a large initial value. A large initial VEV for $\phi$ is an essential component of the constructions we consider and will be justified in Sec.~\ref{sec:initial conditions (NQ)}. It has so far been realistically justified only in the context where the $P$ field acquires a VEV before or during inflation. This means that the axions we are considering fall into the so-called category of {\it ``pre-inflationary"} axions.
\end{description}
A model with the above two components can start a rotation of the PQ field. The resulting $\dot\theta$ right after the kick can be found by solving the equation of motion, which predicts~\cite{Dine:1995uk,Co:2020dya}
\begin{gather}
	\dot{\theta}_{\rm kick} \simeq 2^{1-\frac{n}{2}}\frac{A  \phikick^{n-2} \sin(n\thetakick)}{ 
 M^{n-3} \mphiVprime|_{\phi=\phikick} }, \label{eq:kick size}
\end{gather}
where $ \thetakick $ is the initial misalignment angle. Likewise, $ \phikick $ is the amplitude of the radial mode at the time of the kick and $ \dot{\theta}_{\rm kick} $ is the angular velocity produced by the kick. The potential for the radial mode is denoted by $V$ while the potential for the angular mode is denoted $V_a$.
We are working with large field values, so the usual mass defined by $ V'' $ is not the relevant parameter as this is defined at the potential minimum. Instead, we define the effective radial mode mass
\begin{gather}
\mphiVprime \equiv \sqrt{ \frac{V'}{\phi} }, \label{eq:effective mphi definition}
\end{gather}
as this is the parameter that enters the equation of motion and thus the relevant parameter far from the minimum~\cite{Co:2020jtv}.
This result is independent of the mechanism which sets the large VEV.

It is useful to parametrize the angular velocity inherited from the kick with the parameter $ \epsilon $~\cite{Co:2019jts}, which gives the angular velocity at the apoapsis of the field orbit as a fraction of the velocity required to maintain a circular orbit. If we consider the equation of motion (EOM) for the radial mode,
\begin{gather}
	\ddot{\phi}+3H\dot{\phi} + V' = \dtheta^2 \phi,
\end{gather}
we see that the radial potential competes with the centrifugal term $ \dtheta^2\phi $. 
This is analogues to the usual Kepler problem: The centrifugal term is exactly balanced against the potential when $ \dtheta = \sqrt{\phi^{-1} V'} $, which would then correspond to a perfectly circular orbit. Generically, when a rotation is started with any other value than $\dtheta = \sqrt{\phi^{-1} V'}$, the radial mode will oscillate about the equilibrium value $\phi \sim V' / \dtheta^2$ which corresponds to an elliptic orbit. Because of Hubble friction, radial oscillations about $\phi \sim V' / \dtheta^2$  decrease in amplitude and the equilibrium point decreases towards $f_a$ as $\dtheta$ is depleted.
The link to the potential parameters will be given in the next sections.
We observe that $ \epsilon = \dthetaap / m_\phi  $ is a useful parametrization, where $ \dthetaap $ is the angular velocity at the apoapsis of an orbit. This is equivalent to normalising the PQ charge to $ n_{\phi} = \rho_{\phi,\rm pot} / m_\phi $, i.e.,
\begin{gather}
	\epsilon = \frac{\nPQ /2}{n_\phi} =  \frac{\nPQ/2 }{\rho_{\phi,\rm pot}/m_\phi} = \frac{\dthetaap}{m_\phi}, \label{eq:epsilon equivalence}
\end{gather}
where $\rho_{\phi,\rm pot} = m_\phi^2 \phi^2 / 2$ is the potential energy density and we assumed that the radial potential is a pure quadratic potential. Since the centrifugal term in the $ \phi $-EOM is balanced with the potential exactly for $ \epsilon=1 $, this value corresponds to a perfectly circular orbit. $ \epsilon=0 $ corresponds to purely radial oscillations. Since $\epsilon$ is defined at the apoapsis of the orbit, and $\epsilon>1$ would imply that the measured point is not the apoapsis, we by construction have $0\leq \epsilon \leq 1$. Co et al.~\cite{Co:2019jts,Co:2020dya,Co:2020jtv} note that an insufficiently circular orbit can lead to a parametric resonance in the radial mode. Throughout this work, we assume that $ \epsilon $ is large enough that no such resonance takes place.

\subsection{Evolution after the kick and the need for radial motion damping} \label{sec:evolution after the kick}
After the kick, the radial mode is not in its ground state. It has a large VEV, $ \phikick \gg f_a $, and unless the kick is perfectly circular, $ \phi $ will be oscillating, meaning that $ P $ will be in an elliptic orbit. 
The energy density of radial oscillations behaves as cold dark matter. It will therefore dominate over the angular mode energy density. To realize axion dark matter from kinetic misalignment, these oscillations must be damped. We could in principle also consider the case where the radial mode plays the role of dark matter instead, but this is another project which we leave for future investigation. In this paper we focus on the case where the axion is dark matter.  Defining $\rho_r$ as the radiation energy density from the Standard Model (SM) thermal bath, if damping takes place when $ \rho_\phi > \rho_r $, then it significantly impacts the SM plasma. We denote the temperature at which $ \rho_\phi = \rho_r $ as $ \Tdom $ and we denote the damping temperature as $ \Tdamp $. The yield and the cosmology are significantly impacted by entropy injection only if $ \Tdom > \Tdamp $.
We will discuss damping in the following order:
\begin{enumerate}[label=\arabic*)]
	\item \textbf{Early damping}: $\Tdamp > \Tdom $. In this case, the exact value $ \Tdamp $ does not impact the cosmology or the DM relic directly and thus does not need to be specified beyond the assumption $ \Tdamp > \Tdom $. This analysis will be discussed in Sec.~\ref{sec:Nearly quadratic early damping results}. The relevant figures are Figs.~\ref{fig:nearlyQuadraticEarlyDamping1} and \ref{fig:nearlyQuadraticEarlyDamping2} as well as \ref{fig:quarticHubble} and \ref{fig:quarticHubble13WithAxionDamping}.
	\item \textbf{General damping}: $ \Tdamp > \Tdom $ or $ \Tdamp < \Tdom $.  We will discuss general damping while remaining agnostic about the damping mechanism in Sec.~\ref{sec:nearly quadratic late damping}. The relevant figures are Figs.~\ref{fig:scenarios}, \ref{fig:constantfaPlotn13Mmp}, and \ref{fig:Nearly quadratic M18n13With}. The summary plot is the top panel of Fig.~\ref{fig:summaryPlotNQYukawaAndHiggs}.
	\item \textbf{Thermal damping}: The damping mechanism is implemented by specifying the coupling of the radial mode to the SM plasma. This realizes a subset of the possible solutions identified in 2). This analysis will be discussed in Sec.~\ref{sec:Nearly quadratic Yukawa section} and~\ref{sec:NQHiggs}. The relevant figures are Figs.~\ref{fig:TdampContoursMmpn13} to \ref{fig:minMixAnglePlotNQHiggsMmpn13} and \ref{fig:maxFlucPlotQHigherDimYukawaMmpn13}-\ref{fig:maxFlucPlotMmpn10n7}. The summary plots are the lower two panels of Fig.~\ref{fig:summaryPlotNQYukawaAndHiggs}.
\end{enumerate}

\paragraph{Evolution of the radial and angular modes:}
After the kick and the onset of the elliptic orbit of the field, the equilibrium value $ \expval{\phi} $, about which $ \phi $ oscillates, continuously decreases. This relaxation can be characterized by PQ charge conservation. Once radial oscillations are damped, the generally elliptic orbit with $ 0\leq \epsilon \leq 1 $ is relaxed to a circular orbit with $ \epsilon=1 $, such that $ \dtheta \approx \mphiVprime $ after damping. Furthermore, it was shown in~\cite{Gouttenoire:2021jhk} that even before damping we have $ \expval{\dtheta}\approx \mphiVprime $.
Furthermore, PQ charge conservation implies $\expval{\nPQ} \propto a^{-3}$, such that $ \mphiVprime \phi^2 \propto a^{-3}$, which ensures that 
\begin{gather}
	\expval{\phi} \propto \mphiVprime^{-1/2} a^{-3/2},
\end{gather}
which is the same evolution as that of a homogeneously oscillating scalar field with constant mass, having $\rho \propto a^{-3}$. If the radial mass is time-dependent as a consequence of either the radial-mode dynamics or the thermal effects,  this impacts the relaxation of the radial mode. Let us denote $\amin$ the value of the scale factor of the universe when the energy density of the axion is dominated by its kinetic energy. This corresponds to the time when $ \phi $ reaches the minimum at $ f_a $. 
Expressing the possible time-dependence of the radial mass as $ \mphiVprime\propto a^{-c_{\mphi}} $, we have
\begin{gather}
	\frac{\phikick}{f_a} \approx \left(\frac{\amin}{\akick}\right)^{(3-c_{\mphi})/2}.\label{eq:Tmin definition, general}
\end{gather}
The analyses of axion fragmentation and kinetic misalignment \cite{Eroncel:2022vjg,Co:2019jts,Co:2020dya,Co:2020jtv} are only applicable for $ \Tmin > \Tfrag $. If this condition is violated, then a fragmentation or kinetic misalignment analysis must be carried out in the presence of radial dynamics. Such an analysis is beyond the scope of this work, and we display the parameter space where this assumption is violated even though there is in principle no reason why fragmentation could not take place before the relaxation of the radial mode.

\label{sec:general evolution description}
After $ \phi $ reaches $ f_a $, the angular mode does not follow the equilibrium value $ \dtheta \approx \mphiVprime$. Applying again PQ charge conservation, $ \nPQ = \dtheta \phi^2 \propto a^{-3}$ we observe that once $ \expval\phi = f_a $ the angular mode redshifts as $ \dtheta \propto a^{-3} $. Therefore,
\begin{gather}
	\expval{\dtheta} \propto a^{-c_\theta} \qq{where} c_\theta = \begin{cases}
		c_{\mphi} &\text{for } T > \Tmin, \\
		3 &\text{for } T < \Tmin. \\
	\end{cases}
\end{gather}
As suggested by the subscript, $ \Tmin $ marks the onset of kination-like scaling of the energy density of $ \rho_a \propto \dtheta^2 \propto a^{-6} $ such that a period of kination\footnote{\textit{Kination} is a period of cosmological history characterized by a $ a^{-6} $ evolution of dominant component of the energy density of the universe. We use the term \textit{kination-like} when a specific component of the energy density undergoes $ a^{-6} $ evolution regardless of whether that component dominates the total energy density.} is triggered if $ \rho_a(\Tkin) $ dominates the energy density or not.

Before implementing KMM within specific models, we now discuss in the next section all the constraints that we will have to apply on this scenario and that will eventually translate into bounds on the model parameters.


\section{Phenomenological constraints}
\label{sec:Constraints (general)}

\paragraph{Big Bang Nucleosynthesis (BBN):}
$ \Tfrag $ is bounded by BBN. This limit arises because the energy density of the rotating axion scales as $ \rho_a \propto a^{-6} $ until trapping, so that the energy density in the angular mode will compete with the radiation density at BBN if fragmentation takes place too late. Usually, additional energy density in light fields at BBN is parametrized and constrained through the change in the effective number of neutrinos, $ \Delta N_{\rm eff} $. However, since the energy density in the rotating axion does not scale like radiation, $N_{\rm eff} $ is not an appropriate parametrization. Instead, we apply the numerical code \verb|AlterBBN 2.2|~\cite{Arbey:2018zfh}, which constrains a dark component scaling as $ a^{-6} $ to less than $ \rho_a / \rho_r < 0.19 $ at $ T=1 $ MeV. To be compatible with BBN, $ \Tfrag $ then has to satisfy
\begin{gather}
\Tfrag \gtrsim 20 \text{ keV}.\label{eq:BBN bound Tstar}
\end{gather}
Discussion of this constraint can be found in \cite{Eroncel:2022vjg}.

Furthermore, BBN also constrains damping of radial motion. We do not perform a detailed analysis of how the entropy injection associated with damping interferes with the observed $ ^4 $He production, and we instead simply require damping to take place before BBN. Therefore, in addition to Eq.~\eqref{eq:BBN bound Tstar}, we also demand that
\begin{gather}
\Tdamp \gtrsim \text{ MeV}. \label{eq:BBN bound Tdamp}
\end{gather}
In regimes in which the energy in the radial oscillations is small compared to the energy in radiation, this constraint may be too strong, as it is conceivable that dilution would be small enough not to disturb BBN. Nevertheless, to be conservative we impose this constraint regardless of how significant the entropy injection is. 

\paragraph{Structure formation:} To satisfy the bounds on hot relics and from structure formation, we impose that at radiation-matter equality the axions produced from fragmentation must have a velocity  $ v_a (\Teq) \lesssim 10^{-3}$.
The derivation of the precise constraint from structure formation requires knowledge of the spectrum of fluctuations produced by fragmentation. These spectra were calculated in  \cite{Eroncel:2022vjg,miniclusterpaper}, which showed that the structure formation constraint is subdominant to the BBN constraint discussed above. 
Ref.~\cite{Co:2021lkc} considered a similar structure formation constraint, somewhat stronger than the one implied by our condition. Since either condition is at most comparable with the BBN constraint we do not devote further attention to the matter.
\paragraph{Perturbativity:}
For the physics to remain perturbative, we impose that all dimensionless couplings in the Lagrangian be less than $ 4\pi $. 
This applies in particular to the self-interactions.
This condition typically excludes large radial mode masses, and therefore large axion masses, see for instance  the relation (\ref{eq:mphi nearly quadratic, without injection}) between the two.

\paragraph{Homogeneity:} \label{sec:Cems condition}
Primordial density fluctuations source fluctuations in the axion field (as mentioned earlier, we work with {\it pre-inflationary axions}). At early times, these fluctuations are massless and their energy density redshifts as $ a^{-4} $. However, if the axion homogeneous mode goes through a period of $ a^{-6} $ kination-like redshifting, the fluctuations grow in comparison and may come to dominate over the homogeneous mode. As we here seek to provide model implementations that lead to a rotating, homogenous axion field we treat this condition as a constraint. Nevertheless, such an inhomogeneous state may lead to dark matter with interesting phenomenology, which we leave for future work~\cite{Eroncel:2022b}.

The homogeneity condition was studied in detail in our model-independent paper~\cite{Eroncel:2022vjg}.
For an order-of-magnitude estimate, assume that the energy density of the fluctuations before kination-like scaling can be described by $ \rho_{\rm fluc} \sim \mathcal{P}_{\mathcal{R}} \rho_{a} $, where $ \rho_a $ and $ \rho_{\rm fluc} $ are the energy densities of the homogeneous axion mode and the fluctuations respectively and $ \mathcal{P}_{\mathcal{R}} $ is the amplitude of the primordial power spectrum. After the onset of kination-like scaling at $ \amin $ the energy densities redshift as $ \rho_{\rm fluc} \propto a^{-4} $ and $ \rho_a \propto a^{-6} $. Therefore, if the evolution is not interrupted, the fluctuations will eventually dominate at a time $ a_{\rm NL} $, where $ _{\rm NL} $ signifies the onset of non-linear dynamics. These assumptions imply that
\begin{gather}
\frac{\amin}{a_{\rm NL}} \sim \sqrt{\mathcal{P}_{\mathcal{R}}(\kmin)}.
\end{gather}
This rough estimate reproduces the result in \cite{Eroncel:2022vjg} up to an $ \mathcal{O}(1) $ factor. Here $ \mathcal{P}_{\mathcal{R}} $ is evaluated at the comoving momentum scale $ \kmin $, which is set by $ \kmin = \amin H(\Tkin) $.
$ \mathcal{P}_{\mathcal{R}} $ is on the other hand constrained by Planck~\cite{Planck:2018vyg} to $ A_s \approx 2.1 \times 10^{-9} $ at a pivot scale of 0.05 Mpc$ ^{-1} $. It is usually assumed to be a flat spectrum although it is only tightly constrained around this pivot scale. We expect $ \kmin $ to be many orders of magnitude larger than the pivot scale, such that $ \mathcal{P}_{\mathcal{R}}(\kmin) $ might differ significantly from $A_s$. If it does not differ significantly, then the growth of fluctuations leads to a loss of homogeneity. For instance, if the amplitude of curvature fluctuations of the modes which enter the horizon at $ \amin $ is $ \mathcal{P}_{\mathcal{R}}(\kmin) = 10^{-10} $ then to avoid domination by axion fluctuations the period of $ a^{-6} $ evolution has to be shorter than
\begin{gather}
\frac{\afrag}{\amin} \lesssim 10^{5}.\label{eq:cems condition}
\end{gather}
Conversely, if we leave $ \mathcal{P}_{\mathcal{R}}(\kmin) $ as a free parameter, then the maximally permissible amplitude of fluctuations at the onset of $ a^{-6} $ fluctuations is
\begin{gather}
\mathcal{P}_{\mathcal{R}}(\kmin) \lesssim \left(\frac{\amin}{\afrag}\right)^2. \label{As max}
\end{gather}
If $ \mathcal{P}_{\mathcal{R}} $ has a flat spectrum, the above constraint translates into $ (\amin / \afrag)^2 >  A_s = 2.1\times 10^{-9} $. We can avoid this constraint by assuming a non-trivial spectrum for $\mathcal{P}_{\mathcal{R}}$. 
It will be interesting to identify inflationary models that naturally avoid this constraint. 
A suppressed spectrum of primordial fluctuations also changes the efficiency of fragmentation which would shift the fragmentation lines reported in Fig.~\ref{fig:moneyPlotTwoCol}. However, this dependence is logarithmic~\cite{Eroncel:2022vjg} and we therefore do not take it into account here.
We sketch the needed ingredients on the inflationary model  to lead to the desired suppression of $\mathcal{P}_{\mathcal{R}}(\kmin)$ while satisfying Planck constraints in Section \ref{sec:power}. The details will be presented in a separate paper.

\paragraph{Thermal $ \phi $ relics:}
An interaction that is efficient enough to damp the radial oscillations will also bring the radial mode into thermal equilibrium. This is potentially dangerous for BBN and CMB. The thermal relic will decouple from the plasma if the interaction freezes out at a temperature $ \Tfo > \mphi $.
After decoupling from the plasma, the thermal relic will have an energy density of
\begin{gather}
	\rho_{\phi,\rm thermal}(\Tfo) \sim \frac{\rho_r}{g_*}\Big\vert_{\Tfo}.
\end{gather}
 It will be further diluted by reheating in the SM sector, i.e. by changes in $ g_* $ and $ g_{*s} $. Therefore, as long as the relic remains hot, it will not be problematic unless it freezes out below $ \sim 200 $ MeV. However, if the relic becomes cold, it will grow relative to the radiation density and can become problematic. 

At sufficiently large $ \mphi $, such constraints are absent because the thermal relic decays into axions as soon as it is no longer relativistic, see Appendix \ref{app:Thermal relics}. 
At sufficiently low $ \mphi $ the constraints are absent because the relic remains relativistic until after $ \Teq $ such that it both underproduces dark matter and does not disturb the CMB. The intermediate range is excluded by $ N_{\rm eff} $ constraints from either BBN or CMB. Therefore, $ \phi $ must satisfy 
\begin{gather}
	m_\phi < \mathcal{O}(1)\times\Teq \qq{or} \mathcal{O}(1)\times\frac{f_a^2}{\mplanck} < \mphi, \label{eq:thermal condition}
\end{gather}
where the values of the prefactors depend on the number of relativistic degrees of freedom. These constraints are discussed in more detail in Appendix \ref{app:Thermal relics}.

\paragraph{Fifth-force searches:}
The radial mode couples to photons through $ F_{\mu\nu}F^{\mu\nu} $ rather than $ F_{\mu\nu} \tilde{F}^{\mu\nu} $. We express the scalar radial mode-photon interaction and pseudo-scalar axion-photon interactions as
\begin{gather}
	\mathcal{L}_{\phi\gamma} = \frac{1}{4}g_{\phi\gamma\gamma} \phi F_{\mu\nu} F^{\mu\nu}    \ , \ \ \ \mathcal{L}_{\theta\gamma} = \frac{1}{4}g_{\theta\gamma\gamma} a F_{\mu\nu} \tilde{F}^{\mu\nu} .\label{eq:scalar EM coupling}
\end{gather}
where $ a = \theta f_a $  is the axion field.
We generally expect $ g_{\phi\gamma\gamma} $ to be of the order of the axion-photon coupling $ g_{\theta\gamma\gamma} $.  Scalar couplings 
are subject to much stronger constraints than those from pseudoscalar interactions. Although such constraints are not relevant for theories in which $ \mphi \sim f_a $, they are significant for theories with light radial modes such as those considered in this work. For a recent review of constraints on light scalar fields, see e.g.~\cite{Antypas:2022asj}.

Long-range forces~\cite{Damour:2010rp} arise from scalar-mediated interactions  through couplings like \eqref{eq:scalar EM coupling}.
 This leads to apparent violations of the equivalence-principle (EP) and deviations from the inverse-square law, which can be tested in fifth-force searches such as Eöt-Wash~\cite{Schlamminger:2007ht} and lunar ranging~\cite{Williams:2004qba}.
Searches for long-range forces are sensitive to electromagnetically-coupled scalar fields even if such fields are not DM. If the scalar were to be dark matter, which is not the case studied here, it would also be constrained from atomic clocks through the oscillating electromagnetic perturbations \cite{Hees:2018fpg}. 

The most relevant constraints are summarized by Hess et al.~\cite{Hees:2018fpg} from torsion balance data~\cite{Smith:1999cr,Schlamminger:2007ht,Wagner:2012ui} as well as data from the MICROSCOPE space mission~\cite{Touboul:2017grn}. Assuming a radial mode-photon coupling that is comparable in strength to the axion-photon coupling, such data constrain values of $ f_a $ beyond the range considered here. Therefore, constraints from EP violation in our context appear as a lower bound on $\mphi$:
\begin{gather}
	\mphi \gtrsim 10^{-5}\text{ eV}. \label{eq:EP constraint}
\end{gather}
The constraint is driven by the Eöt-Wash experiments on Cu and Pb test bodies~\cite{Smith:1999cr} which constrains $ g_{\phi\gamma\gamma} < 10^{-20} $ GeV$ ^{-1} $, so if we do not consider $ f_a > 10^{13} $ GeV the $\phi$-photon coupling can deviate from a KSVZ-like coupling by up to four orders of magnitude before exact coupling dependence has to be taken into account in Eq.~\eqref{eq:EP constraint}.

It is possible to use data from ALP searches to constrain radial mode-photon interactions. This was studied by Flambaum et al.~\cite{Flambaum:2022zuq}, who derived scalar-photon constraints from CAST and ADMX. However, scalar-photon constraints from EP violation dominate those from ALP searches by many orders of magnitude~\cite{Flambaum:2022zuq}.

\paragraph{Axion and radial mode mass hierarchy:}
In addition to being spontaneously broken, the PQ symmetry has a small explicit breaking leading to the small axion mass $m_a \ll f_a$.
When $m_\phi \sim f_a$ one thus naturally expects $m_a \ll m_\phi$.
In the case where the radial mode is also light, one may wonder if it is possible to have the opposite hierarchy, i.e., $m_a \gg m_\phi$.
We can address this question with an effective potential for the axion and the radial mode. By integrating out all of the other particles, this potential is of the form
\begin{align}
	V_{\rm eff}(P) = V_{\rm PQinv}(\phi) + V_{\rm PQviol}(\phi,\theta).
\end{align}
$V_{\rm PQinv}$ is invariant under the PQ symmetry, but $V_{\rm PQviol}$ is not.
To realize $m_\phi \ll m_a$, we need to satisfy
\begin{align}
	\frac{\partial^2 V_{\rm PQinv}}{\partial \phi^2} +
	\frac{\partial^2 V_{\rm PQviol}}{\partial \phi^2}
	\ll \frac{1}{\phi^2} \frac{\partial^2 V_{\rm PQviol}}{\partial \theta^2}
\end{align}
at the potential minimum.
This requires some fine-tuning.
First, it is natural to have
$\partial^2 V_{\rm PQinv}/\partial \phi^2 \gg \partial^2 V_{\rm PQviol}/\partial \phi^2,~(1/\phi^2) \partial^2 V_{\rm PQviol}/\partial \theta^2$
since the PQ-violating terms are subdominant.
Also, it is natural to have
$\partial^2 V_{\rm PQviol}/\partial \phi^2 \sim (1/\phi^2)\partial^2 V_{\rm PQviol}/\partial \theta^2$.
In this paper, we do not pursue the possibility of having $m_\phi \ll m_a$
and focus on the region with $m_a < m_\phi$.

\paragraph{Impact of CP-violating term on the axion quality:}
In this work, we are studying both QCD axions and general ALPs. The QCD axion is particularly sensitive to the higher-dimensional PQ-violating operator Eq.~\eqref{eq:higher dimensional PQ breaking potential}. In particular, this term should not introduce $ \theta_{\rm QCD} \gtrsim 10^{-10} $ \cite{Baker:2006ts,Pendlebury:2015lrz}, where $ \theta_{\rm QCD}=0 $ is the CP-conserving minimum of the QCD potential. To estimate this shift, we can take $ \delta_\theta = \pi / 2$ to maximally misalign the higher-dimensional potential from the QCD potential. In this case, in the late-time potential, where $ \phi=f_a $, is
\begin{gather}
V_{\rm today} \approx -2^{1-\frac{n}{2}} \frac{A f_a^n}{M^{n-3}} \theta+\frac{1}{2}  m_a^2 f_a^2\theta^2 +\mathcal{O}\left(\theta ^3\right).
\end{gather}
The CP-violating VEV induced by this potential is then
\begin{gather}
\theta_{\rm CPV} \approx 2^{1-\frac{n}{2}} \frac{A M^3}{m_a^2f_a^2}\left(\frac{ f_a}{ M}\right)^{n}.\label{eq:axion quality}
\end{gather}
For a QCD axion to successfully solve the strong CP problem, we require $ \theta_{\rm CPV} \ll 10^{-10} $. We generally display this condition by shading the otherwise yellow QCD line of $ [m_a,f_a] $ parameter space in blue in the range where it cannot be satisfied.

\paragraph{$\phi$-to-axion decay:}\label{sec:overview - phi-to-axion decay}
The radial and angular modes are coupled.  At the very least, the Lagrangian contains the kinetic term
\begin{gather}
\vert \pd_\mu P \vert^2 = \frac{1}{2}(\pd_\mu \varphi)^2+\frac{1}{2}(\pd_\mu a)^2+ \frac{1}{2}\frac{\varphi^2}{\langle \phi \rangle^2}(\pd_\mu a)^2 + \frac{\varphi}{\langle \phi \rangle}(\pd_\mu a)^2,
\end{gather}
where $P=\frac{\langle \phi \rangle +\varphi(x)}{2}e^{ia(x)/\langle\phi\rangle}$ and $ a = \theta f_a $  is the axion field. The latter of these terms induces an $ \varphi\to aa $ decay with the rate
\begin{gather}
\Gamma_{\varphi a}\sim \frac{1}{64\pi} \frac{m_\phi^3}{\langle \phi \rangle^2}.\label{eq:radial-to-axion rate}
\end{gather}
Although the radial mode also experiences EM couplings, the corresponding rate is always less efficient because EM decay is suppressed by a factor $(f_a / f_{a\gamma})^2\approx 5\times 10^{-6}$ relative to Eq.~\eqref{eq:radial-to-axion rate}, where $f_{a\gamma}=g_{\theta \gamma\gamma}^{-1}$ is the dimension-full axion-photon coupling which we assume to be KSVZ-like.

If the radial mode decayed into axions, it would produce a hot axion relic which could dominate the plasma. To avoid this, we impose the following constraint:
\begin{gather}
\Gamma_{\varphi a} < H \qq{while} \rho_{\rm kin}^\phi > \rho_r \qquad \text{(hot axions dominate the plasma)},\label{eq:hot axions dominate the plasma}
\end{gather}
where $ \rho_{\rm kin}^\phi $ is the energy in radial oscillations. Furthermore, even if the hot axion relic does not dominate the plasma, it may still lead to DM overproduction once it has cooled. Solving for the cold DM relic generated by radial-to-axion decay, we impose the following constraint
\begin{gather}
	\Gamma_{\varphi a} < H \qq{while} \rho_{\rm kin}^\phi \frac{2m_a}{\mphi}\frac{g_*(\Teq)}{g_*(T)}\left(\frac{\Teq}{T}\right)^3>\rho_{r,\rm eq} \qquad \text{(DM overproduction)}, \label{eq:axion decay relic DM overproduction}
\end{gather}
where the subscript $ _{\rm eq} $ refers to quantities measured at radiation-matter equality. 
These constraints are avoided if the radial oscillations are damped before $\phi$-to-axion decay becomes efficient. We therefore impose them only in the model implementation where the damping mechanism is specified.

We are now ready to move to the model implementations of KMM. The early cosmology of the models under consideration have been scrutinised also in \cite{Gouttenoire:2021jhk} with a different purpose,  to determine the parameter region leading to kination eras in the early universe.
Moreover, while the KMM has been pioneered in \cite{Co:2019jts}, our motivation is to provide a synthesis and a status of the mechanism when putting together all requirements, by providing a systematic scan of the parameter space, in particular mapping out observable quantities of interest in the $(m_a,f_a)$ plane, such as the value of the radial-mode mass, the damping temperature, and the needed suppression of the primordial spectrum of fluctuations.


\section{Shape of the Peccei-Quinn potential}\label{sec:potential}

The shape of the radial-mode potential is crucial in determining whether or not KMM can take place. 
A generic mexican-hat potential with quartic coupling has the form
\begin{gather}
	V \supset \lambda \left(\abs{P}^2-\frac{f_a^2}{2}\right)^2. \label{eq:quartic potential}
\end{gather}
As justified below, such generic potential leads to 
KMM only at the cost of fine-tuning $\lambda$ to small values.
We will therefore instead mainly focus on nearly-quadratic potentials a la Affleck-Dine which can be motivated in SUSY constructions where the PQ symmetry is spontaneously broken by radiative corrections. These were originally proposed in Ref.~\cite{Moxhay:1984am}, where soft SUSY-breaking generates a nearly-quadratic potential of the form
\begin{gather}
	V_{\rm late} = m_\phi^2 |P|^2 \left(\frac{1}{2}\ln(\frac{2|P|^2}{f_a^2})-\frac{1}{2}\right)+\frac{1}{4}\mphi^2 f_a^2. \label{eq:Nearly quadratic potential, late time}
\end{gather}
Note that $(dV_{\rm late}/d\phi)|_{\phi=f_a} = 0$ and $(d^2V_{\rm late}/d\phi^2)|_{\phi=f_a} = m_\phi^2$ are satisfied and that we throughout this work assume $N_{\rm DW}=1$ for simplicity, where $N_{\rm DW}$ is the domain wall number.
The radial mass $ \mphi $ may then be naturally light as it is protected by SUSY. In this framework,
PQ is spontaneously broken by the logarithmic running of the potential, which gives the radial mode a VEV at $ \phi = f_a $ once the radial mode is fully relaxed. Such models were first studied in the context of kinetic misalignment in \cite{Co:2019jts}. 
Note that the mass of the radial mode, which around the minimum is $ m_\phi $, receives a logarithmic correction at high field values. However, this logarithmic correction is at most $ \mathcal{O}(1)$. We therefore neglect the correction at large field amplitudes where we then take $ V \approx m_{\phi}^2|P|^2=\frac{1}{2}\mphi^2\phi^2 $. Thus, the fundamental parameter $ m_\phi $ here agrees with $ \mphiVprime \equiv \sqrt{\phi^{-1}V'} $ as defined in Eq.~\eqref{eq:effective mphi definition}, such that there is no distinction between the effective mass at large field values and the fundamental parameter.

In the following sections \ref{sec:initial conditions (NQ)},
\ref{sec:Nearly quadratic early damping results},
\ref{sec:nearly quadratic late damping},
\ref{sec:Nearly quadratic Yukawa section},
\ref{sec:NQHiggs}, we will investigate how radial dynamics in such a model can provide the initial conditions for fragmentation and kinetic misalignment. And in Section \ref{sec:quartic model}, we will present analogous results for the quartic potential. In the next sub-section, we summarise the issues associated with the generic quartic potential \ref{eq:quartic potential} as it has been implemented in literature \cite{Co:2019jts,Co:2020dya,Co:2020jtv}.

\subsection{Hubble-induced mass term versus de Sitter fluctuations}

As discussed in Section \ref{subsection:starting}, a crucial ingredient of the KMM mechanism is an initial field value for the radial mode that is much larger than the value at the minimum $f_a$ of the PQ potential.
There are two main ways to realise this. The first way, which we discuss in more detail later, is a large negative Hubble-induced mass term for the radial mode, $-c_H H^2 |P|^2$, which generates a  minimum at large field values for a large initial Hubble expansion rate value $H$. The second way is a very light radial mode, which via de Sitter fluctuations during inflation drives the field to the value \cite{Co:2020dya}
\begin{gather}
\Skick \sim \sqrt[4]{\frac{3}{8\pi^2}}\frac{H_I}{\sqrt[4]{\lambda}}.\label{eq:phi kick}
\end{gather}  
This second possibility, applied to the quartic potential \ref{eq:quartic potential}, has been considered for KMM in Ref.~\cite{Co:2019jts,Co:2020dya,Co:2020jtv}. 
 However, we find that this scenario faces challenges. 
Note that our notation in Eq.~\eqref{eq:quartic potential} differs from the notation used in Co et al
in that $\lambda=\lambda^2_{\rm{Co \ et \ al}}$.
 An important point is  that the effective radial  depends on the field value
\begin{gather}
	\mphiVprime^2=\frac{1}{\phi}V_{\rm late }'\approx \lambda \phi^2. \label{eq:quartic radial mass}
\end{gather}
This is a key difference  for the phenomenology of kinetic misalignement with respect to the models with a nearly-quadratic potential 
for which the effective radial mass $\mphiVprime$ is constant in $\phi$ up to logarithmic effects. 
The main issue with this scenario is the  formation of domain walls  at the time of trapping, which can be circumvented if~\cite{Co:2020dya}
\begin{gather}
	\frac{H_I}{f_a} \lesssim \mathcal{O}(0.1) .\label{eq:simple DW constraint}
\end{gather}
 Since we are considering pre-inflationary scenario, we do not expect cosmic strings to be present, and the domain walls would therefore be stable and pathological.
Co et al.~\cite{Co:2020dya} propose to resolve this domain wall problem by invoking parametric resonance to non-thermally restore PQ-symmetry, generate cosmic strings and thereby render the string-wall network unstable. However, there is no such analysis to date. We have looked in detail at this case and we do not find any viable  KMM implementations relying on de Sitter fluctuations  with a quartic potential. In the rest of this section, we explain the difficulties. 
To understand this problem, consider the yield produced in this class of models,
\begin{gather}
	Y_{\rm kick} \approx \frac{n_{PQ}}{s}\Big\vert_{\rm kick} \approx \epsilon\frac{\lambdaOLD\Skick^3}{\frac{2\pi^2}{45} g_{*s}\Tkick^3},
\end{gather}
which in terms of model parameters corresponds to
\begin{gather}
	Y_{\rm kick} \approx 0.2	\times\frac{\epsilon}{\lambda^{5/8}}\left(\frac{H_I}{\mplanck}\right)^{3/2},\label{eq:Ykick quartic}
\end{gather}
where the prefactor has a mild dependence on $ g_*(\Tkick) $.
To realize solutions corresponding to sub-eV axions masses it is therefore necessary to consider either large values of the Hubble scale during inflation $ H_I $ or low values of $ \lambda $. Large values of $H_I$ are in tension with Eq.~\eqref{eq:simple DW constraint} and imply domain wall (DW) constraints. Therefore, setting aside concerns of naturalness for the moment, we are led to consider low values of $ \lambda $. Specifically, we can consider values of $ \lambda $ which go below the range constrained by thermal $ \phi $ relics, Eq.~\eqref{eq:radial thermal lower bound}, which corresponds to
\begin{gather}
\lambdaOLD<\mathcal{O}(1)\frac{\Teq}{f_a}.
\end{gather}
However, this range of $ \lambda $ implies a low radial mode mass. This low radial mode mass cannot compete with the thermal mass $\mth $ generated by the interactions introduced for the thermal damping that will be discussed in sections \ref{sec:Nearly quadratic Yukawa section} and \ref{sec:NQHiggs}. This means that the efficiency of the kick, i.e. $\epsilon$, tends to be suppressed in this regime. To avoid this, we must demand $ \mphiz>\mth $ at the time of the kick. In the case where thermal damping is induced by a Higgs portal interaction $ \xi$, as discussed in Section \ref{sec:NQHiggs}, this  corresponds to the condition
\begin{gather}
\xi\lesssim 2 \sqrt{\frac{H_I}{\mplanck}}\lambda^{1/8}.
\end{gather}
A similar condition applies for Yukawa interactions. Such low values of the coupling  does not permit damping. For instance, even with the most efficient Higgs portal interaction, Eq.~\eqref{eq:most important Higgs interaction},
\begin{gather}
	\Gamma_\phi\approx 2 \frac{\xi^4\phi^2}{\pi^2\alphaweak T},
\end{gather}
taken at the most competitive conditions, $ \phi=f_a $ and $ T\sim \mHiggs $, efficient damping as defined by $ \Gamma > H $ requires
\begin{gather}
H_I \gtrsim 3 \times10^{12}\text{ GeV}\times\left(\frac{10^{8}\text{ GeV}}{f_a}\right),
\end{gather}
which is in contradiction with the simple domain wall constraint Eq.~\eqref{eq:simple DW constraint}. If we consider instead solutions above the range constrained by thermal $ \phi $ relics, i.e. 
\begin{gather}
 \mathcal{O}(1)\times\frac{f_a}{\mplanck} < \lambdaOLD ,
\end{gather}
then the large values $ \lambda $ can only correspond to scenarios of eV-scale axions with values of  $ H_I $ that are also constrained by the domain wall problem.

We solved numerically the Boltzmann equations and we found no solutions without DW problems for neither Yukawa-type interactions nor with the Higgs portal interaction. If a solution to the DW problem is assumed, then some parameter space becomes viable.
We therefore discard implementations where the large initial field value is due to de Sitter fluctuations.
In Sections \ref{sec:initial conditions (NQ)},
\ref{sec:Nearly quadratic early damping results},
\ref{sec:nearly quadratic late damping},
\ref{sec:Nearly quadratic Yukawa section},
\ref{sec:NQHiggs},  we will focus on potentials with nearly-quadratic potentials. We also look at quartic potentials in the case where the large initial VEV is due to a negative Hubble induced mass term in Section \ref{sec:quartic model}, for which we find viable parameter space,  a case that was not considered by Co et al.


\section{Initial conditions in a nearly-quadratic potential}\label{sec:initial conditions (NQ)}
We rely on the potential used in
Affleck-Dine Baryogenesis in \cite{Dine:1995kz} and later exploited for KMM in \cite{Co:2019jts}:
\begin{gather}
	V_{\rm early} = (m_\phi^2 - c_H H^2)|P|^2 + \frac{A + c_A H}{n} \frac{P^n}{M^{n-3}}+h.c.+\frac{|P|^{2n-2}}{M^{2n-6}}, \label{eq:potential for kick}
\end{gather}
where we neglected the logarithmic correction to the late-time mass $ m_\phi $, $ H $ is the Hubble parameter, $ M $ is a suppression scale which we typically take to be $ \mplanck \approx 2.4 \times 10^{18} $ GeV, $ A $ is a dimension-full constant which we typically take to be $ \mathcal{O}(1)\times m_\phi $ and $ c_A $ is an $ \mathcal{O}(1) $ constant. The negative Hubble-induced mass can be naturally generated in models where $ P $ couples to fields which dominate the energy density of the universe~\cite{Dine:1995kz,Asaka:1999yc,Kawasaki:2011zi,Kawasaki:2012qm,Kawasaki:2012rs}. 
If at early times $ \mphi\ll H $, then this negative mass contribution drives up the VEV. To stabilize against this negative potential contribution, one can introduce a higher-dimensional term in the superpotential~\cite{Dine:1995kz} of the form
\begin{gather}
	W = \frac{1}{n}\frac{P^n}{M^{n-3}}.
\end{gather}
This superpotential sources a scalar potential corresponding to the last term in Eq.~\eqref{eq:potential for kick}. The A-term, i.e. the middle term of Eq.~\eqref{eq:potential for kick}, is generated from $ W $ by soft SUSY breaking. In addition to the discussion in \cite{Dine:1995kz}, a summary of the generation of these terms can be found in \cite{Gouttenoire:2021jhk}.

At early times when $ H \gg m_\phi $, i.e. before the kick, the $ H^2 $ term then drives $ \phi $ to a large VEV stabilized by the $ |P|^{2n-2} $ term. The VEV at the minimum is
\begin{gather}
	\phi_{\rm early} = \left(\frac{2^{n-2}}{n-1}\right)^{\frac{1}{2 n-4}} \left(H M^{n-3}\right)^{\frac{1}{n-2}}. \label{eq:phi minimum early}
\end{gather}
Around this VEV, both the radial and angular modes acquire large masses,
\begin{gather}
	m_{\phi,\rm early} = \sqrt{2(n-1)}H \qq{and} m_{a,\rm early} = \frac{\sqrt{2 |c_A|n}}{\sqrt[4]{n-1}}H,\label{eq:mphi early masses}
\end{gather}
such that both fields are heavy during inflation and can track the minimum until $ m_\phi \propto 3H $, where the late-time mass becomes important. This large early mass naturally suppresses inflationary fluctuations and precludes the commonly associated domain wall and isocurvature problems. We assume that the higher-dimensional operators become irrelevant once $ m_\phi > 3H $ so that we can take the potential to be $ V_{\rm late} $, given by Eq.~\eqref{eq:Nearly quadratic potential, late time}, after this transition. This implies that the PQ field starts evolving in $ V_{\rm late} $ with an initial radial amplitude of $ \phikick = \phi_{\rm early}(H_{\rm kick}) $, where $ H_{\rm kick} $ is the Hubble parameter measured at the time of the transition.

During the transition, the PQ violating terms $ P^n+P^{*n} $  can drive rotation. This is possible because the angular mode transitions from a minimum set by the $ c_A H$ term into a minimum set by the $ A $ term. The angular minima of these two contributions are determined by the complex phase of $ c_A $ and $ A $, respectively. If the minima of these two contributions are not accidentally aligned, the field experiences a torque as the minimum shifts~\cite{Dine:1995uk} and Eq.~\eqref{eq:kick size} applies with $ \theta_{\rm kick}\neq 0 $. With $ A = \mathcal{O}(1)\times \mphi $ this naturally yields $ \epsilon = \mathcal{O}(1) $ such that 
\begin{gather}
	\dtheta_{\rm kick} = \mathcal{O}(1)\times \mphi,
\end{gather}
where $ \mphi $ is the potential parameter appearing in Eq.~\eqref{eq:potential for kick} and not the much larger early-time mass given by \eqref{eq:mphi early masses}.
This kick is delivered when $ m_\phi \approx 3H $, which implies a kick temperature of
\begin{gather}
	\Tkick \approx \sqrt[4]{\frac{10}{g_*(\Tkick)\pi^2}} \sqrt{m_\phi \mplanck}.\label{eq:Tkick nearly quadratic}
\end{gather}
We assume that soon after the kick the PQ-violating potential becomes irrelevant so that we can disregard the higher-dimensional terms after the first rotation. This is confirmed by the numerical evaluation of the equations of motion~\cite{Dine:1995kz}.

In the nearly-quadratic model, the radial mass $ \mphiVprime $ is approximately constant in time. Therefore, applying Eq.~\eqref{eq:Tmin definition, general} we see that $ \phi $ reaches the minimum at $ f_a $ at $ \amin $ given by
\begin{gather}
	\frac{\phikick}{f_a} \approx \left(\frac{\amin}{\akick}\right)^{3/2}.
\end{gather}
If no significant entropy injection takes place, then this corresponds to a temperature of
\begin{gather}
	\Tmin \approx \Tkick \left(\frac{g_{*s}(\Tkick)}{g_{*s}(\Tmin)}\right)^{1/3}\left(\frac{f_a}{\phikick}\right)^{2/3}.\label{eq:Tkin nearly quadratic}
\end{gather}
After this time, $ \dtheta $ transitions to a kination-like scaling as described in Section \ref{sec:general evolution description}.

\paragraph{Estimate of the yield:}
After the kick is complete and the PQ-violating terms have become subdominant, the comoving PQ-charge is conserved,  $ n_{\rm PQ} \propto a^{-3}$. The resulting yield is constant in the absence of entropy injection:
\begin{gather}
	\Ykick = \frac{n_{PQ}}{s} = \epsilon \frac{m_{\phi}\phikick^2}{\frac{2\pi^2}{45}g_{*s}\Tkick^3}\approx 0.8 \times\epsilon\left(\frac{M}{\mplanck}
	\right)^{\frac{2n-6}{n-2}}\left(\frac{\mplanck}{m_\phi}
	\right)^{\frac{n-6}{2n-4}},\label{eq:Ykick nearly quadratic}
\end{gather}
where the prefactor has a mild dependence on $ n $, $ g_{*}(\Tkick) $, and on $ g_{*s}(\Tkick) $. 

\paragraph{Dark matter solution in the absence of yield dilution:}
If no entropy injection takes place, the yield Eq.~\eqref{eq:Ykick nearly quadratic} eventually sources axion dark matter. The observed DM abundance,  $ \Ydm \approx2\Ykick  $ is achieved for the value $ \mphi=m_{\phi,\rm DM,not\ diluted} $ given by
\begin{align}
	m_{\phi,\rm DM,not\ diluted} \approx\ & \xi_n \times g_*(\Tosc)^{\frac{3 (n-2)}{2 (n-6)}}\left(\frac{g_{*s}(\Teq)  }{g_{*s}(\Tosc) g_{*}(\Teq) }\right)^{\frac{2 (n-2)}{n-6}} \times \mplanck   \left(\frac{M}{\mplanck}\right)^{\frac{4 (n-3)}{n-6}}  \left(\frac{ m_a \epsilon }{ \Teq}\right)^{\frac{2 (n-2)}{n-6}}\notag\\
	&\text{where}\quad \xi_n=(n-1)^{-\frac{2}{n-6}} \pi ^{\frac{2-n}{n-6}} 2^{\frac{9 (n-2)}{2 (n-6)}} 3^{\frac{2 (n-4)}{n-6}} 5^{\frac{n-2}{2 (n-6)}}.
\end{align}
For a few representative choices of $ n $,
\begin{gather}
	m_{\phi,\rm DM,not\ diluted} \approx \begin{cases}
		70 \left(\frac{M}{\mplanck}\right)^{16} \left(\frac{m_a}{10^{-2}\text{ eV}}\right)^{10} \left(\frac{\epsilon}{0.9}\right)^{10} &\qq{if} n=7,\\
		2\times 10^{12} \left(\frac{M}{\mplanck}\right)^{7}  \left(\frac{m_a}{10^{-2}\text{ eV}}\right)^{4}\left(\frac{\epsilon}{0.9}\right)^{4}  &\qq{if} n=10,\\
		5 \times 10^{13} \left(\frac{M}{\mplanck}\right)^{\frac{40}{7}} \left(\frac{m_a}{10^{-2}\text{ eV}}\right)^{\frac{22}{7}} \left(\frac{\epsilon}{0.9}\right)^{\frac{22}{7}}  &\qq{if} n=13.
	\end{cases}\label{eq:mphi nearly quadratic, without injection}
\end{gather}
The yield has a somewhat shallow dependence on the radial mass, i.e. 
\begin{equation}
\Ykick \propto \mphi^{-l} 
\label{eq:yieldkick}
\end{equation}
where $ l\approx $ 0.1, 0.25 or 0.3 for $ n= $ 7, 10 or 13 respectively. Therefore, small changes in various parameters require a large change in $ \mphi $ to compensate, especially for lower values of $ n $, see the very large powers in Eq.~\eqref{eq:mphi nearly quadratic, without injection}. Correspondingly, the numerical prefactor of Eq.~\eqref{eq:mphi nearly quadratic, without injection} is sensitive to small changes in other parameters.

Although $\mphi $ impacts both $\phikick$ and $\Tkick$, its  dominant impact  is through $\Tkick$ where lower values of $\mphi$ act to reduce the amount of redshift and thus act to increase the yield. Therefore, smaller values of $\mphi$ are required when smaller values of $m_a$ are considered.


\section{Early damping in a nearly-quadratic potential}\label{sec:Nearly quadratic early damping results}
The simplest scenario we consider is when the radial mode is assumed to be damped through some unspecified mechanism early enough that no yield dilution takes place. In this scenario, the yield generated by the kick, as given by Eq.~\eqref{eq:Ykick nearly quadratic}, remains constant until late times, and the above results apply without modifications. Because the kick is controlled by the higher-dimensional terms, the viable parameter space is sensitive to the choice of $ n $.

\begin{figure}
	\centering
	\includegraphics[width=0.8 \textwidth]{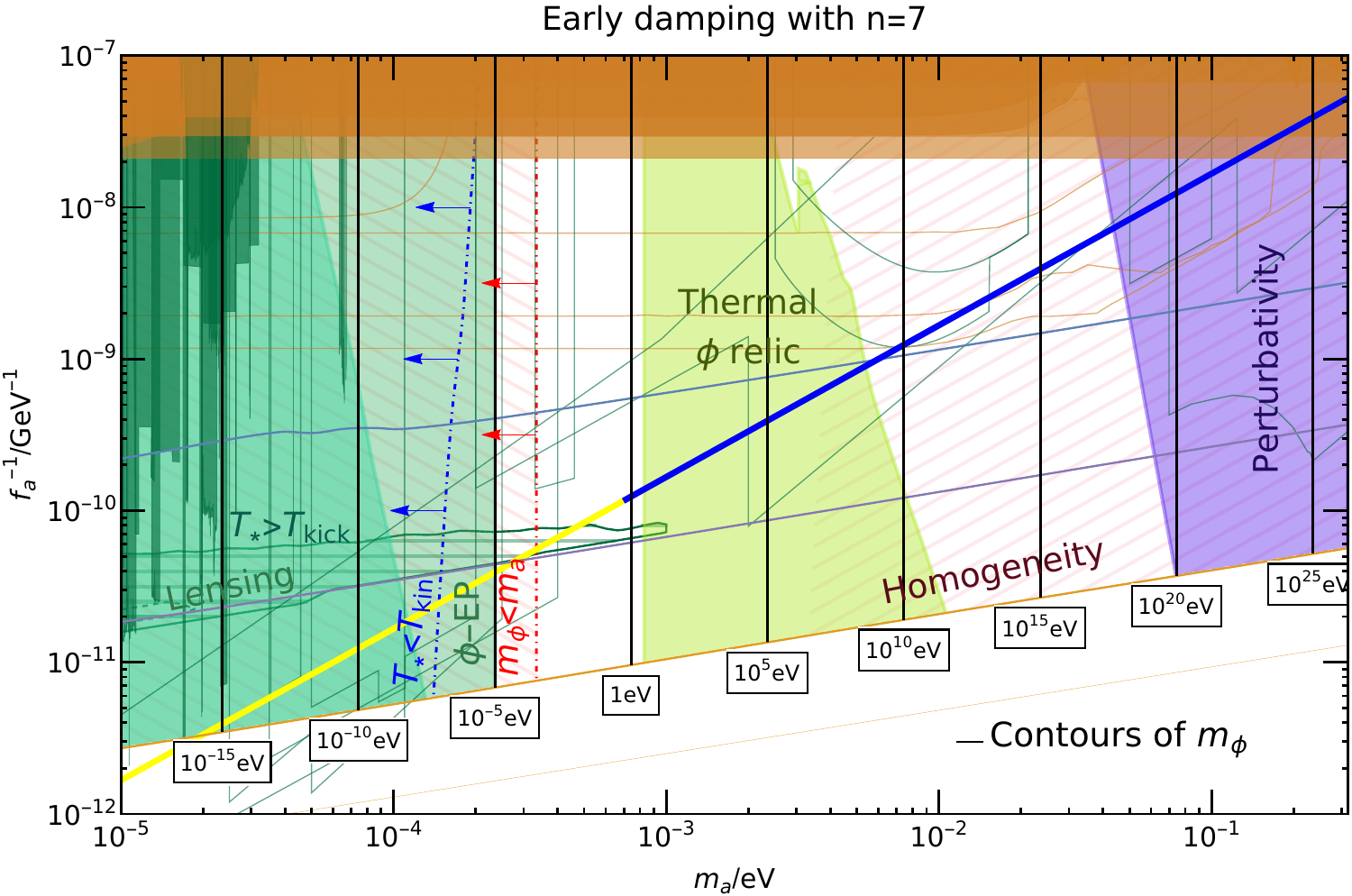}

    \caption{\small \it ALP DM parameter space (in white)  and constraints  for $n=7$ 
    in models with nearly-quadratic potentials Eq.~\eqref{eq:potential for kick}, when early damping without entropy dilution is assumed. The constraints from thermal $ \phi $ relics, perturbativity, homogeneity, and EP violation are given by Eqs.~\eqref{eq:thermal condition}, \eqref{eq:perturbativity bound NQ}, \eqref{eq:homogeneity constraint, early NQ}, and \eqref{eq:EP constraint}, respectively.  The temperatures $ \Tkick,\Tmin, $ and $ \Tfrag $ are given by Eqs.~\eqref{eq:Tkick nearly quadratic}, \eqref{eq:Tkin nearly quadratic} and \eqref{eq:thermal potential}, respectively. The yellow QCD axion line is shaded blue when the axion quality is insufficient to solve the strong CP problem, see Eq.~\eqref{eq:axion quality}. The vertical contours indicate values of the radial mode mass $ m_\phi $, which are fixed by Eq.~\eqref{eq:mphi nearly quadratic, without injection} to obtain the observed axion DM relic. For the labels of experimental sensitivity lines and constraints, see Fig.~\ref{fig:moneyPlotTwoCol}. }
    \label{fig:nearlyQuadraticEarlyDamping1}
\end{figure}

\begin{figure}
	\centering
	\includegraphics[width=0.8 \textwidth]{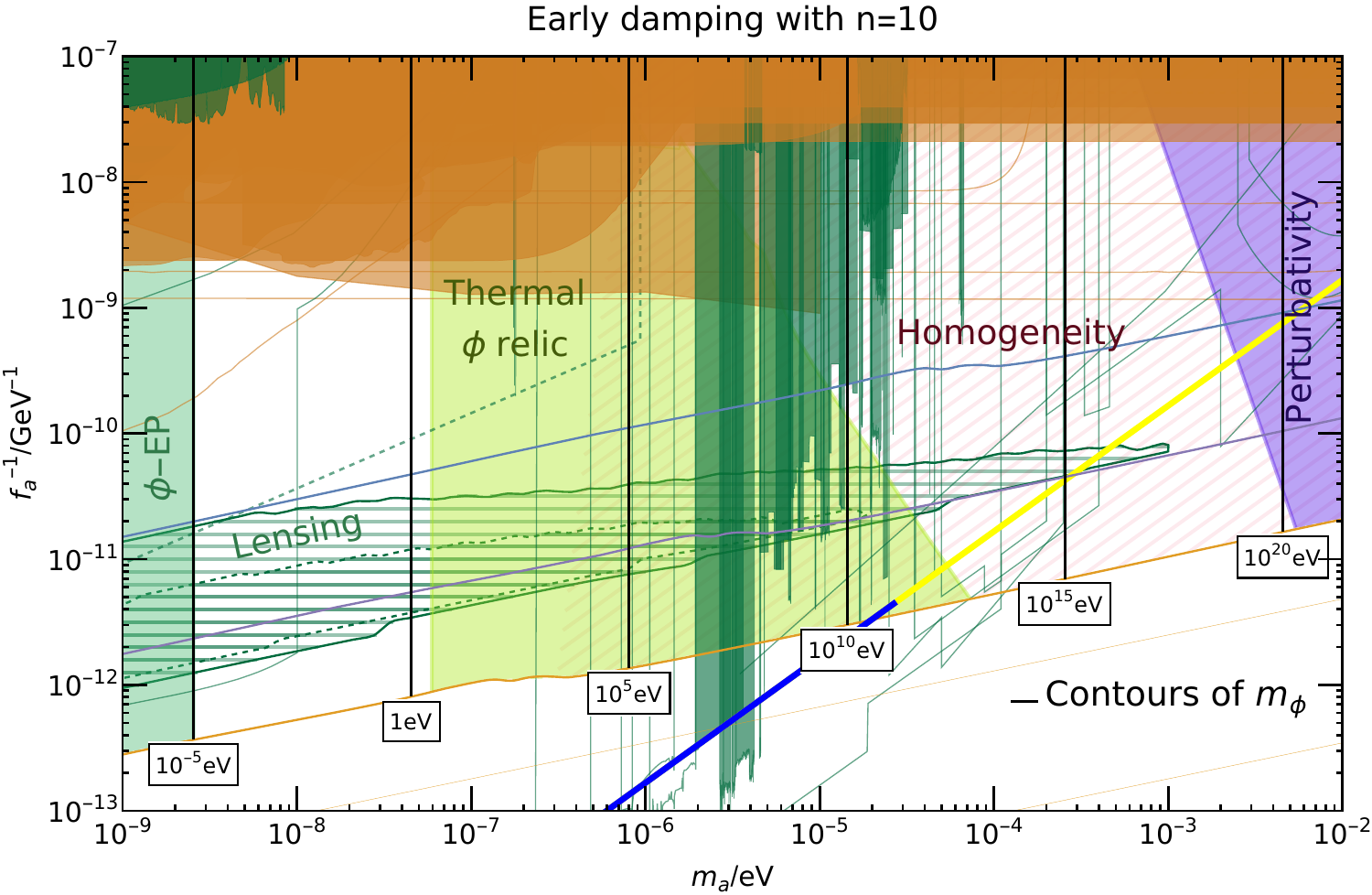}
	
	\vspace{0.5cm}
	\includegraphics[width=0.8 \textwidth]{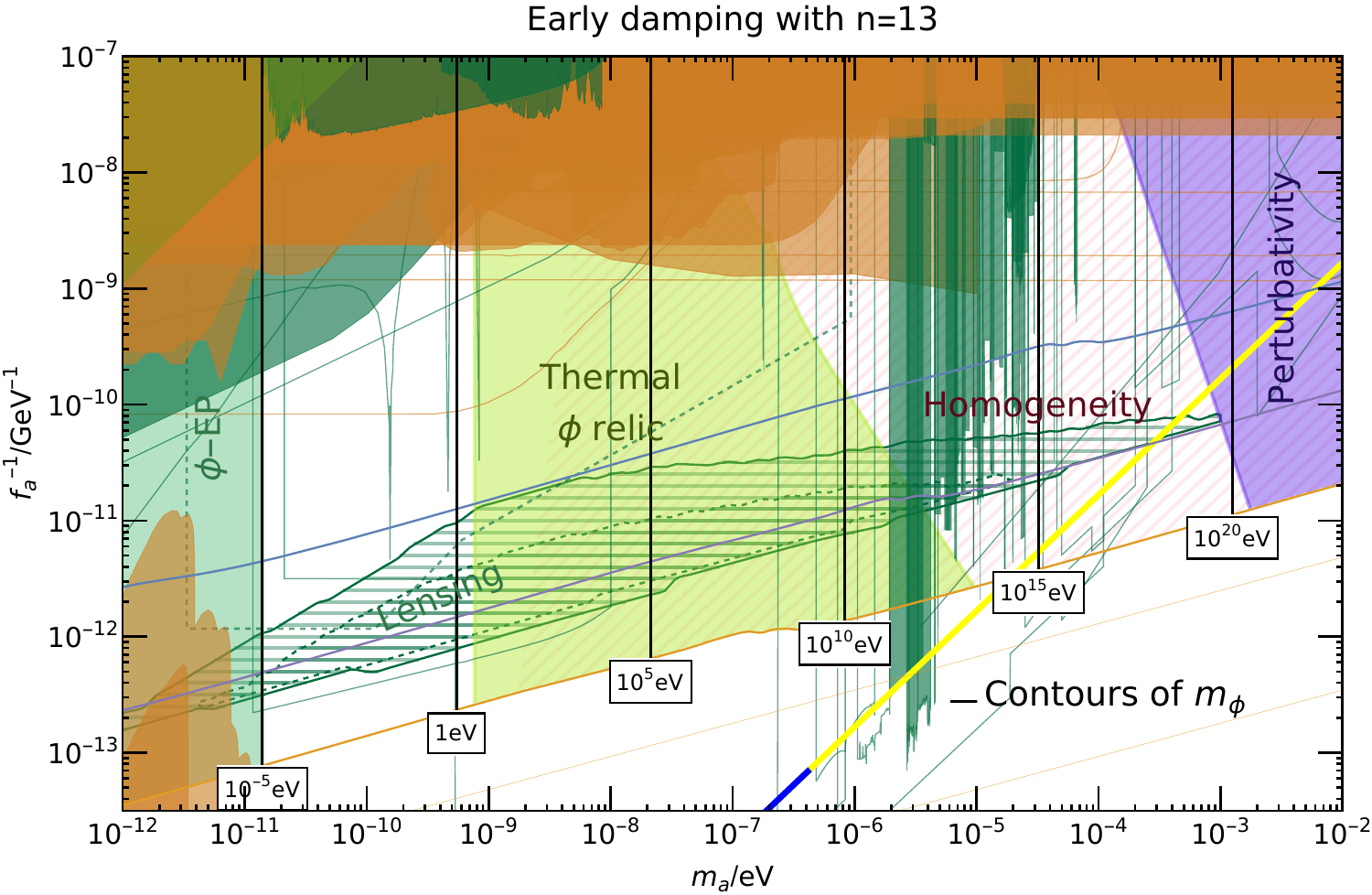}
 
	\caption{\small\it Same as Fig.~\ref{fig:nearlyQuadraticEarlyDamping1} but for $ n=10 $ and $ n=13 $. Note that each of the three plots displays a different  $ m_a $ range to focus on the viable regions arising from each choice of $ n $. To understand the qualitatively different behaviour of the axion quality constraint (blue line), see Eq.~\eqref{eq:theta CPV qualitiative}.}
	\label{fig:nearlyQuadraticEarlyDamping2}
\end{figure}

This scenario is subject to the constraints described in Section \ref{sec:Constraints (general)}. The most significant are the experimental EP constraint, the consistency conditions to ensure that the kick takes place before $ \Tfrag $, that the field remains homogeneous until $ \Tfrag $, and that a thermal $ \phi $ relic does not spoil CMB or BBN.
The thermal $\phi$ relic constraint from Eq.~\eqref{eq:thermal condition} and the EP constraint from Eq.~\eqref{eq:EP constraint} can be directly applied to the $ \mphi $ solution Eq.~\eqref{eq:mphi nearly quadratic, without injection}. 
Below we apply the remaining constraints. We show the viable parameter space in Figs. \ref{fig:nearlyQuadraticEarlyDamping1} and \ref{fig:nearlyQuadraticEarlyDamping2}.

\paragraph{Perturbativity}
 To extract the self-interactions, we define the radial excitation around $ f_a $ as $ S = \phi-f_a $ and expand the potential Eq.~\eqref{eq:Nearly quadratic potential, late time} around $ S=0 $ in powers of $ S/f_a $:
\begin{gather}
	V_{\rm late} = m_\phi^2 |P|^2 \left(\frac{1}{2}\ln(\frac{2|P|^2}{f_a^2})-\frac{1}{2}\right) \approx \frac{1}{2}\mphi^2S^2+\frac{\lambda_{\rm ln}}{3!}f_aS^3-\frac{\lambda_{\rm ln}}{4!}S^4+ \mathcal{O}\left(\lambda_{\rm ln}\frac{S^5}{f_a}\right),
\end{gather}
where 	$\lambda_{\rm ln}={m_\phi^2}/{f_a^2}$.
Therefore, perturbativity of the self-interactions implies the  bound 
\begin{gather}
	\frac{m_\phi^2}{f_a^2} < 4 \pi. \label{eq:perturbativity bound NQ}
\end{gather}
This form of the constraint applies for all scenarios based on the nearly-quadratic potential, Eq.~\eqref{eq:potential for kick}. When we also assume early damping, then the constraint Eq.~\eqref{eq:perturbativity bound NQ} is applied to Eq.~\eqref{eq:mphi nearly quadratic, without injection}.

\paragraph{Early enough kick:}
For the scenario to be consistent we have to impose the condition $ \Tosc > \Tfrag $. Applying Eq.~\eqref{eq:mphi nearly quadratic, without injection} to Eq.~\eqref{eq:Tkick nearly quadratic} and comparing with Eq.~\eqref{eq:Tfrag} we find the following lower bound on the viable axion masses:
\begin{gather}
	m_a \gtrsim \begin{cases}
		9.0 \times 10^{-5}\text{ eV } \left(\frac{f_a}{10^{10}\text{ GeV}}\right)^{0.13} \left(\frac{M}{\mplanck}\right)^{1.80} &\text{for } n=7\\
		1.3 \times 10^{-12}\text{ eV } \left(\frac{f_a}{10^{10}\text{ GeV}}\right)^{0.40} \left(\frac{M}{\mplanck}\right)^{2.45} &\text{for } n=10\\
		1.6 \times 10^{-17}\text{ eV } \left(\frac{f_a}{10^{10}\text{ GeV}}\right)^{0.57} \left(\frac{M}{\mplanck}\right)^{2.86} &\text{for } n=13\\
	\end{cases}
\end{gather}
While significant, this constraint is generally weaker than the constraint set by EP violation.

\paragraph{Homogeneity:}
As described in Section \ref{sec:Constraints (general)}, the growth of fluctuations limits the duration of kination-like scaling. Using Eq.~\eqref{eq:Tkin nearly quadratic}, we can write the condition Eq.~\eqref{eq:cems condition} as
\begin{gather}
\mathcal{P}_{\mathcal{R}}(\kmin) \lesssim \left(\frac{g_{*s}(\Tfrag)}{g_{*s}(\Tosc)}\right)^{2/3}\left(\frac{\Tfrag}{\Tosc}\right)^2 \left(\frac{\phikick}{f_a}\right)^{4/3}.
\end{gather}
Plugging in Eqs.~\eqref{eq:Tfrag}, \eqref{eq:phi minimum early}, and \eqref{eq:Tkick nearly quadratic}  implies an upper bound on $ m_a $:
\begin{gather}
	m_a < \mathcal{O}(10-100) \times \left(\mathcal{P}_{\mathcal{R}}(\kmin)\right)^{-\frac{177 (n-6)}{8 (19 n+4)}} f_a^{-\frac{17 (n-6)}{76 n+16}} \left(\frac{M}{\mplanck}\right)^{-\frac{59 (n-3)}{19 n+4}} \mplanck^{-\frac{59 (n-6)}{76 n+16}} \left(\frac{\Teq}{\epsilon }\right)^{\frac{59 (3 n-10)}{76 n+16}},
\end{gather}
where the exact value of the prefactor depends on $ n $ and the number of relativistic degrees of freedom. Equivalently, for representative choices of $ n $, the condition is
\begin{gather}
	m_a \lesssim \begin{cases}
		2\times 10^{-2}\text{ eV } \left(\frac{\mathcal{P}_{\mathcal{R}}(\kmin)}{2.1\times 10^{-9}}\right)^{-0.16} \left(\frac{f_a}{10^9\text{ GeV }}\right)^{-0.031} \left(\frac{M}{\mplanck}\right)^{-1.7} \epsilon^{-1.2} &\text{ if } n = 7 \\
		5\times 10^{-6}\text{ eV } \left(\frac{\mathcal{P}_{\mathcal{R}}(\kmin)}{2.1\times 10^{-9}}\right)^{-0.46} \left(\frac{f_a}{10^9\text{ GeV }}\right)^{-0.088} \left(\frac{M}{\mplanck}\right)^{-2.1} \epsilon^{-1.5} &\text{ if } n = 10 \\
		5\times 10^{-8}\text{ eV } \left(\frac{\mathcal{P}_{\mathcal{R}}(\kmin)}{2.1\times 10^{-9}}\right)^{-0.62} \left(\frac{f_a}{10^9\text{ GeV }}\right)^{-0.12}  \left(\frac{M}{\mplanck}\right)^{-2.3} \epsilon^{-1.7} &\text{ if } n = 13
	\end{cases}\label{eq:homogeneity constraint, early NQ}
\end{gather}
Furthermore, together with the constraint from the thermal radial relic, the condition $ \Tfrag < \Tkick $ most strongly restricts this scenario. 
The resulting parameter space is displayed for $ n=7 $, 10 and 13 in Figs. \ref{fig:nearlyQuadraticEarlyDamping1} and \ref{fig:nearlyQuadraticEarlyDamping2}. The shallow dependence of $ \Ykick $ 
on $ \mphi $ leaves only a narrow range of parameter space viable for $ n=7 $. For larger $ n $, $ \Ykick $ becomes more sensitive to $ \mphi $ such that a larger range of parameter space can be accommodated, although this space will also be shifted to lower $ m_a $.

\paragraph{Axion quality:}
The PQ-violating potential, which drives the kick, also threatens the axion quality by generating the VEV given by Eq.~\eqref{eq:axion quality}. For this scenario, we evaluate Eq.~\eqref{eq:axion quality} with $A=m_\phi$ and apply Eq.~\eqref{eq:mphi nearly quadratic, without injection}. 
We have 
\begin{gather}
    \theta_{\rm CPV} \propto f_a^{n-2\frac{n-2}{n-6}} \label{eq:theta CPV qualitiative}
\end{gather} 
where the second term in the power is from the implicit dependence in the DM solution, Eq.~\eqref{eq:mphi nearly quadratic, without injection}. Note that lower values of $f_a$ are constrained for $n\leq 7$ while larger values of $f_a$ are constrained for $8\leq n$, which is the reason why the axion quality constraint (blue line) appears in different regimes between Fig.~\ref{fig:nearlyQuadraticEarlyDamping1} and \ref{fig:nearlyQuadraticEarlyDamping2}.



\section{General damping in a nearly-quadratic potential} \label{sec:nearly quadratic late damping}
\begin{figure}
	\centering
	\hspace{-1.5cm}\includegraphics[width=0.8\textwidth]{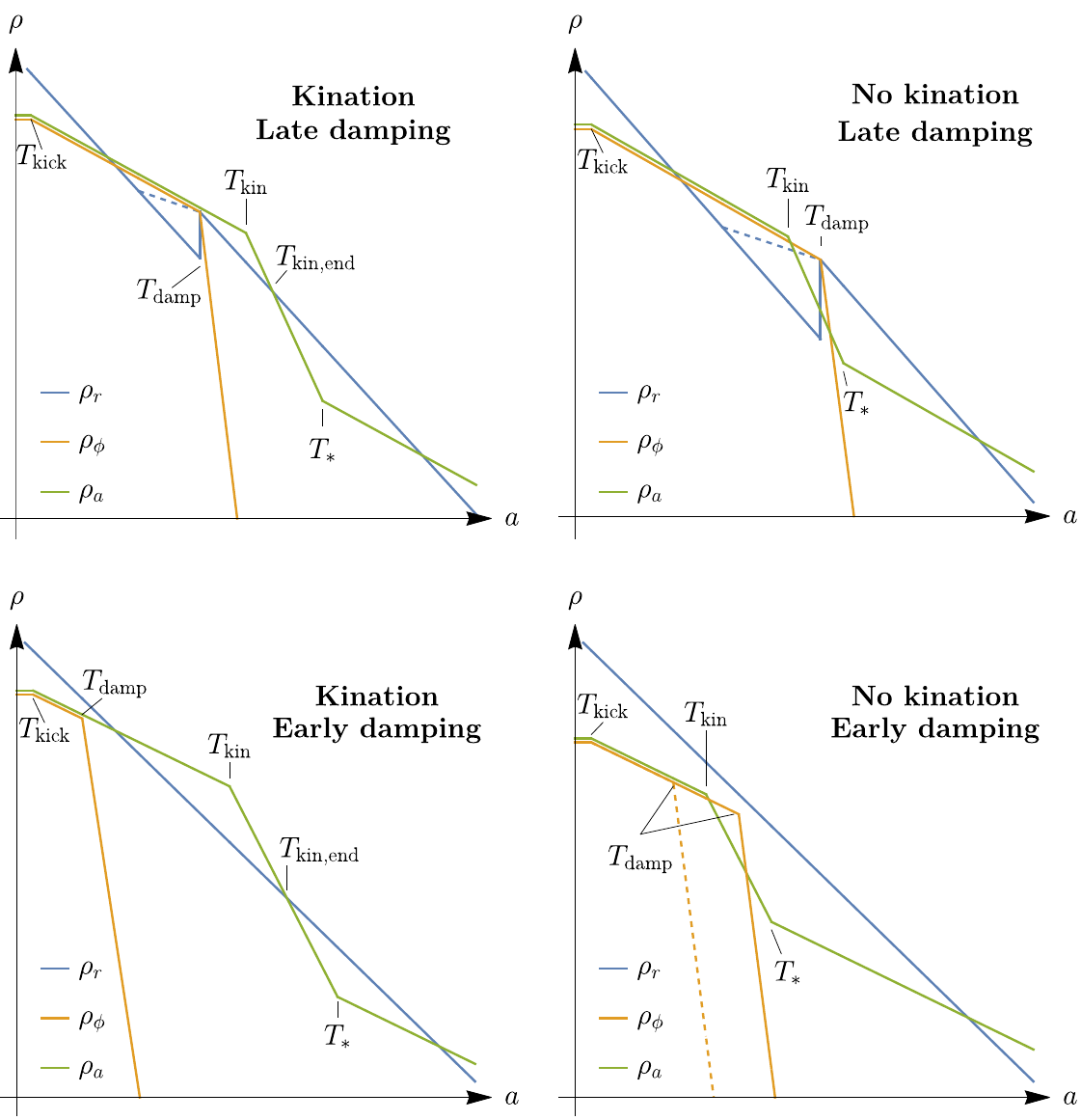}
	
	\caption{\small\it Overview of possible scenarios in models with nearly-quadratic potentials. The energy densities of the axion (orange) and radial mode (green) are comparable until either damping takes place ($\Tdamp)$ or the radial mode reaches its minimum at $f_a$, ($\Tkin)$. The top examples feature entropy injection (late damping) while the bottom examples do not (early damping). The left examples feature kination, while the right examples do not. Regardless of whether a period of kination is triggered or not, the temperature at which the radial mode is relaxed to $ f_a $ is labelled $ \Tkin $. The end of kination is labelled by $ \Tkend $ if it does take place. Physically, damping may not be instantaneous, and the energy density of the plasma could reheat smoothly, as indicated by the blue dashed lines. However, as only the initial and final states matter for yield dilution, this question does not impact the damping-mechanism-agnostic solution presented in this section. The physical evolution of $ T  $ does impact the more realistic damping considered in later sections of this work, since the damping rates depend on $T$. The evolutions displayed here are sketches intended to highlight the characteristic differences between scenarios and do not represent actual solutions. They are to be compared with the full numerical solutions shown at the end of Appendix \ref{app:boltzmann equations}.}
	\label{fig:scenarios}
\end{figure}

We now generalize our analysis to include the case when damping occurs after the energy in radial oscillations has become dominant. This leads to richer cosmologies. Fig.~\ref{fig:scenarios} shows sketches of the different possibiliities for the evolution of energy densities. In some cases, kination takes place, which can have interesting phenomenological consequences~\cite{Gouttenoire:2021wzu,Gouttenoire:2021jhk,Co:2021lkc}.

\paragraph{Estimate of yield dilution for late damping:}  The radial mode begins to dominate over radiation at the temperature $ \Tsd $ for which $ \rho_\phi \approx \rho_r $:
\begin{gather}
	\Tsd \approx 0.8 \left(\frac{M}{\mplanck}\right)^{\frac{n-3}{2(n-2)}} \mplanck^{\frac{n-6}{8(n-2)}} \mphi^{\frac{n+2}{8(n-2)}}\Tth^{3/4},
\end{gather}
where the prefactor has a mild dependence on $ g_* $ and on $n$.
If  $ \Tdamp<\Tsd  $,  the associated entropy injection is significant. 
If all energy in the radial oscillations is instantaneously transferred to the SM plasma at some temperature $ \Tth $ and the entropy injection increases the plasma temperature to $ \TSth $, then the yield is diluted to
\begin{gather}
	\Ydiluted = \Ykick \frac{g_{*s}(\Tth)\Tth^3}{g_{*s}(\TSth)\TSth^3}. \label{eq:Ydiluted nearly quadratic}
\end{gather}
where the reheated temperature $ \TSth $ is determined by
\begin{gather}
	\frac{1}{2}m_\phi^2 \phikick^2 \frac{g_{*s}(\Tth)\Tth^3}{g_{*s}(\Tosc)\Tosc^3} = \frac{\pi^2}{30}g_*(\TSth) \TSth^4.
\end{gather}
The observed DM yield is realized for radial masses significantly different from the scenario without entropy injection:
\begin{gather}
	m_{\phi,\rm DM, diluted} \approx  5\left(\epsilon \frac{m_a}{\Teq}\right)^{\frac{8(n-2)}{7n-18}}\left(\mplanck^{3(n-2)}M^{4(n-3)}\Tth^{6(n-2)}\right)^{\frac{1}{7n-18}},
\end{gather}
where the exact value of the $ \mathcal{O}(1) $ factor depends on $ n $ and the number of relativistic degrees of freedom. For specific choices of $ n $, this corresponds to
\begin{gather}
	m_{\phi,\rm DM, diluted} \approx \begin{cases}
		2\times 10^{3}\text{ GeV }  \left(\frac{m_a}{\text{ eV }}\right)^{40/31}\left(\frac{M}{\mplanck}\right)^{16/31} \left(\frac{\Tth}{10^{2}\text{ GeV}}\right)^{30/31}\epsilon^{40/31} &\qq{if} n=7,\\
		1\times 10^{4} \text{ GeV } \left(\frac{m_a}{\text{ eV }}\right)^{16/13}\left(\frac{M}{\mplanck}\right)^{7/13} \left(\frac{\Tth}{10^{2}\text{ GeV}}\right)^{12/13}\epsilon^{16/13} &\qq{if} n=10,\\
		2\times 10^{4} \text{ GeV }  \left(\frac{m_a}{\text{ eV }}\right)^{88/73}\left(\frac{M}{\mplanck}\right)^{40/73} \left(\frac{\Tth}{10^{2}\text{ GeV}}\right)^{66/73}\epsilon^{88/73} &\qq{if} n=13.
	\end{cases}\label{eq:mphi nearly quadratic, with injection}
\end{gather}
Note in particular that the $ m_a $ dependence in $ \mphi $ is much shallower in the presence of entropy injection than in the undiluted case, see Eq.~\eqref{eq:mphi nearly quadratic, without injection}. This allows the solutions to span over a larger range of axion masses without requiring unreasonable radial masses.

\begin{figure}
	\centering
	\includegraphics[width=0.8 \textwidth]{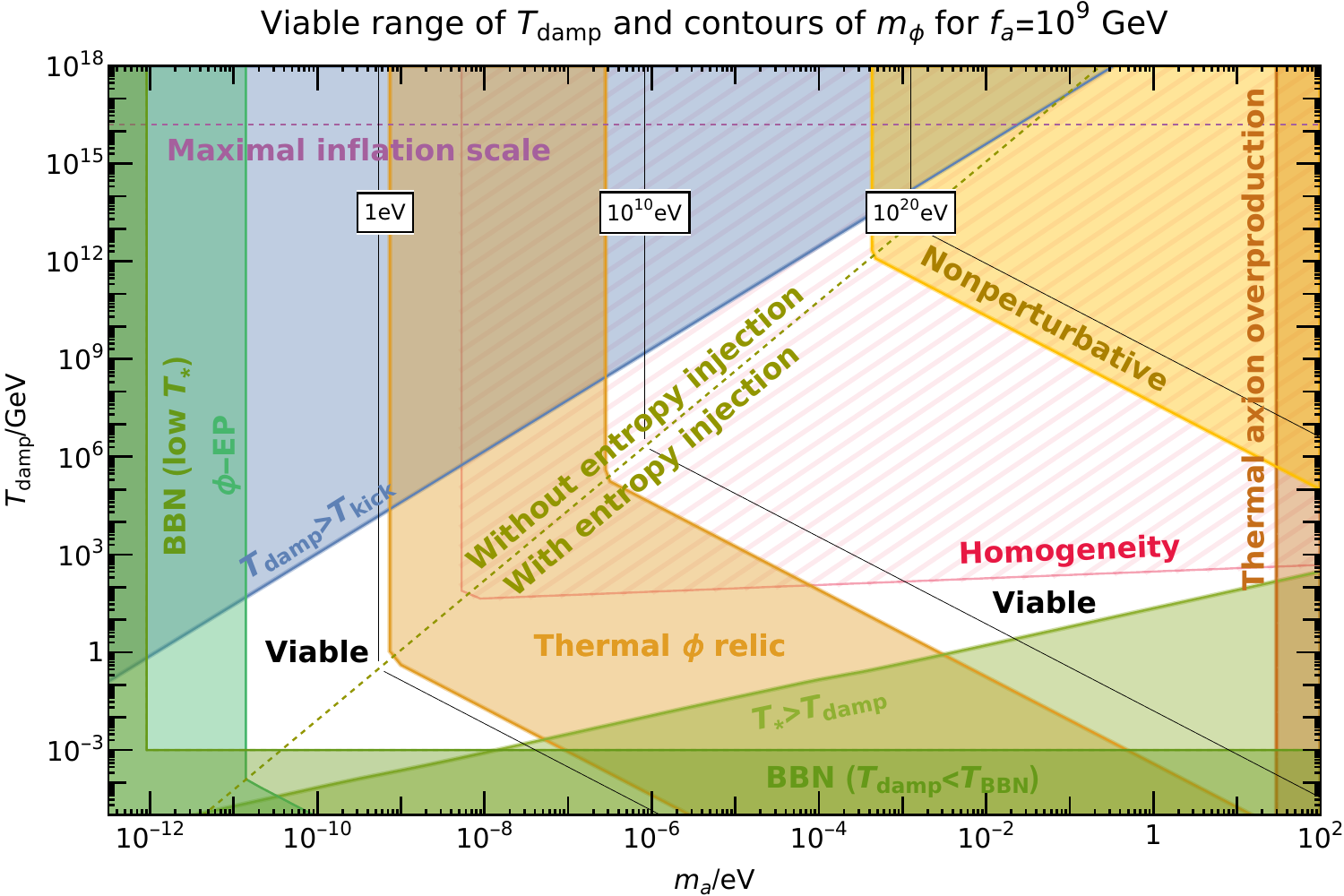}

	\caption{\small\it Constraints on the [$ \Tdamp $, $ m_a $] parameter space for fixed $ f_a = 10^9 $ GeV, $ n=13 $ and $ M=\mplanck $. In addition to the constraints presented in Fig.~\ref{fig:nearlyQuadraticEarlyDamping1}, the damping temperature $ \Tdamp $ is constrained by the BBN bound \eqref{eq:BBN bound Tdamp} and by \eqref{eq:range of validity}. Also shown is the upper bound $ \Tdamp\lesssim E_{I,\rm max} $ where $ E_{I,\rm max} \sim 1.6\times 10^{16} $ is the upper bound on the maximal energy scale of inflation set by Planck 2018~\cite{Planck:2018jri}. The plot corresponds to a horizontal slice of Fig.~\ref{fig:Nearly quadratic M18n13With}.}
	\label{fig:constantfaPlotn13Mmp}
\end{figure}

\begin{figure}
	\centering
    \vspace{-5mm}
	\includegraphics[width=0.8 \textwidth]{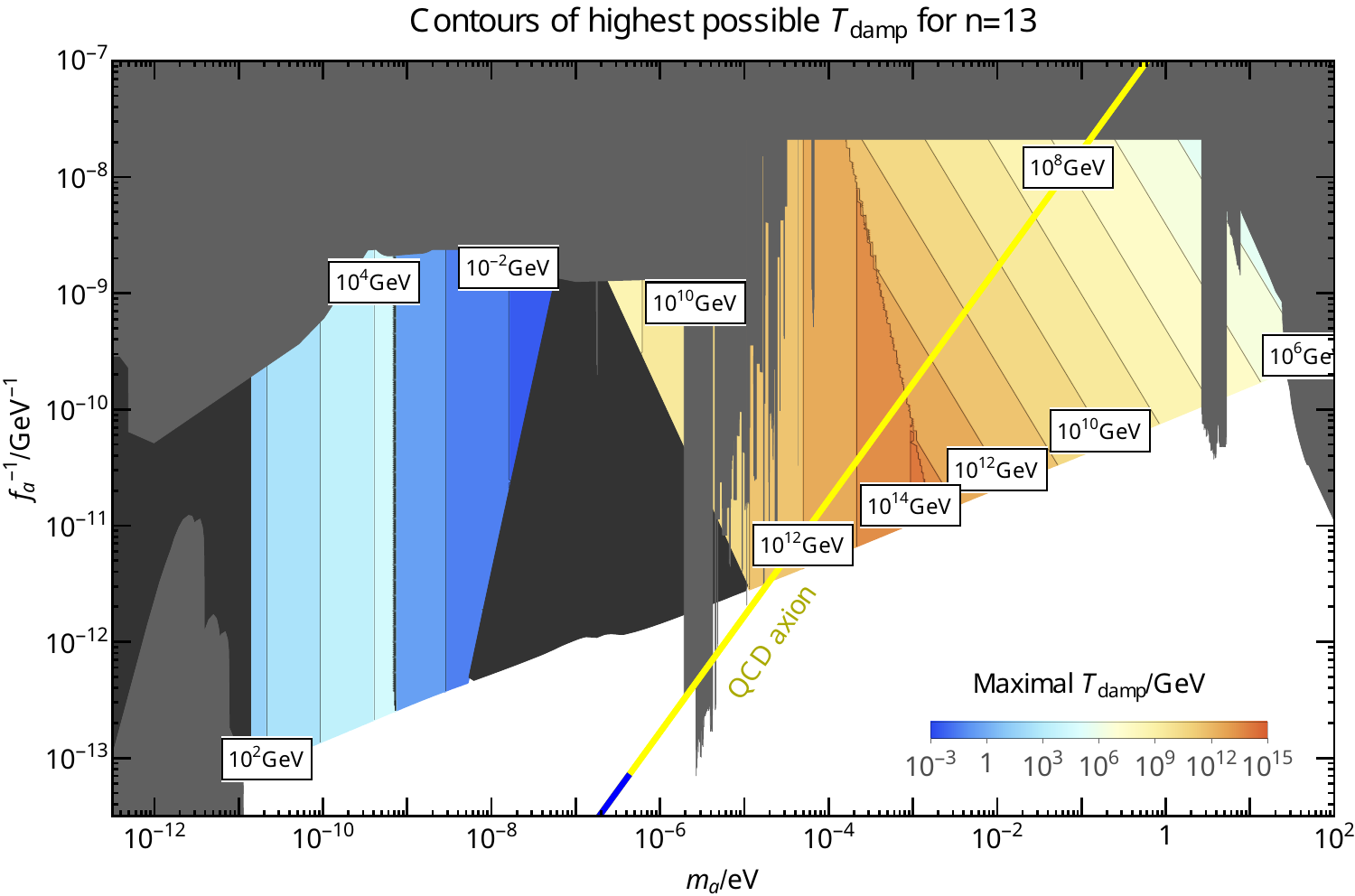}\\
	\vspace{2mm}
	\includegraphics[width=0.8 \textwidth]{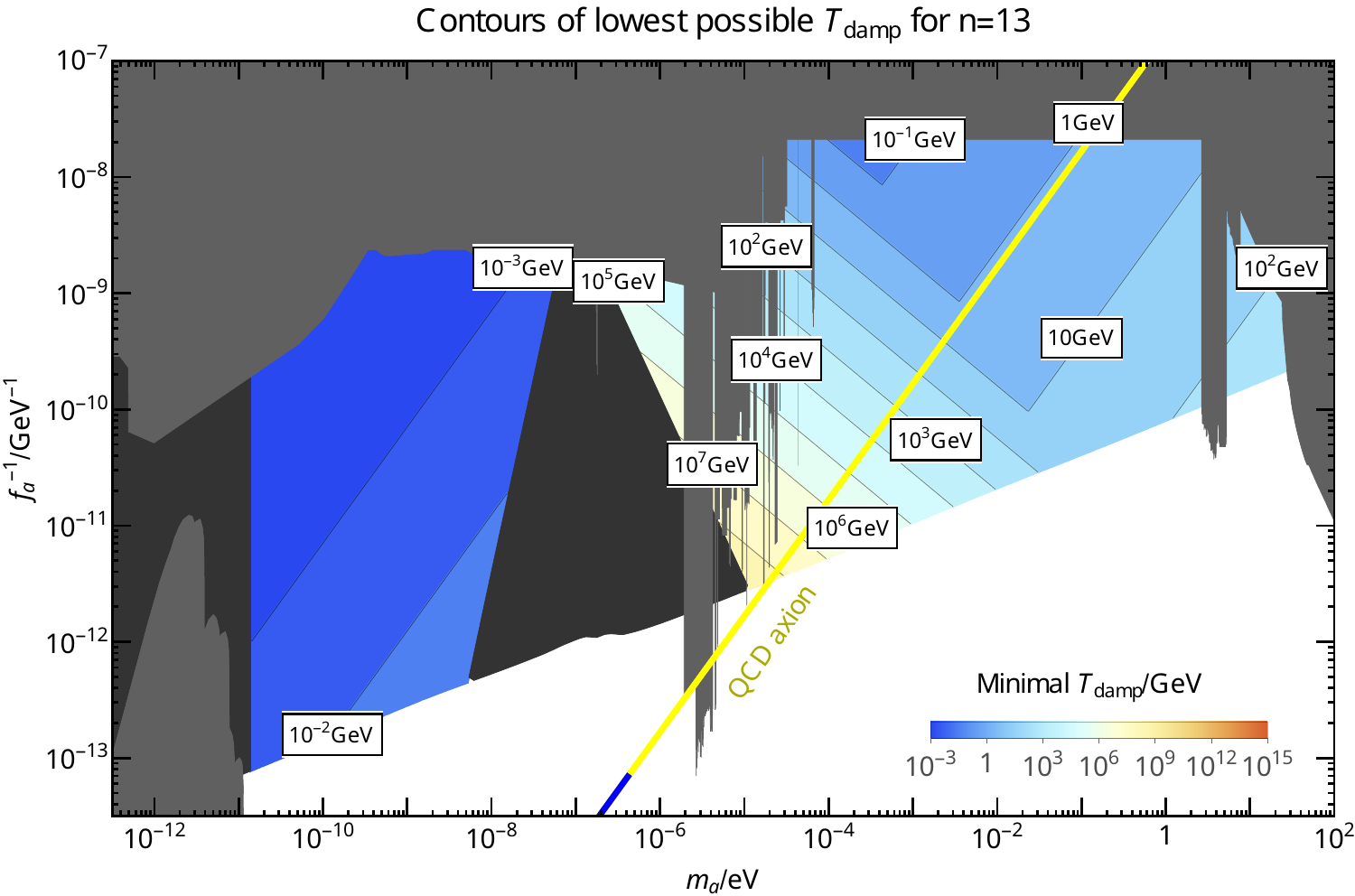}

	\caption{\small\it Survey of viable damping temperatures. Fig.~\ref{fig:constantfaPlotn13Mmp} serves as a guide to the structure of this plot. We assume that the homogeneity condition can be relaxed such that the dashed region in Fig.~\ref{fig:constantfaPlotn13Mmp} can be made viable. The temperature range can be extended further in scenarios in which the thermal relic constraint is resolved. \textbf{Top:} Contours of the highest possible damping temperature for which the scenario can be realized. The discontinuities seen at $ m_a \sim 10^{-8} $ eV and $ m_a\sim 10^{-3.5} $ eV can be understood from the onset of thermal relic and perturbativity constraints, respectively. The regions to the right of these steps only feature solutions with late damping. \textbf{Bottom:} Similar plot of the lowest viable damping temperature. }
	\label{fig:Nearly quadratic M18n13With}
\end{figure}

\paragraph{Viable range of $ \Tdamp $}
For the analysis of \cite{Eroncel:2022vjg} to apply, the radial oscillations must be damped before fragmentation.  Besides,  if efficient at $ \Tkick $, the damping would interfere with the kick. Therefore, $ \Tdamp $ is ultimately limited by\footnote{It is possible to consider scenarios in which damping takes place after fragmentation. However, the fragmentation analysis would have to be redone in the presence of radial oscillations. We leave such studies to future work.}
\begin{gather}
	\Tkick > \Tdamp > \Tfrag. \label{eq:range of validity}
\end{gather}
This condition gives the wedge-shaped envelope of the [$ \Tth,m_a $] parameter space, which is displayed for constant $ f_a $ in Fig.~\ref{fig:constantfaPlotn13Mmp}.
The damping temperature is further restricted by constraints on the thermal relic of the radial mode. These constraints arise when $ \mphi $, as determined by either Eq.~\eqref{eq:mphi nearly quadratic, without injection} or Eq.~\eqref{eq:mphi nearly quadratic, with injection}, falls in the range given by Eq.~\eqref{eq:thermal condition} where ALP thermal relics are constrained by CMB or BBN~\cite{Fields:2019pfx,Zyla:2020zbs}. 
The constraint corresponds to the orange band labelled "Thermal $ \phi $ relic" in Fig.~\ref{fig:constantfaPlotn13Mmp} and to the central dark grey constraint in Fig.~\ref{fig:Nearly quadratic M18n13With}.

Yield dilution also modifies the homogeneity condition. Combining the solution Eq.~\eqref{eq:mphi nearly quadratic, with injection} with the condition Eq.~\eqref{eq:cems condition} yields an upper bound on $ \Tdamp $:
\begin{gather}
	\Tdamp \lesssim \begin{cases}
		2\times 10^{1} \text{ GeV }  \left(\frac{m_a}{\text{eV}}\right)^{0.27}\left(\frac{\mathcal{P}_{\mathcal{R}}(\kmin)}{2.1\times 10^{-9}}\right)^{-1.4} \left(\frac{f_a}{10^9\text{ GeV }}\right)^{-0.27} \left(\frac{M}{\mplanck}\right)^{0.97} \epsilon^{4/3} &\text{ if } n = 7 \\
		4\times 10^{2} \text{ GeV }  \left(\frac{m_a}{\text{eV}}\right)^{0.15}\left(\frac{\mathcal{P}_{\mathcal{R}}(\kmin)}{2.1\times 10^{-9}}\right)^{-1.3} \left(\frac{f_a}{10^9\text{ GeV }}\right)^{-0.25} \left(\frac{M}{\mplanck}\right)^{0.93} \epsilon^{4/3} &\text{ if } n = 10 \\
		1\times 10^{3} \text{ GeV }  \left(\frac{m_a}{\text{eV}}\right)^{0.10}\left(\frac{\mathcal{P}_{\mathcal{R}}(\kmin)}{2.1\times 10^{-9}}\right)^{-1.3} \left(\frac{f_a}{10^9\text{ GeV }}\right)^{-0.24} \left(\frac{M}{\mplanck}\right)^{0.92} \epsilon^{4/3} &\text{ if } n = 13 \\
	\end{cases}\label{eq:Tdamp max for homogeneity, NQ general}
\end{gather}
Interestingly, this condition permits entropy injection to circumvent the homogeneity condition such that a large-$ m_a $ window opens up in parameter space. This window is visualized in both Fig.~\ref{fig:constantfaPlotn13Mmp}, which displays the structure of the constraints along a line of constant $ f_a $, and in Fig.~\ref{fig:Nearly quadratic M18n13With}, which displays the viable range of $ \Tdamp $ across the [$ m_a,f_a $] parameter space.

We next seek to implement specific realizations of the damping mechanism. In Section \ref{sec:Nearly quadratic Yukawa section}, we  realise it through a coupling of the PQ field to a fermion while in Section \ref{sec:NQHiggs} we consider a Higgs-portal interaction.


\section{Thermal damping via Yukawa interactions in nearly-quadratic potential}\label{sec:Nearly quadratic Yukawa section}

A well-motivated damping of radial oscillations is through a Yukawa interaction of the form
\begin{gather}
	\mathcal{L}_{\rm int} = \sqrt{2}y P\bar{\chi} \left( \frac{1-\gamma^5}{2}\right) \chi+h.c.,\label{eq:yukawa coupling}
\end{gather}
where $ \chi $ is a  fermion that is efficiently coupled to the SM thermal bath~\cite{Mukaida:2012qn}. 
Such a Yukawa coupling is an integral part of KSVZ-type axion models. In this framework, the $ \chi $-fermions are KSVZ fermions which are coupled to the SM via SU(3)$_{\rm QCD}$, and damping  would be a natural consequence of the KSVZ construction. 
This construction deviates from a conventional KSVZ model as we must assume $y\ll 1$ for reasons discussed later in this section. Nevertheless, we take this as our benchmark case and assume the $\chi$ fermions to be charged under SU(3)$_{\rm QCD}$. 
Schematically, we have
\begin{gather}
	\phi\hspace{3mm}\xleftrightarrow[\text{Yukawa}]{y} \hspace{3mm}\chi\hspace{3mm} \xleftrightarrow[\text{Gauge}]{\gchi}\hspace{3mm} \text{SM plasma with gluons }g.
\end{gather} 
The $\chi$ fermion mass receives a contribution from the $ \phi $-interaction and at high temperature from the SM gauge interaction. We neglect any other sources such that the effective $ \chi $ mass is
\begin{gather}
	\mchieff^2\approx \mchiS^2+\mchiT^2 \approx y^2 \phi^2 + \gchi^2 T^2 .
\end{gather}
We define  $ \alphachi \equiv \gchi^2 / 4\pi $. We assume that the Yukawa coupling is weak such that 
$y < \alphachi$.

\subsection{Thermal effects and damping rates}
\label{sec:why no light dm}
Although the above structure appears rather simple, thermal effects lead to a very rich dynamics. For a detailed study of such thermal effects we refer to Mukaida et al.~\cite{Mukaida:2012qn,Mukaida:2012bz,Mukaida:2013xxa}. They are also crucial for an analysis of kination as discussed extensively and reviewed in \cite{Gouttenoire:2021jhk}.
In particular, a Yukawa coupling of the type considered here generates thermal contributions to the radial potential. If $T > y \phi$, $\chi$ particles  in the thermal plasma give a thermal correction $\approx y^2 T^2$ to $m_\phi^2$. On the other hand, if $T < y \phi$, although $\chi$ is absent from the thermal bath, it affects the effective gauge coupling via logarithmic running effect. This provides a thermal correction to the potential energy as $\delta V\approx \alphachi^2 T^4 \log(y^2 \phi^2 / T^2)$ \cite{Anisimov:2000wx, Mukaida:2012qn}. The second derivative of this term can be interpreted as a thermal mass correction to $\phi$.  The full effective mass $ \meff $ is thus of the form~\cite{Anisimov:2000wx, Mukaida:2012qn}
\begin{gather}
	\begin{gathered}
		\meff^2 \approx \begin{cases}
			m_{\phi,0}^2 + \mth^2 &\text{if}\quad T > y \phi, \\
			m_{\phi,0}^2 + \mln^2 &\text{if}\quad T<y \phi,
		\end{cases} \\
		\text{where}\quad \mth^2 \approx y^2 T^2 \qq{and} \mln^2 \approx \alphachi^2 \frac{T^4}{\phi^2}.
	\end{gathered}
\end{gather}
Here $ \mphiz $ is the zero-temperature mass considered in the previous section, $ \mth $ is the thermal mass and $ \mln $ is the thermal-log mass. The thermal-log mass replaces the ordinary thermal mass when the $ \chi $-fermions are Boltzmann suppressed, i.e. for $ T < y \phi $. 
If either thermal potential is non-negligible, then it can significantly alter the dynamics of the kick. Furthermore, when the Yukawa interaction damps the radial oscillations, this energy is transferred to the SM plasma. This can further modify the parameter space if the yield dilution is significant.

\label{sec:Thermal damping and yield dilution}
The Yukawa coupling introduced in Eq.~\eqref{eq:yukawa coupling} gives rise to thermal damping through a variety of processes. The most relevant rates are the following: 

\noindent - Perturbative decay:
\begin{gather}
	\Gamma_{\phi\bar{\chi}\chi} \approx \frac{y^2 \mphiz}{8\pi} \qq{if} m_{\phi,0}<g T \qq{and} m_{\phi,0} < y \phi, \\
	\Gamma_{\phi gg} \approx \alphachi^2 \frac{\mphiz^3}{\phi^2} \qq{if}m_{\phi,0}<g T \qq{and} m_{\phi,0} > y \phi \\
	\qq{where} b = \frac{T(r)}{16\pi^2}\frac{(12\pi)^2}{\ln \alphachi^{-1}},\label{eq:gauge scattering}\notag
\end{gather}
- Scattering processes with $ \chi $-fermions:
\begin{gather}
	\Gamma_{\chi-\text{scat.}}\approx \begin{cases}
		y^2\alphachi T &\text{if}\quad y \phi < \alphachi T, \\
		y^4 \frac{\phi^2}{\alphachi T} &\text{if}\quad \alphachi T < y \phi < T,
	\end{cases}\label{eq:fermion scattering rate}
\end{gather}
- Loop-induced scattering processes with gluons $ g $:
\begin{gather}
	\Gamma_{\text{gluon scat.}} \approx \frac{b \alphachi^2T^3}{\phi^2}.
\end{gather}
These damping rates enter as a friction term in the equation of motion
\begin{gather}
	\ddot{\phi}+3H\dphi+\Gammatot\dphi+\meff^2\phi = \dtheta^2\phi,
\end{gather}
where the last term is the centrifugal term connecting the radial dynamics to the angular dynamics and $ \Gammatot $ is the sum of all damping rates. A summary of all relevant damping rates as well as references can be found in Appendix \ref{app:damping rates}.

\subsection{Boltzmann equations}
The energy lost through damping is transferred to the radiation bath. Since several of the damping rates depend on $ \phi $ and/or $ T $, the dynamics of this system can be non-trivial. The PQ fields may dominate the energy density and thus control $ H $. Therefore, in Appendix \ref{app:boltzmann equations}, we derive a set of Boltzmann equations to be solved numerically. The system of equations is
\begin{align}
	\dot{\rho}_{\rm kin}^\phi &= -(3H +\Gammatot + c_{\mphi}H) \rho_{\rm kin}^\phi,\label{eq:boltzmann1} \\
	\dot{\rho}_{\rm circ} &= -(3H + c_{\mphi}H) \rho_{\rm circ}, \\
	\dot{\rho}_r &= -4H\rho_r +\Gammatot\rho_{\rm kin}^\phi,\\
	\dot{\rho}_a &= - c_{\rho_a}H\rho_a, \label{eq:boltzmann4}
\end{align}
where $ \rho_{\rm kin}^\phi $ is the kinetic energy of the radial oscillations, $ \rho_{\rm circ} $ is the potential energy associated with the circular orbit about which the radial mode is oscillating and $ \rho_r $ is the energy density of the SM plasma. We model a possible temperature dependence of the radial mass with a parameter $ c_{\mphi} $ defined as $ \mphi \propto a^{-c_{\mphi}} $, such that $ c_{\mphi}=1 $ if thermal effects dominate the potential and $ c_{\mphi}=0 $ otherwise. Similarly, we parametrize the axion redshift with $ \rho_a\propto a^{-c_{\rho_a}} $ where $ c_{\rho_a}=3 $ when $ \phi \gg f_a $ and $ c_{\rho_a} = 6 $ when $ \phi \sim f_a $. The initial conditions for the system, as well as the $ \chi $-fermion relativistic/non-relativistic transitions, are also discussed in Appendix \ref{app:boltzmann equations}. 

\subsection{Consequences of thermal effects}\label{sec:Yukawa discussion} 

\paragraph{Thermal potential and the absence of low-$ m_a $ solutions:}
One interesting consequence of the thermal potential is that it precludes the low$-m_a $ class of solutions seen e.g. on the left-hand-side of Figs. \ref{fig:constantfaPlotn13Mmp} and \ref{fig:Nearly quadratic M18n13With}. 
To understand this, recall from Eq.~\eqref{eq:Ykick nearly quadratic} that $ \Ykick\propto\epsilon\times \meff^{-k} $, where $ k $ is a positive constant depending on $ n $. In the absence of thermal effects, the parameter $ \epsilon $ can be assumed to be of $ \mathcal{O}(1) $. This is a result of the balance between $ A\sim \mathcal{O}(1) m_{\phi,0} $ in the PQ-breaking angular potential and the radial potential.
Therefore, a thermal mass contribution will suppress $ \epsilon $. From Eq.~\eqref{eq:kick size} and Eq.~\eqref{eq:epsilon equivalence} we find
\begin{gather}
	\epsilon \approx \mathcal{O}(1)\times \frac{m_{\phi,0}}{m_{\phi,\rm eff}}. \label{eq:epsilon supression}	
\end{gather}
This suppression disfavors\footnote{Note that if $ \epsilon $ is suppressed then the radial mode may also experience parametric resonance which is not taken into account here. Since we favor solutions with $ \meff \sim \mphiz $, this does not have a major impact on our results.} kicks with $ m_{\phi,\rm eff}\gg \mphiz $, and introduces a lower bound on $ \mphiz $. 
Because a larger yield corresponds to a lower $ \mphi $, this competition sets a lower bound on $ m_a $ for which sufficient DM can be accounted for, see Eq.~\eqref{eq:mphi nearly quadratic, without injection} and Figs. \ref{fig:nearlyQuadraticEarlyDamping1} and \ref{fig:nearlyQuadraticEarlyDamping2} for the relation between $m_a$ and $\mphi$. As discussed in the next paragraph, this problem can be ameliorated by considering scenarios with only intermediate relativistic phases. Nevertheless, this precludes solutions with axion/$\phi$ masses below the thermal $ \phi $ relic bound, Eq.~\eqref{eq:thermal condition}.

\paragraph{Intermediate relativistic phases:} 
 In solutions with an intermediate relativistic phase, fermions are initially non-relativistic because of a large $ \phi $-induced mass and turn relativistic once the $ \phi $-induced mass falls below $ T $. This phenomenon allows efficient damping through scattering with relativistic fermions without the problems associated with a large thermal mass at the time of the kick. 

In particular, if $ y \Skick > \Tkick $ then fermions are non-relativistic and the lower limit on the $ \mphiz $ is relaxed to $ \mphiz> \mln $.
Because $ \phi\propto a^{-3/2} $ if $ \phi $ is not dominated by the thermal potential, it is possible for the fermions to enter an intermediate relativistic phase with efficient damping.
We find that this method of evading $ \epsilon $-suppression leads to solutions with DM for $ m_a \gtrsim 10^{-6} $ eV. An explicit example of this evolution can be found among the example solutions in the Appendix \ref{app:boltzmann equations}, see Fig.~\ref{fig:niceBoltzmannExample}. 

Solutions where a thermal contribution is evaded by an early non-relativistic phase are only consistent for some values of the reheating temperature after inflation, $ T_{\rm reheat} $. In particular, some choices of $ y $ impose upper bounds on $ T_{\rm reheat} $ below which a kick in a non-relativistic phase is inconsistent. See Appendix \ref{app:TReheat} for a discussion of such potential conflicts. We generally allow $ T_{\rm reheat} $ to take any value between the inflation scale and $ \Tkick $, except for regimes in which a too-large $ T_{\rm reheat} $ would contradict a kick with non-relativistic fermions. For simplicity we always assume $ T_{\rm reheat} > \Tkick $. Our results have no further dependence on $ T_{\rm reheat} $.

\paragraph{Constraints from $ \chi $ fermions:}
 Since the $ \chi $-fermions are efficiently interacting with the SM and cannot be SM fermions,  to ensure compatibility with collider data, we must impose that
\begin{gather}
	\mchitoday\approx y f_a \gtrsim \text{1 TeV}. \label{eq:collider constraint}
\end{gather}
 Thermal relics from $\chi$ freeze-out are not problematic because they can always be made to decay before BBN, see e.g. \cite{Ibe:2021ctf}.

\subsection{Parameter space}\label{sec:Visualization of constraints}
Even when we fix the $
\chi$-SM interaction to be SU(3)$_{\rm QCD}$, the model we are considering has six parameters: $ m_a, f_a,y,M,n, $ and $ \mphiz $. This makes the visualization of the viable parameter space challenging. For simplicity, we only consider $ M=\mplanck $ and only show results for $ n=13 $, although alternative choices of $ n $ can be found in Appendix \ref{app:parameter values}. The dimensionality is further reduced to 3 by enforcing that the axion relic matches the observed DM density. This remaining free parameter means that for any given choice of $ [m_a,f_a] $ the correct DM relic can be realized by a family of solutions with varying $ \mphiz $ and $ y $. We display the viable  parameter in the two following ways: 
\begin{enumerate}[label=\alph*)]
	\item To visualize the impact of the most important constraints, we show the parameter space that  becomes viable if a given constraint is \textit{relaxed}. Any such region is indicated by colored hatching. Parameter space that cannot be made viable by the relaxation of any single condition is shown in solid color.
	\item To visualize parameter values, we plot contours of the highest and/or lowest possible values at any given point $ [m_a,f_a] $.
\end{enumerate}
In particular, $ a) $ implies that if two constraints overlap in our visualization, then that point of $ [m_a,f_a] $ space becomes viable if \textit{either} constraint is relaxed. This is quite different from usual visualizations of two-dimensional parameter spaces and is a consequence of our three-dimensional parameter space being projected down onto the two-dimensional $ [m_a,f_a] $ plane. 
 
By numerically solving the Boltzmann equations~\eqref{eq:boltzmann1}-\eqref{eq:boltzmann4} we can sample across the parameter space $ [f_a,y,M,n, \mphiz] $ and  identify solutions that correspond to the correct DM relic for a given $ m_a $. Solutions with and without yield dilution can both be found with all solutions belonging to the large-mass window of the thermal $\phi$ relic constraint, Eq.~\eqref{eq:thermal condition}. A summary of regions of $ [m_a,f_a] $ space that are viable given $ n=7,8,10,13$ is given in Fig.~\ref{fig:summaryPlotNQYukawaAndHiggs}.
Among these choices, $ n=13 $ allows the largest parameter space, and we therefore focus on this case. Interestingly, there are regions of parameter space in which the radial mass can be large enough to saturate the perturbativity condition $ \mphiz^2 \leq 4 \pi f_a^2 $. The full range of parameter values  are detailed in Appendix \ref{app:parameter values}.
We highlight the following:

\paragraph{Dominant damping rates:} We find that the process most commonly responsible for damping is scattering with relativistic $ \chi $-fermions as given by Eq.~\eqref{eq:fermion scattering rate}. Damping tends to take place at temperatures near either the end or the beginning of the phase of relativistic $ \chi $-fermions. Damping near the beginning of a relativistic phase is common because of the lifting of the $ \chi $-fermion Boltzmann suppression rapidly increases the damping rate. Damping near the end of a relativistic phase is common because the $ \Gamma_{\chi-\text{scat.}}\propto T^{-1} $ behavior rapidly out-competes $ H\propto T^2 $ in the short regime, $ \alphachi T < y\phi <T $, in which the damping rate is relevant. Damping near the end of relativistic phase can be problematic if the damping does not complete before Boltzmann suppression closes the process down. 
See Appendix \ref{app:boltzmann equations} for explicit examples of these scenarios.

Aside from $ \chi $-scattering, $\phi$-to-axion decay can also play an important role in the damping of radial oscillations. As discussed in Section \ref{sec:overview - phi-to-axion decay}, this mechanism is constrained by the danger of producing a hot axion relic in excess of constraints on dark radiation. This limits $\phi$-to-axion decay as a stand-alone damping mechanism.
Interestingly, we find that when $\phi$-to-axion decay is considered in conjunction with $ \chi $-scattering then the decay can account for the remaining energy density which is left behind after $ \chi $-scattering has damped most, but not all, of energy out of radial oscillations. The synergy in such two-step damping lies in the fact that after $ \chi $-scattering has reduced the energy in radial oscillations to a fraction of $ \rho_r $ $\phi$-to-axion decay can take place without the resulting hot axion being a significant part of the overall energy density.
An explicit example of such two-step damping can be found in Appendix \ref{app:boltzmann equations}. 

\begin{figure}
	\centering
	\includegraphics[width=0.8 \textwidth]{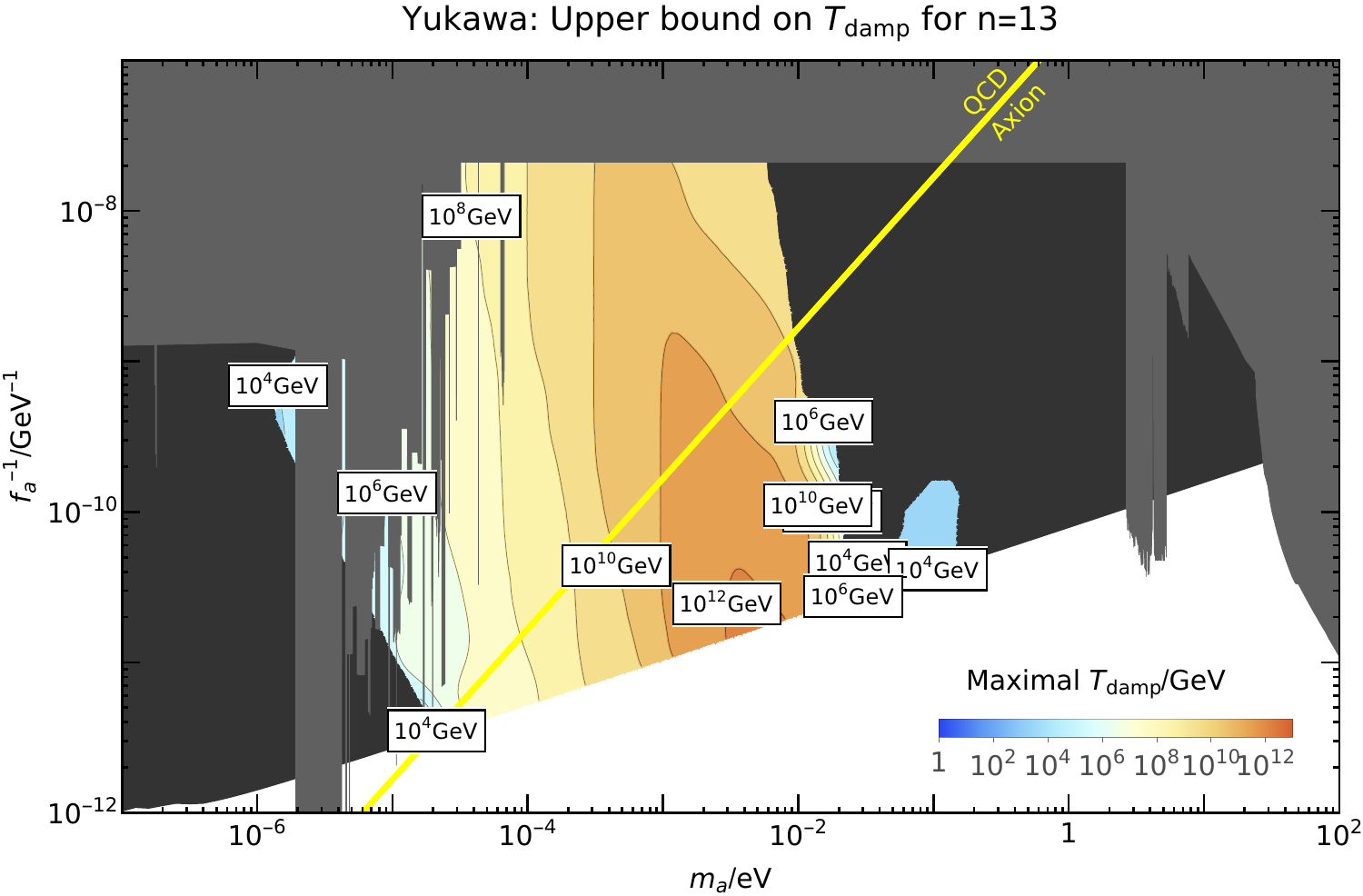}
	
	\vspace{0.5cm}
	\includegraphics[width=0.8 \textwidth]{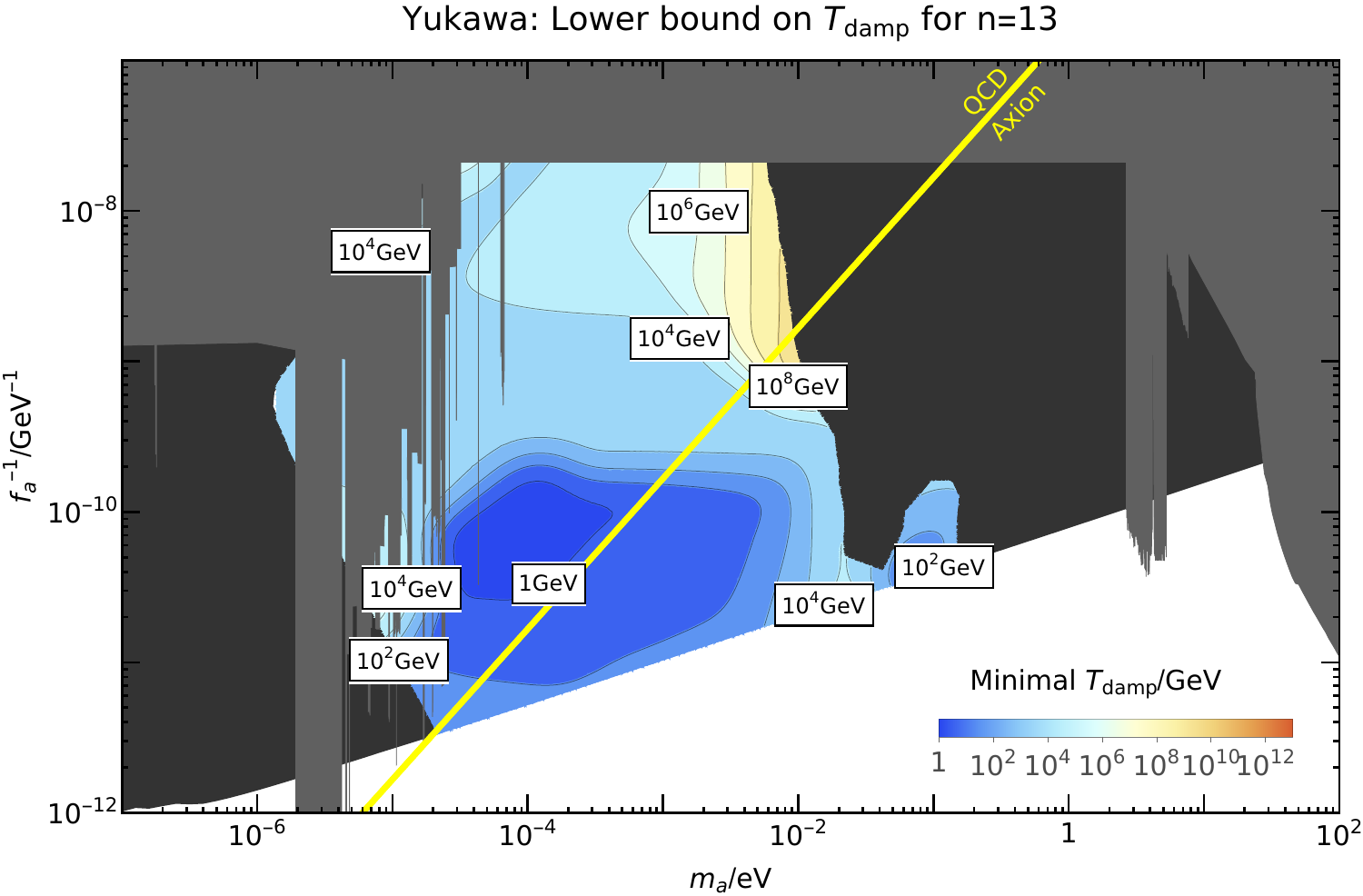}
	
	\caption{\small\it Realized range of damping temperatures. Here the homogeneity condition \eqref{eq:cems condition} has been relaxed by assuming that $ \mathcal{P}_{\mathcal{R}}(\kmin) $ is sufficiently suppressed. \textbf{Top:} Contours of the highest possible thermalization temperature. \textbf{Bottom:} Contours of the lowest possible thermalization temperature}
	\label{fig:TdampContoursMmpn13}
\end{figure}

\paragraph{Realized range of damping temperatures:}

For scenarios in which $ \chi $-scattering is the sole relevant damping rate, we find  scenarios for $ n=13 $  with damping temperatures which generally lie in the range $ \Tdamp \sim 10^4 $ GeV to $ 10^{12} $ GeV.
When $\phi$-to-axion decay is also considered, then scenarios can be found in which some energy remains in radial oscillations down to $ \Tdamp\gtrsim $ GeV. 
This may appear to be in conflict with the results seen in Section \ref{sec:nearly quadratic late damping}. However, because damping through $\phi$-to-axion decay does not establish thermal contact with the plasma, damping through this channel avoids thermal relic constraints given by Eq.~\eqref{eq:thermal condition}. Therefore, lower damping temperatures than those suggested by e.g. Fig.~\ref{fig:Nearly quadratic M18n13With} can be achieved without conflict.

Furthermore, the definition of $ \Tdamp $ is made more complex by the possibility of two-step damping. For the analytic approach of Section \ref{sec:nearly quadratic late damping} we considered an instantaneous damping in which $ \Tdamp $ is unambiguously defined.
For the numerical solutions, damping generally does not take place at any single time, so we must be more careful with the definition of $ \Tdamp $.
We define $ \Tdamp $ as the temperature at which the energy in radial oscillations is damped below all other energy densities, i.e. we demand $ \rho_{\rm kin}^\phi < \rho_r,\rho_a $ for all $ T<\Tdamp $ in addition to the usual condition $ \Gammatot > H $. Even with this refinement, the possibility of two-step damping implies that $ \Tdamp $ is not necessarily anywhere near the temperature at which the majority of energy is transferred from the PQ field to the plasma. Crucially, the bulk of the energy in two-step solutions is transferred by $ \chi $-scattering much earlier than $ \Tdamp $. In such scenarios, the behavior is quite different from the simple scenario of instantaneous damping and the analysis of Section \ref{sec:nearly quadratic late damping} should not be expected to apply.

With the knowledge that the very different behavior of $ \Tdamp $ in two-step scenarios in some circumstances circumvent some lower bounds discussed in Section \ref{sec:nearly quadratic late damping}, our numerical solution otherwise realizes a subset of the temperature range derived in that section.
This range is visualized in Fig.~\ref{fig:TdampContoursMmpn13}, which can be compared to the general case visualized in Fig.~\ref{fig:Nearly quadratic M18n13With}.
Again, note that each point in the [$ m_a, f_a] $ plane may be covered by a range of solutions with different combinations of $ \mphi $ and $ y $, such that each point in our [$ m_a, f_a $] parameter space is associated with a range of possible temperatures.  However, note that the color gradient between the two plots cannot be compared directly because the range has been chosen to best fit each plot individually.

The low damping temperatures permitted around $ m_a \sim 10^{-4}$ eV and $ m_a\sim 10^{-1} $ eV correspond to two-step damping in which the damping is finalized by $\phi$-to-axion decay without violation of the thermal $ \phi $ relic constraints that enforced a higher lower bound in Fig.~\ref{fig:Nearly quadratic M18n13With}.

\paragraph{Radial mode mass range:}

\begin{figure}
	\centering
	\includegraphics[width=\standardwidth \textwidth]{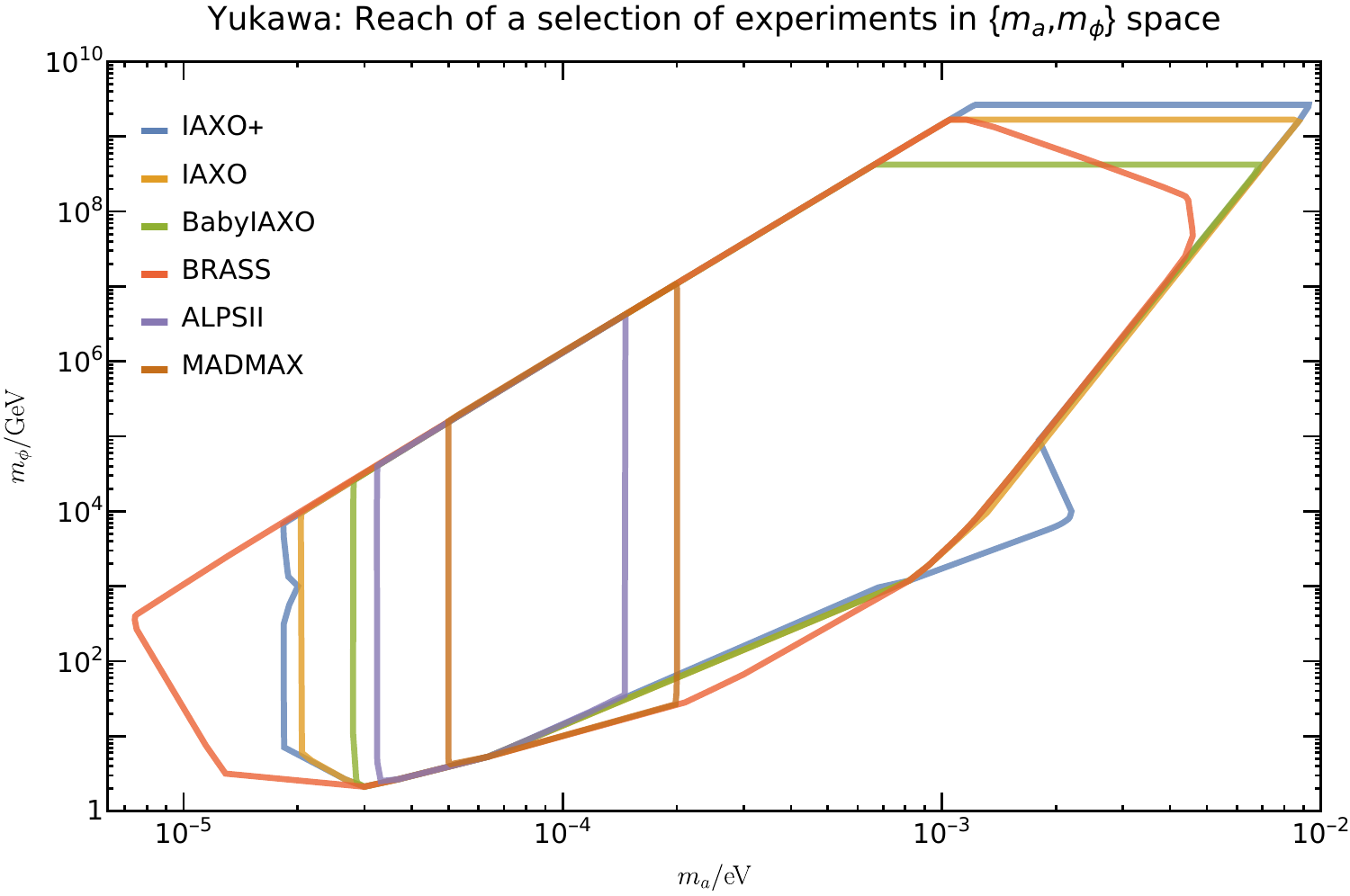}
	
	\caption{\small\it Range of $[m_a,\mphi]$ parameter space  associated with a selection of experiments. The regions inside the contours feature axion DM solutions that are within the projected reach of the indicated experiments. Given an axion discovery by a given experiment, this figure gives the prediction for the associated radial mass range. We assumed $n=13$.}
	\label{fig:HamburgExperimentReachPlotYukawa}
\end{figure}

We stress the relation between the axion mass and the radial mode-mass. For every single point in the  $(m_a, f_a)$ plane, we can predict the range of the radial mode mass. In some cases, $\mphi$ can be very light, which should be kept in mind for experimental searches. We show in Fig.~\ref{fig:HamburgExperimentReachPlotYukawa} the specific reach for a selection of experiments. Assuming a given axion  is detected in one of the axion experiments, the range of masses for the radial-mode partner can be inferred and potentially searched for in a different type of experiment.

\paragraph{Necessity of $ \mathbf{\mathcal{P}_{\mathcal{R}}(\kmin)} $ suppression:}
All solutions found involve an extended duration of $ a^{-6} $ evolution of the energy density of the angular mode. This leads to significant constraints from the homogeneity condition Eq.~\eqref{eq:cems condition}, which applies if the spectrum of density perturbations from inflation $ \mathcal{P}_{\mathcal{R}} $ is assumed to be a flat spectrum. 
If the bound on $ \mathcal{P}_{\mathcal{R}} $ from CMB measurements, i.e. $ A_s\sim 2.1 \times 10^{-9} $~\cite{Planck:2018vyg}, is assumed to apply on all scales then Yukawa damping is not viable.
However, the relevant scale $ \kmin^{-1} $ is many orders of magnitude smaller ($ \kmin^{-1} \ll 10^{-10} $ Mpc) than those constrained by CMB measurements ($ k\sim 0.05  $ Mpc$ ^{-1} $). In principle, $ \mathcal{P}_{\mathcal{R}}(\kmin) $ is allowed to differ by several orders of magnitude from the value suggested by CMB measurements without tension.

\begin{figure}
	\centering
	\vspace{-1cm}
	\includegraphics[width=0.8 \textwidth]{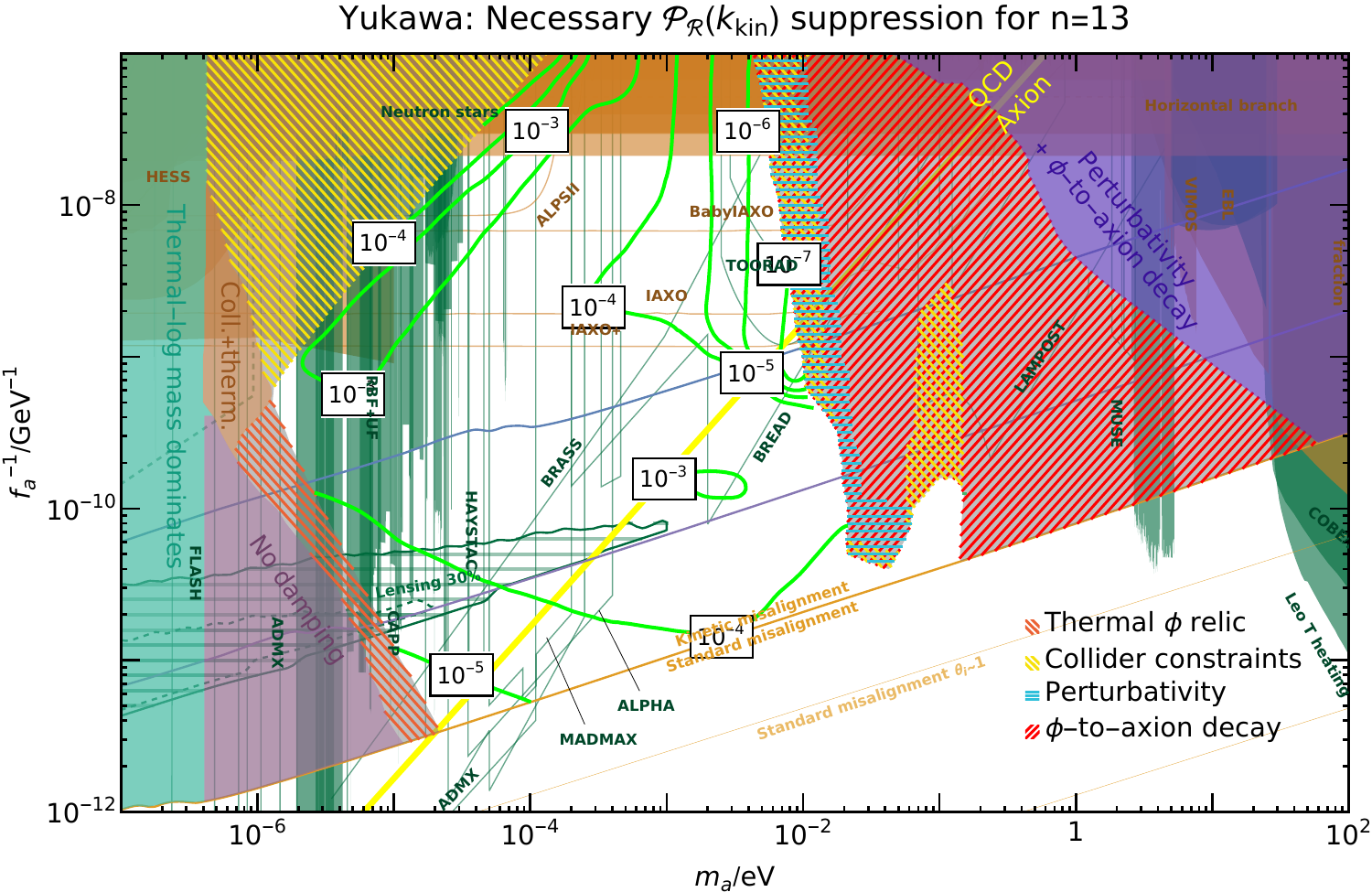}
	
	\caption{\small\it Green contours: Necessary suppression of $ \mathcal{P}_{\mathcal{R}}(\kmin) $ relative to $ A_s $, where $ \mathcal{P}_{\mathcal{R}} $ is the amplitude of the primordial power spectrum, see Eq.~\eqref{eq:cems condition}. 
    The dominant constraints are from the thermal $ \phi $ relic \eqref{eq:thermal condition}, collider bounds \eqref{eq:collider constraint}, and hot axion relics produced by $\phi$-to-axion decay~\eqref{eq:hot axions dominate the plasma}. As each point in $ [m_a,f_a] $ space may be supported by a range of solutions, we illustrate the constraints by highlighting regions that would become viable if a given constraint is lifted.}
	\label{fig:maxFlucPlotMmpn13}

	\vspace{0.5cm}
	
    \includegraphics[width=0.8 \textwidth]{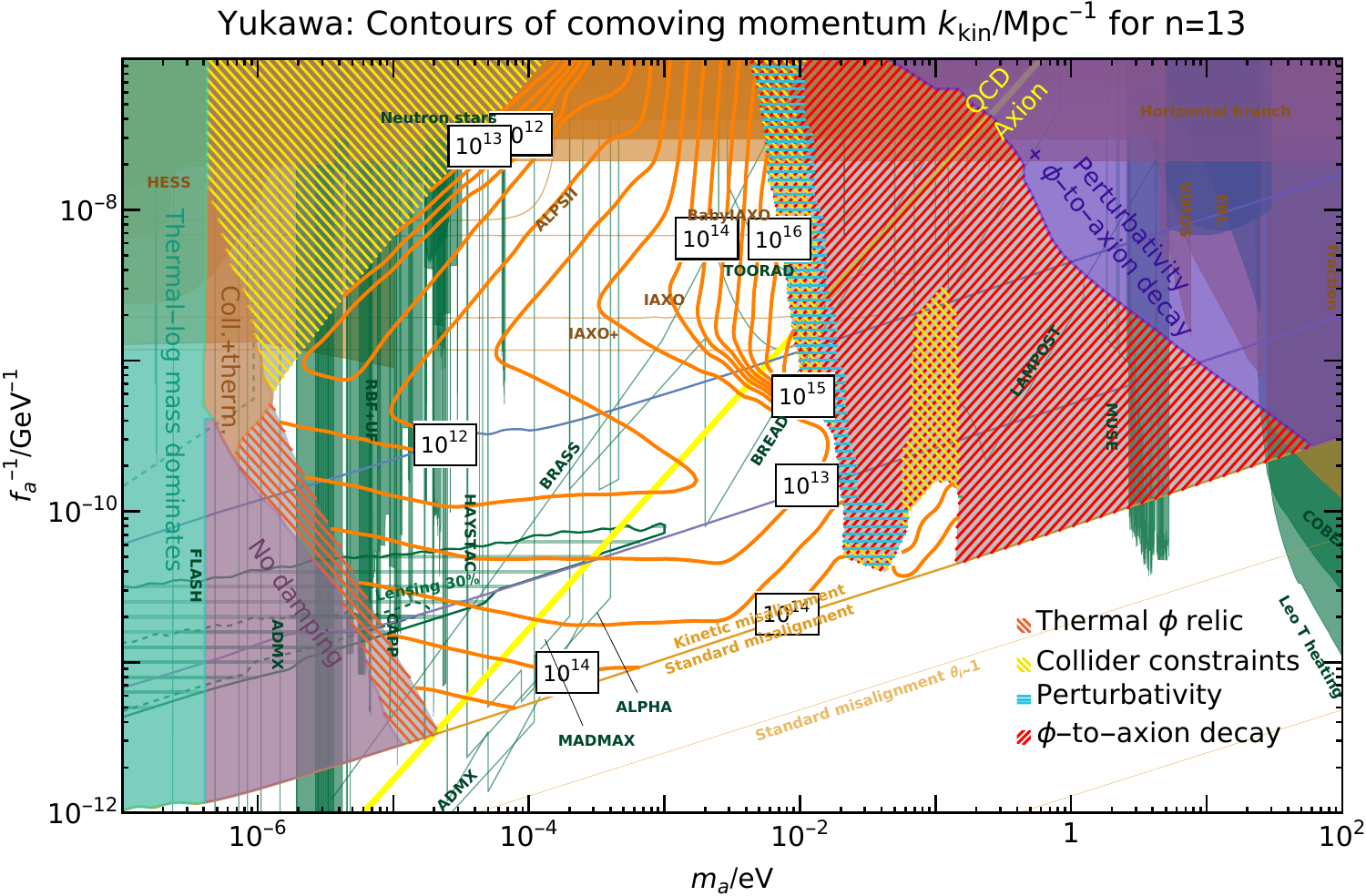}
    \caption{\small\it Orange contours:  Comoving momentum scale $ \kmin$ at which $ \mathcal{P}_{\mathcal{R}}(\kmin) $ should be suppressed to satisfy the homogeneity constraint. The solutions shown here correspond to those shown in Fig.~\ref{fig:maxFlucPlotMmpn13}. This clearly 
    indicates that $ \kmin \gg 0.05 $ Mpc$ ^{-1} $, which is the scale probed by Planck.}
    \label{fig:kminPlotMmpn13}
\end{figure}

If $ \mathcal{P}_{\mathcal{R}}(\kmin) $ is assumed to be sufficiently smaller than $ \mathcal{P}_{\mathcal{R}}\sim 2.1 \times 10^{-9} $, then a region around $ m_a \sim 10^{-6} $ eV to $ 10^{-2} $ eV becomes viable for a benchmark value of $ n=13 $. We display this region in Fig.~\ref{fig:maxFlucPlotMmpn13} along with the necessary $ \mathcal{P}_{\mathcal{R}}(\kmin) $ suppression and the surrounding constraints. We display the associated values of $ \kmin $ in Fig.~\ref{fig:kminPlotMmpn13}. Assuming $ \mathcal{P}_{\mathcal{R}}(\kmin) \ll 2.1 \times 10^{-9} $ corresponds to relaxation of the homogeneity constraint seen e.g. in Fig.~\ref{fig:constantfaPlotn13Mmp}.

\subsection{Kination era}
\label{subseckination}
\begin{figure}
    \centering
\includegraphics[width=0.8 \textwidth]{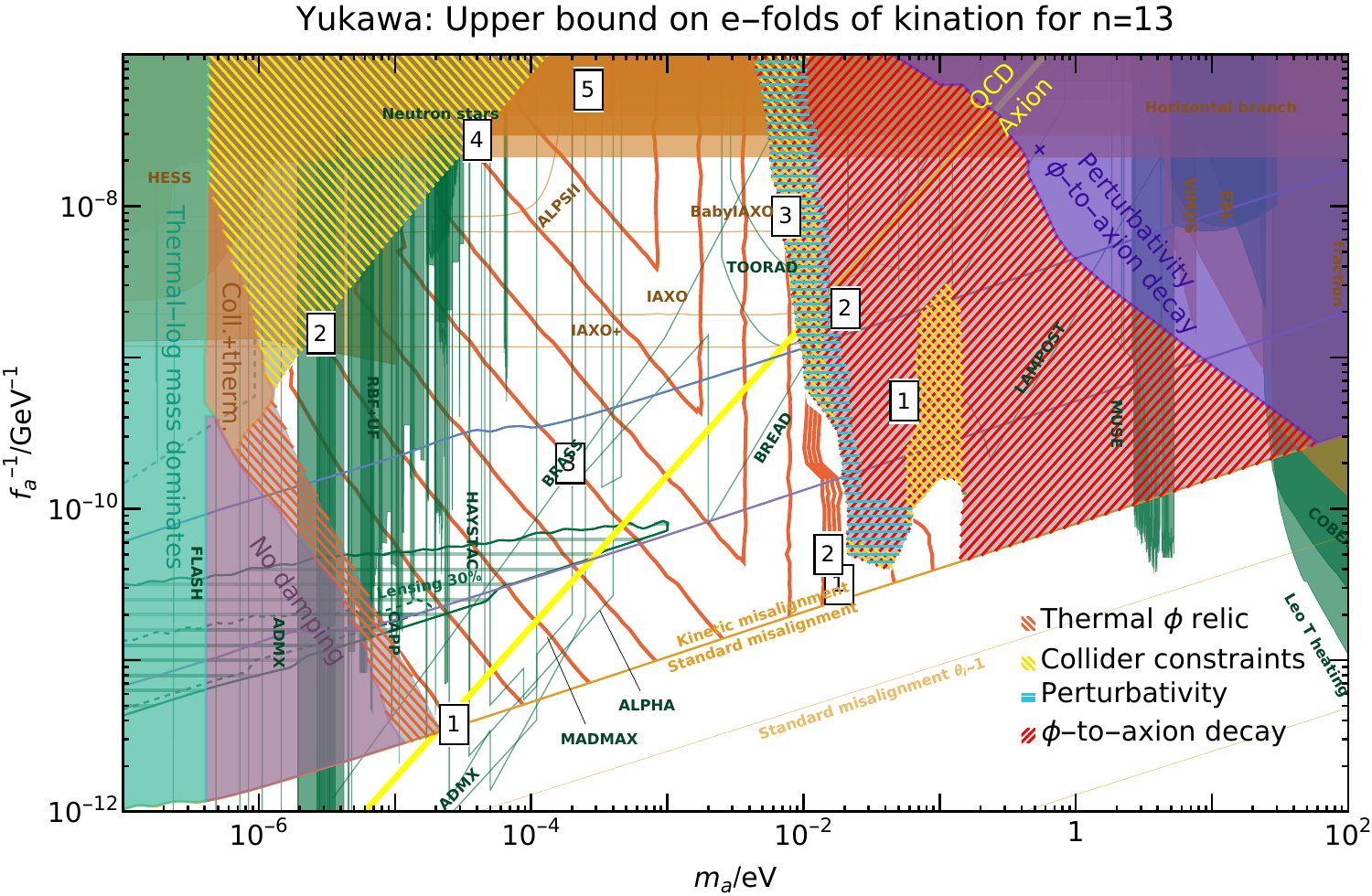}
	
	\caption{\small\it Map of the largest number of e-folds of kination which can be realized if it is assumed that the curvature fluctuations can be sufficiently suppressed. Each point in the white [$ m_a,f_a $] parameter space, in general, has a range of axion DM solutions with varying radial mode mass $ m_\phi $ and the Yukawa coupling $ y $ specified by Eq.~\eqref{eq:yukawa coupling}. The contours given here indicate the largest possible number of e-folds at each point.}
	\label{fig:EFoldsContoursMmpn13}
	
\end{figure}

To conclude this section, we discuss the potential for an era of kination and associated gravitational waves (GWs). 
This scenario involves such a period of kination, i.e., the total energy density starts to scale $a^{-6}$,
if the kinetic energy of the angular mode dominates when $\phi$ reaches the minimum at $f_a$.
In Fig.~\ref{fig:EFoldsContoursMmpn13} we show the duration of kination that can arise in this setup. 

A period of kination can amplify GWs from inflation. This effect was studied extensively in \cite{Gouttenoire:2021jhk} where it was found that a period of kination leads to GWs with a peak amplitude of
\begin{gather}
	\begin{gathered}
		\Omega_{\rm GW-peak}(f_{\rm GW-peak})h^2 \approx 2.8 \times 10^{-13}\\ \times\left(\frac{g_*(\Tkend)}{106.25}\right)\left(\frac{g_{*s}(\Tkend)}{106.25}\right)^{-4/3} \left(\frac{E_I}{10^{16}\text{ GeV}}\right)^4 \left(\frac{e^{2\Nkin}}{e^{10}}\right),\label{eq:GW peak}
	\end{gathered}
\end{gather}
and a peak frequency of
\begin{gather}
	f_{\rm GW-peak} \approx 1.1\times 10^{-3} \text{ Hz } \left(\frac{g_*(\Tkend)}{3.37}\right) \left(\frac{g_{*s}(\Tkend)}{3.91}\right)^{-4/3} \left(\frac{E_{\rm kin}}{10 \text{ TeV}}\right)\left(\frac{e^{\Nkin /2}}{10}\right).
\end{gather} 
Here $ \Tkend $ is the temperature at the end of kination domination, $ \Nkin $ is the number of e-folds of inflation and $ E_I $ is the energy scale of inflation which is constrained to $ E_I < 1.6 \times 10^{16} $ GeV by Planck \cite{Planck:2018vyg}. 
The peak frequency is controlled by the energy scale of kination,
\begin{gather}
	E_{\rm kin}=\rho_\theta(\Tkin)=\frac{1}{2}\mphi^2f_a^2.
\end{gather}
Therefore, the frequency of the GW peak depends on the radial mass $ \mphi $. 

The suppression of $ \mathcal{P}_{\mathcal{R}}(\kmin) $, which is required to avoid the homogeneity constraints and which was discussed towards the end of the previous subsection, will nevertheless also typically suppress the stochastic GW background from inflation, as
it is usually assumed that the tensor-to-scalar ratio be cannot be varied much (but  this may be challenged by proper inflation model building, see discussion in Sec.~\ref{sec:power}).
We therefore do not show in this paper the associated GW predictions.  Note however that there is another source of 
 GW at very low frequencies from the axion fragmentation itself~\cite{Eroncel:2022vjg}.

\section{Thermal damping via Higgs portal in a nearly-quadratic potential}\label{sec:NQHiggs}
We now work out an alternative damping mechanism relying on a Higgs interaction of the form
\begin{gather}
	\mathcal{L}\supset \xi^2 (\phi^2-f_a^2) \left(H^\dagger H- \frac{v^2}{2}\right), \label{eq:phi-higgs coupling}
\end{gather}
where $ H $ is the SM Higgs doublet with  a VEV $ \expval{H^\dagger H} = v_{\rm EW}^2/2 $ where $ v_{\rm EW} \approx 246 $ GeV. Such an interaction was also considered for the quartic model by  Co et al.~\cite{Co:2020dya,Co:2020jtv} and the thermal effects were  studied by Mukaida et al.~\cite{Mukaida:2012qn,Mukaida:2012bz,Mukaida:2013xxa}. We refer in particular to~\cite{Mukaida:2013xxa}.

The interaction term is tuned to cancel a contribution to the Higgs mass arising from the late-time VEV of $ \expval{\phi}\sim f_a $, so as to avoid generating a Higgs mass larger than the observed Higgs mass $ \xi f_a \gg \mHiggs $. This is the usual Higgs hierarchy problem which we do not discuss further here. Assuming the cancellation, the effective Higgs mass is
\newcommand{\mHiggseff}{m_{\rm H,eff}}
\begin{gather}
\mHiggseff^2 \approx \mHiggs^2+\frac{1}{6}y_t^2 T^2 + \xi^2(\phi^2-f_a^2), \label{eq:Higgs effective mass}
\end{gather}
where $ y_t \sim \mathcal{O}(1) $ is the Higgs-top Yukawa coupling, which we assume to be the dominant source of the Higgs thermal potential.

\subsection{Thermal effects and damping rates}
As was the case with the Yukawa interaction, the Higgs portal coupling Eq.~\eqref{eq:phi-higgs coupling} gives rise to a thermal potential for $ \phi $. If $T^2 > \mHiggs^2 + \xi^2 (\phi^2-f_a^2)$, the SM Higgs boson in the thermal plasma gives thermal corrections $\approx \xi^2 T^2$ to $m_\phi^2$. On the other hand, if $T^2 <  \mHiggs^2 + \xi^2 (\phi^2-f_a^2)$, although the SM Higgs boson is absent in the thermal bath, it affects the effective gauge coupling via a logarithmic running effect and the potential energy obtains a thermal correction $\delta V \approx \alphaweak^2 T^4 \log [ (\mHiggs^2 + \xi^2 (\phi^2-f_a^2)) / T^2]$ \cite{Anisimov:2000wx, Mukaida:2012qn}. The effective mass of $ \phi  $ is then
\begin{gather}
\begin{gathered}
	\meff^2 \approx \begin{cases}
		m_{\phi,0}^2 + \mth^2 &\text{if}\quad T > \xi \phi, \\
		m_{\phi,0}^2 + \mln^2 &\text{if}\quad T<\xi \phi,
	\end{cases} \\
	\text{where}\quad \mth^2 \approx \xi^2 T^2 \qq{and} \mln\approx \alphaweak^2 \frac{\xi^2 T^4}{\xi^2\phi^2+\mHiggs^2}\label{eq:Higgs thermal log},
\end{gathered}
\end{gather}
where $ \alphaweak \sim 1/30 $\footnote{A contribution also arises from the top quark Yukawa coupling. 
For a simple order-of-magnitude estimate, we consider only the contribution from the weak gauge coupling here.
}
and the thermal-log contribution is suppressed by the zero-temperature Higgs mass $ \mHiggs $. 
In the absence of this suppression, much of the otherwise viable parameter space would feature $ \mln $ domination and Q-ball formation. 

The Higgs portal coupling introduces a number of dissipation channels which are summarized in Appendix \ref{app:damping rates}. The  most relevant interaction for damping is the counterpart of vacuum decay via the effective three-point interaction $ (\xi^2\expval{\phi}H^\dagger H) $, which in the thermal environment takes the form~\cite{Mukaida:2013xxa},
\begin{gather}
		 \Gamma_\phi\approx 2 \frac{\xi^4\phi^2}{\pi^2\alphaweak T} \qq{for} 
			 \mphi < \alphaweak T \text{ and } 
			 \mHiggs< T \text{ and }\xi \phi < T .
         \label{eq:most important Higgs interaction}
\end{gather}
This interaction can be an effective source of friction in the regime where $ H $ is relativistic.

In addition, the Higgs portal coupling also leads to interactions induced by $ \phi $-$ H $ mixing  which can be important at low temperatures~\cite{Co:2020dya,Co:2020jtv}. By diagonalizing the $ \phi $-Higgs mass mixing matrix, which takes the form
\begin{gather}
	M_{\rm mix} = \begin{pmatrix}
		m_\phi^2	&	2\xi^2 f_a v_{\rm EW}	\\
		2\xi^2 f_a v_{\rm EW}	& m_h^2			\\
	\end{pmatrix},
\end{gather}
and expanding in the $ v_{\rm EW}/f_a \ll 1 $ and $ |\vartheta_{\phi H}| \ll 1 $ limits, the coupling induces a mixing angle 
\begin{gather}
	\vartheta_{\phi H}\approx-2\xi^2\frac{f_a v_{\rm EW}}{m_h^2-m_\phi^2}.\label{eq:mixing angle}
\end{gather}
which leads to a decay of the form
\begin{gather}
\Gamma_{\rm mix-decay}\approx \vartheta_{\phi H}^2\times  \Gamma_{\rm SM-h-decay}(m_\phi), \label{eq:lepton decay}
\end{gather}
where $\Gamma_{\rm SM-h-decay}$ is the standard model decay rate of the Higgs. Depending on kinematics, controlled by $m_\phi$, this can take place either though leptons, quarks/mesons or gauge bosons. The full set of decay channels we take into account is listed in Appendix \ref{app:damping rates}.

\subsection{Discussion}
For thermal damping though Eq.~\eqref{eq:phi-higgs coupling}, many of the same considerations as in damping through a Yukawa interaction apply. In particular, thermal effects compete with the zero-temperature mass in a way that rules out scenarios with radial mode masses below the thermal $ \phi $ relic bound. Furthermore, scenarios with kination are possible and because the parameter space is still 6-dimensional we employ the same visualization approach as introduced in the Yukawa scenario, see Section \ref{sec:Yukawa discussion}.
The main differences with the Yukawa case include limits arising from radial mode-Higgs mixing and the Higgs mass contribution.
As SM Higgs interactions are known to the percent-level~\cite{CMS:2022dwd}, 
we implement the upper bound on the radial mode-Higgs mixing
\begin{gather}
    \vartheta_{\phi H}<0.1 
\end{gather}
A consequence of the cancellation in Eq.~\eqref{eq:phi-higgs coupling}, which is necessary to prevent the contribution $ \delta m_{\rm H}^2\sim \xi^2 \phi^2$ from the Higgs portal coupling, is a rapid decrease in the $ \phi $ contribution to the Higgs mass as $ \phi $ approaches $ f_a $. This rapid change is accounted for in our numerical analysis. It can trigger efficient damping by removing Boltzmann suppression and permitting scattering processes.
If no such cancellation is assumed, then the observed Higgs mass imposes a strict upper bound of
\begin{gather}
	\xi \ll \frac{\mHiggs}{f_a}.
\end{gather}
While some parameter space remains viable even when asuming this bound, it implies fine-tuned very low coupling constants\footnote{Some authors have proposed that DFSZ-like axion models could feature naturally small Higgs-portal couplings as a result of an enhanced Poincaré symmetry in the $ \xi \to 0 $ limit. For a discussion of this approach, see \cite{Foot:2013hna}.}.

\subsection{Numerical Results}
\begin{figure}
	\centering
	\includegraphics[width=0.8 \textwidth]{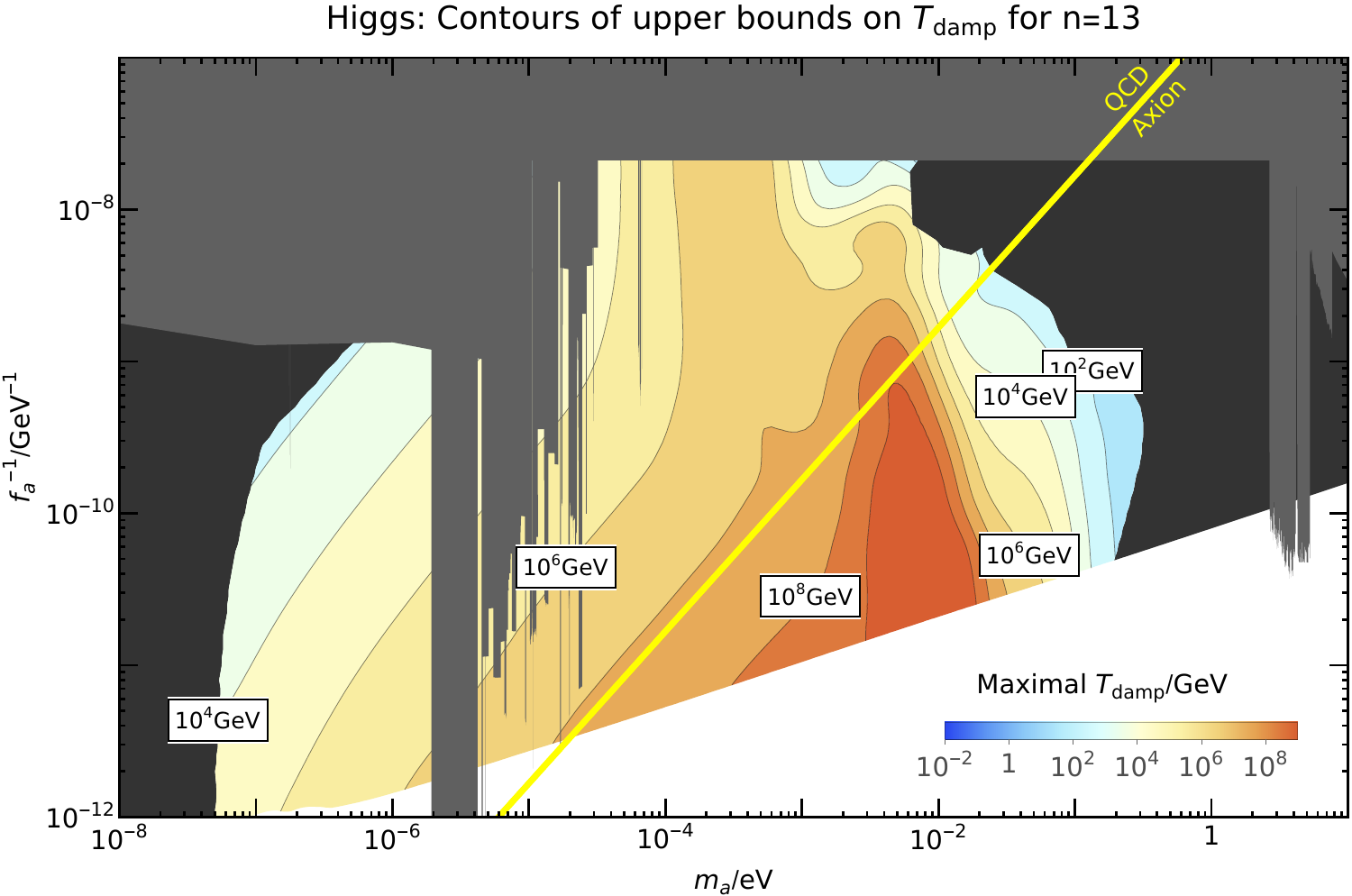}
	
	\vspace{0.5cm}
	\includegraphics[width=0.8 \textwidth]{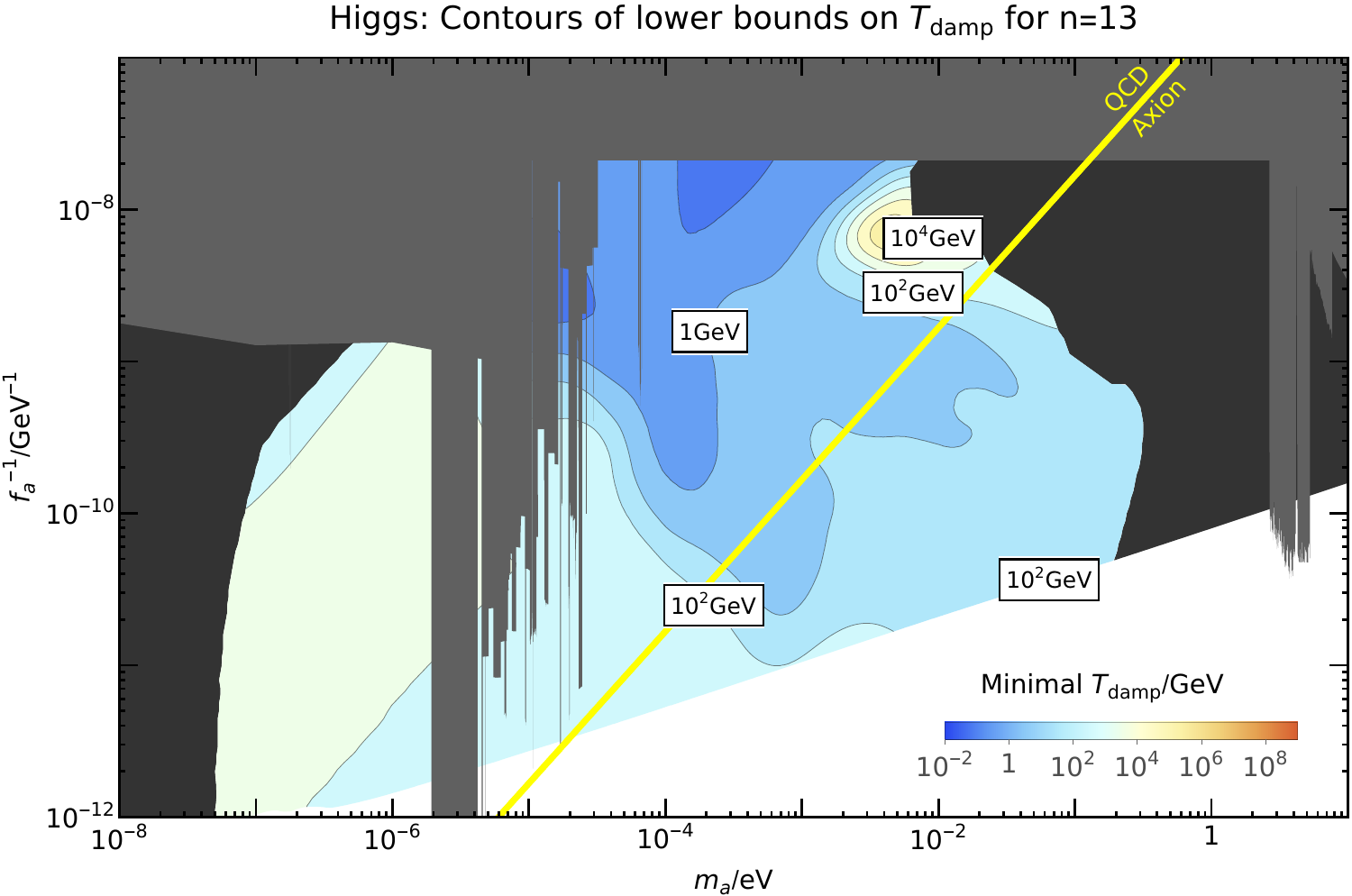}
	
	\caption{\small\it Realized range of damping temperatures similar to Fig.~\ref{fig:TdampContoursMmpn13} but for Higgs portal damping with $ n=13 $. \textbf{Top:} Contours of the highest possible damping temperature. \textbf{Bottom:} Contours of the lowest possible damping temperature}
	\label{fig:TdampContoursMmpn13NQHiggs}
\end{figure}

\begin{figure}
	\centering
	\includegraphics[width=\standardwidth \textwidth]{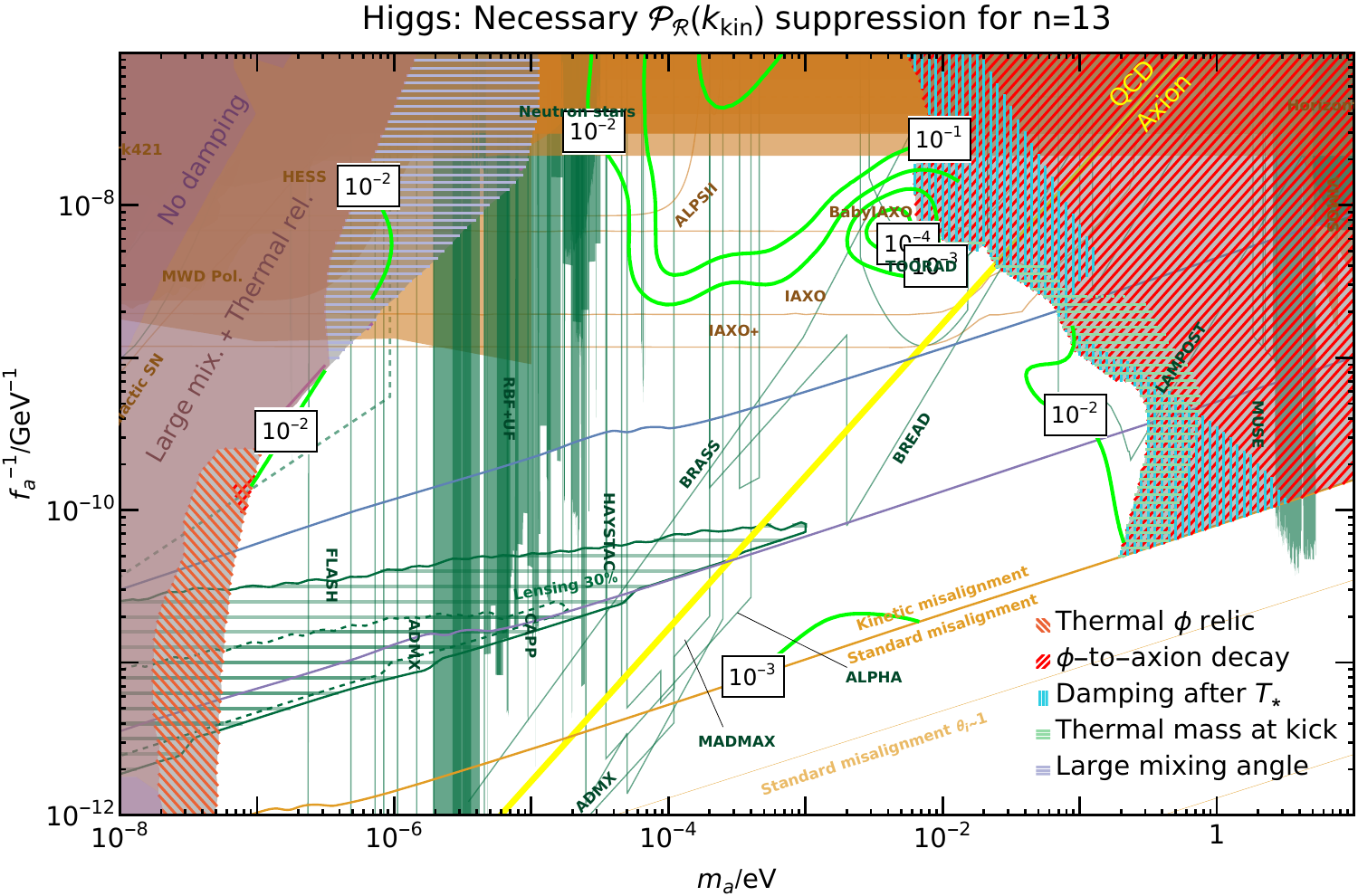}
	\caption{\small\it Green contours: Necessary suppression of $ \mathcal{P}_{\mathcal{R}}(\kmin) $ relative to $ A_s $. The plot is similar to Fig.~\ref{fig:maxFlucPlotMmpn13} except that damping is through the Higgs portal with $ n=13 $.}
	\label{fig:maxFlucPlotMmpn13NQHiggs}
\end{figure}

Applying the same numerical integration of Boltzmann equations as described in Section \ref{sec:Nearly quadratic Yukawa section}, we map out the ALP and QCD axion parameter space for which the axion DM can be accounted for by KMM in models with nearly-quadratic potentials and radial oscillations damped though the Higgs portal coupling (Eq.~\eqref{eq:phi-higgs coupling}). 
We have scanned the parameter space for the benchmark choices $ n= 7$, 10 and 13. Again, we find that $ n=13 $ offers the largest parameter space, with ALP masses in the range $ 5\times 10^{-8} $ eV $ \lesssim m_a  \lesssim  0.3$ eV including QCD axion solution with up to $ m_a \lesssim 5 \times 10^{-3} $. The viable range of damping temperatures is summarized in Fig.~\ref{fig:TdampContoursMmpn13NQHiggs} and the relevant constraints are visualized in Fig.~\ref{fig:maxFlucPlotMmpn13NQHiggs}. In most of the viable parameter space the mixing angle, Eq.~\eqref{eq:mixing angle}, can be made much smaller than the $\vartheta_{\rm mix} \lesssim 0.1 $ limit, see Fig.~\ref{fig:minMixAnglePlotNQHiggsMmpn13}.  We provide a full map of the required model parameters in Appendix \ref{app:parameter values}. We highlight the following:

\paragraph{Relaxation of the thermal relic constraint:}
The Higgs portal coupling leads to damping rates that are efficient even in the low-temperature regime, i.e. mixing-induced decay of the form of Eq.~\eqref{eq:lepton decay}.
Similarly to Co et al. in \cite{Co:2020dya,Co:2020jtv}, we  find that the thermal relic constraint is relaxed by the mixing-induced decay. Therefore, solutions are extended to  axion masses lower than permitted by the Yukawa interaction. Specifically, the parameter space around $ m_a\sim 10^{-6}$ eV that appears to be inaccessible in Fig.~\ref{fig:Nearly quadratic M18n13With} can be made viable with such couplings. With the Higgs-portal coupling, the lower bound on $ m_a $ is relaxed to $ 5\times 10^{-8}$ eV $ \lesssim m_a$.

\paragraph{Necessity of $ \mathcal{P}_{\mathcal{R}} $ suppression:} Similar to the Yukawa case, the viability of damping through the Higgs-portal interaction requires that $ \mathcal{P}_{\mathcal{R}}(\kmin) $ differs from the value $ A_s $ measured on CMB scales. If this is not the case, the condition discussed in Section \ref{sec:Cems condition} is violated and homogeneity is lost during kination-like scaling of the axion.
As can be seen in Fig.~\ref{fig:maxFlucPlotMmpn13NQHiggs}, a suppression of $ \mathcal{P}_{\mathcal{R}}(\kmin) $ relative to $ A_s $ on the order of two to three orders of magnitude is sufficient.


\begin{figure}
    \centering
    \includegraphics[width=\standardwidth \textwidth]{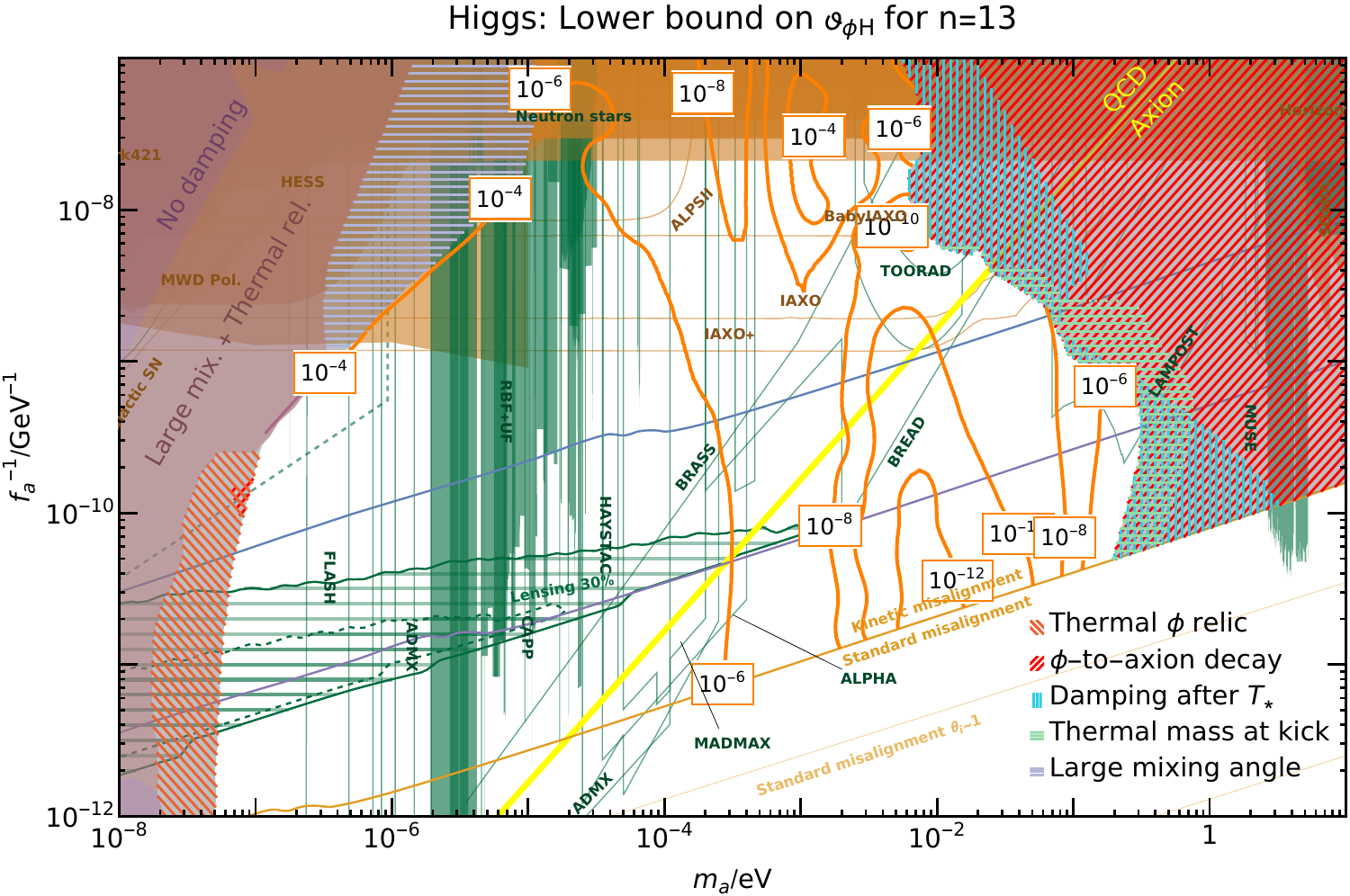}
    \caption{\small\it Contours of the smallest values of the mixing angle $\vartheta_{\rm mix}$ as given by \eqref{eq:mixing angle}. The mixing angle can take values between this lower bound and an upper bound that in most of the parameter space saturates the enforced limit of $\vartheta_{\rm mix}\lesssim 0.1$.}
    \label{fig:minMixAnglePlotNQHiggsMmpn13}
\end{figure}


\FloatBarrier
\section{KMM in a quartic potential with a Hubble-induced mass term}
\label{sec:quartic model}
\begin{figure}
	\centering
	\includegraphics[width=0.68 \textwidth]{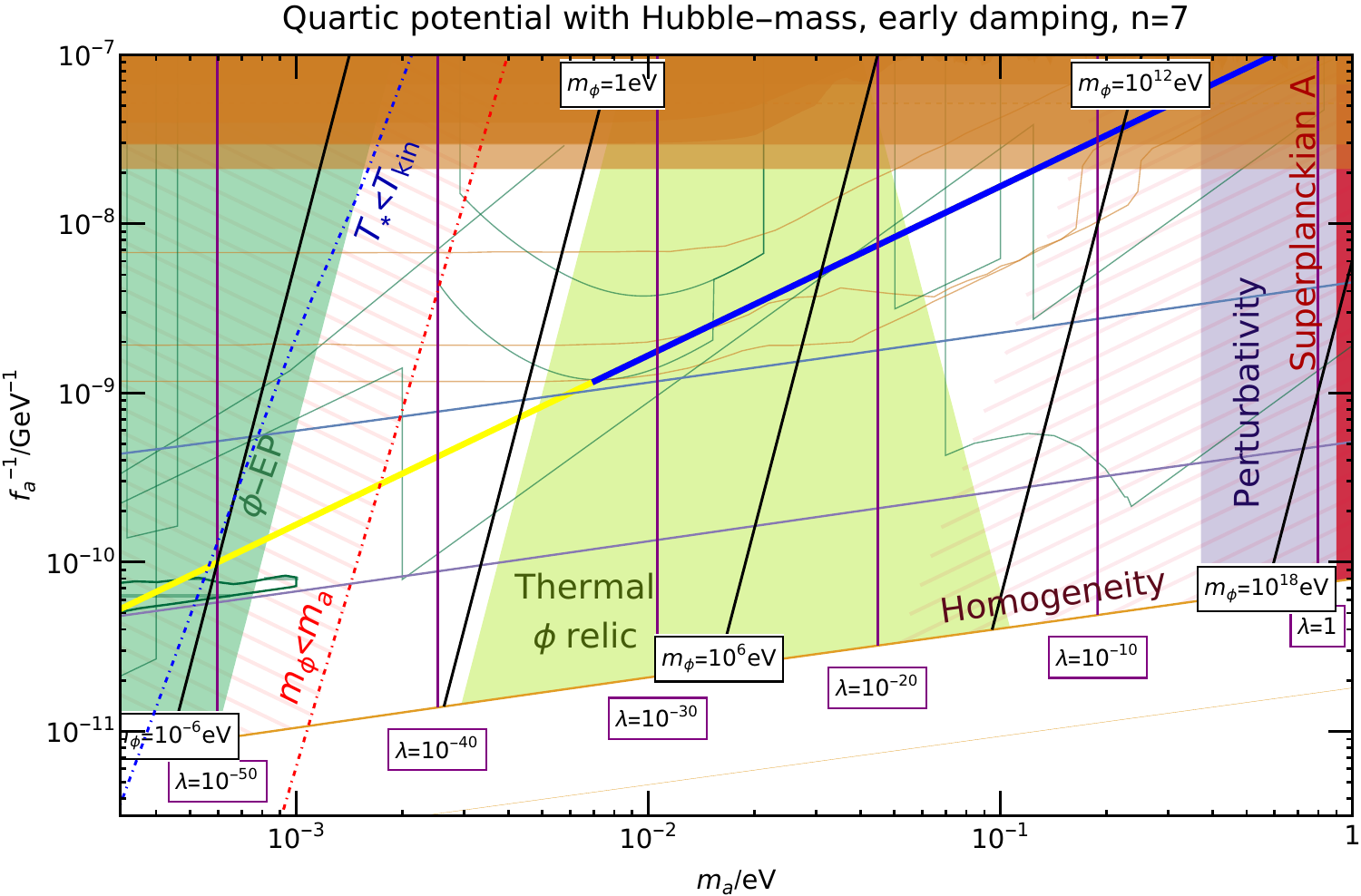}
 \includegraphics[width=0.68 \textwidth]{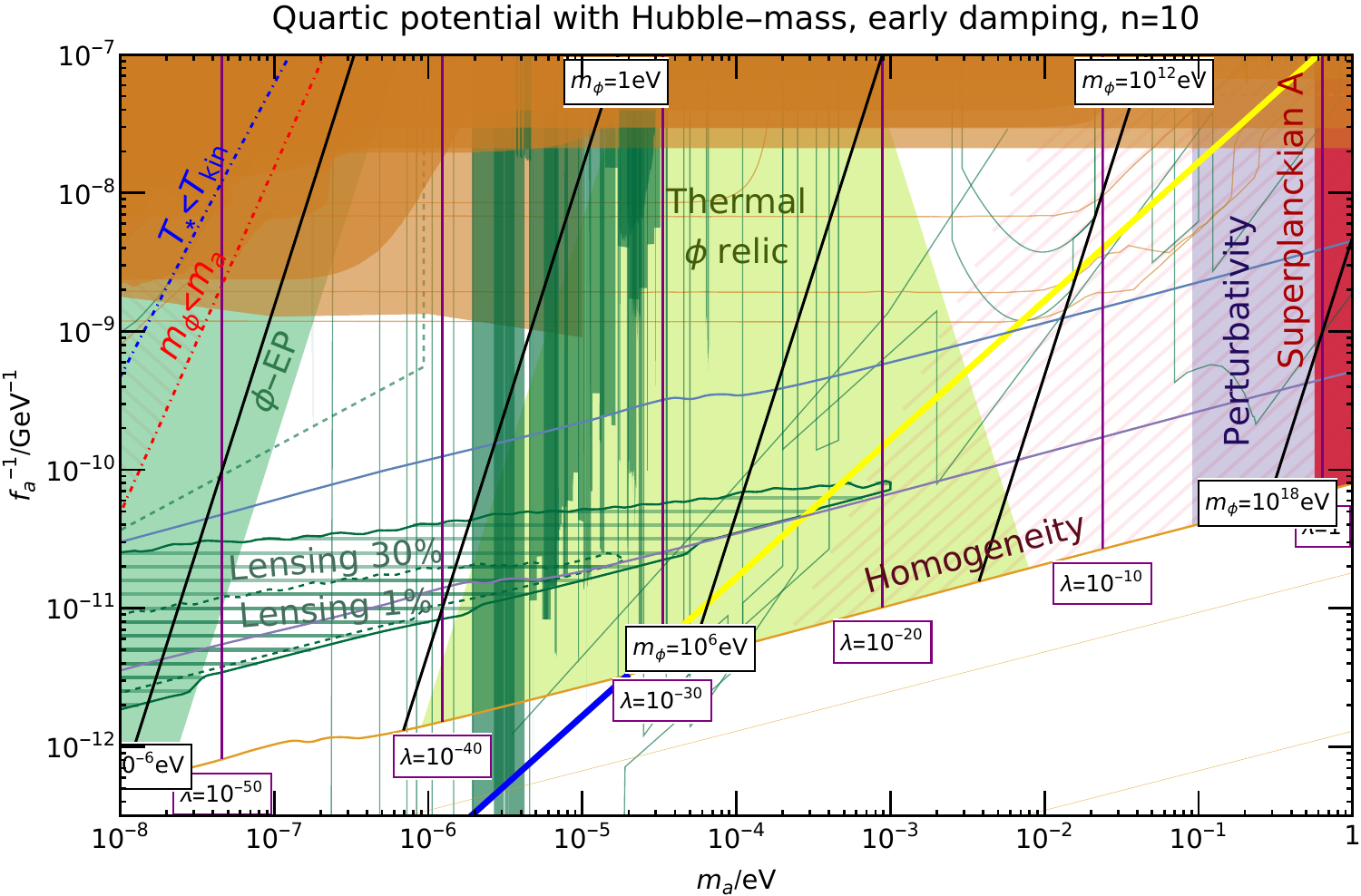}
 \includegraphics[width=0.68 \textwidth]{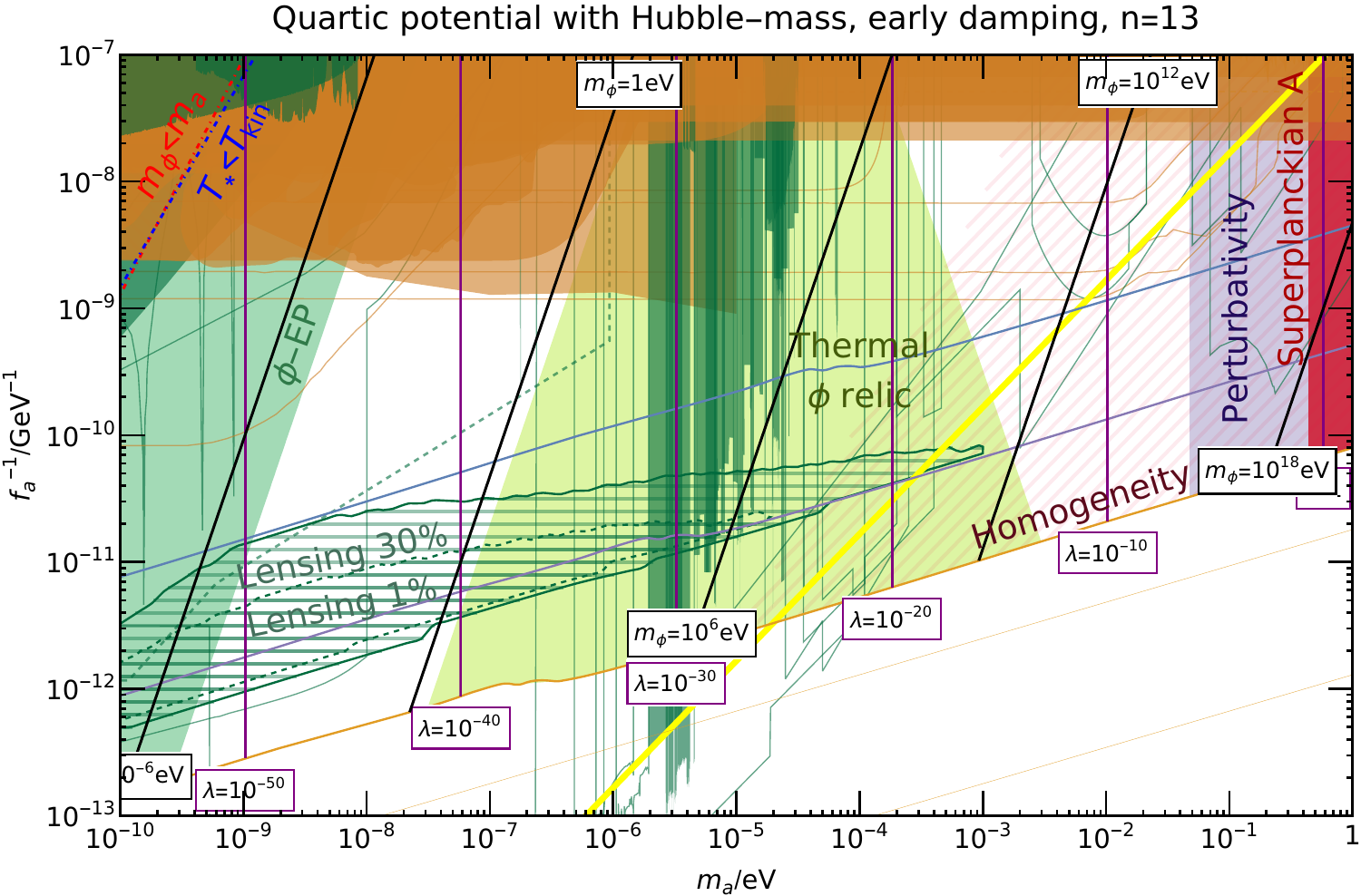}
	\caption{\small \it Viable parameter space and constraints for models with quartic potentials 
when assuming early damping for $n=7$ (upper panel), $n=10$ (middle panel) and  $n=13$ (bottom panel). See Figs. \ref{fig:nearlyQuadraticEarlyDamping1} and \ref{fig:nearlyQuadraticEarlyDamping2} for the analogous plots for models with nearly-quadratic potentials.
}
	\label{fig:quarticHubble}
\end{figure}

As an alternative to the implementations described above, we now investigate how higher-dimensional terms can drive a large VEV in a model with a quartic potential. The potential under consideration is
\begin{gather}
V =  \frac{\lambda}{4} \left(\phi^2-f_a^2\right)^2 - c_H H^2\phi^2 + \frac{A + c_A H}{n} \frac{P^n}{M^{n-3}}+h.c.+\frac{|P|^{2n-2}}{M^{2n-6}}, \label{eq:quartic higher-dimensional}
\end{gather}
which is analogous to Eq.~\eqref{eq:potential for kick}. As with the nearly-quadratic potential, the Hubble-induced terms generate an early large mass at the VEV 
\begin{gather}
\phi_{\rm early} = \left(\frac{2^{n-2}}{n-1}\right)^{\frac{1}{2 n-4}} \left(H M^{n-3}\right)^{\frac{1}{n-2}}.\label{eq:time-dependent VEV, quartic}
\end{gather}
This time-dependent VEV is the same as in the nearly-quadratic model since the VEV is set purely by the higher dimensional terms. It makes both the angular and radial modes heavy during inflation which resolves the isocurvature and domain wall problems associated with models using instead de Sitter fluctuations to drive the large initial value of the field. 
Although the early time-dependent minimum given by Eq.~\eqref{eq:time-dependent VEV, quartic} is the same as in models with nearly-quadratic potentials, the VEV at the time of the kick differs. This is because the kick takes place at $ \mphi\sim \lambdaOLD \phi \sim 3 H $ which implies
\begin{gather}
\Tkick \approx 0.3\times  \sqrt{M \mplanck} \lambda^{\frac{n-2}{4n-12}}, \\
\phikick \approx 1.\times  M \lambda^{\frac{1}{2n-6}}
\end{gather}
where the exact value of the prefactors depends on $ n $ and $ g_*(\Tkick) $. Such a kick implies a yield of
\begin{gather}
\Ykick \approx 0.8 \times \epsilon\left(\frac{M}{\mplanck}\right)^{3/2}\lambda^{\frac{n-6}{4n-12}}.
\end{gather}
The observed DM yield is achieved for
\begin{gather}
\lambdaOLD \sim \mathcal{O}(1)\times \left(\frac{M}{\mplanck}\right)^{\frac{3(n-3)}{n-6}}\left(\frac{\epsilon m_a}{\Teq}\right)^{\frac{2(n-3)}{n-6}},\label{eq:qDM in quartic with Hubble}
\end{gather}
which determines the late-time radial mass $ m_{\phi-\rm late}\approx \lambdaOLD f_a $. With this solution, we can repeat the analysis discussed in Section \ref{sec:Nearly quadratic early damping results} and describe the viable parameter space under the simplifying assumption that damping takes place early enough so that entropy injection can be ignored. This assumption leads to the parameter space described by Fig.~\ref{fig:quarticHubble}. Similar to the models with nearly-quadratic models, we observe two main families of solutions divided into lower and higher axion/$\phi$ masses by the thermal $\phi$ relic constraint. As with models with nearly-quadratic, we expect the lower-mass window to be closed once thermal potentials are considered.

\paragraph{CP violation and the $A$ parameter:} As in the previous models, the amplitude of the CP-violating potential responsible for driving the kick is controlled by the parameter $A$. However, in models based on a quartic potential, contrary to the SUSY-motivated nearly-quadratic potential, this parameter is not naturally linked to a known scale. Furthermore, the CP-violating term is competing against an effective radial mass, Eq.~\eqref{eq:quartic radial mass}, which depends on the field amplitude. This has an important consequence:  $A$ no longer naturally yields a  nearly circular kick, i.e. $\epsilon\sim\mathcal{O}(1)$. In models with quartic potentials, $A$ has to be set to a particular value by hand, in order to generate a kick with $\epsilon \sim \mathcal{O}(1)$. To see this, consider the approximate angular velocity generated by the kick as given by Eq.~\eqref{eq:kick size}, which can be solved for $A$ after plugging the effective radial mass Eq.~\eqref{eq:quartic radial mass}. This yields
\begin{gather}
A \sim \mathcal{O}(1)\times \left(\frac{m_a}{\Teq}\right)^{\frac{2(n-2)}{n-6}} \mplanck, \label{eq:value of A required by quartic models}
\end{gather}
which means that $A$ necessarily exceeds the Planck scale if masses above $m_a\sim 1$ eV are considered.  Note that perturbativity already excludes such values of $m_a$. Nevertheless, the necessity of introducing an additional scale $A$ at an un-motivated value weakens the appeal of models based on quartic potentials.

We now check whether the axion can be the QCD axion. By plugging the required value of $A$, Eq.~\eqref{eq:value of A required by quartic models} into Eq.~\eqref{eq:axion quality}, we find that the axion quality behaves in a way that is qualitatively similar to models with nearly-quadratic potentials, i.e. with differing behaviour for $n>7$ and $n\leq 7$.
The resulting axion quality constraint is displayed with a blue line in Fig.~\ref{fig:quarticHubble}, as explained in the paragraph after Eq.(~\ref{eq:axion quality}).

\begin{figure}
    \centering
    \includegraphics[width=0.8 \textwidth]{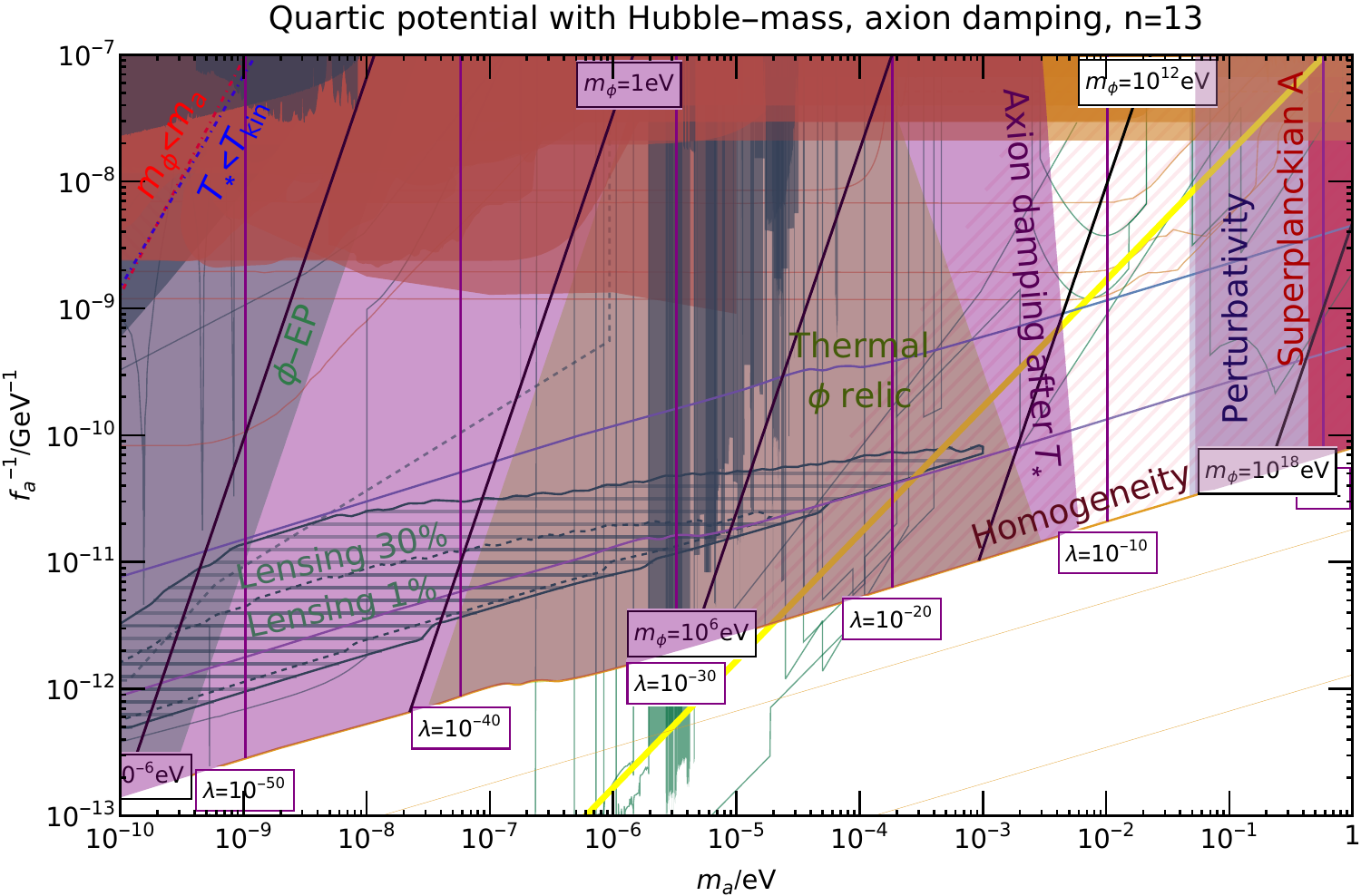}
    \caption{\small \it Extends Fig.~\ref{fig:quarticHubble} by assuming that early damping is implemented entirely through $\phi$-to-axion decay, see Eq.~\eqref{eq:phi-to-axion decay, internal damping}. The unshaded region is viable even without the introduction of an additional SM coupling to implement thermal damping. Nonetheless, suppression of $\mathcal{P}_{\mathcal{R}}(\kmin)$ is still required, see Fig.~\ref{fig:maxFlucPlotQHigherDimYukawaMmpn13} of which this scenario represents a subset.}
    \label{fig:quarticHubble13WithAxionDamping}
\end{figure}

\subsection{Internal damping in quartic potential}
An important characteristic of the quartic potential is that since $\phi \propto a^{-1}$ and $\mphiVprime \propto a^{-1}$, $\rho_\phi \propto a^{-4}$. This means that the energy in the PQ field remains a constant fraction of the energy in radiation. While this means that no kination can arise in this class of models \cite{Gouttenoire:2021jhk}, it also makes $\phi$-to-axion decay less dangerous. 
This is because $\rho_\phi$ never redshifts slower than radiation and therefore can never dominate $\rho_r$. This ensures that dark radiation constraint \eqref{eq:hot axions dominate the plasma} is always avoided. The faster redshift of $\rho_\phi$, relative to nearly-quadratic models, also weakens the DM overproduction constraint Eq.~\eqref{eq:axion decay relic DM overproduction}.
This allows $\phi$-to-axion decay to play a larger role in models with quartic potentials, and we find that radial damping can happen entirely within the PQ field without the need for any additional coupling to the SM. The energy in radial fluctuations is transferred to a hot axion relic which constitutes a dark radiation far subdominant to the SM plasma.

To derive the region in which $\phi$-to-axion decay can lead to damping, we rely on the decay rate Eq.~\eqref{eq:radial-to-axion rate}
\begin{gather}
\Gamma_{Sa}\sim \frac{1}{64\pi} \frac{\mphiVprime^3}{\phi^2}. \label{eq:phi-to-axion decay, internal damping}
\end{gather}
Comparing this to Hubble, we find that this will lead to efficient damping around the temperature
\begin{gather}
    \Tdamp\sim \mathcal{O}(1)\times 10^{-2} \left(\frac{M}{\mplanck}\right)^{\frac{7n-18}{n-6}}\left(\frac{\epsilon m_a}{\Teq}\right)^{\frac{5n-14}{n-6}},
\end{gather}
where the exact value of the $\mathcal{O}(1)$ prefactor depends on the choice of $n$.
At the very least, our scenario assumes that this damping takes place before $\Tfrag$. We find that this condition is consistent with $m_a\sim 10^{-2}$ eV, see Fig.~\ref{fig:quarticHubble13WithAxionDamping}. As the PQ field cannot dominate the energy density, this decay mode implements early damping.
Interestingly, we find that this scenario not only avoids the isocurvature and domain wall problems inherent to the original models~\cite{Co:2019jts,Co:2020dya,Co:2020jtv}, but also requires  less tuning in the quartic coupling. Co et al. found that the isocurvature bound generically required the quartic coupling to be less than $\lambda \lesssim 10^{-22}$ while we find solutions with quartic couplings up to approx. $\lambda \lesssim 10^{-6}$.
We report the full range of viable values of $\lambda$ in Appendix \ref{app:parameter values}.

\begin{figure}
	\centering
	\includegraphics[width=0.8 \textwidth]{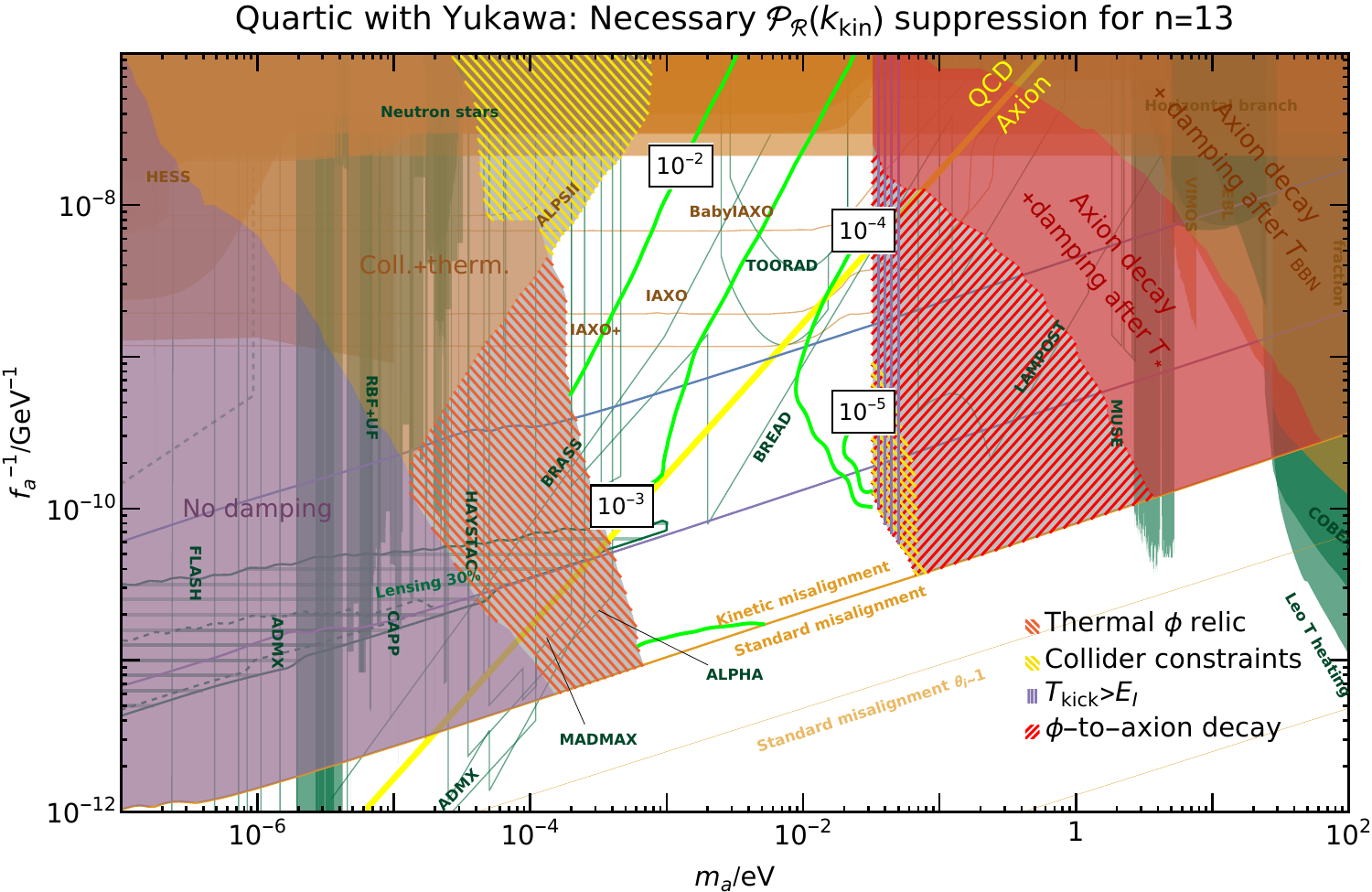}
 \includegraphics[width=0.8 \textwidth]{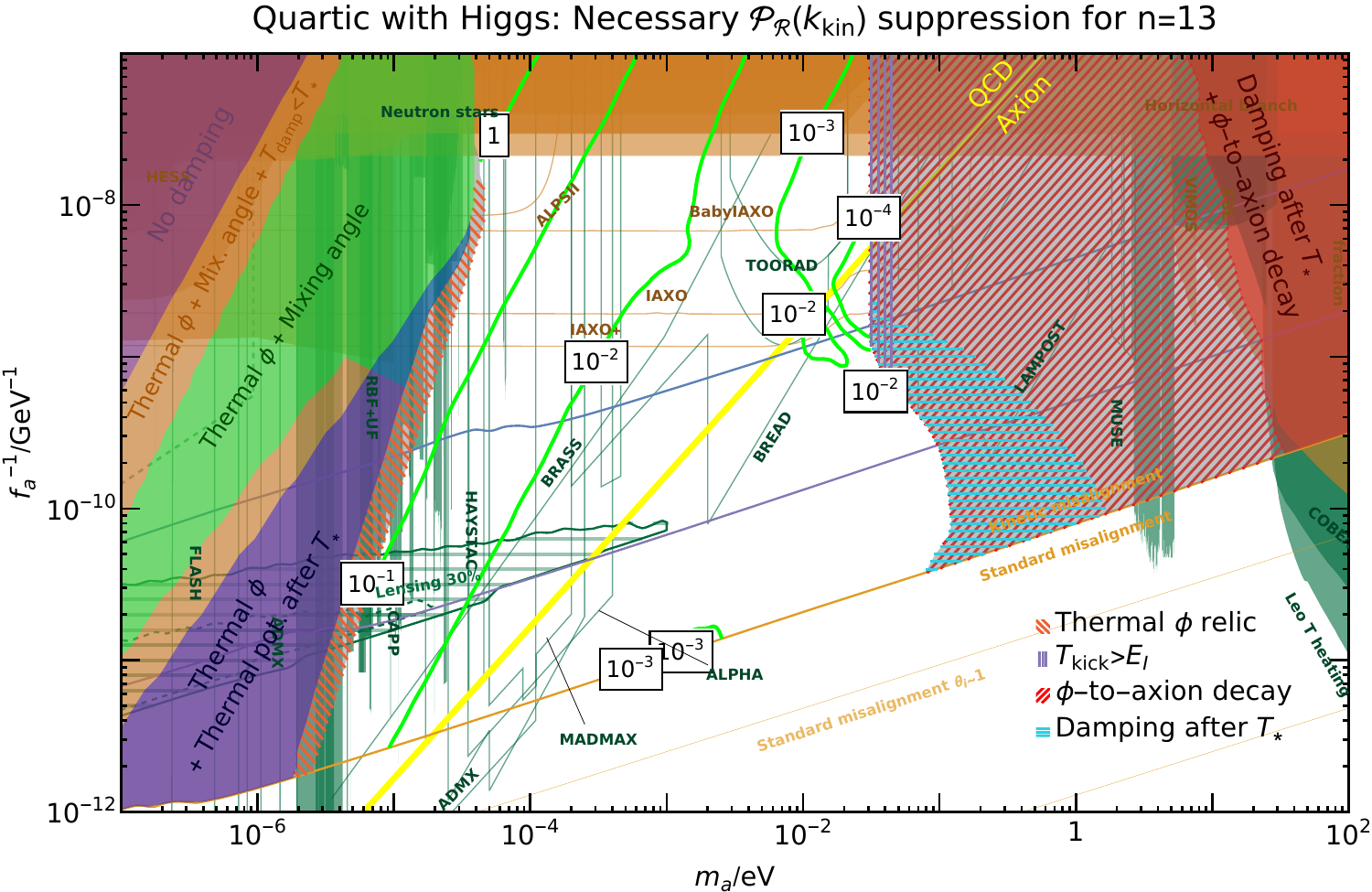}
	
	\caption{\small\it Green contours: Necessary suppression of the amplitude of the primordial power spectrum,  $ \mathcal{P}_{\mathcal{R}}(\kmin) $ relative to $ A_s $, see Eq.~\eqref{eq:cems condition}, for models with quartic potentials and higher dimensional operators with $n=13$, $M=\mplanck$. Solutions at slightly higher $m_a$ are possible when considering lower values of $n$. Upper panel: Damping with a Yukawa interaction; Lower panel: Damping with a Higgs-portal interaction;} 
	\label{fig:maxFlucPlotQHigherDimYukawaMmpn13}
\end{figure}

\subsection{Thermal damping in quartic potential}

To see if more parameter space can be opened in models with quartic potentials by introducing thermal damping we now consider again the Yukawa coupling \eqref{eq:yukawa coupling} and the Higgs portal interaction  \eqref{eq:phi-higgs coupling}. By repeating the analysis of Section \ref{sec:Nearly quadratic Yukawa section} and \ref{sec:NQHiggs}, we find that these interactions extend the parameter space. 
The numerical solutions with {radial masses}
%
above the bound from the thermal $\phi$ relic, Eq.~\eqref{eq:thermal condition}, largely reproduce the results expected from the early damping analysis, see Fig.~\ref{fig:maxFlucPlotQHigherDimYukawaMmpn13}. 
Similar to models  with a nearly-quadratic potential, the  thermal potential rules out radial masses in the low-mass window, i.e. the solutions left of the thermal $\phi$ relic bound on Fig.~\ref{fig:quarticHubble} are incompatible with a coupling to the thermal plasma.

Both the Yukawa and the Higgs interactions extend the parameter space to lower axion masses thanks to additional damping channels. The Yukawa scenario is limited to $m_a\gtrsim $  $ 10^{-4}$ eV by the thermal $\phi$ relic bound. The Higgs portal scenario circumvents this thermal $\phi$ relic bound as $\phi$ can decay into the SM plasma through $\phi$-Higgs mixing. This allows for solutions down to $m_a\gtrsim 10^{-5}$ eV.


\section{Suppressed primordial power spectrum at the kination scale}
\label{sec:power}

\begin{figure}
    \centering
    \includegraphics[width=0.8\textwidth]{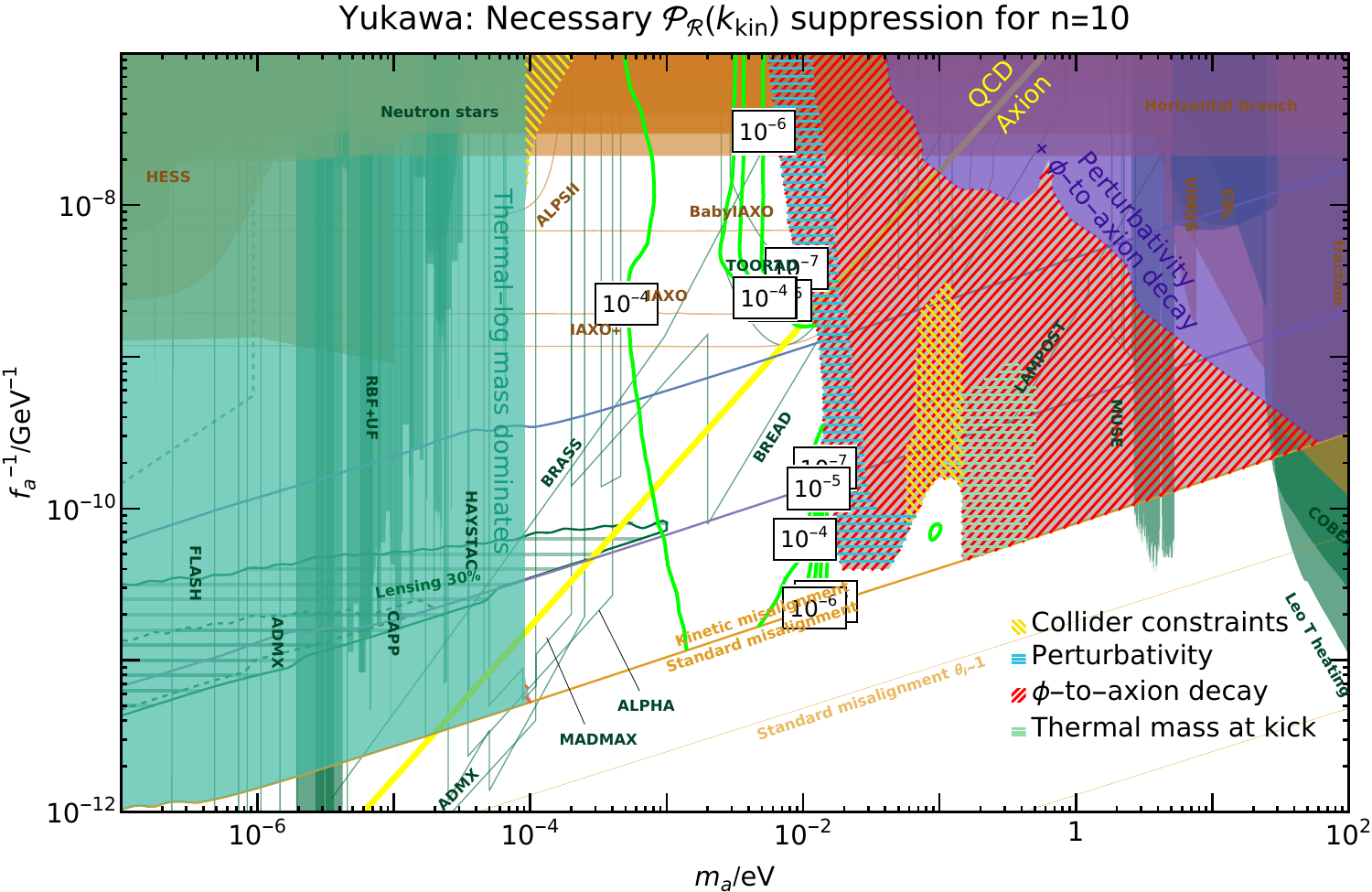}
    \includegraphics[width=0.8\textwidth]{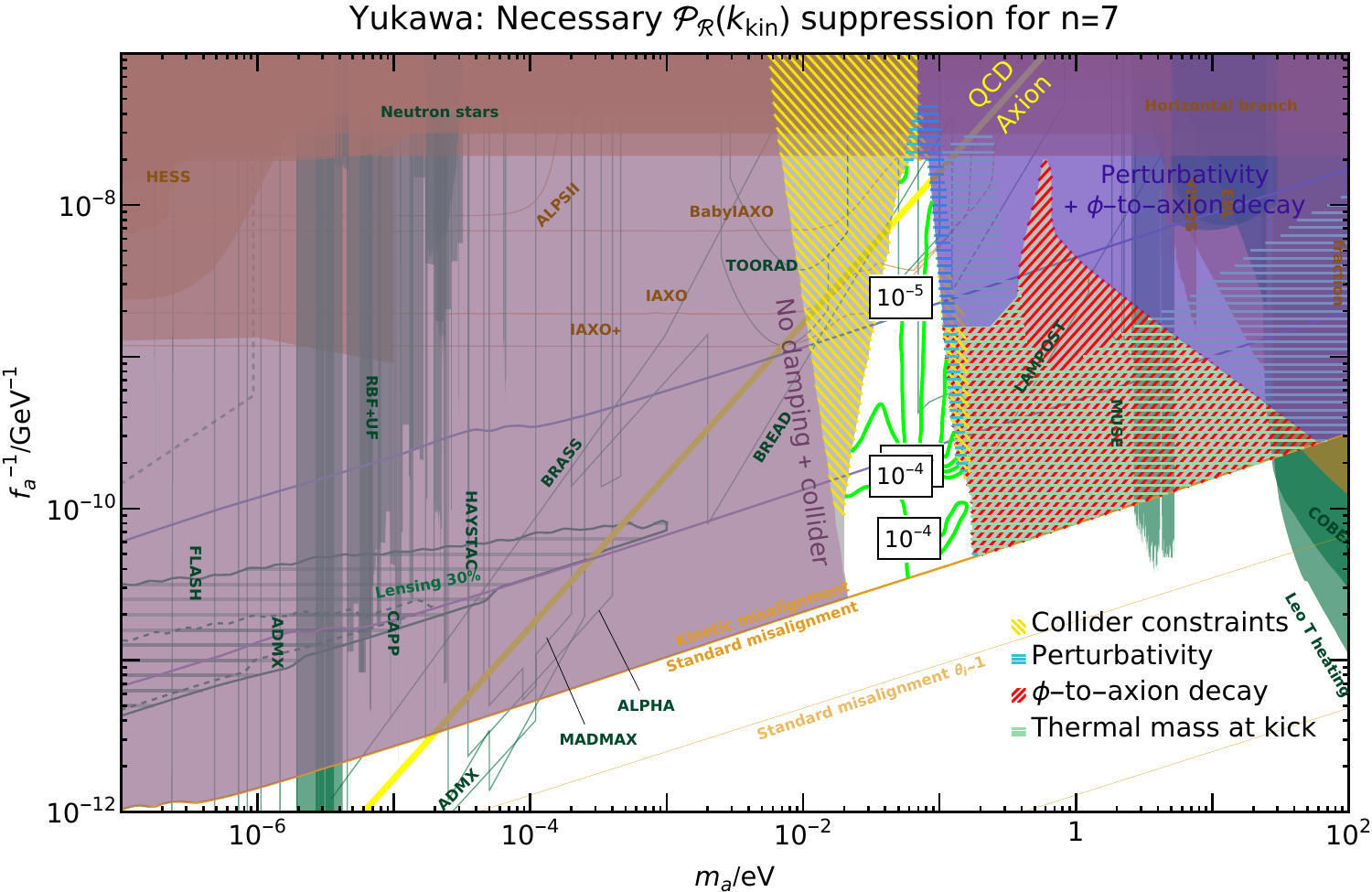}
    \caption{\small\it Green contours: Necessary suppression of $ \mathcal{P}_{\mathcal{R}}(\kmin) $ relative to $ A_s $ for alternative choices of $ n $. Both $ n=10 $ and $ n=7 $ support less parameter space than $ n=13 $ and both still require $ \mathcal{P}_{\mathcal{R}}(\kmin) $ suppression.  For $ n=10 $, the lower bound on $ m_a $ is set by thermal-log contributions to $ m_\phi $. We do not consider parameter space with thermal-log domination as such may feature oscillon formation~\cite{Mukaida:2012qn}. }
    \label{fig:maxFlucPlotMmpn10n7}
\end{figure}

In this work, we impose that dark matter is constituted by the (fragmented) zero-mode energy of the axion field. This implies that in essentially all the relevant parameter space, there is a stage in the early universe (when the axion is still effectively massless) during which the energy density of the scalar field has a kination equation of state. This suppresses the zero-mode energy component (which scales as $a^{-6}$) with respect to the energy in the primordial fluctuations of the axion field  inherited from inflation (which scales as $a^{-4}$), such that the later could come to dominate at late times and be the dark matter. Such scenario is also of high interest and will be treated in a separate article~\cite{Eroncel:2022b}. In this paper, we focus on the case where dark matter is due to the primordial homogeneous component of the  axion field. For this to happen, the primordial power spectrum of fluctuations $ \mathcal{P}_{\mathcal{R}}$ has to be suppressed at the scale $\kmin$ relative to that at the pivot scale $ k_{\rm pivot} = 0.05 $ Mpc$ ^{-1}$, which is constrained by Planck~\cite{Planck:2018vyg}: $ \mathcal{P}_{\mathcal{R}}(k_{\rm pivot}) = A_s\approx 2.1 \times 10^{-9}$.

The needed amount of suppression  $\mathcal{P}_{\mathcal{R}}(\kmin) / A_s$ is shown in Fig.~\ref{fig:maxFlucPlotMmpn13} (nearly-quadratic potential with Yukawa damping),
Fig.~\ref{fig:maxFlucPlotMmpn13NQHiggs} (nearly-quadratic potential with Higgs damping), Fig.~\ref{fig:maxFlucPlotQHigherDimYukawaMmpn13} (quartic potential with Yukawa and Higgs damping), 
and Fig.~\ref{fig:maxFlucPlotMmpn10n7} (nearly-quadratic potential with Yukawa damping for $n=7$ and $n=10$). 
It is clear  that in all the parameter space that we have scanned in detail, $\mathcal{P}_{\mathcal{R}}(\kmin)$ has to be suppressed with respect to $A_s$.
The typical needed suppression is in the range $  [10^{-7}, 10^{-3}]$. $\mathcal{P}_{\mathcal{R}}(\kmin) / A_s$ can be larger, as large as $10^{-2}$ in the case of
a nearly quadratic potential with Higgs damping or 
for a quartic potential with Yukawa damping. For a (tuned) quartic potential and Higgs damping there is a small region of parameter space where nearly no suppression is needed. However, for most of parameter space, a significant suppression of the power spectrum is needed at $ \kmin $.

The comoving momentum scale $ \kmin $ that is relevant 
for the kinetic misalignement mechanism,  is many orders of magnitude apart from the Pivot scale as shown in Fig.~\ref{fig:kminPlotMmpn13}, such that $ \mathcal{P}_{\mathcal{R}}(\kmin) $ may very well deviate from the value constrained by Planck. On such scales, LISA might set constraints on enhanced $\mathcal{P}_{\mathcal{R}}$ the order of $\mathcal{P}_{\mathcal{R}}(\kmin) / A_s \lesssim 10^6$~\cite{Byrnes:2018txb}. On the other hand, suppressed $\mathcal{P}_{\mathcal{R}}$ values are unconstrained.

Even though the suppression of the primordial power spectrum at scales far smaller than the CMB scales are unconstrained, it is important to keep in mind that axion dark matter from the kinetic misalignement mechanism as presented in this paper and in all the earlier literature on KMM 
rests on an inflationary model which can account for this kind of suppression while  also producing the correct spectral index $n_s$ at the pivot scale, and obeying the constraints on the tensor-to-scalar ratio $r$. An existence proof with a toy model of inflation is provided in Ref.~\cite{SuppressedPowerPaper}. We sketch the main requirements below.

Let $N\equiv \ln(a_{\rm end}/a)$ denote the number of efolds, counting from the end of inflation.
Also, we defined $N_{\pvt}$ and $N_{\rm kin}$ as the efold times at which the pivot scale $k_{\pvt}$ and $k_{\rm kin}$ exit the horizon respectively. Then, by approximating the Hubble scale during inflation as a constant we can write
$$   k/ {k_{\pvt}}=\frac{a(N_k) H(N_k)}{a(N_{\pvt}) H(N_{\pvt})}\sim \frac{a(N_k)}{a(N_{\pvt})}=e^{N_{\pvt} - N_{k}}.
$$
From Fig.~\ref{fig:kminPlotMmpn13}, we can read that
\begin{equation}
    \label{eq:kkin_range}
    \frac{k_{\rm kin}}{k_{\pvt}} = 2\times 10^{13} \divisionsymbol 2\times 10^{17},
\end{equation}
which  implies that
\begin{equation}
   \label{eq:nkin_range}
    N_{\pvt} - N_{\rm kin} \approx 
    30\divisionsymbol 40.
\end{equation}
From Fig.~\ref{fig:maxFlucPlotMmpn10n7}, we can also read that we need at least $4$ to $7$ orders of magnitude of suppression of the primordial power spectrum at $k_{\rm kin}$ relative to $\mathcal{P}_{\mathcal{R}}$. 

\begin{figure}
    \centering
    \includegraphics[width=0.8\textwidth]{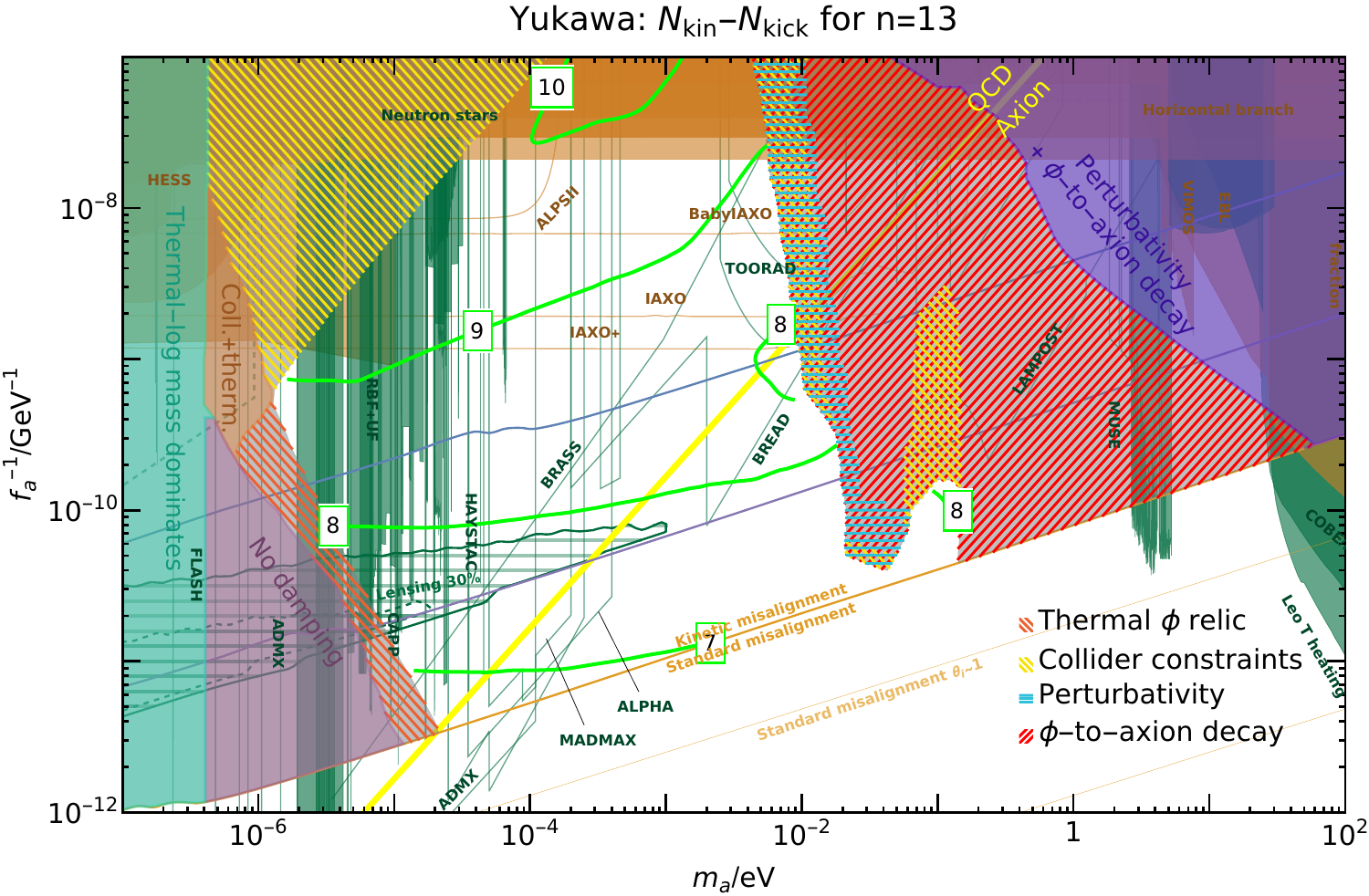}
    \caption{\small\it Contours of the number of e-folds of inflation between the times when the scales $
    k_{\rm kin}$ and $k_{\rm kick}$ leave the horizon as defined by Eq.~\eqref{eq:number of e-folds during inflation}. The solutions plotted here correspond to the same solutions shown in Figs. \ref{fig:maxFlucPlotMmpn13} and \ref{fig:kminPlotMmpn13}.}
    \label{fig:maxminefolds}
\end{figure}
In most inflation models, the curvature power spectrum is approximately flat for the modes that exit the horizon during inflation. A suppression only occurs for the modes that exit the horizon during the last few efolds of inflation. Naively, we might think that by using an inflation model where $k_{\rm kin}$ exits the horizon at the last few efolds, we can achieve the suppression that we want. However, this is problematic. The reason is that we have assumed that the kick happens after reheating. Therefore, the mode $k_{\rm kick}\equiv a_{\rm kick} H_{\rm kick}$ which re-enters the horizon at the time  of kick should leave the horizon before the end of inflation. By denoting $N_{\rm kick}$ as the efold time at which $k_{\rm kick}$ exits the horizon during inflation, we can write
\begin{equation}
    N_{\rm kin}-N_{\rm kick}=\ln\qty(\frac{a(N_{\rm kick})}{a(N_{\rm kin})})\sim \ln\qty(\frac{k_{\rm kick}}{k_{\rm kin}})=\ln\qty(\frac{a_{\rm kick}H_{\rm kick}}{a_{\rm kin}H_{\rm kin}}).
\end{equation}
We now show that there is a lower bound on this value. For simplicity, let us assume that the reheating is instantaneous and the kick happens right after reheating. The universe between the kick and the kination-like scaling is either dominated by radiation all the time, or there is an early matter dominated era that starts at $a_{\rm dom}$ and ends at $a_{\rm kin}$. Let $N_{\rm EMD}=\ln (a_{\rm kin}/a_{\rm dom})$ denote the number of efolds of such an era if it exists. Then, by using the fact $H\propto a^{-2}$ and $H\propto a^{-3/2}$ during radiation and matter domination respectively, we can show that
\begin{equation}
    N_{\rm kin}-N_{\rm kick} \sim \ln\qty(\frac{a_{\rm kin}}{a_{\rm kick}}) - \frac{1}{2}N_{\rm EMD}.\label{eq:number of e-folds during inflation}
\end{equation}
In the extreme case, when the matter domination lasts from the kick to the start of the kination, then $N_{\rm EMD}=\ln\qty(a_{\rm kin}/a_{\rm kick}) $, so we have $N_{\rm kin}-N_{\rm kick} \sim\ln\qty(a_{\rm kin}/a_{\rm kick}) /2$, i.e.  the number of required efolds is reduced by half. 
The range of $N_{\rm kin}-N_{\rm kick}$ realized in our models is given in Fig.~\ref{fig:maxminefolds}, which shows $7
\lesssim N_{\rm kin}-N_{\rm kick}\lesssim10$.

\begin{figure}
    \centering
    \includegraphics[width=\textwidth]{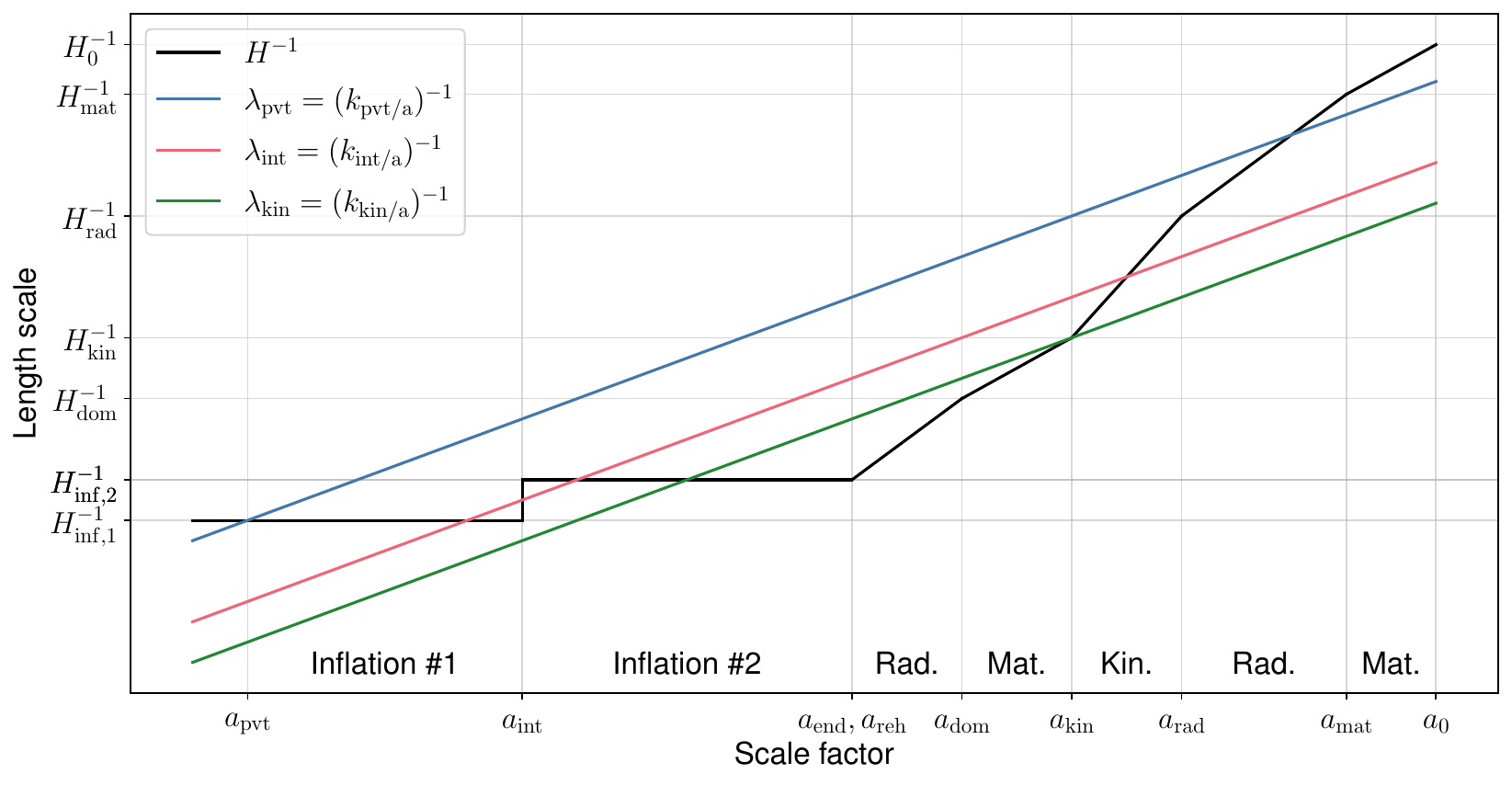}
    \caption{\small \it A sketch showing the evolution of the Hubble radius throughout the cosmological history, starting from the pivot scale exit during inflation until today. Inflation is divided into two stages, where the Hubble scale in the second stage is much smaller compared to the first one. 
    A brief non-inflationary period at $a_{\rm int}$ connects the two stages. 
    We have approximated this transition as instantaneous in the plot. The second stage of inflation ends at $a_{\rm end}$, with the decay of the inflaton instantaneously reheating the universe. At $a_{\dom}$, the PQ field starts to dominate the energy density of the universe producing an early matter era until the radial mode relaxes to its minimum at $a=a_{\rm kin}$, when the universe enters into an kination era due to the kinetic energy stored in the axion field. Radiation takes over again at $a=a_{\rm rad}$, and the usual cosmological history applies after that. Together with the Hubble scale evolution, we also show the evolution of three length scales. First is the pivot scale shown in \textcolor{tolblue}{\textbf{(blue)}}. This mode exits the horizon during the first stage of inflation, remains outside the horizon during the second stage, and re-enters much later, close to the matter-radiation equality. Second is the scale corresponding to $k_{\rm kin}$, shown in \textcolor{tolgreen}{\textbf{(green)}}, where the latter is the mode that re-enters the horizon at the start of the kination. This mode is inside the horizon during the first stage, and exits during the second stage. Hence, its power spectrum amplitude is suppressed. Lastly, we also show a length scale, shown in \textcolor{tolred}{\textbf{(red)}}, which exits the horizon during the first stage, re-enters again during the transition period connecting the two inflationary eras, and re-exits again during the second stage.}
    \label{fig:inverse-hubble}
\end{figure}

This result demonstrates that we need a moderately long period of inflation after the mode $k_{\rm kin}$ exits the horizon during inflation, so we cannot suppress $\mathcal{P}_{\mathcal{R}}(k_{\rm kin})$ just by ending the inflation at $N_{\rm kin}$. We show how to realise this in \cite{SuppressedPowerPaper}. The idea is to invoke two stages of inflation, connected to each other via a brief non-inflationary period. The modes relevant in CMB observations exit the horizon during the first stage of inflation, and remain outside the horizon until much later. We must demand that the potential during this inflationary period obey the measurements and constraints from Planck. The modes $k\sim k_{\rm kin}$ are sub-horizon during the first stage of inflation, and exit the horizon during the second stage of inflation. The inflationary model during this stage is unconstrained except that the power spectrum needs to be suppressed. In the slow-roll approximation, the power spectrum can be expressed analytically as~\cite{Lyth:1998xn}
\begin{equation}
\mathcal{P}_{\mathcal{R}}(k)\approx \frac{H^2(N_k)}{8\pi^2 \mplanck^2 \varepsilon_1(N_k)},
\end{equation}
where $N_k$ is the efold time at which the mode $k$ exits the horizon, and $\varepsilon_1=H'(N)/H$ is the first slow-roll parameter. Therefore, demanding $\mathcal{P}_{\mathcal{R}}(k_{\rm kin})\ll \mathcal{P}_{\mathcal{R}}(k_{\pvt})$ requires that
\begin{equation}
    \frac{\mathcal{P}_{\mathcal{R}}(k_{\rm kin})}{\mathcal{P}_{\mathcal{R}}(k_{\pvt})}\sim \frac{H_{\rm inf, 2}}{H_{\rm inf, 1}}\ll 1,
\end{equation}
where $H_{\rm inf, 1}$ and $H_{\rm inf, 2}$ are the Hubble scales during the first and second stages of inflation respectively. 

We show in Fig.~\ref{fig:inverse-hubble} a sketch of the Hubble radius evolution throughout the whole cosmological history, starting from the exit of the pivot scale until today. For comparison, we also show the evolution of three representative length scales. The pivot scale, \textcolor{tolblue}{\textbf{(blue)}}, exits the horizon during the first stage of inflation, and re-enters much later. The length scale corresponding to $k_{\rm kin}$, \textcolor{tolgreen}{\textbf{(green)}}, exits the horizon during the second stage of inflation, and re-enters the horizon at the start of the kination era. Due to the drop in the Hubble scale when transitioning from the first stage of inflation to the second stage, some modes that exit the horizon during the first stage re-enter the horizon during the transition, and re-exit in the second stage. The length scale shown in \textcolor{tolred}{\textbf{(red)}} is such an example. This figure being a  sketch, the durations of the individual eras do not correspond to a particular benchmark solution. Note that even though we show a kination era in this figure, our discussion remains relevant in the case where there is no kination era. In this case, $k_{\rm kin}$ corresponds to the wavenumber that exits the horizon when the equation of state of the axion is that of kination.

An important implication is that the needed suppression of the power will also typically lead to the suppression of the gravitational wave signal discussed in Sec.~\ref{subseckination} and in \cite{Gouttenoire:2021jhk,Gouttenoire:2021wzu,Co:2021lkc} if one imposes the homogeneous axion zero-mode to be dark matter. However, the amplification of the primordial signals of GW due to a kination era remains in a large class of realisations of ALP dark matter, in particular the one discussed in \cite{Eroncel:2022b} or if the radial mode is dark matter instead of the axion.


\section{Summary and conclusion}\label{sec:summary}
We have re-examined Affleck-Dine-like implementations of the axion dark matter kinetic misalignment mechanism (KMM). By evaluating the models proposed by Co et al in  \cite{Co:2019jts,Co:2020dya,Co:2020jtv}, we have determined precisely which $ [m_a,f_a] $ parameter space can pass all the tests for a successful UV realization.
A successful implementation consists in a nearly-quadratic potential for the Peccei-Quinn scalar field $P$, Eq.~\eqref{eq:Nearly quadratic potential, late time} with an explicit PQ breaking term, Eq.~\eqref{eq:higher dimensional PQ breaking potential}.
 We also analysed models with quartic potentials in Sec.~\ref{sec:quartic model} and found that these could be improved by driving a large radial VEV with a Hubble induced mass term Eq.~\eqref{eq:time-dependent VEV, quartic} instead of de Sitter fluctuations Eq.~\eqref{eq:phi kick}.
The viable $ [m_a,f_a] $ parameter space crucially depends on the axion partner, the radial mode of the PQ field.
To investigate the viable dark matter parameter space, we carried out both a damping-mechanism-independent analysis and analyses of models in which the radial oscillations are damped either through a KSVZ-like Yukawa interaction Eq.~\eqref{eq:yukawa coupling} or through a Higgs portal interaction Eq.~\eqref{eq:phi-higgs coupling}.

\begin{figure}
\centering
\includegraphics[width=0.63 \textwidth]{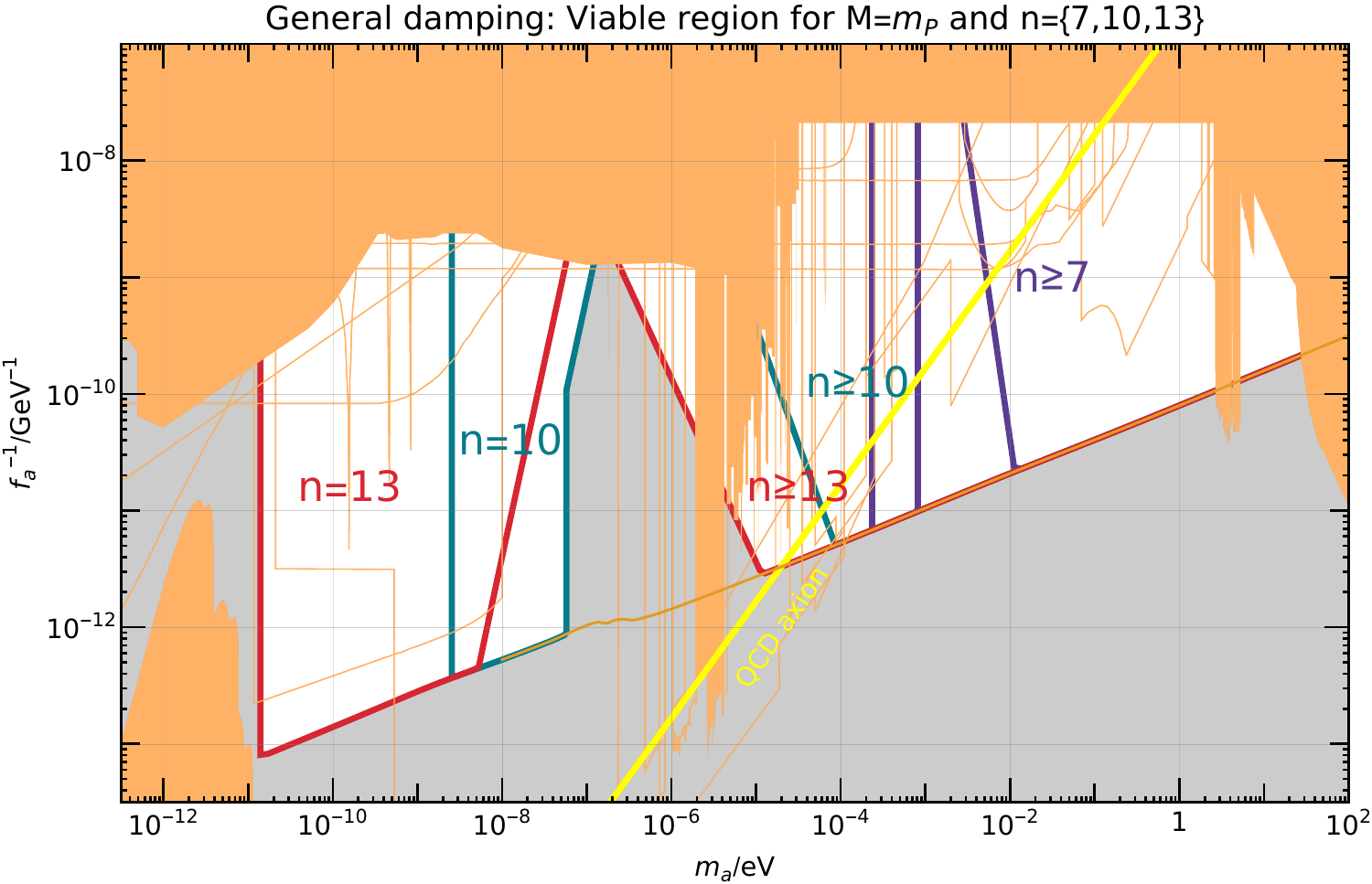}
\includegraphics[width=0.65 \textwidth]{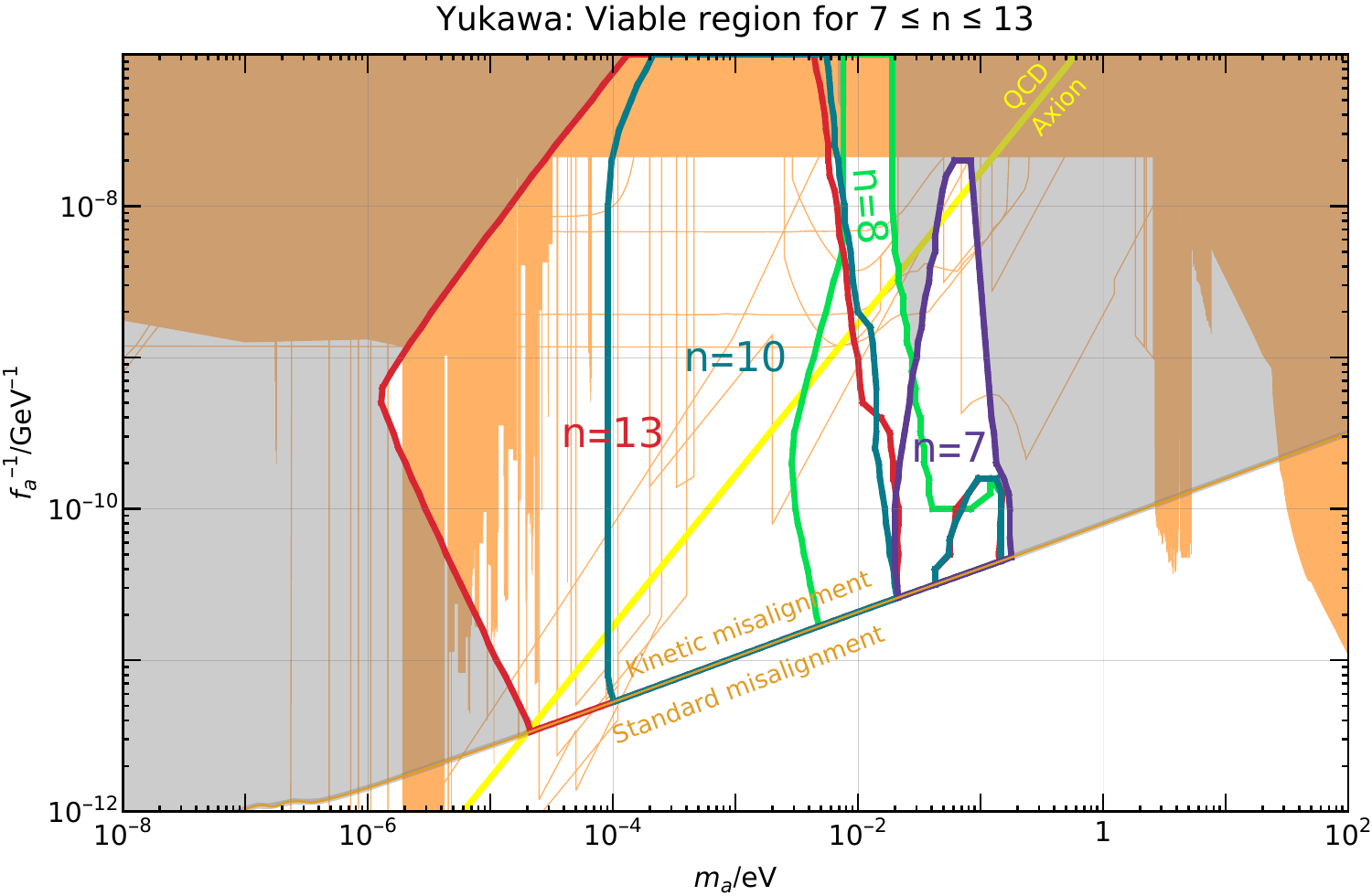}
\includegraphics[width=0.65 \textwidth]{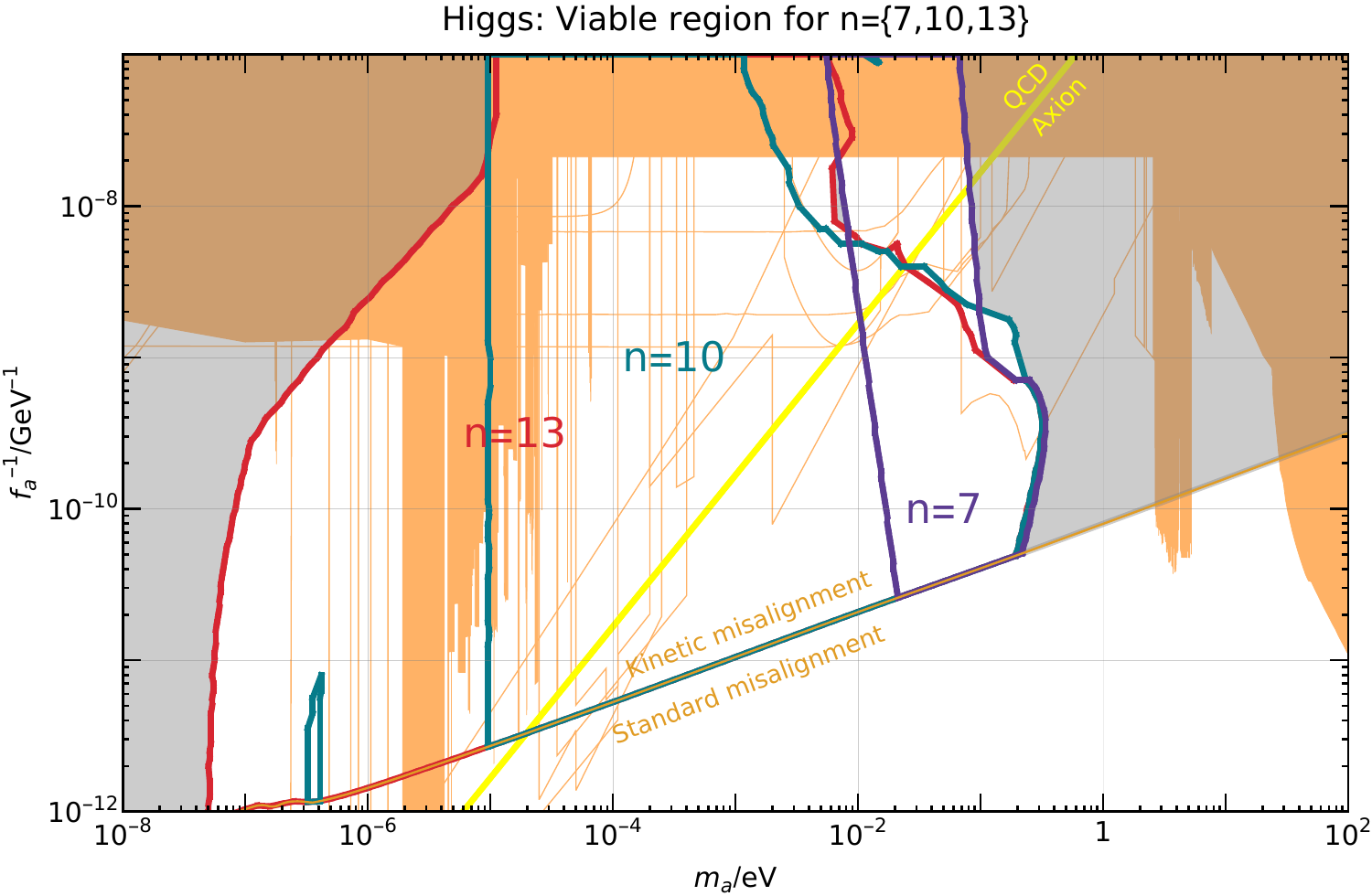}

\caption{ \small \it Summary of the viable parameter space (in white) for axion DM from kinetic misalignment in models with nearly-quadratic potentials.
Upper panel: Assumes that instantaneous damping can be implemented at any $ \Tdamp $.
The region around $ m_a\sim 10^{-6} $ is ruled out by 
 the bound on radial-mode thermal relics unless a specified damping mechanism can relax this constraint.
Intermediate panel: Damping through a Yukawa interaction \eqref{eq:yukawa coupling}. 
Lower panel: Damping through a Higgs portal \eqref{eq:phi-higgs coupling}. 
Compared to the scenario with damping through a Yukawa interaction, lower  $ m_a $ values can be achieved with the Higgs portal as the Higgs mixing relaxes the thermal relic constraint.}

\label{fig:summaryPlotNQYukawaAndHiggs}
\end{figure}

In the damping-channel-independent analysis in Sec.~\ref{sec:Nearly quadratic early damping results} and Sec.~\ref{sec:nearly quadratic late damping}, 
we have mapped out the best-case parameter space that can be realized by any model based on the nearly-quadratic potential.
The results are shown in
Fig.~\ref{fig:nearlyQuadraticEarlyDamping1}, \ref{fig:nearlyQuadraticEarlyDamping2}, \ref{fig:constantfaPlotn13Mmp}, and \ref{fig:Nearly quadratic M18n13With}. The axion mass $m_a$ and the radial mode mass $\mphi$ are related by Eq.~\eqref{eq:mphi nearly quadratic, without injection}.  Generic features of these figures are the large-$m_a$ regime which is constrained by the perturbativity condition, Eq.~\eqref{eq:perturbativity bound NQ}, and the small-$m_a$ regime that is constrained by equivalence-principle constraints on the radial mode, Eq.~\eqref{eq:EP constraint}. In addition, the intermediate  $m_a$ region is constrained by bounds on the thermal radial mode relic, Eq.~\eqref{eq:thermal condition}. As a result, the available parameter space is separated into two disconnected regions; one is the low $m_a$ region with $\mphi < $ eV and another is the high $m_a$ region with $f_a^2/\mplanck < \mphi$.
As discussed in Sec.~\ref{sec:evolution after the kick}, our analysis includes both the simplest scenario in which damping is assumed to take place through an unspecified mechanism early enough to avoid any entropy injection and the more general scenario in which damping takes place at specified temperature $\Tdamp$, which potentially gives rise to significant injection of entropy to the SM plasma. The results for the simplest scenario, early damping, is summarised in Figs.~\ref{fig:nearlyQuadraticEarlyDamping1} and \ref{fig:nearlyQuadraticEarlyDamping2}, which details relevant constraints and viable parameter space for $n=7, 10,$ and 13. Interestingly, the homogeneity condition presented in Sec.~\ref{sec:Cems condition}, which we identified in our previous work \cite{Eroncel:2022vjg}, turns out to severely restrict the high-mass regime unless the power spectrum of primordial density fluctuations $\mathcal{P}_{\mathcal{R}}$ is much smaller on the relevant scale $\kmin$ than the value suggested by Planck on CMB scales. Such suppressed values of $\mathcal{P}_{\mathcal{R}}(\kmin)$ are observationally unconstrained and in our analysis we treat the condition as a constraint on the implementation of inflation which we assume is met.
To generalize the damping-independent analysis to arbitrary damping temperatures, we include entropy injections and evaluate how this shifts DM solutions and constraints. Possible thermal histories are classified in Fig.~\ref{fig:scenarios}. Generically, entropy injection shifts solutions to larger $m_a$ and can thus open up parameter space that would otherwise not have been viable. This analysis is done in Sec.~\ref{sec:nearly quadratic late damping} where we describe the structure of the constraints in Fig.~\ref{fig:constantfaPlotn13Mmp} and report the damping temperatures that can be made viable in Fig.~\ref{fig:Nearly quadratic M18n13With}. If the required damping temperatures can be realized, the parameter space summarized in the top panel of Fig.~\ref{fig:summaryPlotNQYukawaAndHiggs} can be made viable. The figure includes results for $n=7,10,$ and 13.

In Sec.~\ref{sec:Nearly quadratic Yukawa section}, we investigated a KSVZ-like Yukawa coupling as an implementation of a mechanism to damp radial oscillations. The results are shown in Figs.~\ref{fig:TdampContoursMmpn13}-\ref{fig:HamburgExperimentReachPlotYukawa}.
We describe the thermal effect on the potential from this interaction and details on the damping rates. The thermal potential excludes the low-mass regime identified above because of the competition between the $T=0$ light radial masses and the thermal effects, but solutions for the high-mass regime remain. To  describe the non-linear dynamics of this system, we cast the equations of motion into a system of Boltzmann equations which we integrate numerically. 
The damping temperatures that are realized with this damping mechanism are reported in Fig.~\ref{fig:TdampContoursMmpn13}.
As reported in Fig.~\ref{fig:maxFlucPlotMmpn13}, we find that it is necessary to assume that $\mathcal{P}_{\mathcal{R}}(\kmin)$ is smaller than the Planck reference value by a factor of about $10^{-3}$ or $10^{-4}$. Assuming this, a parameter space with $10^{-6}~{\rm eV} \lesssim m_a \lesssim 10^{-1}~{\rm eV}$ can be viable. 
In a large range of parameter space, the QCD axion as well as a general ALP can be dark matter and in the reach of experiments such as ALPS-II, MADMAX, BRASS, BREAD and IAXO.
The parameter space also overlaps with the regime in which there may be an observable signal from gravitational lensing of axion miniclusters.
As shown in Fig.~\ref{fig:EFoldsContoursMmpn13}, our solutions allow to measure the implied duration of kination. Kination generally gives rise to a characteristic amplification of primordial spectra of gravitational waves~\cite{Gouttenoire:2021wzu,Co:2021lkc,Gouttenoire:2021jhk}. 
Unfortunately, this GW amplification appears in tension with the need for the inflationary fluctuations, $\mathcal{P}_{\mathcal{R}}(\kmin)$, to be suppressed on the scales relevant for kination, for the homogeneous axion mode to explain DM. In the absence of knowledge of a realistic model of inflation that suppresses $\mathcal{P}_{\mathcal{R}}(\kmin)$ without  suppressing the primordial GW background on the same scales $\kmin$, we do not show GW predictions in this work.
The parameter space that can be made viable with this implementation is summarized for $n=7,10$ and 13 in Fig.~\ref{fig:summaryPlotNQYukawaAndHiggs}.

In Sec.~\ref{sec:NQHiggs}, we investigated damping through a Higgs-portal interaction as an alternative to damping through a KSVZ-like Yukawa interaction. The results are shown in Figs.~\ref{fig:TdampContoursMmpn13NQHiggs} to \ref{fig:minMixAnglePlotNQHiggsMmpn13}.
Similarly to Sec.~\ref{sec:Nearly quadratic Yukawa section}, we describe the thermal effects and damping rates. 
Assuming that the Higgs mass contribution from this interaction can be canceled (i.e. there exists a solution to the usual hierarchy problem), a region of the $[m_a,f_a]$ parameter space that is excluded in the Yukawa-scenario can be made viable with damping through the Higgs portal. The most important difference is that thermal $\phi$ relic constraints are relaxed because the radial mode decay is mediated by Higgs mixing, see Eq.~\eqref{eq:lepton decay}. As a result, solutions with axion masses down to $m_a\sim10^{-7}$ eV can be realized as DM via KMM, which increases the overlap with projections from haloscopes, the helioscope IAXO(+) as well as the region of interest for minicluster lensing, see Fig.~\ref{fig:maxFlucPlotMmpn13NQHiggs}. Similar to the Yukawa case, $\mathcal{P}_{\mathcal{R}}(\kmin)$ must be suppressed relative to the Planck value by at least two to three orders of magnitude (Fig.~\ref{fig:maxFlucPlotMmpn13NQHiggs}).
We have verified that the mixing angle between the SM Higgs and the radial mode is small enough to evade the current LHC constraint as shown in Fig.~\ref{fig:minMixAnglePlotNQHiggsMmpn13}. The damping temperatures that are realized with this damping mechanism are reported in Fig.~\ref{fig:TdampContoursMmpn13NQHiggs}.
The viable parameter space for $n=7, 10, $ and 13 is summarized in Fig.~\ref{fig:summaryPlotNQYukawaAndHiggs}.

In Sec. \ref{sec:quartic model}, we investigated models based on a quartic potentials. These are modified from the setup outlined in \cite{Co:2019jts,Co:2020jtv,Co:2020xlh} by using a Hubble-induced mass instead of de Sitter fluctuations to drive a large initial radial VEV. The main motivation for this change is to alleviate the isocurvature and domain wall issues which challenge KMM scenarios with light PQ fields during inflation. In our implementation, such problems are avoided because both PQ components  are heavy during inflation compared to the Hubble parameter. 
In this way, the required tuning in the quartic coupling $\lambda $ is strongly reduced from $\lambda < 10^{-22}$, set by isocurvature, to about $\lambda \lesssim 10^{-6}$. Nevertheless, much lower values of $\lambda$ are required to reach lower $m_a$, see Appendix \ref{app:parameter values}. 
Because of these low values of $\lambda$, models with quartic potentials are less natural than SUSY-motivated models based on nearly-quadratic potentials. The case for quartic potentials is further weakened by the need to set the amplitude of the CP-violating potential by hand, see Eq.~\eqref{eq:value of A required by quartic models}, for the kick to yield a roughly circular rotation with $\epsilon\sim\mathcal{O}(1)$. Such problem is not present in models based on nearly-quadratic potentials.
If these assumptions are permitted, then the models studied in Sec.~\ref{sec:quartic model} give rise to similar axion parameter space as in the nearly-quadratic scenario. The results for early damping is given in Fig.~\ref{fig:quarticHubble} and for thermal damping in Fig.~\ref{fig:maxFlucPlotQHigherDimYukawaMmpn13}. Interestingly, this model features a regime in which neither a KSVZ-like Yukawa coupling nor a Higgs portal interaction is required for damping the radial mode, see Fig.~\ref{fig:quarticHubble13WithAxionDamping}, as the decay of the radial mode into axions is enough to damp radial motion. Models based on quartic potentials require a similar amount of $\mathcal{P}_{\mathcal{R}}(\kmin)$ suppression as required for nearly-quadratic potentials, see Fig.~\ref{fig:maxFlucPlotQHigherDimYukawaMmpn13}.

In Appendix \ref{app:parameter values}, we collect a large number of additional  figures scanning  the full parameter space in both nearly-quadratic and quartic models  (radial mode mass, Yukawa coupling, Higgs-portal coupling and Higgs quartic coupling) for completeness.

While all these scenarios only weakly depend on the reheat temperature of the universe, they are not independent of the inflation physics. We showed that 
the primordial power spectrum has to be suppressed by typically 4 to 7 orders of magnitude at high wavenumbers. As discussed in Section \ref{sec:power},
this motivates a class of two-step inflationary models that are discussed in \cite{SuppressedPowerPaper}.

To close this summary, let us comment on possible scenarios which are not covered in this paper. We have assumed that the radial mode is relaxed to the potential minimum $\phi \simeq f_a$ when the axion fragmentation starts. However, in principle, fragmentation could happen even if the radial mode is still rolling or oscillating. Such a scenario requires a detailed analysis of the dynamics of the radial and angular modes together. Another unexplored possibility is a scenario which does not satisfy the homogeneity condition Eq.~\eqref{As max}. This scenario also  has a rich phenomenology and will be presented in a future work \cite{Eroncel:2022b}. Finally, we expect that the discussions in this paper remain relevant in stringy UV completions, as the axion kinetic energy cannot be dealt with independently, and damping the motion of other scalar moduli arising from the string compactififcation also needs to be addressed.

\acknowledgments
We thank Raymond Co, Yann Gouttenoire, Keisuke Harigaya, Hyungjin Kim, Peera Simakachorn, and Pablo Quílez for discussions.
This work is supported by the Deutsche Forschungsgemeinschaft under Germany’s Excellence Strategy - EXC 2121 “Quantum Universe” - 390833306. 
The work of RS is supported in part by JSPS KAKENHI No.~23K03415, No.~24H02236, and No.~24H02244. During the preparation of this article,
C.E. received support from the 2236 Co-Funded Brain Circulation Scheme2 (CoCirculation2) of the Scientific and
Technological Research Council of Türkiye TÜBİTAK (Project No: 121C404).

\clearpage


\appendix
\addtocontents{toc}{\protect\setcounter{tocdepth}{1}}

\section{Table of subscripts, abbreviations and symbols}\label{app:notation}
In this appendix, we provide a list of subscripts, abbreviations and symbols that are used throughout this paper. We here use $ q $ as a placeholder for other quantities.

\paragraph{List of subscripts:}
\begin{description}
	\setlength\itemsep{1mm}
	\item[$ {q_\frag} $ :] Quantity measured at the time where the rotation of the PQ field ends, i.e. the time of trapping. Around this time, the field may either fragment, begin oscillating or become frozen.
	\item[$ {q_0} $ :] Quantity measured at zero-temperature (i.e., not necessarily today).
	\item[$ {q_{\rm ap}} $ :] Quantity measured at the apoapsis (i.e., the top) of the orbit. Only used with $ \dthetaap $.
	\item[$ {q_\damp} $ :] Quantity measured at the time of damping.
	\item[$ {q_{\rm diluted}} $ :] Quantity which has been diluted by the entropy injection associated with radial damping.
	\item[$ {q_\dom} $ :] Quantity measured at the time when the energy density of the PQ field starts to dominate the energy density of the universe if such domination takes place.
	\item[$ {q_\eq} $ :] Quantity measured at the time of radiation-matter equality.
	\item[$ {q_{\rm heated}} $ :] Quantity which has been heated by the entropy injection associated with radial damping.
	\item[$ {q_I} $ :] Quantity measured during inflation.
	\item[$ {q_\kick} $ :] Quantity measured at the time of the kick, i.e. at the onset of rotation of the PQ field.
	\item[$ {q_\kin} $ :] Quantity measured at the time where $ \phi $ reaches the minimum at $ f_a $, which corresponds to the onset of kination-like scaling of the axion.
	\item[$ {q_{\rm today}} $ :] Quantity measured today.
\end{description}

\paragraph{List of symbols:}
\begin{description}
	\setlength\itemsep{1mm}
	\item[$ \alphachi $ :] The SU(3)$_{\rm QCD}$ structure constant $ \alphachi \equiv \gchi^2 / 4\pi $. At the  temperatures relevant for damping we take $\alphachi\sim 0.1$. 
	\item[$ \alphaweak $ :] The $SU(2)$ structure constant. We take $\alphaweak\sim 1/30$.
	\item[$ \gamma $ :] Symbol both used for the photon 
 and for the parameter which quantifies the temperature dependence of the axion mass $ m_a(T) $ (see Eq.~\eqref{eq:thermal potential}).
	\item[$ \Gamma $ :] Damping rate or interaction rate, usually specified by a subscript.
	\item[$ \epsilon $ :] Parameter which measures the ellipticity of the initial rotation. By construction, $ 0\leq \epsilon \leq 1 $ where $ \epsilon=0 $ corresponds to perfectly radial oscillation and $ \epsilon=1 $ corresponds to a perfectly circular rotation, see Eq.~\eqref{eq:epsilon equivalence}.
	\item[$ \theta $ :] The angular mode of the PQ field $ P $, also refereed to as the axion field. Can refer to either a QCD axion or a general axion-like-particle.
	\item[$ \lambda $ :] Quartic coupling in the potential for $P$, see Eq.~\eqref{eq:quartic potential}.
	\item[$ \xi $ :] Coupling between the radial mode $\phi$ and the Higgs, see Eq.~\eqref{eq:phi-higgs coupling}.
	\item[$ \rho_q $] Energy density of q. In particular, $ \rho_a $ is the axion energy density.
	\item[$ \phi $ :] The radial mode of the PQ field $ P $ see Eq.~\eqref{eq:P definition}.
	\item[$ \Omega_{q} $ :] Critical energy density of the universe of component $ q $, i.e. $ \rho_q / \rho_{\rm tot}, $ where $ \rho_{\rm tot} $ is the total energy density of the universe.
	\item[$ \Omega_{\rm GW}(f) $ :] Fraction of energy density in gravitational waves per logarithmic frequency, see Eq.~\eqref{eq:GW peak}.
	\item[$ a $ :] The scale factor of the universe;
	\item[$ A $ :] Dimension-full parameter of the PQ-violating higher-dimensional terms, see Eq.~\eqref{eq:higher dimensional PQ breaking potential}.
	\item[$ \mathcal{P}_{\mathcal{R}}(k) $ :] Amplitude of the primordial power spectrum measured on comoving momentum scale $ k $.
	\item[$ \Asplanck $ :] The value of $ \mathcal{P}_{\mathcal{R}}(k) $ measured by Planck~\cite{Planck:2018vyg} at the comoving momentum scale $ k = 0.05  $ Mpc$^{-1} $, which is referred to as the pivot scale. $ \Asplanck = \approx 2.1\times 10^{-9} $.
	\item[$c_q$ :] Redshift parameter which indicates that quantity q redshifts as $ q\propto a^{-c_q} $.
	\item[$ f_a $ :] Decay constant of the axion. Also the final VEV of the radial mode.
	\item[$ g_* $ :] Number of relativistic degrees of freedom in the SM plasma.
	\item[$ g_{*s} $ :] Number of effective degrees of freedom in entropy.
	\item[$ h $ :] Scaling factor for Hubble expansion rate, $ h\approx 0.674 $.
	\item[$ H $ :] Hubble parameter. Note that in Section \ref{sec:NQHiggs} the symbol $ H $ may also refer to the Higgs doublet. The distinction should be clear from context.
	
	\item[$ m_\phi $ :] Mass of the radial mode. As we are concerned with dynamics far from the minimum, we define this as $ m_\phi\equiv\sqrt{V'\phi^{-1}} $ so that the EOM takes the standard form.
	\item[$ m_{\phi,DM} $ :] Solution for the radial mode mass which ensures that the observed DM relic is produced by KMM.
	\item[$ m_{\phi,\rm eff},m_{\phi,0},m_{\phi,\rm th}\text{, and }m_{\phi,\rm ln} $ :] The contributions to the radial mode mass are specified when thermal effects are involved. These indicate the effective (total), zero-temperature, thermal, and thermal-log masses, respectively.
	\item[$ m_a(T) $ :] Mass of the axion / angular mode as measured at temperature T.
	\item[$ m_a $ :] Zero-temperature mass of the axion / angular mode.
	\item[$ \mplanck $ :] The reduced Planck mass, $ \mplanck \approx 2.4 \times 10^{18} $ GeV.
	\item[$ M $ :] Suppression scale of higher dimensional terms. In this work, we usually assume $ M=\mplanck $ unless otherwise stated.
	\item[$ n $ :] Dimension of the higher dimensional terms.
	\item[$ n_a $ :] Axion number density.
	\item[$ n_\phi $ :] Radial mode number density.
	\item[$ \nPQ $ :] PQ-charge. The conserved charge related to the axion shift symmetry.
	\item[$ N_{\rm eff} $ :] Effective number of neutrinos. See Appendix \ref{app:Thermal relics}.
	\item[$ P $ :] The full complex PQ field, which is parametrized as $ P = \frac{1}{\sqrt{2}}\phi e^{i\theta} $.
	\item[$ y $ :] Yukawa coupling. See, e.g., Eq.~\eqref{eq:yukawa coupling}.
	\item[$ Y $ :] The yield variable $ Y = \nPQ / s $, where $ s $ is the entropy density.
	\item[$ Y_{\rm DM} $ :] The observed DM yield defined from $ n_{\rm DM}/s $ where $ n_{\rm DM} $ is the number density of DM particles and $ s $ is the entropy density. It is conveniently parametrized as $ Y_{\rm DM} \approx 0.64 \Teq / m_a$.
	
\end{description}

\paragraph{List of abbreviations:}
\begin{description}
	\setlength\itemsep{1mm}
	\item[ALP] Axion-like-particle. In this work, the word \textit{axion} is used synonymously with ALP.
	\item[BBN] Big Bang Nucleosynthesis
	\item[CMB] Cosmic Microwave Background
	\item[DW] Domain wall
	\item[EOM] Equation of motion
	\item[GW] Gravitational wave
	\item[KMM] Kinetic misalignment mechanism
	\item[KSVZ] Kim-Shifman-Vainshtein-Zakharov. Our benchmark QCD axion model.
	\item[PQ] Peccei-Quinn, as related to the PQ solution of the strong CP problem. Here used to refer to the PQ field $ P $ or the PQ charge $ \nPQ $ related to the axion shift symmetry. Note that as we are working with general ALP scenarios, the reference to Peccei and Quinn here is often by analogy and goes beyond the QCD axion scenario.
	\item[QCD] Quantum Chromodynamics. Often used in conjunction with the term ``QCD axion", which refers to the axion which solves the strong CP problem of QCD.
	\item[SM] Standard model
	\item[SMM] Standard misalignment mechanism
	\item[SUSY] Supersymmetry
\end{description}

\section{Thermal relics}
\label{app:Thermal relics}
\newcommand{\Tnr}{T_{\rm nr}}
\newcommand{\Tsa}{T_{\rm \phi a}}
In this work, we study the process in which oscillations of the radial mode are damped. Generically, this is achieved by introducing an interaction which efficiently connects the radial mode to the thermal plasma. While this solves the issue of excessive energy stored in radial oscillations, it also introduces a thermal population of $\phi$ particles. In this appendix, we discuss the resulting constraints.

If the radial mode remains in thermal contact with the plasma indefinitely, and $\mphi \gg \TBBN$, then this population will become Boltzmann suppressed without posing a conflict with $N_{\rm eff}$. However, some damping interactions considered in this work are only efficient at high temperatures. Therefore, a thermal $\phi$ relic may freeze out at some temperature $\Tfo \gg \TBBN$.
At $\Tfo$, the thermal relic makes up a fraction $\rho_{\phi,\rm thermal} /\rho_r \sim 1/g_*$. Except for reheating, this ratio will remain constant until the thermal $\phi$ relic becomes non-relativistic at some $\Tnr$. Because $\Tnr$ refers to the temperature of the reheated SM plasma, it differs slightly from $T\sim \mphi$ and is
\begin{gather}
 \Tnr \approx \left(\frac{g_{*s}(\Tfo)}{g_{*s}(\Tnr)}\right)^{1/3}\mphi.
\end{gather}
Below this temperature, the thermal $\phi$ behaves as cold dark matter. If $\mphi$ is light enough for the relic to be stable against decay, then, at radiation-matter equality we find 
\begin{gather}
 \frac{\rho_{\phi,\rm thermal}}{\rho_{\rm DM}}\Big\rvert_{\Teq} \approx \frac{1}{g_*(\Teq)}\frac{g_{*s}(\Teq)}{g_{*s}(\Tfo)}\frac{\mphi}{\Teq}  ,
\end{gather}
where we made use of $\rho_{\rm r}(\Teq) \approx \rho_{\rm DM}(\Teq)$. To avoid overproduction of DM we arrive at
\begin{gather}
 \text{Low mass window:}\quad	\mphi < \mathcal{O}(1) \times \Teq  . \label{eq:radial thermal lower bound}
\end{gather}
This condition defines the low-mass regime of solutions observed e.g. in Fig.~\ref{fig:nearlyQuadraticEarlyDamping1}. If instead $\mphi$ is large then the relic will decay directly into axions. The decay rate $\Gamma_{\varphi a}$ is given by Eq.~\eqref{eq:radial-to-axion rate}, which becomes efficient at a temperature $\Tsa$ defined by $\Gamma_{\varphi a}\sim H(\Tsa)$. The decay product is a hot population of axions which constitutes dark radiation subject to constraints from $\Delta N_{\rm eff}$. 
Assuming that the thermal relic evolves as cold DM between $\Tnr$ and $\Tsa$ and as dark radiation at all other times, one can derive
\begin{gather}
	\Delta N_{\rm eff}(T) \approx 2 \frac{f_a}{\sqrt{\mplanck\mphi}}\frac{g_{*s}^{4/3}(T)}{g_{*s}(\Tfo)g_{*}^{1/12}(\Tsa)}.
\end{gather}
$ N_{\rm eff} $ is constrained directly at BBN to $ \Delta N_{\rm eff}\lesssim 0.4 $ \cite{Fields:2019pfx} and at CMB to $ \Delta N_{\rm eff}\lesssim 0.3 $~\cite{Zyla:2020zbs}. For simplicity, we demand that $ \Delta N_{\rm eff} $ does not exceed 0.3 at either $ \TBBN $ or at around the time of CMB, i.e. $ T\sim $ eV $ \sim \Teq $. At large enough $\mphi$, $\phi$-to-axion decay terminates cold DM evolution early enough so that the mass range
\begin{gather}
	 \text{High mass window:}\quad \mathcal{O}(1)\times\frac{f_a^2}{\mplanck} < \mphi\label{eq:radial thermal upper bound}
\end{gather}
is compatible with $\Delta N_{\rm eff}$ constraints. In summary, if the damping interaction does not maintain $\phi$ in thermal equilibrium with the plasma until $T \sim \mphi$, then a constraint from the thermal $\phi$ relic must be considered. This relic can be avoided through either Eq.~\eqref{eq:radial thermal lower bound}, by making $\mphi$ light enough for the relic to underproduce DM, or through Eq.~\eqref{eq:radial thermal upper bound}, by allowing $\phi$ to decay into a hot but subdominant axion relic.

\section{Summary of damping rates}\label{app:damping rates}
In the following we summarize the processes and associated damping rates which are taken into account in description of thermal damping. As the majority of these are quoted from literature, primarily \cite{Mukaida:2012qn} and \cite{Mukaida:2013xxa}, they are presented without derivation. For more details, we refer to the cited sources. All the rates listed here are taken into account in our numerical analysis.

\subsection{Model-independent process}
Regardless of the coupling added to the PQ field to achieve damping, the radial excitations always have the possibility to decay into the much lighter angular mode. We refer to this process as $\phi$-to-axion decay, which leads to a damping rate of the form:
\begin{table}[ht!]
	\centering
	\begin{tabular}{c|c|c|c}
		Rate  & Assumptions & Reference & Mechanism\\
		\hline
		$ \Gamma_{\varphi a}\approx\frac{1}{64\pi} \frac{m_\phi^3}{\phi^2} $ & \makecell{$ m_\phi > 2 m_a $} & Eq.~\eqref{eq:radial-to-axion rate}, this work  &  $\phi$-to-axion decay  \\
	\end{tabular}
\end{table}

\noindent
In addition to $\phi$-to-axion decay, we always assume that the axion has KSVZ-like EM couplings, which would also generate radial mode EM couplings. However, $\phi$-to-photon decay through the direct EM-coupling is subdominant to $\phi$-to-axion decay, see discussion around Eq.~\eqref{eq:radial-to-axion rate}.

\subsection{Fermion-induced damping rates}
We here consider damping processes arising from the fermion Yukawa Eq.~\eqref{eq:yukawa coupling}, of the form
\begin{gather}
    \mathcal{L}_{\rm int} = \sqrt{2}y P\bar{\chi} \left( \frac{1-\gamma^5}{2}\right) \chi+h.c.
\end{gather}
The damping rates arising from this coupling were derived by Mukaida et al. in \cite{Mukaida:2012qn,Mukaida:2012bz}. The rates we take into account in our study are following:
\begin{table}[h!]
	\centering
	\begin{tabular}{c|c|c|c}
		Rate  & Assumptions & Reference & Mechanism\\
		
		\hline
		$ \Gamma \approx y^2 \alphachi T$ & \makecell{$ y \phi < \alphachi T $} & Eq.~3.35,\cite{Mukaida:2012qn}  & $ \chi $-scattering\\
		
		\hline
		$ \Gamma \approx y^4 \frac{\phi^2}{\alphachi T} $ & \makecell{$ \alphachi T < y \phi < T $} & Eq.~3.35,\cite{Mukaida:2012qn} &$ \chi $-scattering\\
		
		\hline
		$ \Gamma \approx b\alphachi^2 \frac{T^3}{\phi^2} $ & $ T<y \phi $ & Eq.~3.16,\cite{Mukaida:2012qn} & gluon scattering \\
		
		\hline
		$ \Gamma \approx \frac{1}{8\pi}y^2 \mphi $ & $ \phi^2>4(y^2\phi^2+g_\chi^2T^2) $ & Eq.~3.37,\cite{Mukaida:2012qn} & decay into $ \chi $\\
		
		\hline
		$ \Gamma \approx b \alphachi^2 \frac{\mphi^3}{\phi^2} $ & $ 4(y^2\phi^2+g_\chi^2T^2)>\mphi^2>4g_\chi^2T^2 $ & Eq.~5.7,\cite{Mukaida:2012qn} & decay into gluons

	\end{tabular}
\end{table}

\noindent
Here $g$ is the gluon.
We here assume $ b\sim 10^{-5} $ and $ \alphachi\sim 0.1 $, from Eq.~3.30 of \cite{Co:2020jtv}, as these value are representative for strong interactions around the relevant temperatures.

\newpage
\subsection{Higgs-induced damping rates}
We here consider interactions arising from the Higgs portal interaction, \eqref{eq:phi-higgs coupling} of the form
\begin{gather}
    \mathcal{L}\supset \xi^2 (\phi^2-f_a^2) \left(H^\dagger H- \frac{v^2}{2}\right).
\end{gather}
In the limit $ \phi \gg f_a $, this scenario corresponds to the scenario studied by Mukaida et al. in \cite{Mukaida:2013xxa} from which we quote the following damping rates:

\begin{table}[h!]
	\hspace{-0.45cm}	\begin{tabular}{c|c|c|c}
		Rate  & \makecell{Assumptions / \\ Range of validity} & Reference & Mechanism\\
		\hline
		
		$ \Gamma\approx \frac{b \alphaweak^2 T^3}{\phi^2}$ & \makecell{ $\phi \gg f_a $\\ $ \mphi \ll \alphaweak T $ \\
			$ \xi \phi \gg \max\left[T,m_{H,0}\right] $
		} & \makecell{Eq.~(2.9),\cite{Mukaida:2013xxa} \\ Eq.~(3.16),\cite{Mukaida:2012bz}} &  Loop decay ($ \phi $-mass)\\
		
		\hline
		$ \Gamma\approx N_{\rm dof} \frac{\xi^4\phi^2}{\pi^2\alphaweak T} $ & \makecell{
			$ \mphi \ll \alphaweak T $ \\
			$ \xi^2 \phi^2-\xi^2f_a^2+\mHiggs^2 \ll y_t^2 T^2 $ \\
		} & \makecell{Eq.~(2.11),\cite{Mukaida:2013xxa} \\ Eq.~(3.35),\cite{Mukaida:2012qn}}&  Loop decay (T-mass) \\
		
		\hline
		$ \Gamma\approx N_{\rm dof} \frac{\xi^4 \phi^2}{8\pi \mphi} $ & \makecell{$m_\phi > 2 m_{H,\rm eff}$}\tablefootnote{This range of validity is not explicitly stated in \cite{Mukaida:2012qn,Mukaida:2013xxa}.} & Eq.~(2.13),\cite{Mukaida:2013xxa} &  Perturbative $ \phi\to H H $ decay \\
		
		\hline
		$ \Gamma\approx N_{\rm dof} \frac{\xi^4 T^3}{12\pi \mHiggsth^2} $ & \makecell{
			$ \mphi \ll \alphaweak T $ \\
			$ \xi^2 \phi^2-\xi^2f_a^2+\mHiggs^2 \ll y_t^2 T^2 $
		} & Eq.~(2.12),\cite{Mukaida:2013xxa} &  $ \phi H \to \phi H $ (condensate$ \to $particle)\tablefootnote{Our model does not account for a scenario with significant $ \phi $-particle production, so we do not allow for this process to be efficient while  $ \phi $ dominates the energy density.}\\
		
		\hline
		$ \Gamma\approx N_{\rm dof} \frac{\xi^4 T^2}{48\pi \mphi}$ & \makecell{$ \alphaweak T \ll m_\phi \ll T $ \\$ \xi^2 \phi^2-\xi^2f_a^2+\mHiggs^2 \ll y_t^2 T^2 $ } & Eq.~(2.14),\cite{Mukaida:2013xxa} & $ H $-scattering \\
		
	\end{tabular}
\end{table}

\noindent
We take $ b\sim 10^{-3} $,  $\alphaweak\sim 1/30 $, $ N_{\rm dof}=2 $ and $ y_t\sim 1 $. The ranges of validity have been modified to account for the non-zero late-time Higgs mass. Mukaida et al. also describe a non-perturbative decay process which arises from a parametric resonance, but this does not apply to our scenario because the $ \phi $ oscillates around the equilibrium point at $ \expval{\phi} \gg f_a $ and not $ \phi \sim 0 $. Here, the labels "$T$-mass" and "$\phi$-mass" indicate how the behavior of the loop decay process changes according to whether the second or third term of Eq.~\eqref{eq:Higgs effective mass} dominates, respectively.

As discussed in Section \ref{sec:NQHiggs}, the Higgs portal interaction induces a mixing angle 
\begin{gather}
	\vartheta_{\phi H}\approx-2\xi^2\frac{f_a v_{\rm EW}}{m_h^2-m_\phi^2}.
\end{gather}
Here we assumed a small mixing angle limit $|\vartheta_{\phi H}| \ll 1$.
This radial mode-Higgs mixing induces damping through a number of decay channels. For radial mode masses lower than the GeV scale the relevant decays are~\cite{Bezrukov:2009yw}
\begin{gather}
    \text{Photons:}\quad \Gamma_{\phi \gamma\gamma} =  \vartheta_{\phi H}^2 \frac{F_{\rm photon}^2(\mphi)}{16\pi}\frac{\mphi^3}{v_{\rm EW}^2} \left(\frac{\alphaEM}{4\pi}\right)^2, \\
    \text{Pions:}\quad \Gamma_{\phi\pi\pi} = \vartheta_{\phi H}^2 \frac{3}{16\pi}\frac{\mphi^3}{v_{\rm EW}^2}\left(\frac{2}{9}+\frac{11}{9}\frac{m_\pi^2}{\mphi^2}\right)^2\sqrt{1-\frac{4m_\pi^2}{\mphi^2}}, \\
    \text{Leptons:}\quad \Gamma_{\phi f f} = \vartheta_{\phi H}^2\frac{1}{8\pi}\frac{m_f^2}{v_{\rm EW}^2} \left( 1-\frac{4 m_f^2}{\mphi^2} \right)^{3/2} \qq{for} \mphi>2m_f,
\end{gather}
where $F_{\rm photon}$ is a loop factor $F$ given in Eq.~(3.4) of \cite{Bezrukov:2009yw}.
For radial mode masses above the GeV scale, in addition to decays into photons and leptons, we have~\cite{Sato:2011gp}:
\begin{gather}
    \text{Quarks:}\quad \Gamma_{\phi q q} = \vartheta_{\phi H}^2\frac{3}{8\pi}\frac{m_q^2}{v_{\rm EW}^2} \left( 1-\frac{4 m_q^2}{\mphi^2} \right)^{3/2} \qq{for} \mphi>2m_q.
\end{gather}
and decays into the remaining SM gauge bosons become relevant:~\cite{Bezrukov:2009yw,Sato:2011gp}
\begin{gather}
    \text{Gluons:}\quad \Gamma_{\phi gg} =  \vartheta_{\phi H}^2 \frac{F_{\rm gluon}(\mphi)}{8\pi}\frac{\mphi^3}{v_{\rm EW}^2} \left(\frac{\alpha_s}{4\pi}\right)^2, \\
    \text{$W$ bosons:}\quad \Gamma_{\phi WW} = \vartheta_{\phi H}^2\frac{1}{16\pi}\frac{\mphi^3}{v_{\rm EW}^2} \left( 1-\frac{4 m_W^2}{\mphi^2} +\frac{12 m_W^4}{\mphi^4} \right)\sqrt{1-\frac{4 m_{W}^2}{\mphi^2}} \qq{for} \mphi>2m_W, \\
    \text{$Z$ bosons:}\quad \Gamma_{\phi ZZ} = \vartheta_{\phi H}^2\frac{1}{32\pi}\frac{\mphi^3}{v_{\rm EW}^2} \left( 1-\frac{4 m_Z^2}{\mphi^2} +\frac{12 m_Z^4}{\mphi^4} \right)\sqrt{1-\frac{4 m_{Z}^2}{\mphi^2}} \qq{for} \mphi>2m_Z,
\end{gather}
where $F_{\rm gluon}$ is the loop factor  $F$ given in Eq.~(3.9) of \cite{Bezrukov:2009yw} for colored degrees, which is analogous to $F_{\rm photon}$.
Finally, the decay into Higgs bosons also contributes:~\cite{Sato:2011gp}
\begin{gather}
    \text{Higgs bosons:}\quad\Gamma_{\phi hh} = \vartheta_{\phi H}^2\frac{1}{32\pi}\frac{\mphi^3}{v_{\rm EW}^2} \left( 1 + \frac{2 \mHiggseff^2}{\mphi^2}  \right)^2\sqrt{1-\frac{4 \mHiggseff^2}{\mphi^2}} \qq{for} \mphi>2\mHiggseff.
\end{gather}
We include all of the above damping channels in our numerical solution of the Boltzmann equations. In the latter equation, the physical Higgs mass $\mHiggseff$ is given by Eq.~\eqref{eq:Higgs effective mass}.

\section{Restrictions on the reheating temperature}\label{app:TReheat}

In this work, we generally allow for any reheating temperature between $ \Tkick $ and the energy scale of inflation $ E_I $. However, there is a regime in which the choice of reheating temperatures is restricted by the consistency of the scenario, which we explore in this appendix. This discussion is specific to the nearly-quadratic models.

Because thermal effects respect the PQ symmetry and only enhance the radial potential and not the angular potential, $ \epsilon $ and $Y$ are suppressed in scenarios where the radial mode is dominated by the thermal mass contributions at the time of the kick. We therefore favour solutions where $ \mphiz $ dominates at the time of the kick. Such $ \mphiz $ dominated scenarios can be realized in the regime where the $ \chi $-fermions are Boltzmann suppressed, and the thermal potential is reduced from $ \mth $ to the much smaller $ \mln $ at the time of the kick. However, this scenario is only consistent if there is no earlier time in which an early phase of relativistic $ \chi $-fermions contribute a thermal mass large enough to start a kick. 

\begin{figure}
	\centering
	\includegraphics[width=\standardwidth \textwidth]{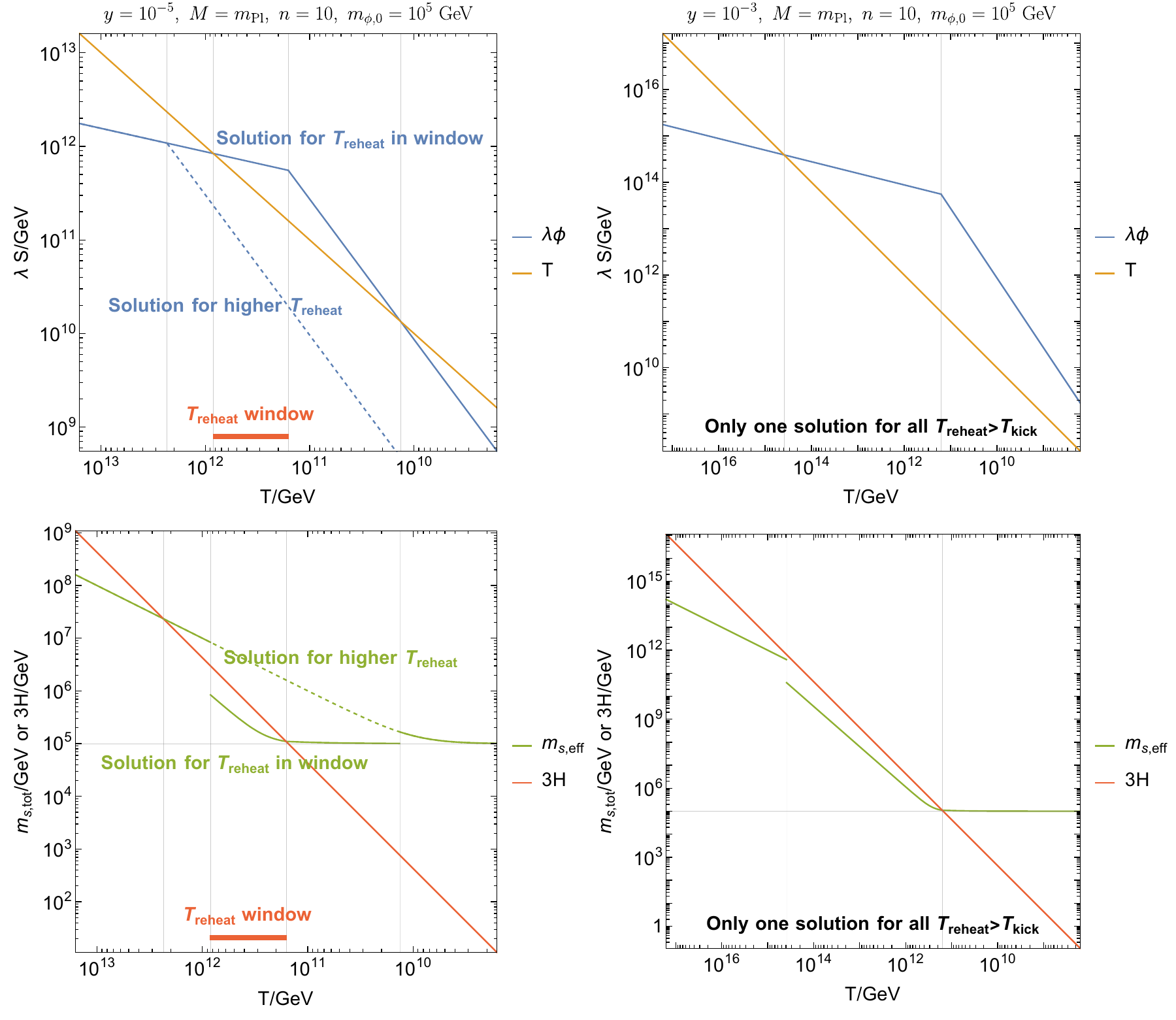}
	\caption{\small\it This figure illustrates how $ T_{\rm reheat} $ has an impact on which type of kick is realized. \textbf{Left:} If $ T_{\rm reheat} $ falls in the indicated window then a kick with non-relativistic fermions and $ m_{\phi,\rm eff} \approx m_{\phi,0} $ will be realized. If $ T_{\rm reheat} $ is larger than this window, then a kick will be given with $ m_{\phi,th}\gg m_{\phi,0} $ instead. The evolution corresponding to this early kick is indicated with the dashed lines. \textbf{Right:} For this choice of parameters the kick is given with $ m_{\phi,\rm eff} \approx m_{\phi,0} $ for any $ T_{\rm reheat} > \Tkick $ because the fermions become non-relativistic before a $ \mth $-dominated kick can take place, even if the $ T_{\rm reheat} $ is assumed to be arbitrarily high. The transition between scenarios in which either type of kick is possible depending on $ T_{\rm reheat} $ and scenarios in which a $ \mphiz $-dominated kick is realized regardless of $ T_{\rm reheat} $ leads to the discontinuity is seen around $ y\approx 10^{-3.5} $ in Fig.~\ref{fig:baseVersion}.}
	\label{fig:discontinuityExplaination}
\end{figure}

\begin{figure}
	\centering
	\includegraphics[width=0.7 \textwidth]{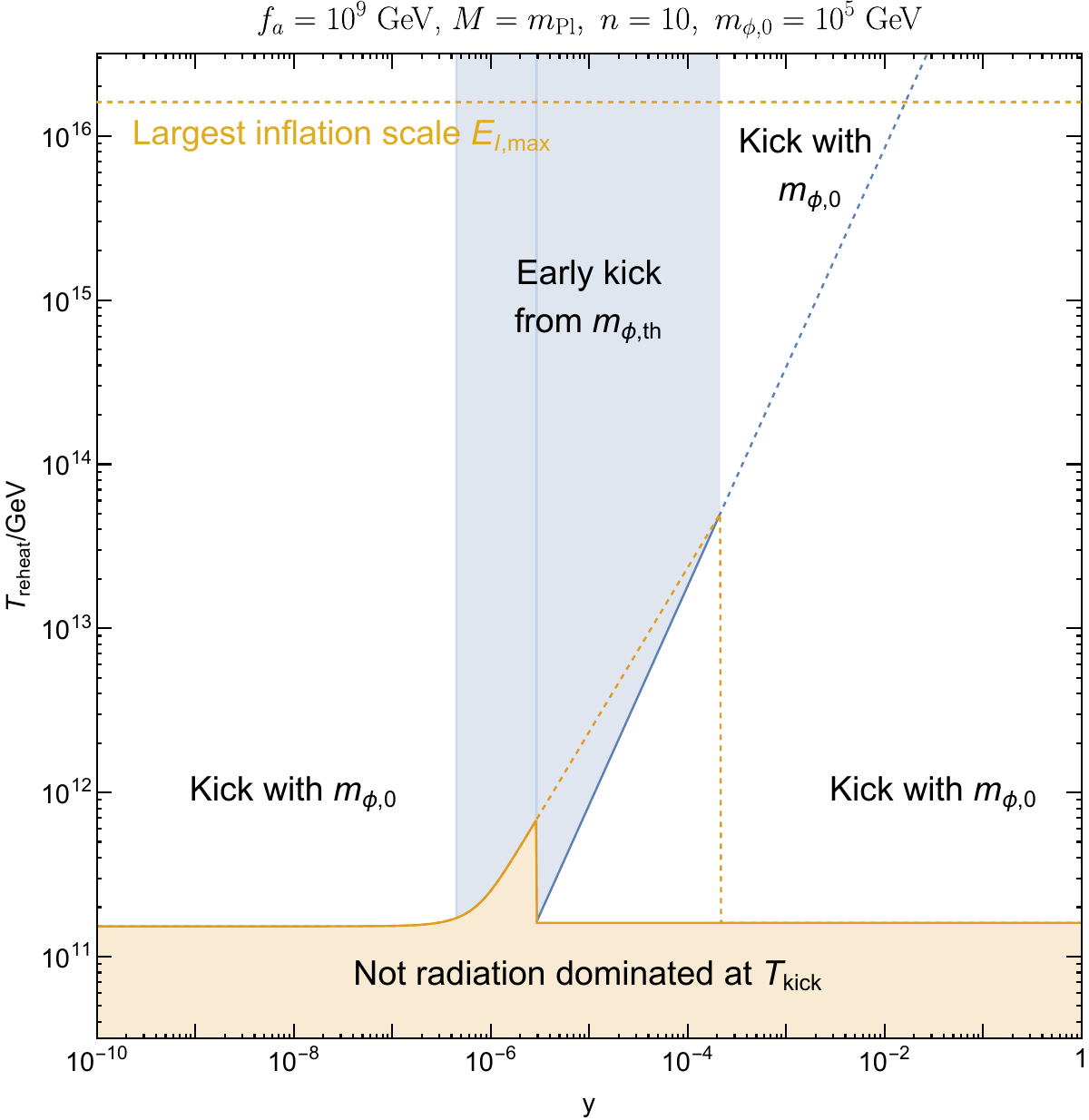}
	\caption{\small\it Impact of $ T_{\rm reheat} $ on the choice of kick type. In the white region the kick is realised with $ m_{\phi,\rm eff}\approx m_{\phi,0} $ while in the blue region the kick is realised with $ m_{\phi,\rm eff}\approx m_{\phi,\rm th} $. The orange range corresponds to $ T_{\rm reheat}<\Tkick $, which we do not consider in this work. \textbf{In the central range of }$ \mathbf{y} $, approximately $ 10^{-5.5} \lesssim y \lesssim 10^{-3.5} $, the choice of $ T_{\rm reheat} $ can determine which type of kick is realized. This $ T_{\rm reheat}$-dependent region corresponds to the left column of Fig.~\ref{fig:discontinuityExplaination}. \textbf{The high-$ \mathbf{y} $ region}, approximately $ 10^{-3.5} \lesssim y $, in which only the late $ m_{\phi,\rm eff}\approx m_{\phi,0} $ kick can be realized corresponds to the right column of Fig.~\ref{fig:discontinuityExplaination}. \textbf{In the low-$ \mathbf{y} $ region}, i.e. $ y \lesssim 10^{-6.5} $, the kick can only take place with $ \mphiz $ because a $ \mth $-dominated kick fails to take place before the zero-temperature mass becomes dominant. \textbf{Lines:} The orange dashed line corresponds to $ \Tkick $ in the $ \mth $-dominated solution and lowest temperature for which $ T > y\phi $ corresponds to blue line. \textbf{Other remarks:} Note that the parameters here are fixed such that the correct DM-yield is realized with the $ \mphiz $-dominated kick, i.e. in the white parameter space. For simplicity, we have not imposed any other constraints on this plot. Therefore, the white parameter space needs \textit{not} correspond to physically viable scenarios.}
	\label{fig:baseVersion}
\end{figure}

To see how a would-be $ \mphiz $-dominated kick can be spoiled by an earlier phase of relativistic $ \chi $-fermions, note that prior to the kick $ \phi $ will track the early potential minimum given by Eq.~\eqref{eq:phi minimum early}, which evolves as
\begin{gather}
\phi \propto T^{\frac{2}{n-2}} \quad \text{before }\Tkick.
\end{gather}
Thus, $ \phi $ redshifts slower than $ T $ for any choice of $ n>4 $. Therefore, any era in which $ y \phi > T $ will always be preceeded by an era in which $ y \phi < T $ if $ T_{\rm reheat} $ is arbitrarily high (see first plot of Fig.~\ref{fig:discontinuityExplaination}). In such an earlier phase of relativistic $ \chi $-fermions, $ \phi $ acquires the usual thermal mass $ \mth $, which can be much larger than $ \mln $ and which can therefore lead to an early kick. For some choices of parameters, this phenomenon makes solutions in which the kick is given in the non-relativistic phase, i.e. with $ y \phikick > \Tkick $, inconsistent unless reheating after inflation takes place in the window between the end of any would-be early relativistic phase with $ m_{\phi,\rm th} > 3H $ and $ \Tkick $ itself. A specific example of how different choices $ T_{\rm reheat} $ can choose between two different families of solutions is illustrated in Fig.~\ref{fig:discontinuityExplaination}. An exemplary map of the $ [T_{\rm reheat},y] $ parameter space is given in Fig.~\ref{fig:baseVersion}, where the blue region corresponds to parameter space in which the kick takes place with $ \mth $ and the triangular region around $ y\sim 10^{-5} $ corresponds to region in which an early $ \mth $ kick can be avoided by postulating a sufficiently low $ T_{\rm reheat} $.

In the regime for which either a $ \mphiz $-dominated kick or a $ \mth $-dominated kick is possible depending on $ T_{\rm reheat} $ we  assume that $ T_{\rm reheat} $ is chosen such that the $ \mphiz $ kick is realized. We do not make any further assumptions on $ T_{\rm reheat} $ other than $ E_I> T_{\rm reheat} > \Tkick $. In most of our parameter space, our solutions are compatible with any $ E_I> T_{\rm reheat} > \Tkick $.

\section{Boltzmann equations and solutions of thermal damping}\label{app:boltzmann equations}
In this appendix, we derive the Boltzmann equations, which describe thermal damping. We then describe the numerical scheme we use to solve these systems of equations. We first consider the nearly-quadratic potential and then generalize to the quartic potential. Finally, in Subsection \ref{sec:examples of solutions} we present examples of solutions.

\subsection{Derivation}
We first consider the case of the nearly-quadratic potential. We will, as usual, neglect the logarithmic correction and restrict ourselves to the regime in which higher-dimensional terms are suppressed such that the potential is
\begin{gather}
V\sim \frac{1}{2} \meff^2 \phi^2,
\end{gather}
where we have included the thermal correction in $ \meff $.
The equations of motion for the radial and angular degrees of freedom of the PQ field are (see also Appendix G of \cite{Gouttenoire:2021jhk} for an extensive discussion)
\begin{gather}
\ddot{\phi}+3H\dphi+\Gamma\dphi+\meff^2\phi = \dtheta^2\phi, \\
\ddot{\theta} + 3H\dtheta = -2\frac{\dtheta\dphi}{\phi},
\end{gather}
where $ \Gamma $ is the total dissipation rate. The total energy density of the field is
\begin{gather}
\rho_{\rm tot} = \rho_{\rm kin}^\phi + \rho_{\rm kin}^\theta + \rho_{\rm pot} = \frac{1}{2}\dphi^2 + \frac{1}{2}\phi^2\dtheta^2 + \frac{1}{2}\meff^2\phi^2.\label{eq:rho tot}
\end{gather}
To express the EOM in terms of Boltzmann equations, we consider the derivative of the energy density. If we allow for a time-dependence in the mass term from, e.g., thermal corrections, then we find
\begin{gather}
\dot{\rho}_{\rm tot} = \dphi\ddot{\phi}+\phi\dphi\dtheta^2+\phi^2\dtheta\ddot{\theta}+\meff\dmeff\phi^2+\meff^2\phi\dot{\phi}.
\end{gather}
We then apply the EOM and reduce: 
\begin{gather}
\dot{\rho}_{\rm tot} = -(3H+\Gamma)\dphi^2-3H\dtheta^2\phi^2 + \meff\dmeff\phi^2. \label{eq:equation with m mdot}
\end{gather}
It is convenient to parameterize the time-dependence of $ \meff $ in terms of the parameter $ c_{\mphi} $,
\begin{gather}
\meff \propto a^{-c_{\mphi}}. \label{eq:parameter for mass time-dependence}
\end{gather}
In terms of this parameter $ c_{\mphi} $ the last term of Eq.~\eqref{eq:equation with m mdot} becomes
\begin{gather}
\meff \dmeff\phi^2 = -c_{\mphi} H \meff^2\phi^2,
\end{gather}
so that $ \dot{\rho}_{\rm tot} $ can be written as
\begin{gather}
\dot{\rho}_{\rm tot} = -(6H+2\Gamma)\rho_{\rm kin}^\phi-6H\rho_{\rm kin}^\theta -2c_{\mphi}H\rho_{\rm pot}.
\end{gather}
\paragraph{Virial theorem:} To understand the relation between the potential and kinetic energy of the system, we apply the virial theorem, which for a monomial potential $ \propto \phi^n $ states that
\begin{gather}
\frac{n}{2}\expval V = \expval T \qq{where} n=2 \text{ for a quadratic potential},
\end{gather}
where $ V $ is the total potential energy, and $ T $ is the total kinetic energy. We then have
\begin{gather}
\expval{\rho_{\rm pot}}= \expval{\rho_{\rm kin}^\theta+\rho_{\rm kin}^\phi}.\label{eq:virial theorem for nearly quadratic}
\end{gather}
We henceforth drop the brackets and understand all energy densities as time-averaged quantities. Combining Eq.~\eqref{eq:virial theorem for nearly quadratic} with Eq.~\eqref{eq:rho tot} we conclude that
\begin{gather}
\rho_{\rm tot} = 2\rho_{\rm kin}^\phi + 2\rho_{\rm kin}^\theta,
\end{gather}
so that
\begin{gather}
\dot{\rho}_{\rm tot}= 2\dot{\rho}_{\rm kin}^\phi + 2\dot{\rho}_{\rm kin}^\theta = -(6H+2\Gamma+2c_{\mphi}H)\rho_{\rm kin}^\phi-(6H+2c_{\mphi}H)\rho_{\rm kin}^\theta .
\end{gather}
If we consider $ \rho_{\rm kin}^\phi $ and $ \rho_{\rm kin}^\theta $ to be independent we can separate this equation into two Boltzmann equations:
\begin{gather}
\dot{\rho}_{\rm kin}^\phi = -(3H +\Gamma + c_{\mphi}H) \rho_{\rm kin}^\phi, \label{nearly quadratic boltzmann 1} \\
\dot{\rho}_{\rm kin}^\theta = -(3H + c_{\mphi}H) \rho_{\rm kin}^\theta. \label{nearly quadratic boltzmann 2}
\end{gather}
Let us denote the energy of the circular orbit corresponding to a given angular velocity $ \dtheta $ as $\rho_{\rm circ}$. Since for a circular orbit we have $ \dtheta=\meff $ it follows that $ \rho_{\rm circ} = \rho_{\rm kin}^\phi $. We can therefore express the Boltzmann equations as a Boltzmann equation for the circular orbit and a Boltzmann equation for the oscillations about this orbit:
\begin{gather}
\dot{\rho}_{\rm kin}^\phi = -(3H +\Gamma + c_{\mphi}H) \rho_{\rm kin}^\phi, \label{eq:Boltzmann 1}\\
\dot{\rho}_{\rm circ} = -(3H + c_{\mphi}H) \rho_{\rm circ}. \label{eq:Boltzmann 2}
\end{gather}
With a total energy density of $ \rho_{\rm tot} = 2 \rho_{\rm kin}^\phi+2 \rho_{\rm circ}  $. By conservation of energy this can be linked to the energy density of the plasma:
\begin{gather}
\dot{\rho}_r = -4H\rho_r +\Gamma\rho_{\rm kin}^\phi .\label{eq:Boltzmann 3}
\end{gather}
The set of equations Eqs.~(\ref{eq:Boltzmann 1}-\ref{eq:Boltzmann 3}) are then the equations we will solve numerically to describe the thermal damping of the PQ field. To complete the numerical setup, we also need to determine the initial conditions and specify how to describe the transition between phases of relativistic/non-relativistic fermions, where we do not have a description of the mass change of the form $ m\propto a^{-c_m} $.

\paragraph{Initial conditions:} After the kick, we assume that the energy density is composed of angular kinetic and potential energy, i.e.
\begin{gather}
\rho_{\rm ini} = \frac{1}{2}\meff^2\phikick^2+\frac{1}{2}\dtheta_{\rm kick}^2\phikick^2 = \frac{1+\epsilon^2}{2}\meff^2\phikick^2.
\end{gather}
After the kick this energy is distributed into $ \rho_{\rm kin}^\phi $, $ \rho_{\rm kin}^\theta $ and $ \rho_{\rm pot} $. To identify the distribution, note that in \cite{Gouttenoire:2021jhk} it was found that after the kick
\begin{gather}
\rho_{\rm kin}^\theta = \epsilon \rho_{\rm pot},
\end{gather}
so that in combination with the virial theorem, this implies
\begin{align}
\rho_{\rm pot} &= \frac{1}{2}\rho_{\rm ini}, \\
\rho_{\rm kin}^\phi &= \frac{1-\epsilon}{2}\rho_{\rm ini} \\
\rho_{\rm kin}^\theta &= \frac{\epsilon}{2}\rho_{\rm ini}.
\end{align}
These equations then specify the initial conditions for the Boltzmann equations.

\paragraph{Relativistic/non-relativistic fermion transition:} The parameter $ k $ takes into account continuous and gradual changes in the mass as with the temperature-dependence of a thermal mass. However, we generally approximate the transition between phases with relativistic and non-relativistic fermions as instantaneous because it is non-trivial to express in greater detail. Therefore we cannot encode the change in radial mass across such a transition in the parameter $ c_{\mphi} $ as in Eq.~\eqref{eq:parameter for mass time-dependence}. To take the transition into account in a physical way, we will instead assume that the transition is gradual enough to be adiabatic such that $ \phi \propto m^{-1/2} $ applies. Across a transition where the mass changes from $ m\to m^* $ this implies that $ \phi $ and an energy density $ \rho \propto m^2\phi^2 $ is modified as
\begin{align}
\phi &\underset{m\to m^*}{\xrightarrow{\hspace*{1cm}}} \left(\frac{m}{m_*}\right)^{\frac{1}{2}}\phi, \\
\rho &\underset{m\to m^*}{\xrightarrow{\hspace*{1cm}}} \left(\frac{m_*}{m}\right)\rho.
\end{align}
We use this solution to account for a change in $ \meff $ across a relativistic/non-relativistic fermion transition. We assume that $ \rho_{\rm pot} $, $ \rho_{\rm kin}^\phi $ and $ \rho_{\rm kin}^\theta $ are all shifted by the same factor and that the energy required for this change is transferred to or from the radiation bath. 

In the numerical  Boltzmann solver, we then apply this transformation to $ \phi $ and the energy densities whenever the fermions transition between their relativistic and non-relativistic phases.

\paragraph{Quartic potential}
For a quartic potential,
\begin{gather}
V\sim \frac{1}{4}\lambda\phi^4,
\end{gather}
we can follow the same steps as above to arrive again at
\begin{gather}
\dot{\rho}_{\rm tot} = -6H\rho_{\rm kin}^\phi-2\Gamma\rho_{\rm kin}^\phi-6H\rho_{\rm kin}^\theta.
\end{gather}
The virial theorem for $ \phi^n $ with $ n=4 $ now states that $ 2\expval{V}=\expval{T} $, so that
\begin{gather}
\rho_{\rm pot} = \frac{1}{2}\rho_{\rm kin}^\phi+\frac{1}{2}\rho_{\rm kin}^\theta.
\end{gather}
The combined Boltzmann equation is then
\begin{gather}
\frac{3}{2}\dot{\rho}_{\rm kin}^\phi+\frac{3}{2}\dot{\rho}_{\rm kin}^\theta= -6H\rho_{\rm kin}^\phi-2\Gamma\rho_{\rm kin}^\phi-6H\rho_{\rm kin}^\theta.
\end{gather}
If we again assume that $ \rho_{\rm kin}^\phi $ and $ \rho_{\rm kin}^\theta $ are independent, then we arrive at the following two Boltzmann equations:
\begin{align}
\dot{\rho}_{\rm kin}^\phi &= -4H \rho_{\rm kin}^\phi -\frac{4}{3}\Gamma \rho_{\rm kin}^\phi, \\
\dot{\rho}_{\rm kin}^\theta &= -4H \rho_{\rm kin}^\theta.
\end{align}
If a thermal potential becomes dominant, then the evolution is instead determined by \ref{nearly quadratic boltzmann 1} and \ref{nearly quadratic boltzmann 2} as in the quadratic case. The initial conditions are modified to
\begin{align}
\rho_{\rm pot} &= \frac{1}{3}\rho_{\rm ini}, \\
\rho_{\rm kin}^\phi &= (1-\epsilon)\frac{2}{3}\rho_{\rm ini} \\
\rho_{\rm kin}^\theta &= \epsilon\frac{2}{3}\rho_{\rm ini}.
\end{align}

\subsection{Examples of solutions}\label{sec:examples of solutions}
\begin{figure}
	\centering
	\includegraphics[height=0.9\textheight]{"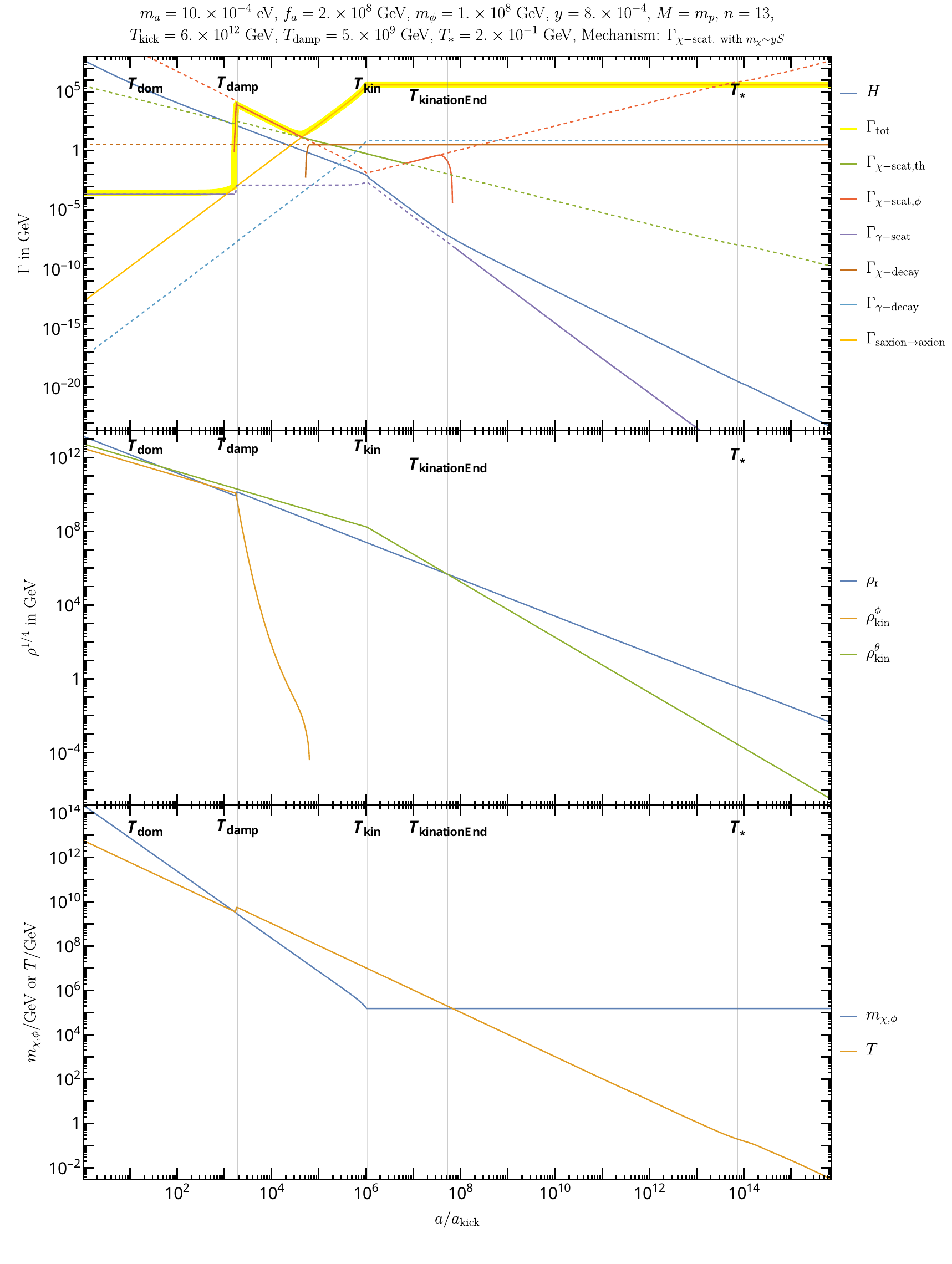"}
	\caption{\small\it Example of thermal damping within the KSVZ-like nearly-quadratic model, which assumes Yukawa damping. The solution is found by numerical solution of eqs. (\ref{eq:boltzmann1}-\ref{eq:boltzmann4}) and damping is triggered by an intermediate relativistic phase. Various events of interest are tagged with vertical lines. \textbf{Top}: Damping rates compared to the Hubble parameter. Dashed lines indicate that the conditions for a given effect are not met. \textbf{Middle:} Evolution of the energy densities. Note the transfer of kinetic energy from $ \rho_\phi $ to $ \rho_r $. \textbf{Bottom}: Evolution of temperature as compared to the fermion mass.}
	\label{fig:niceBoltzmannExample}
\end{figure}

\begin{figure}
	\centering
	\includegraphics[height=0.9\textheight]{"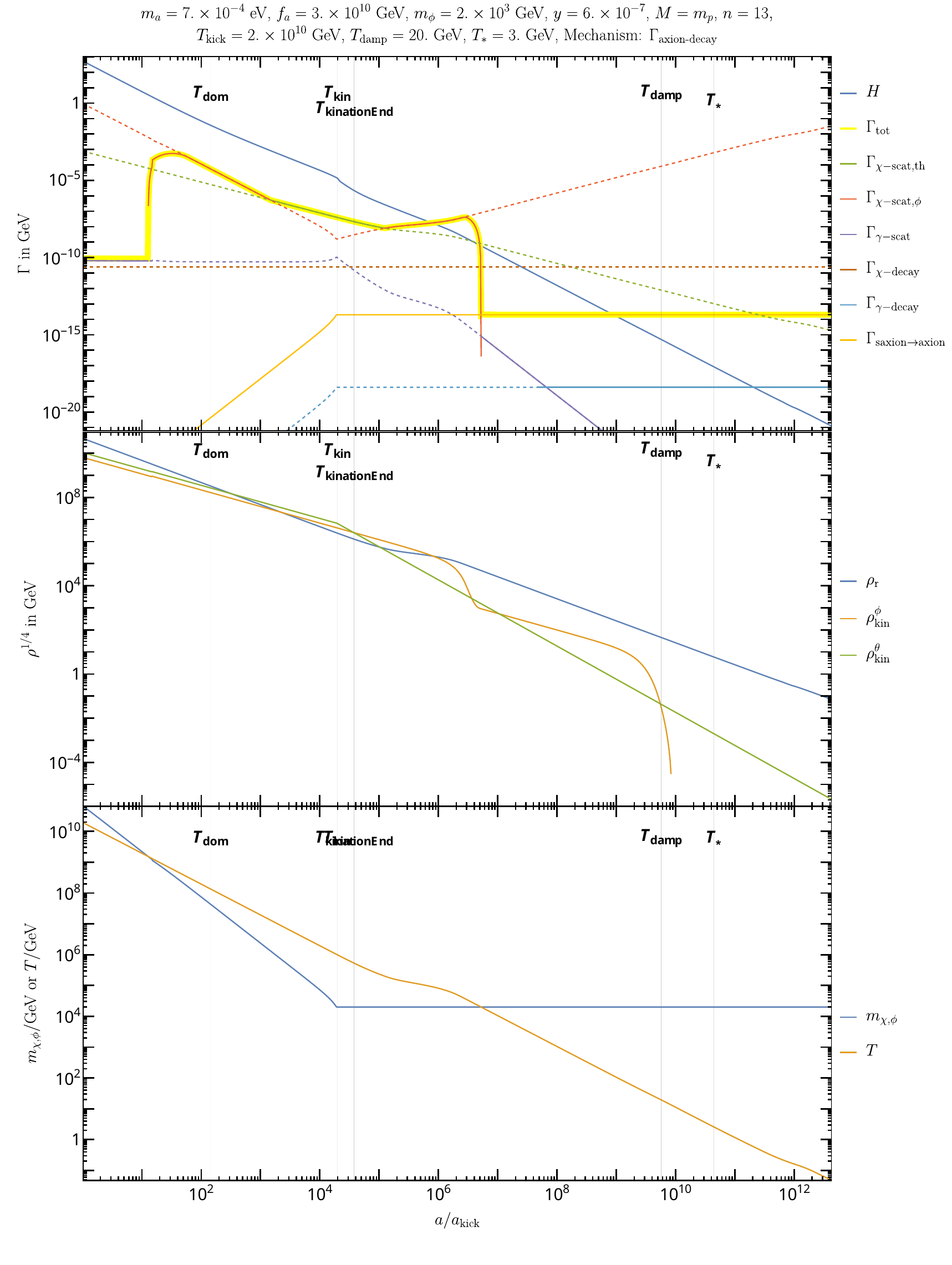"}
	\caption{\small\it Similar to Fig.~\ref{fig:niceBoltzmannExample}, but for another choice of parameters for which damping is not triggered by an intermediate relativistic phase. Instead, late damping is partially realized towards the end of a relativistic phase. Here, $ \chi $-scattering is insufficient to completely damp $ \rho_{\rm kin}^\phi $. Damping is completed by $\phi$-to-axion decay. The initial, incomplete, damping through $ \chi $-scattering prevents the production of a cosmologically dangerous hot axion relic.}
	\label{fig:niceBoltzmannKinationExample}
\end{figure}

\begin{figure}
	\centering
	\includegraphics[height=0.9\textheight]{"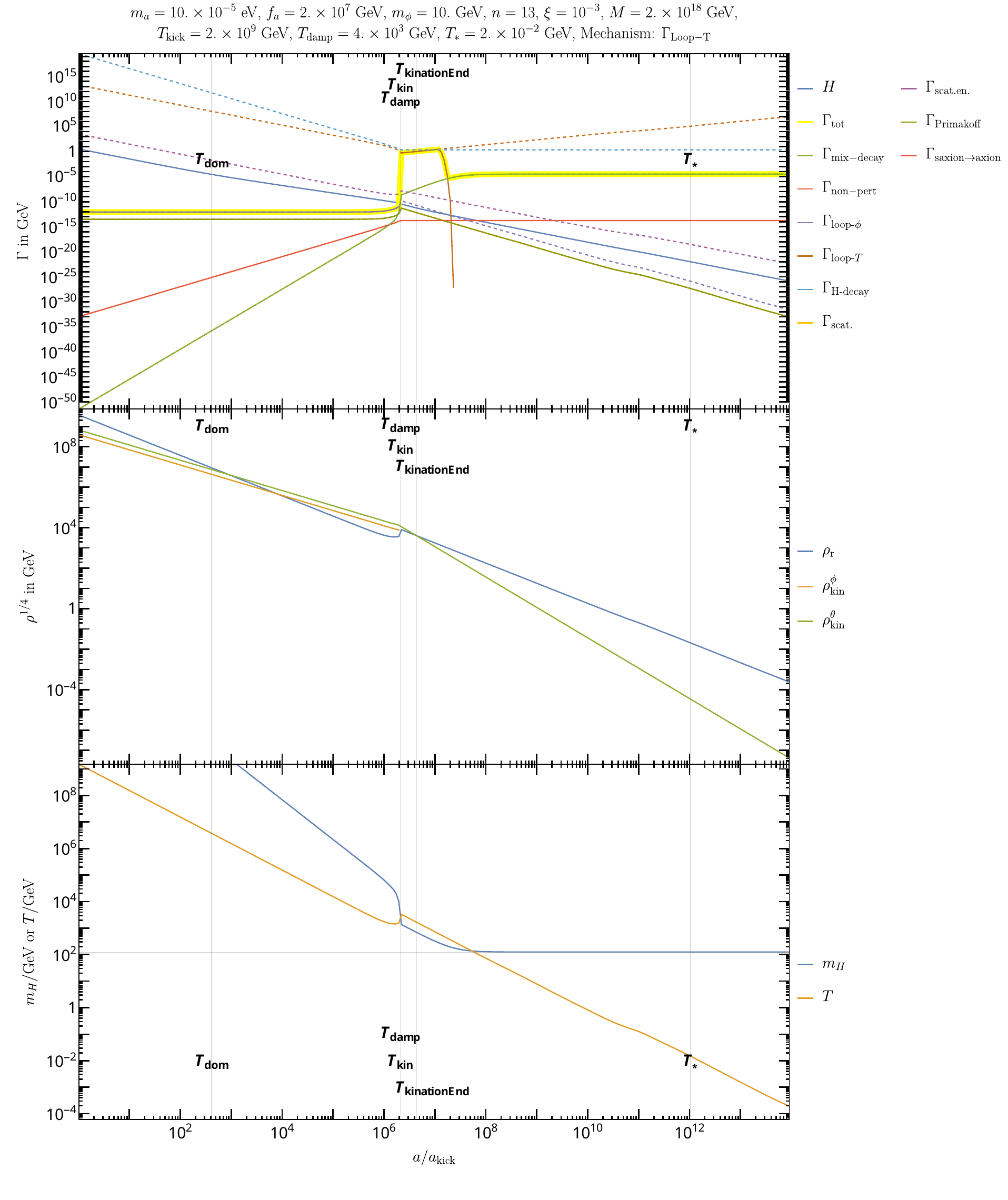"}
	\caption{\small\it Similar to Figs. \ref{fig:niceBoltzmannExample} and \ref{fig:niceBoltzmannKinationExample} except that we consider a Higgs portal coupling instead of the $ \chi $-fermion coupling. Damping takes place in an intermediate relativistic phase triggered by cancellation of the Higgs mass.}
	\label{fig:plotNQHiggsSharpStepCausedByHiggsMassCancellation}
\end{figure}

We provide three examples of solutions produced with the Boltzmann prescription detailed above:
\begin{description}
	\item[Fig.~\ref{fig:niceBoltzmannExample}:] Model with a nearly-quadratic potential and damping through $ \chi- $fermion scattering triggered by the onset of an intermediate relativistic phase.
	\item[Fig.~\ref{fig:niceBoltzmannKinationExample}:]  Model with a nearly-quadratic potential and two-step damping through $ \chi- $fermion scattering near the end of a relativistic phase and $\phi$-to-axion decay.
	\item[Fig.~\ref{fig:plotNQHiggsSharpStepCausedByHiggsMassCancellation}:] Model with a nearly-quadratic potential and damping through a Higgs portal coupling in which damping takes place in a relativistic phase triggered by cancellation of the Higgs mass.
\end{description}
In all three figures, the upper frame shows the evolution of the damping rates, the middle frame shows the evolution of the energy densities and the bottom frame shows the $ \chi$-fermion or Higgs mass compared to the temperature. 
Note that the models do not attempt to describe the behavior after $ \Tfrag $ so the results should only be considered correct until that point.

\FloatBarrier
\section{Range of viable parameters}\label{app:parameter values}
In this appendix, we make the  connection between the UV model parameters and the viable range of parameter space in the $ [m_a,f_a] $ plane. For Yukawa damping, these model parameters are the following six variables: $$ m_a, f_a,y,M,n  \mbox{ and }  \mphiz . $$ For Higgs damping, $y$ is traded for $\xi^2$ and $\mphiz$ is set by $\lambda$ so that we still have six model parameters. As stated before, we assume that the axion relic makes up all of the dark matter which reduces the number of free parameters to five. Furthermore,  we restrict ourselves to the well-motivated case $ M=\mplanck $.  Thus, in the $ [m_a,f_a] $ plane, each point may be supported by a family of solutions spanning values of the coupling constant ($y$ or $\xi^2$) and the radial-mode mass parameter ($\mphiz $ or $\lambda$).
In the next pages, we plot the viable range for these parameters for $ n=10$ and $13 $. 
The figures given in this appendix are the following:
\paragraph{Nearly-quadratic potential with Yukawa damping:} \ \\ 
\vspace{0.0cm}

\hspace{0.0cm}
\begin{tabular}{l l l}
	Range of $ \mphiz $		& for $ n=13 $:		&Figure \ref{fig:msPlotNQYukawaMmpn13}	\\
	Range of $ y $ 			&for $ n=13 $: 		&Figure \ref{fig:YukawaPlotNQYukawaMmpn13}	\\
	Range of $ \mphiz $ 	&for $ n=10 $:		&Figure \ref{fig:msPlotNQYukawaMmpn10}	\\
	Range of $ y $ 			&for $ n=10 $: 		&Figure \ref{fig:YukawaPlotNQYukawaMmpn10}	\\
\end{tabular}

\paragraph{Nearly-quadratic potential with Higgs damping:} \ \\ 
\vspace{0.0cm}

\hspace{0.0cm}
\begin{tabular}{l l l}
	Range of $ \mphiz $		& for $ n=13 $:		&Figure \ref{fig:msPlotNQHiggsMmpn13}	\\
	Range of $ \xi^2 $ 			&for $ n=13 $: 		&Figure \ref{fig:HiggsCouplingPlotNQHiggsMmpn13}	\\
	Range of $ \mphiz $ 	&for $ n=10 $:		&Figure \ref{fig:msPlotNQHiggsMmpn10}	\\
	Range of $ \xi^2 $ 			&for $ n=10 $: 		&Figure \ref{fig:HiggsCouplingPlotNQHiggsMmpn10}	\\
\end{tabular}

\paragraph{Quartic potential with Hubble-induced mass and Yukawa damping:} \ \\ 
\vspace{0.0cm}

\hspace{0.0cm}
\begin{tabular}{l l l}
	Range of $ \lambda $		& for $ n=13 $:		&Figure \ref{fig:QuarticPlotQHigherDimYukawaMmpn13}	\\
    Range of $\mphiz$           & for $ n=13 $:     &Figure \ref{fig:msPlotQHigherDimYukawaMmpn13}      \\
	Range of $ y $ 			&for $ n=13 $: 		&Figure \ref{fig:YukawaPlotQHigherDimYukawaMmpn13}	\\
	Range of $ \lambda $ 	&for $ n=10 $:		&Figure \ref{fig:QuarticPlotQHigherDimYukawaMmpn10}	\\
    Range of $\mphiz$           & for $ n=10 $:     &Figure \ref{fig:msPlotQHigherDimYukawaMmpn10}      \\
	Range of $ y $ 			&for $ n=10 $: 		&Figure \ref{fig:YukawaPlotQHigherDimYukawaMmpn10}	\\
\end{tabular}

\paragraph{Quartic potential with Hubble-induced mass and Higgs damping:} \ \\ 
\vspace{0.0cm}

\hspace{0.0cm}
\begin{tabular}{l l l}
	Range of $ \lambda $		& for $ n=13 $:		&Figure \ref{fig:QuarticPlotQHigherDimHiggsMmpn13}	\\
    Range of $\mphiz$           & for $ n=13 $:     &Figure \ref{fig:msPlotQHigherDimHiggsMmpn13}      \\
	Range of $ \xi^2 $ 			&for $ n=13 $: 		&Figure \ref{fig:HiggsPlotQHigherDimHiggsMmpn13}	\\
	Range of $ \lambda $ 	&for $ n=10 $:		&Figure \ref{fig:QuarticPlotQHigherDimHiggsMmpn10}	\\
    Range of $\mphiz$           & for $ n=10 $:     &Figure \ref{fig:msPlotQHigherDimHiggsMmpn10}      \\
	Range of $ \xi^2 $ 			&for $ n=10 $: 		&Figure \ref{fig:HiggsPlotQHigherDimHiggsMmpn10}	\\
\end{tabular}

\vspace{0.5cm}
\noindent
More details on various interesting aspects of these solutions are given in the relevant sections of the the main text. Specifically, the models with Yukawa damping are described in Section \ref{sec:Nearly quadratic Yukawa section} and models with Higgs damping are described in Section \ref{sec:NQHiggs}.


\begin{figure}
	\centering
	\includegraphics[width=\standardwidth \textwidth]{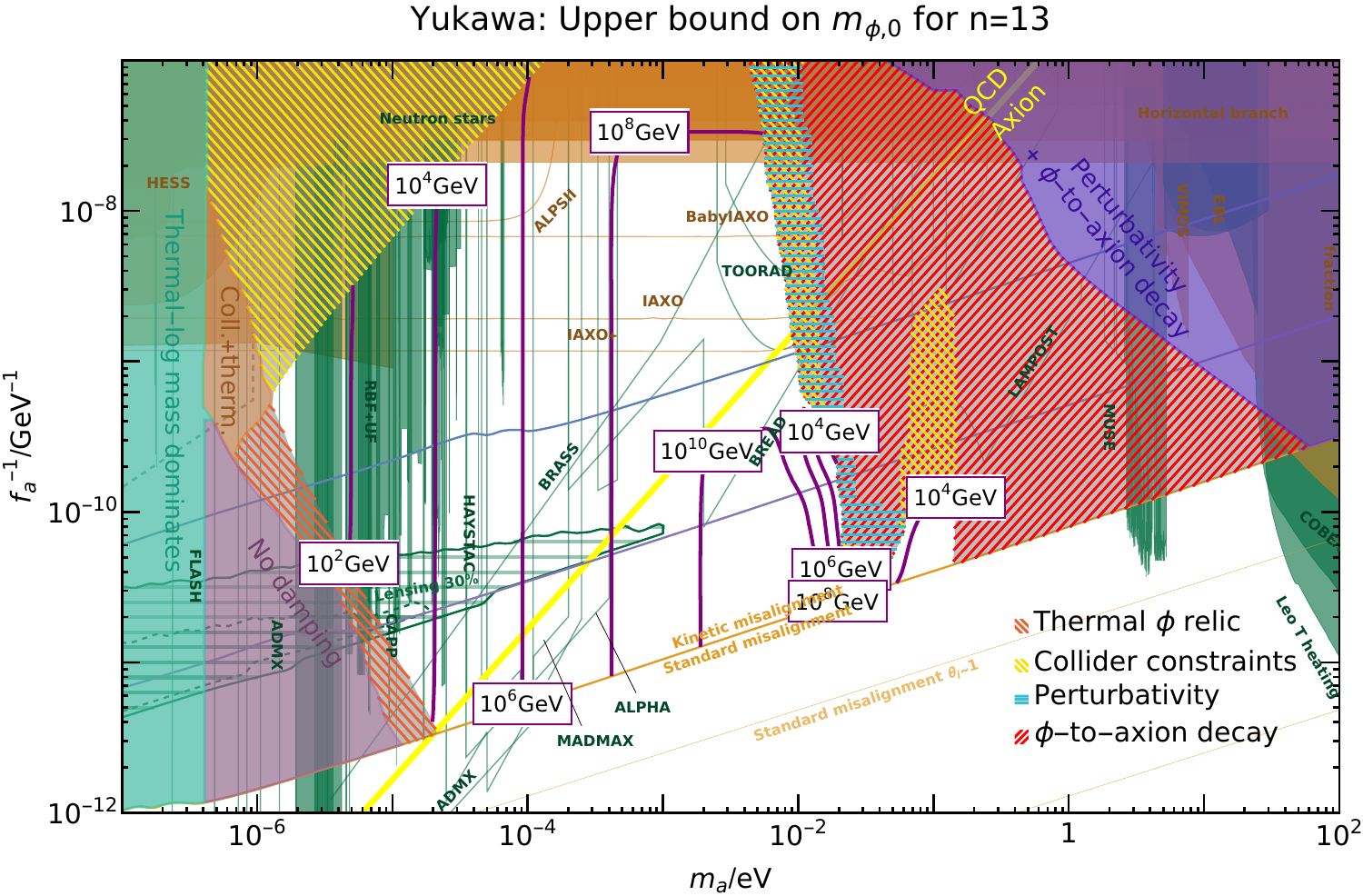}
	
	\vspace{0.5cm}
	\includegraphics[width=\standardwidth \textwidth]{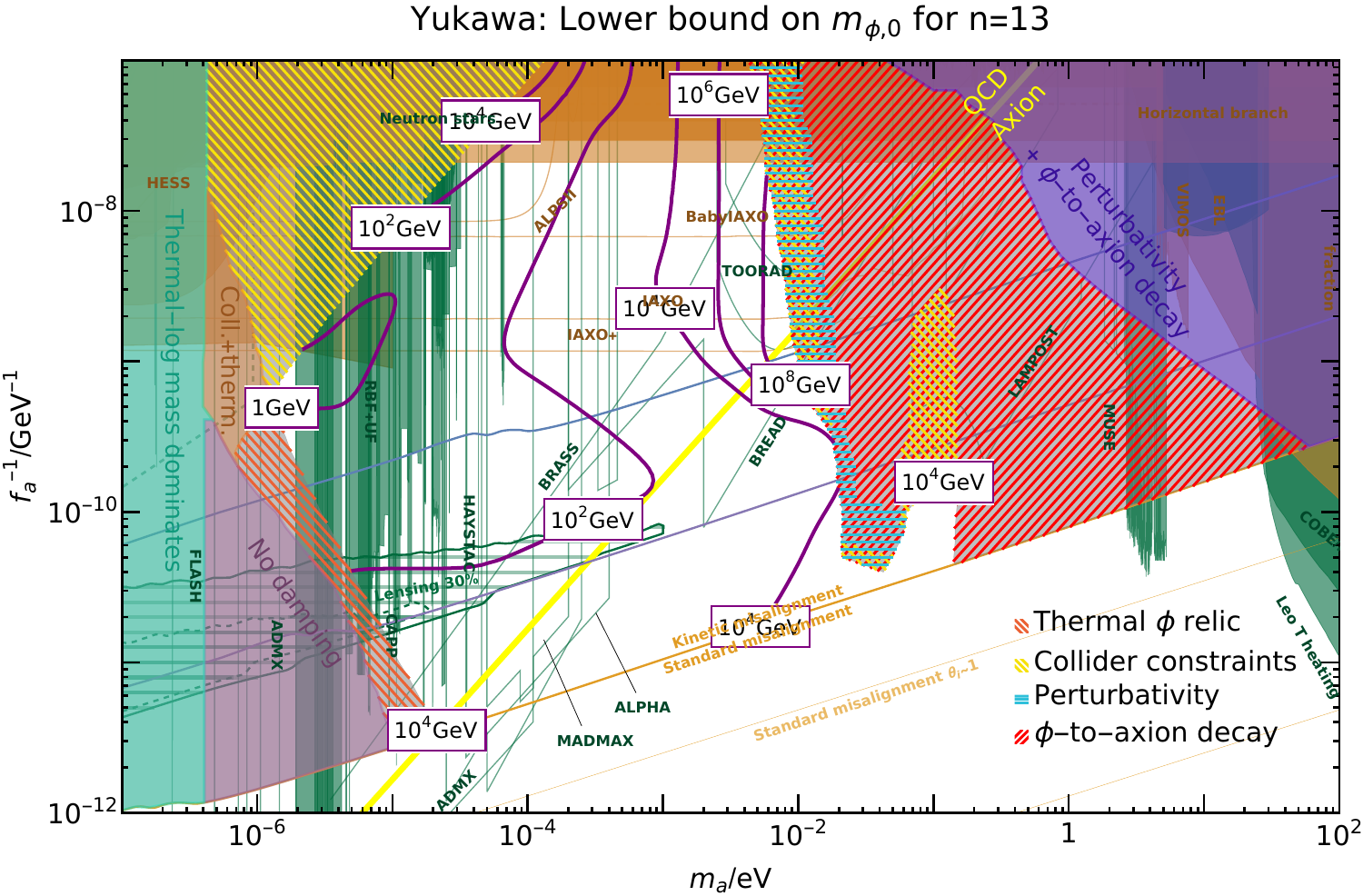}
	
	\caption{\small\it Upper and lower bounds on the radial mass $m_\phi$ in Yukawa damped models with $n=13$.}
	\label{fig:msPlotNQYukawaMmpn13}
\end{figure}

\begin{figure}
	\centering
	\includegraphics[width=\standardwidth \textwidth]{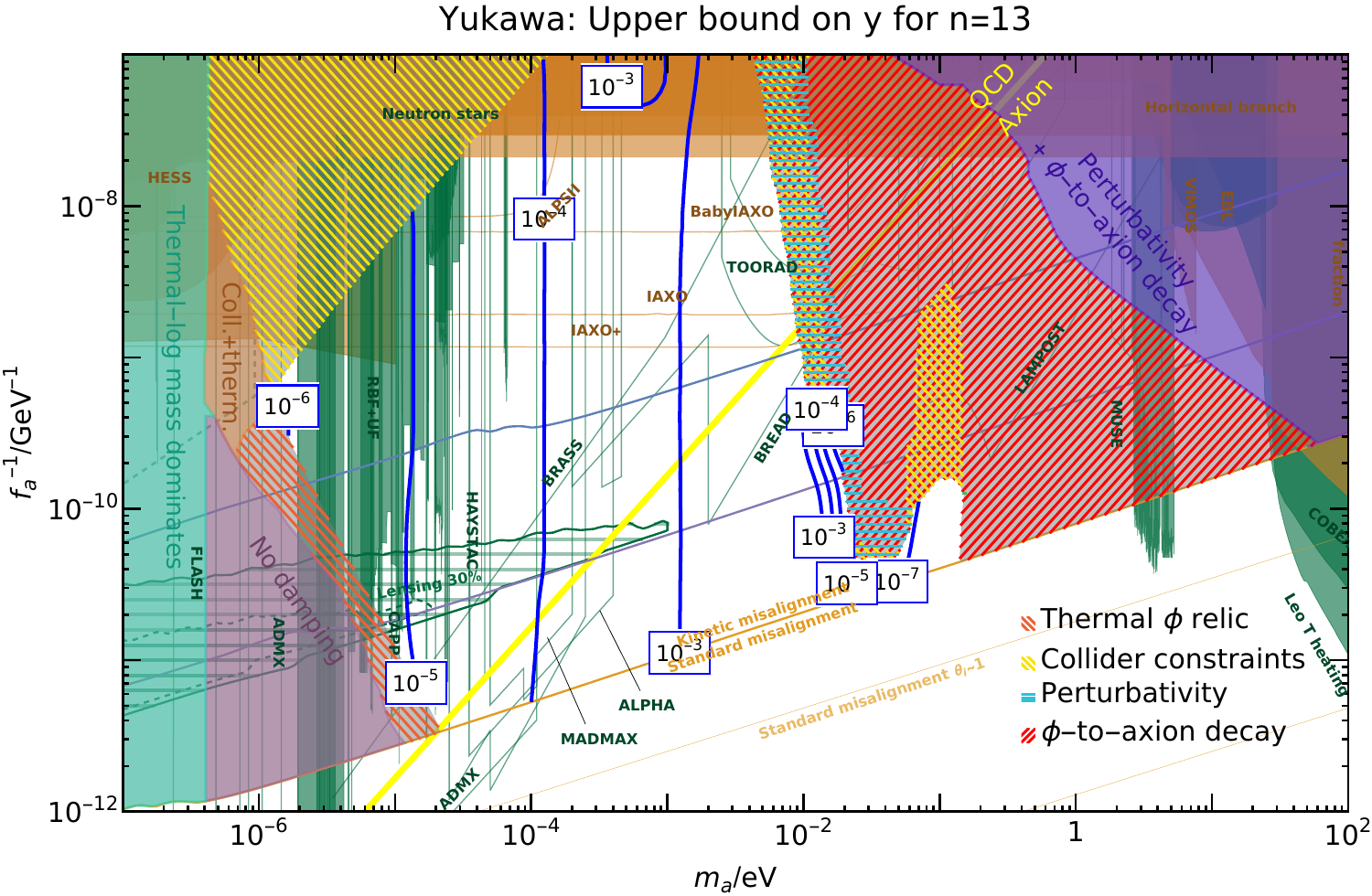}
	
	\vspace{0.5cm}
	\includegraphics[width=\standardwidth \textwidth]{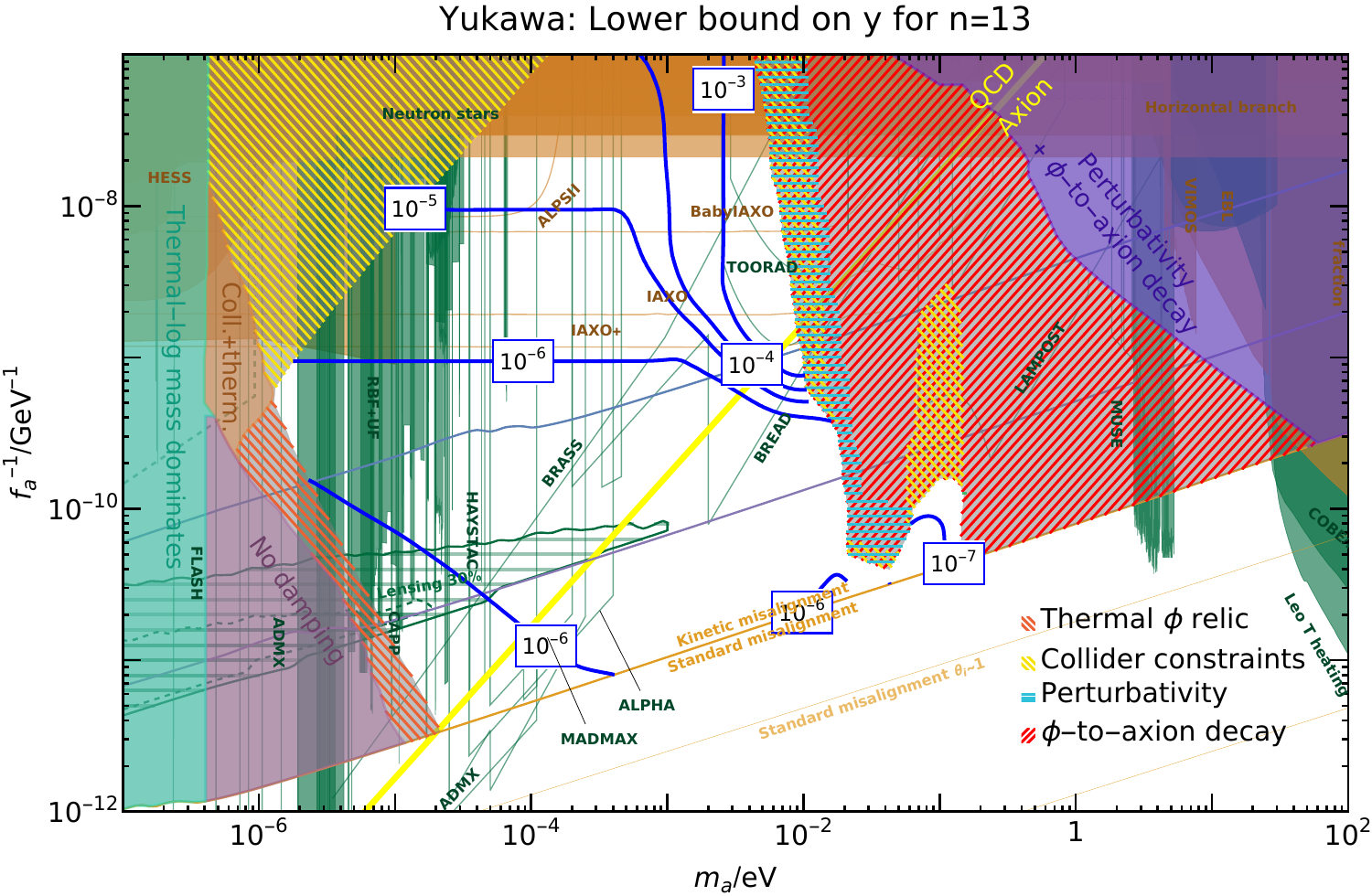}
	
	\caption{\small\it Upper and lower bounds on the Yukawa coupling $y$ in Yukawa damped models with $n=13$.}
	\label{fig:YukawaPlotNQYukawaMmpn13}
\end{figure}


\begin{figure}
	\centering
	\includegraphics[width=\standardwidth \textwidth]{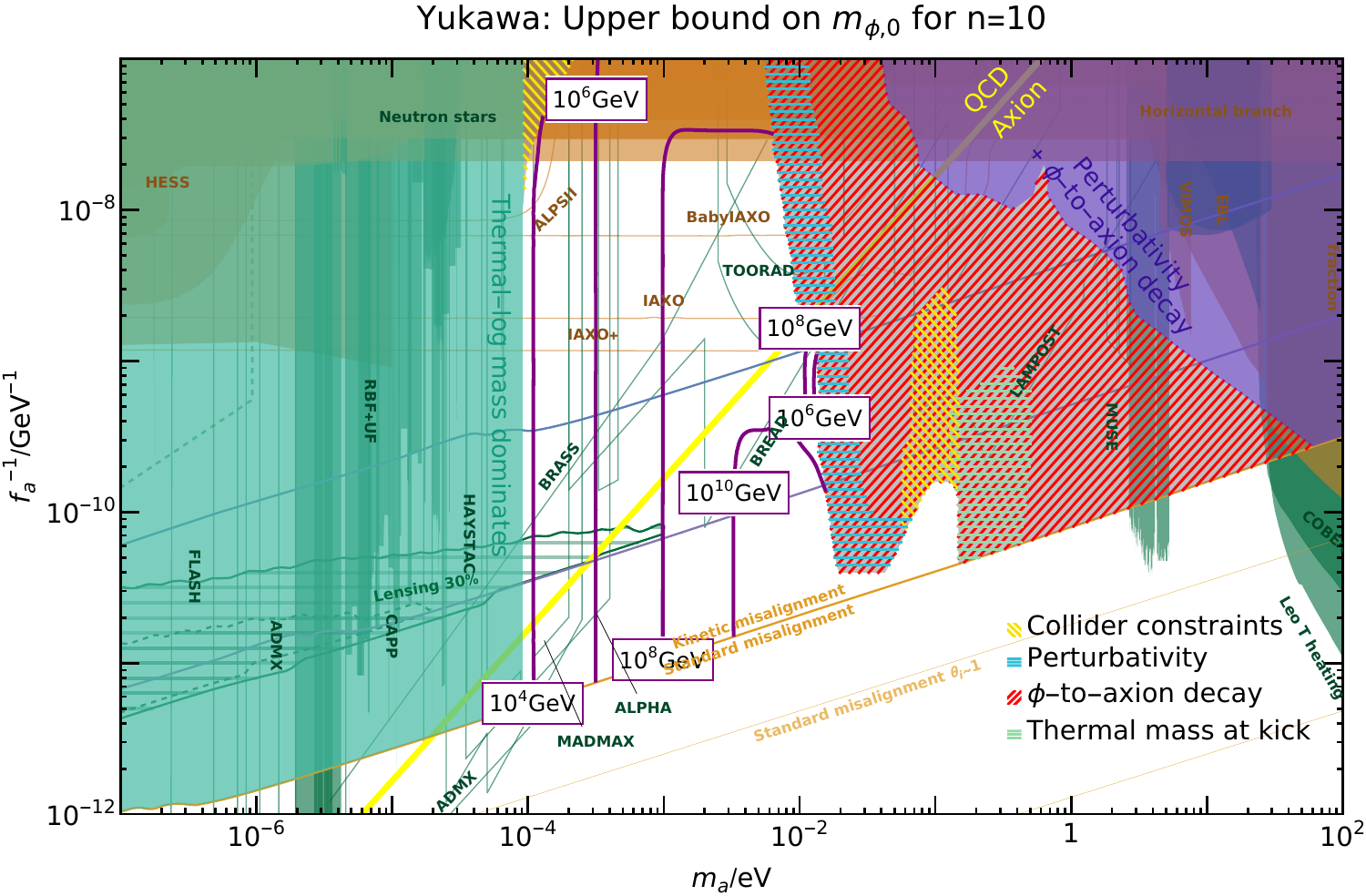}
	
	\vspace{0.5cm}
	\includegraphics[width=\standardwidth \textwidth]{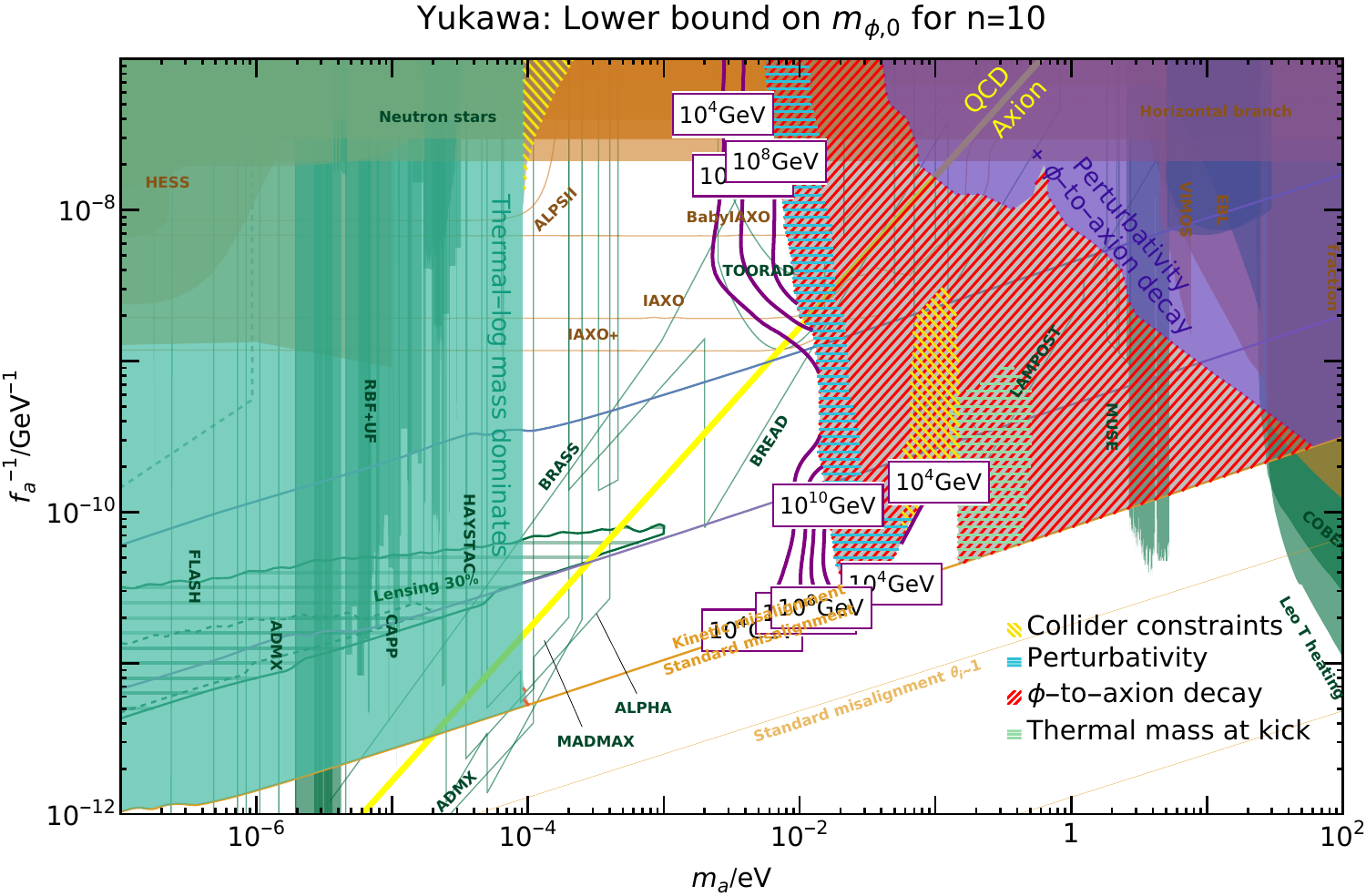}
	
	\caption{\small\it Upper and lower bounds on the radial mass $m_\phi$ in Yukawa damped models with $n=10$.}
	\label{fig:msPlotNQYukawaMmpn10}
\end{figure}

\begin{figure}
	\centering
	\includegraphics[width=\standardwidth \textwidth]{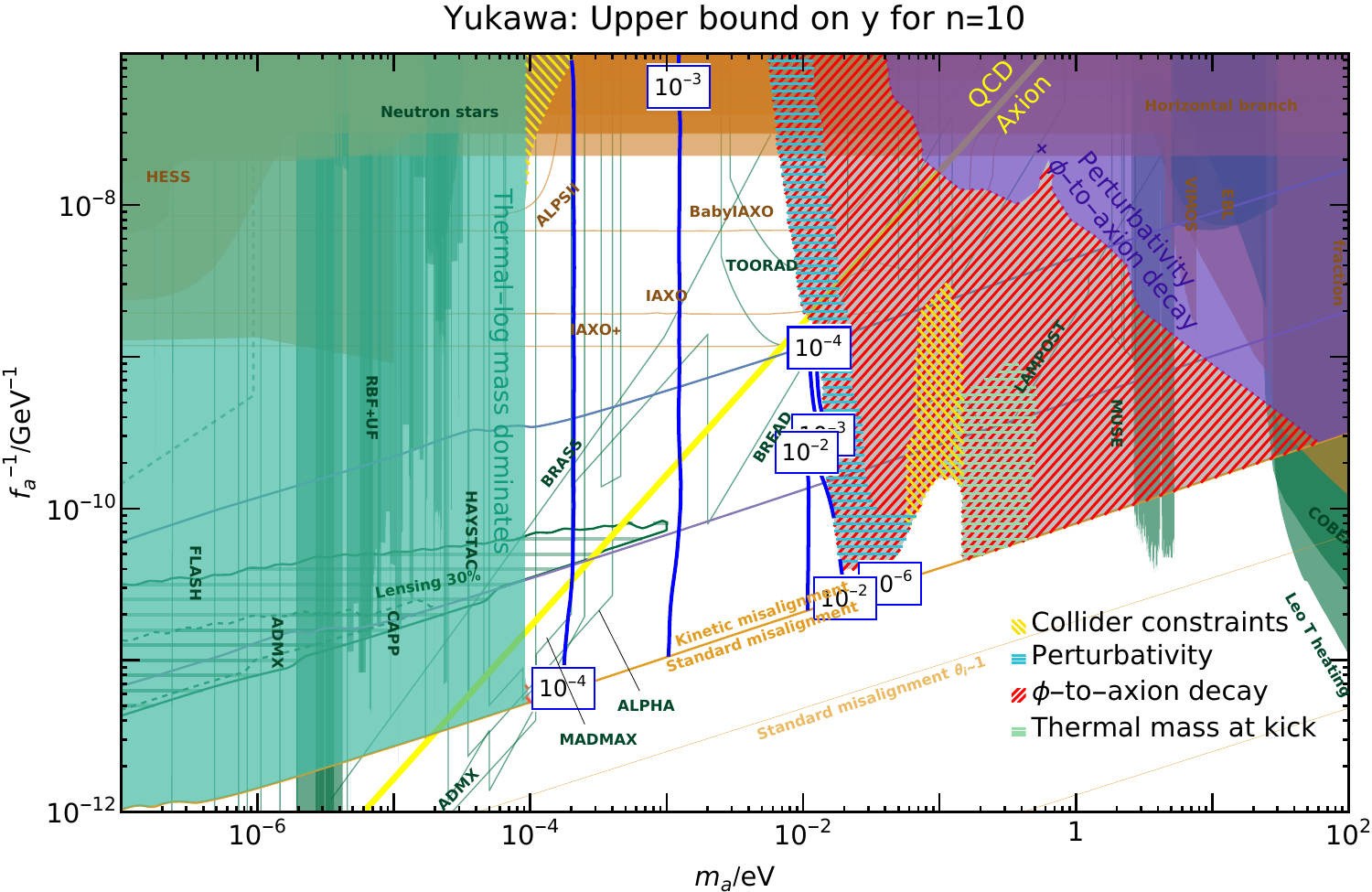}
	
	\vspace{0.5cm}
	\includegraphics[width=\standardwidth \textwidth]{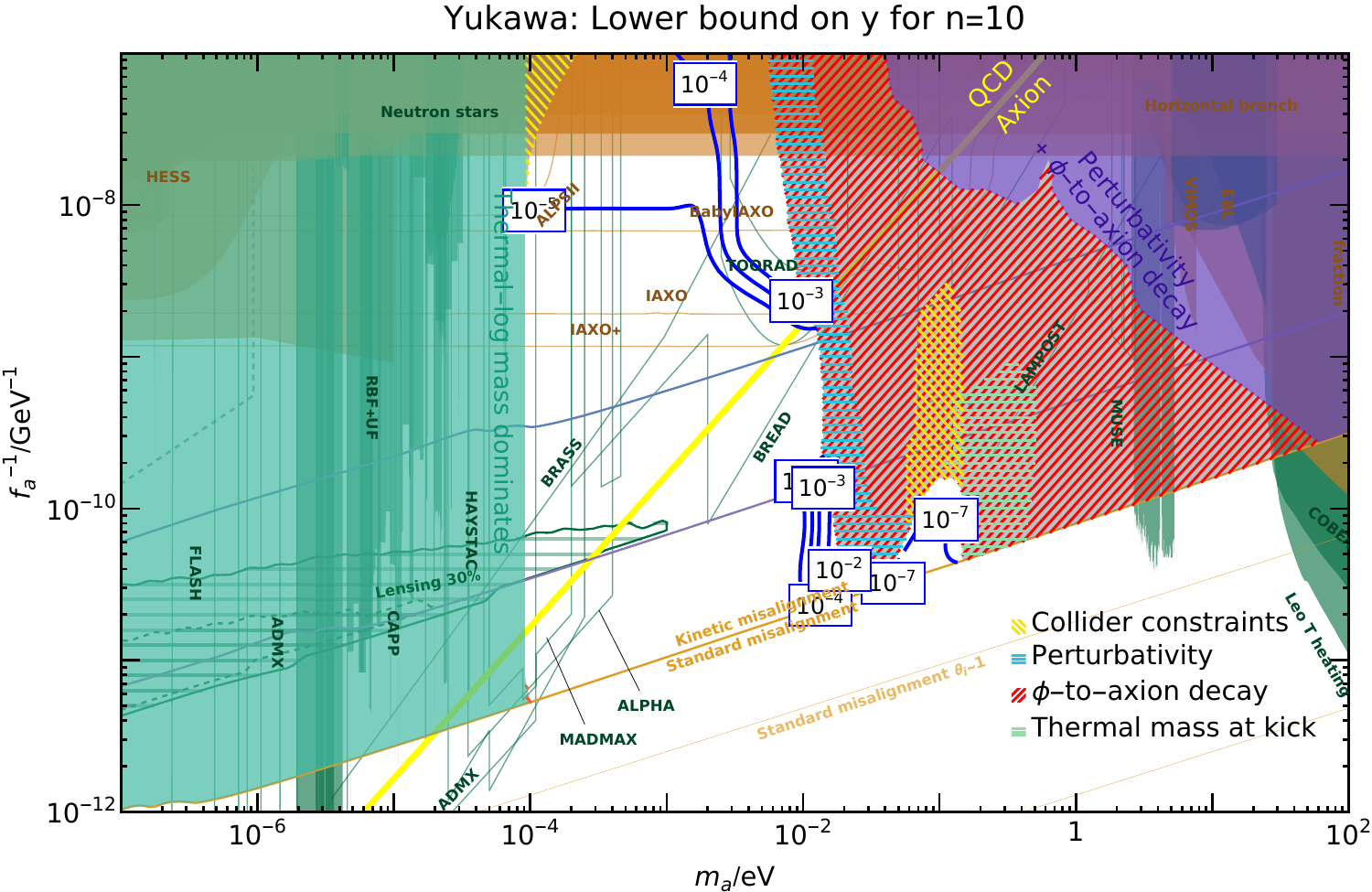}
	
	\caption{\small\it Upper and lower bounds on the Yukawa coupling $y$ in Yukawa damped models with $n=10$.}
	\label{fig:YukawaPlotNQYukawaMmpn10}
\end{figure}


\begin{figure}
	\centering
	\includegraphics[width=\standardwidth \textwidth]{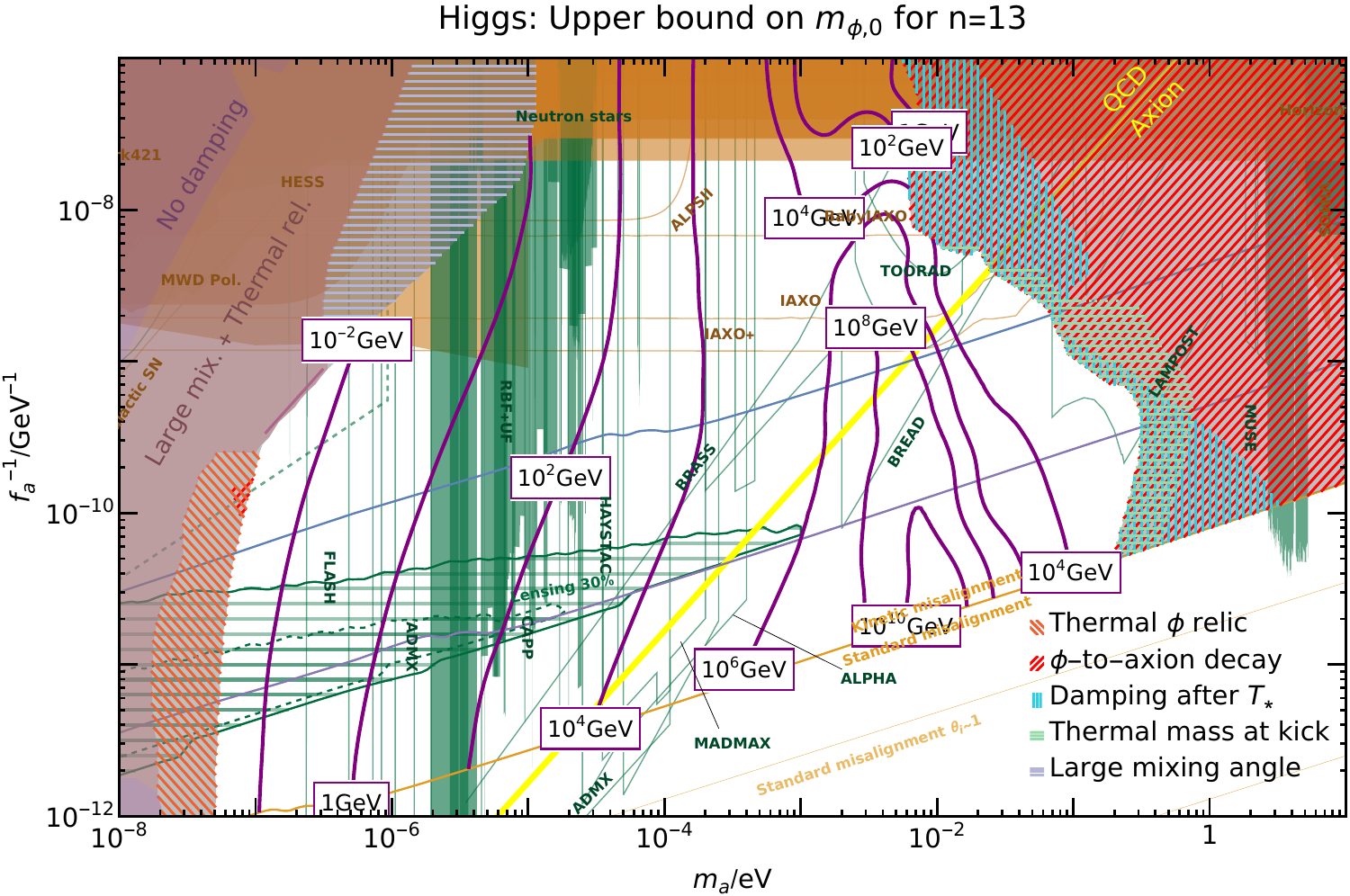}
	
	\vspace{0.5cm}
	\includegraphics[width=\standardwidth \textwidth]{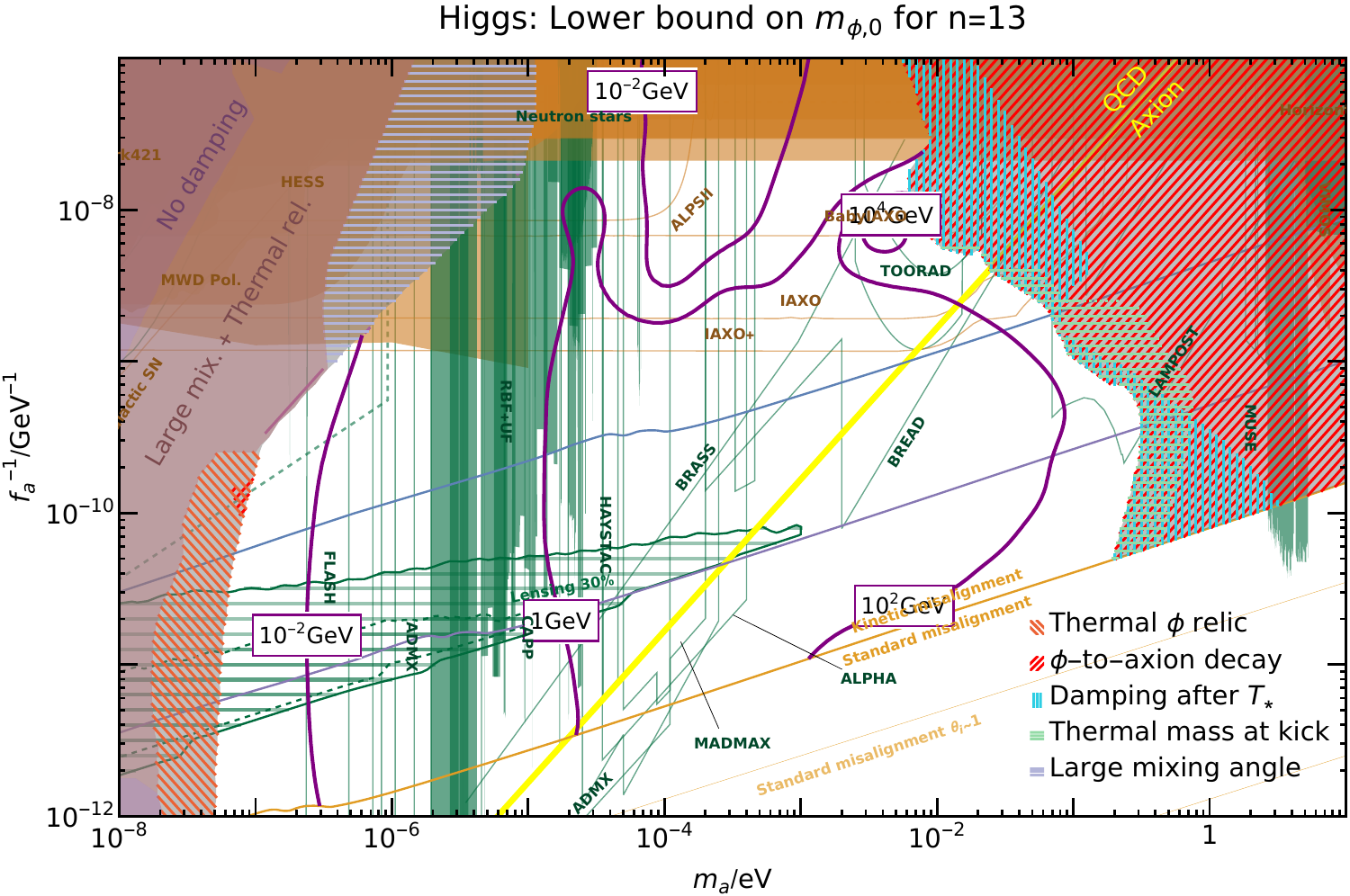}
	
	\caption{\small\it Upper and lower bounds on radial mass $m_\phi$ in Higgs damped models with $n=13$}
	\label{fig:msPlotNQHiggsMmpn13}
\end{figure}

\begin{figure}
	\centering
	\includegraphics[width=\standardwidth \textwidth]{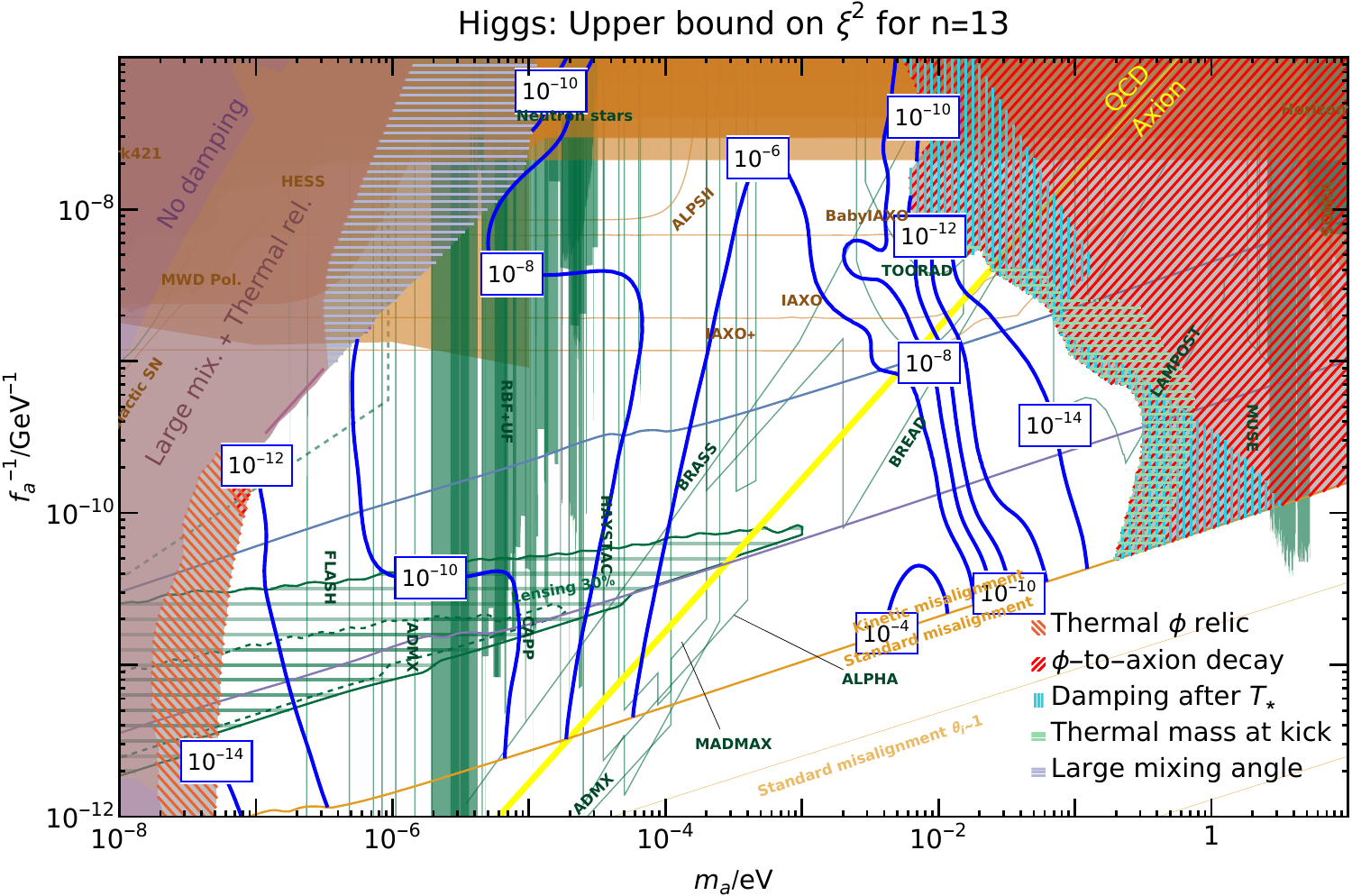}
	
	\vspace{0.5cm}
	\includegraphics[width=\standardwidth \textwidth]{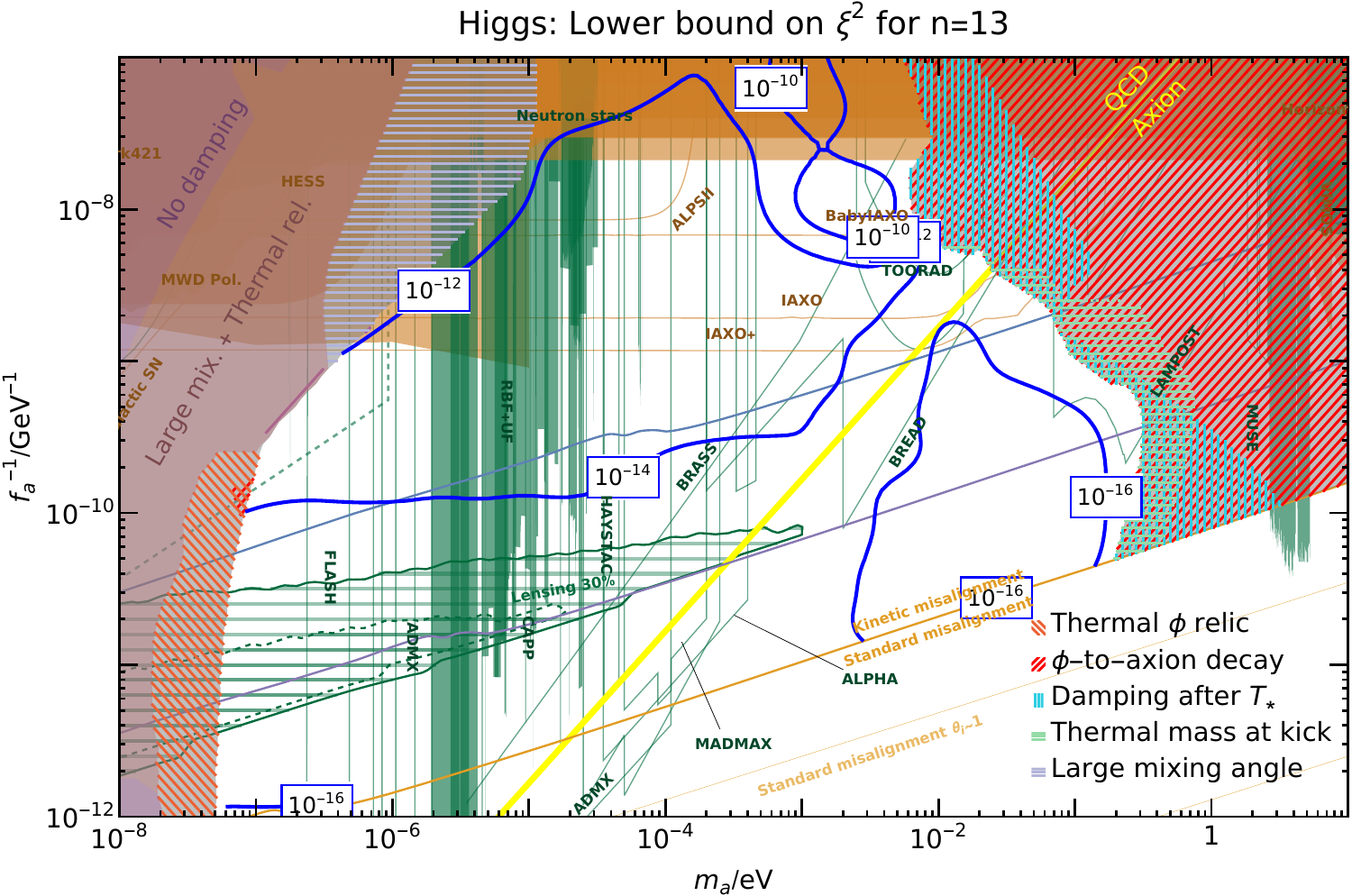}
	
	\caption{\small\it Upper and lower bounds on portal coupling $\xi^2$ in Higgs damped models with $n=13$}
	\label{fig:HiggsCouplingPlotNQHiggsMmpn13}
\end{figure}


\begin{figure}
	\centering
	\includegraphics[width=\standardwidth \textwidth]{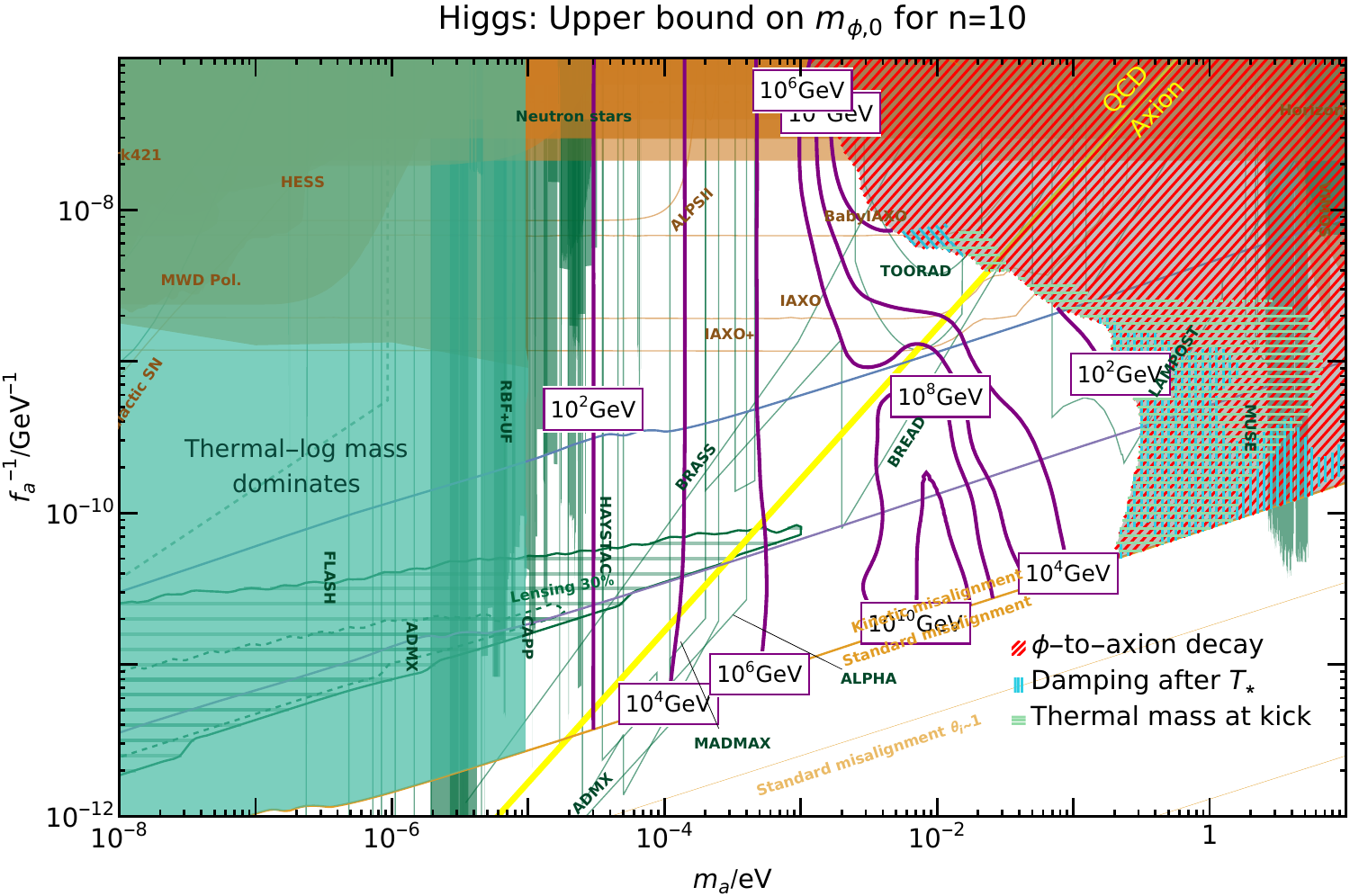}
	
	\vspace{0.5cm}
	\includegraphics[width=\standardwidth \textwidth]{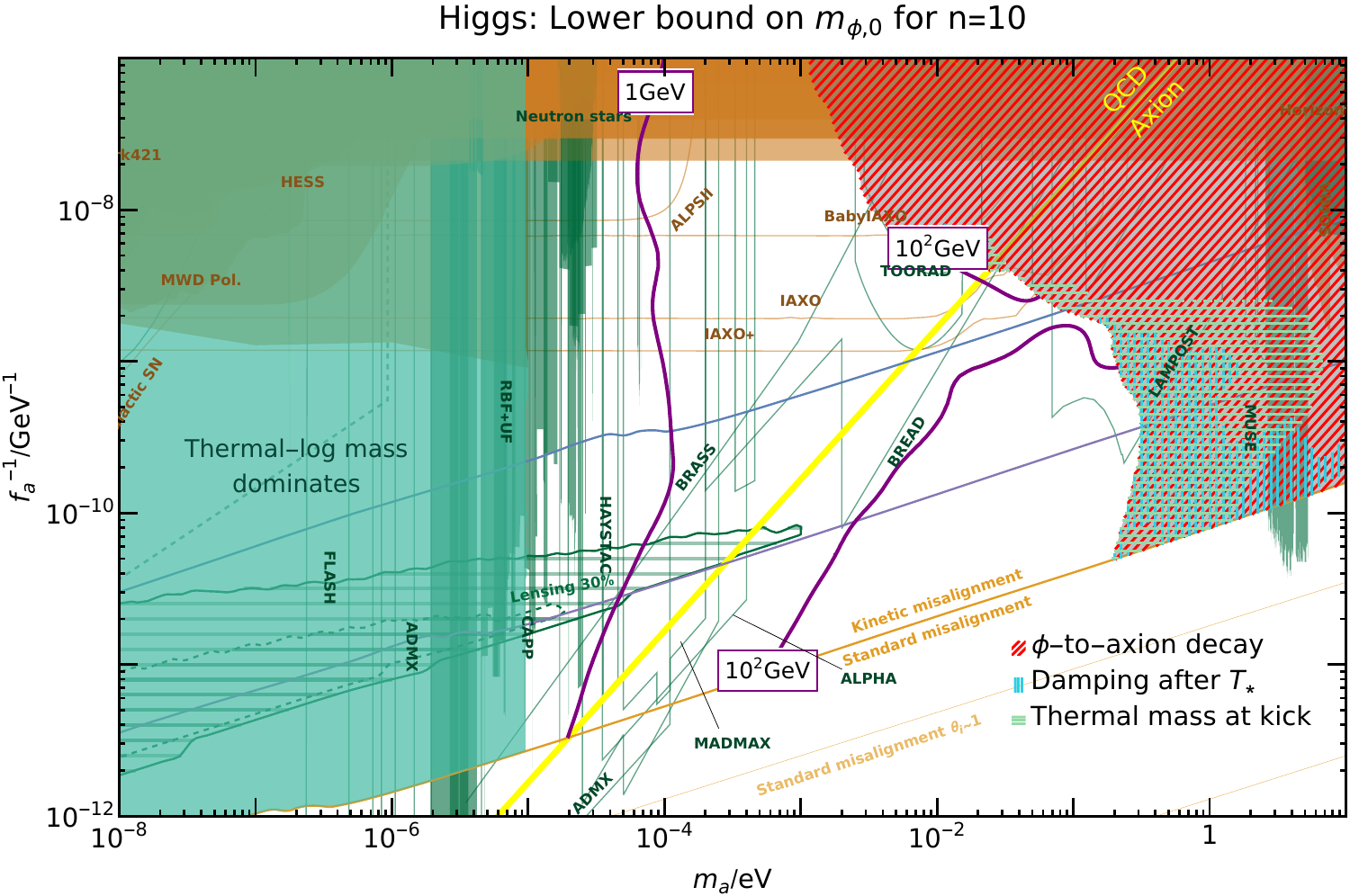}
	
	\caption{\small\it  Upper and lower bounds on radial mass $m_\phi$ in Higgs damped models with $n=10$}
	\label{fig:msPlotNQHiggsMmpn10}
\end{figure}

\begin{figure}
	\centering
	\includegraphics[width=\standardwidth \textwidth]{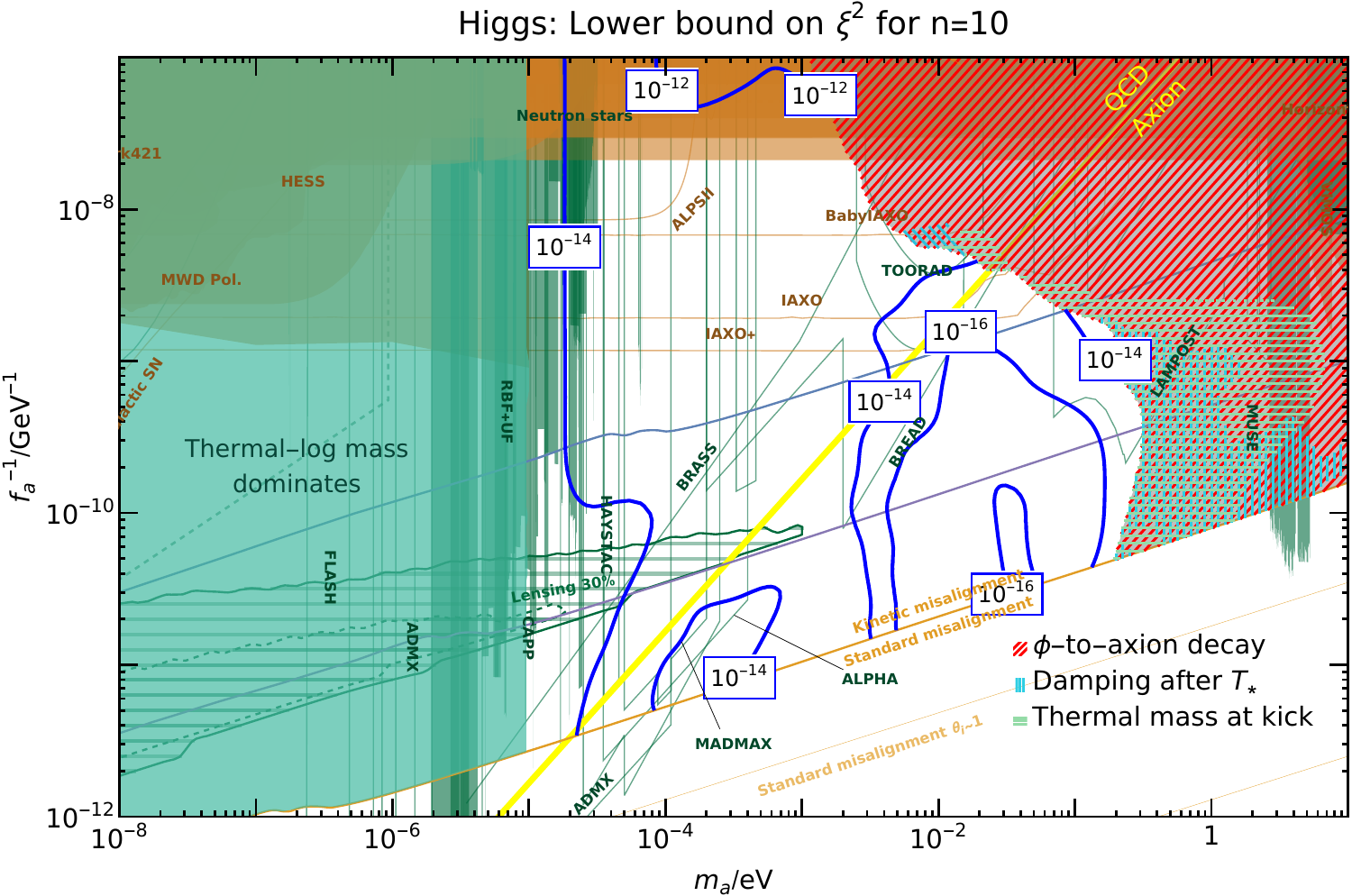}
	
	\vspace{0.5cm}
	\includegraphics[width=\standardwidth \textwidth]{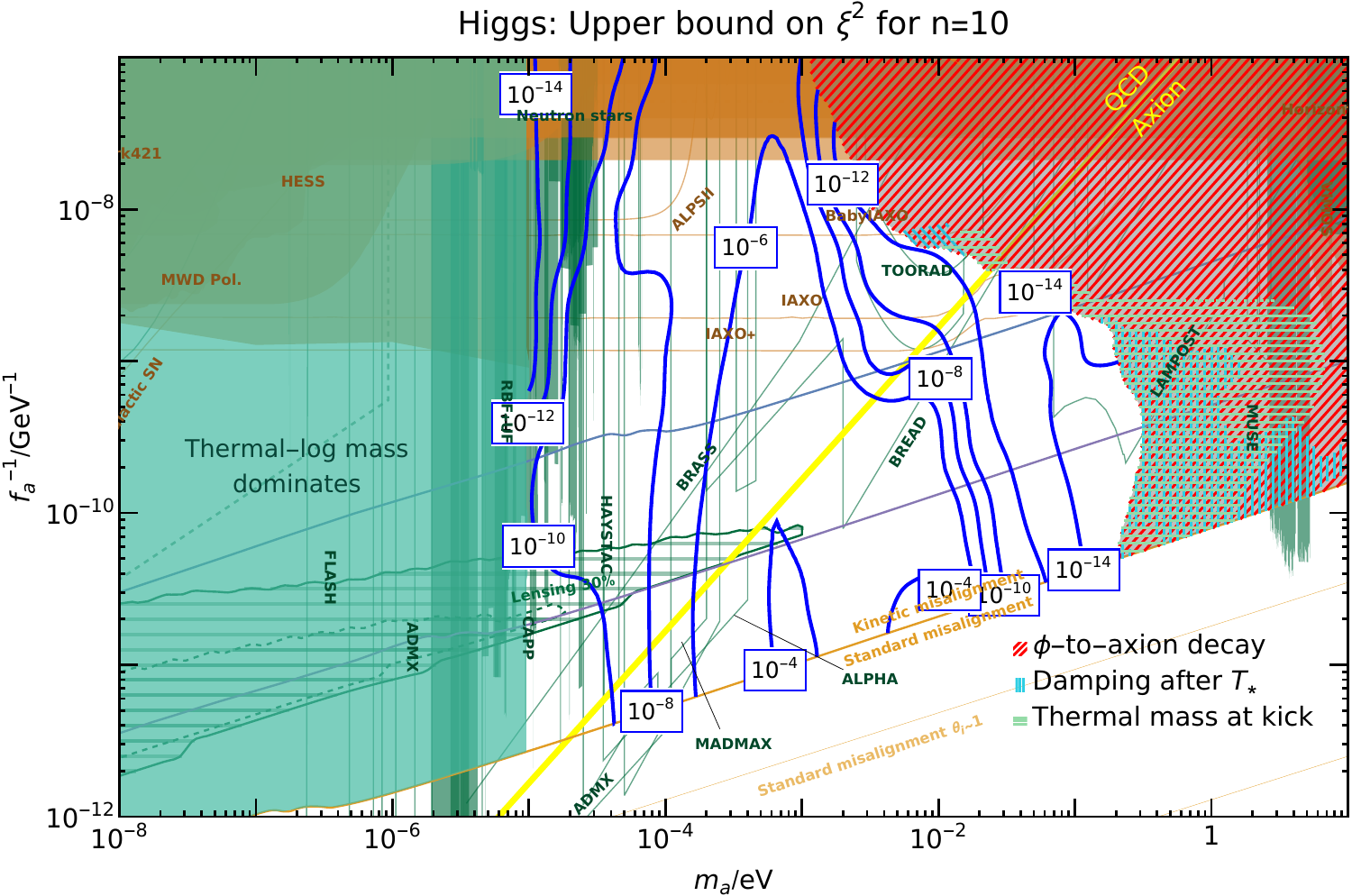}
	
	\caption{\small\it Upper and lower bounds on portal coupling $\xi^2$ in Higgs damped models with $n=10$}
	\label{fig:HiggsCouplingPlotNQHiggsMmpn10}
\end{figure}


\begin{figure}
	\centering
	\includegraphics[width=\standardwidth \textwidth]{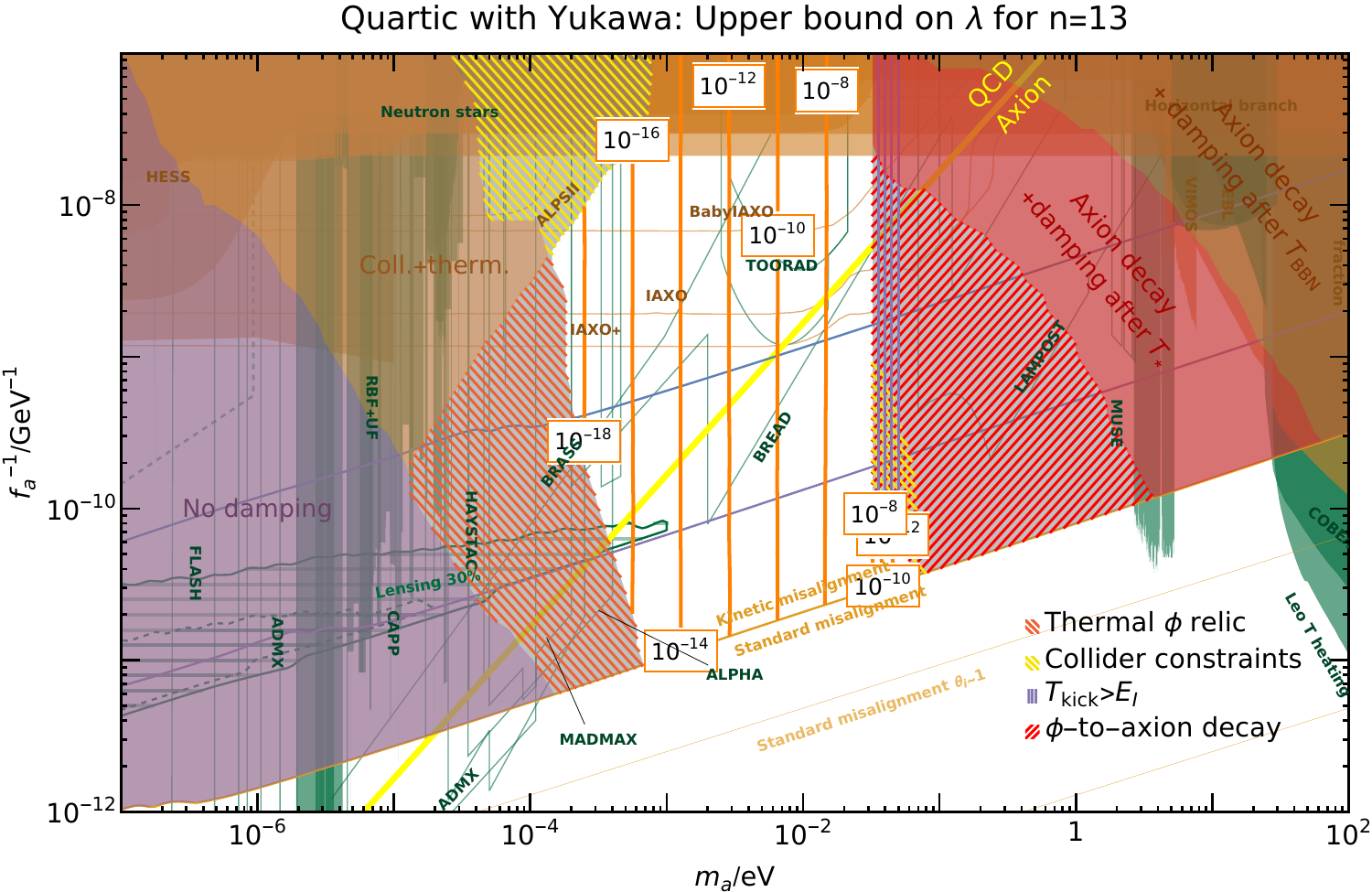}
	
	\vspace{0.5cm}
	\includegraphics[width=\standardwidth \textwidth]{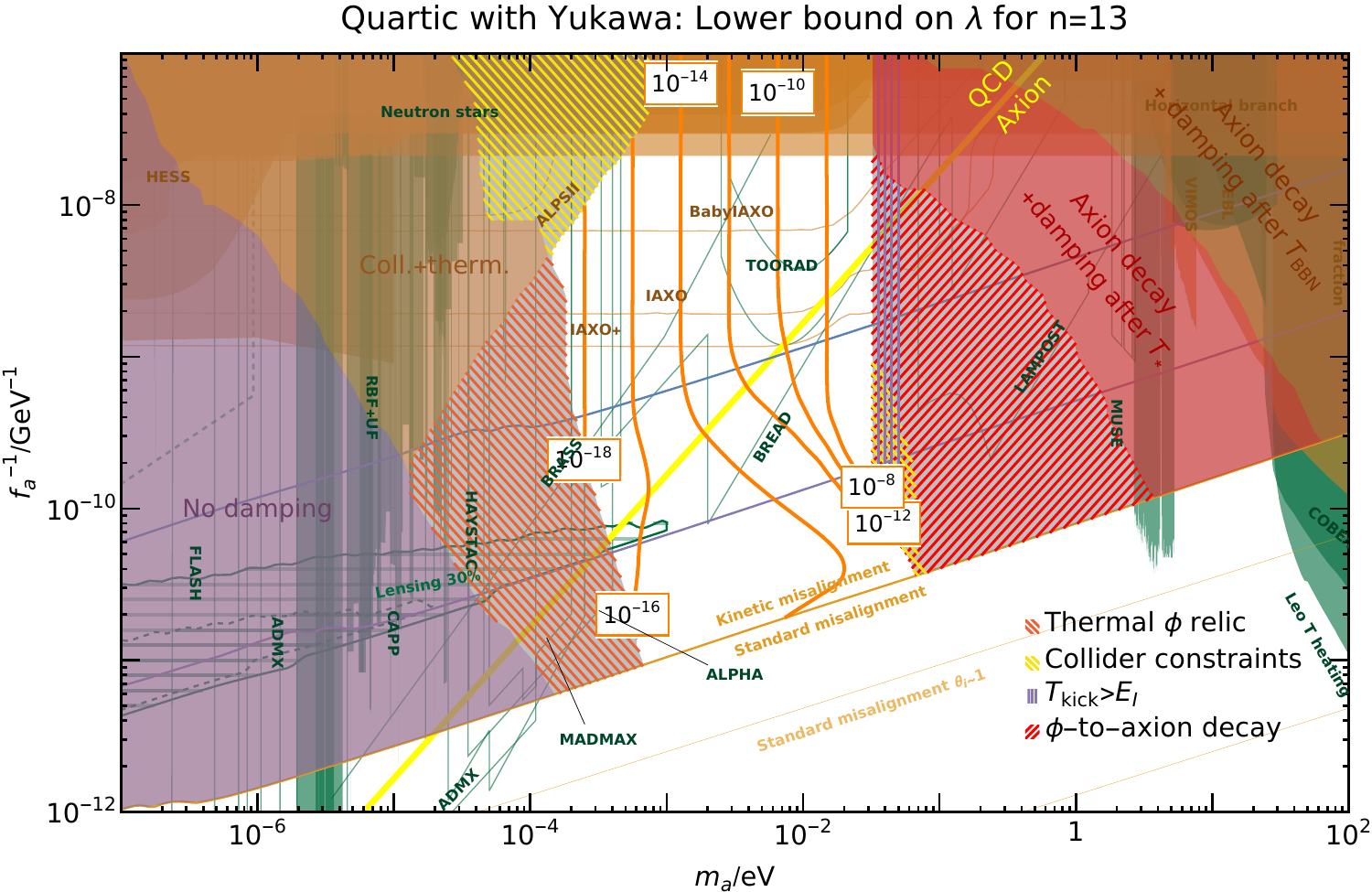}
	
	\caption{\small\it Upper and lower bounds on the quartic coupling $\lambda$ in models with quartic potentials, higher dimensional operators with $n=13$, and damping implemented with a Yukawa interaction.}
	\label{fig:QuarticPlotQHigherDimYukawaMmpn13}
\end{figure}

\begin{figure}
	\centering
	\includegraphics[width=\standardwidth \textwidth]{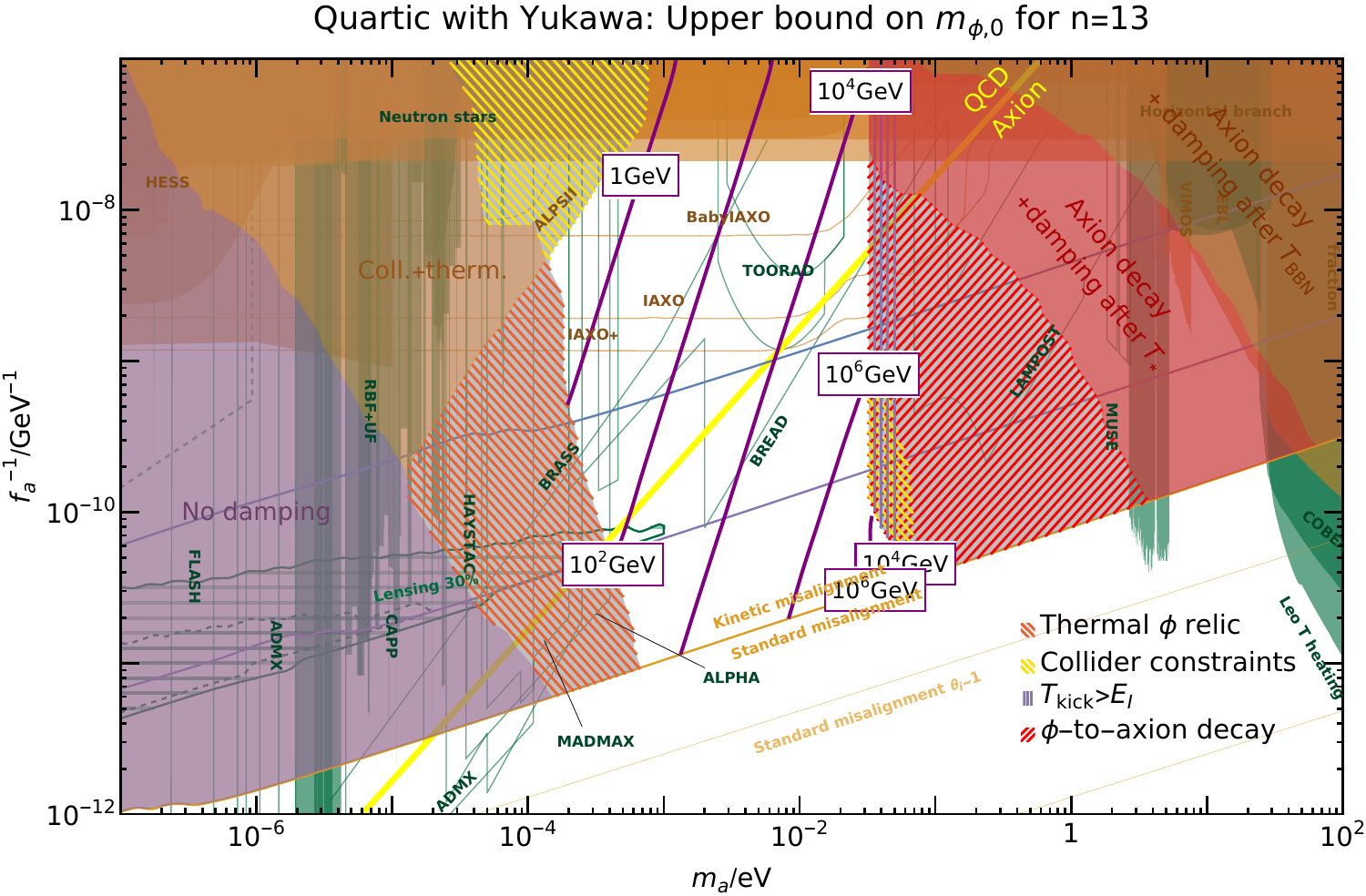}
	
	\vspace{0.5cm}
	\includegraphics[width=\standardwidth \textwidth]{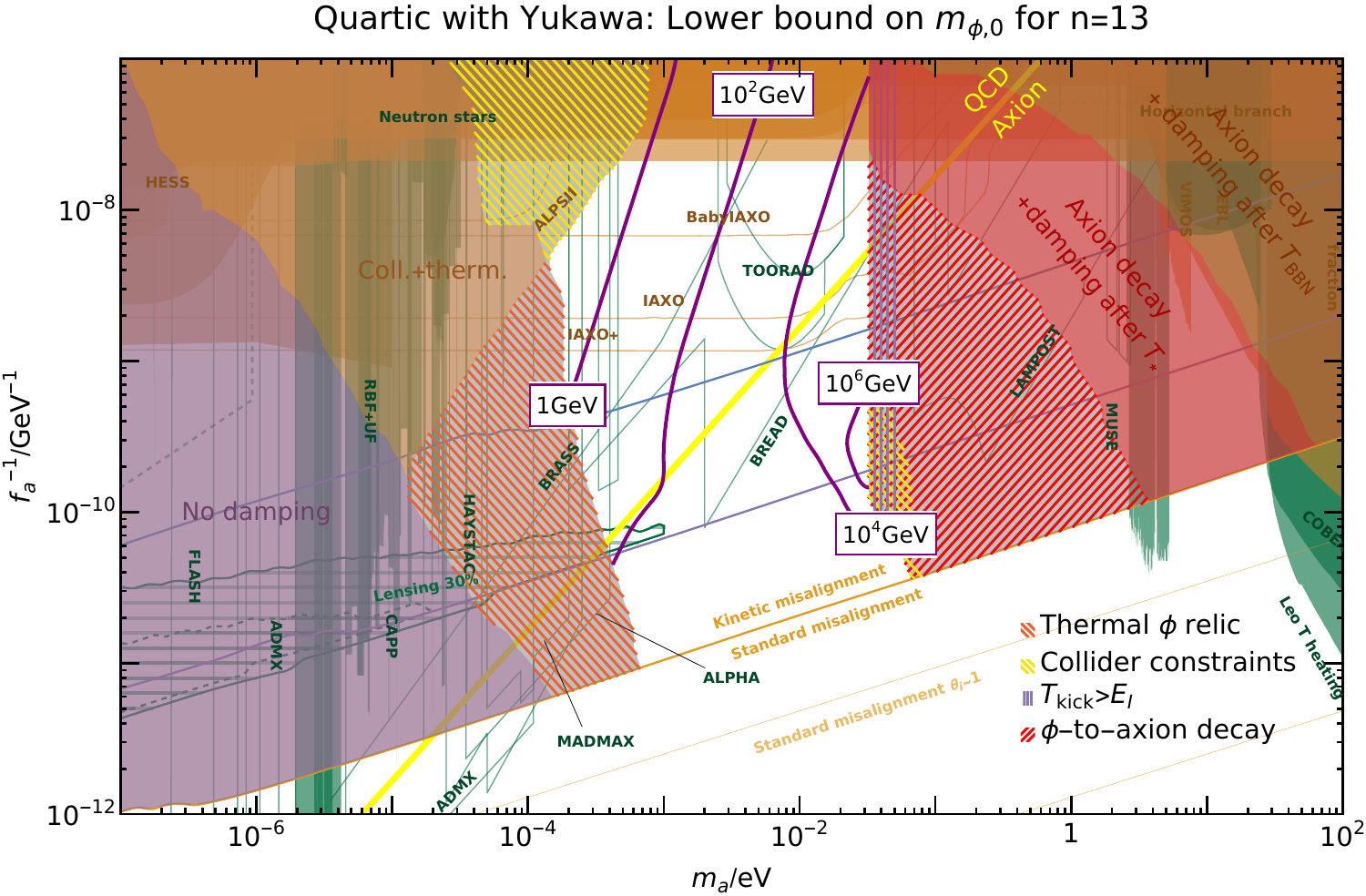}
	
	\caption{\small\it Upper and lower bounds on the radial mass $m_\phi$ in models with quartic potentials, higher dimensional operators with $n=13$, and damping implemented with a Yukawa interaction.}
	\label{fig:msPlotQHigherDimYukawaMmpn13}
\end{figure}

\begin{figure}
	\centering
	\includegraphics[width=\standardwidth \textwidth]{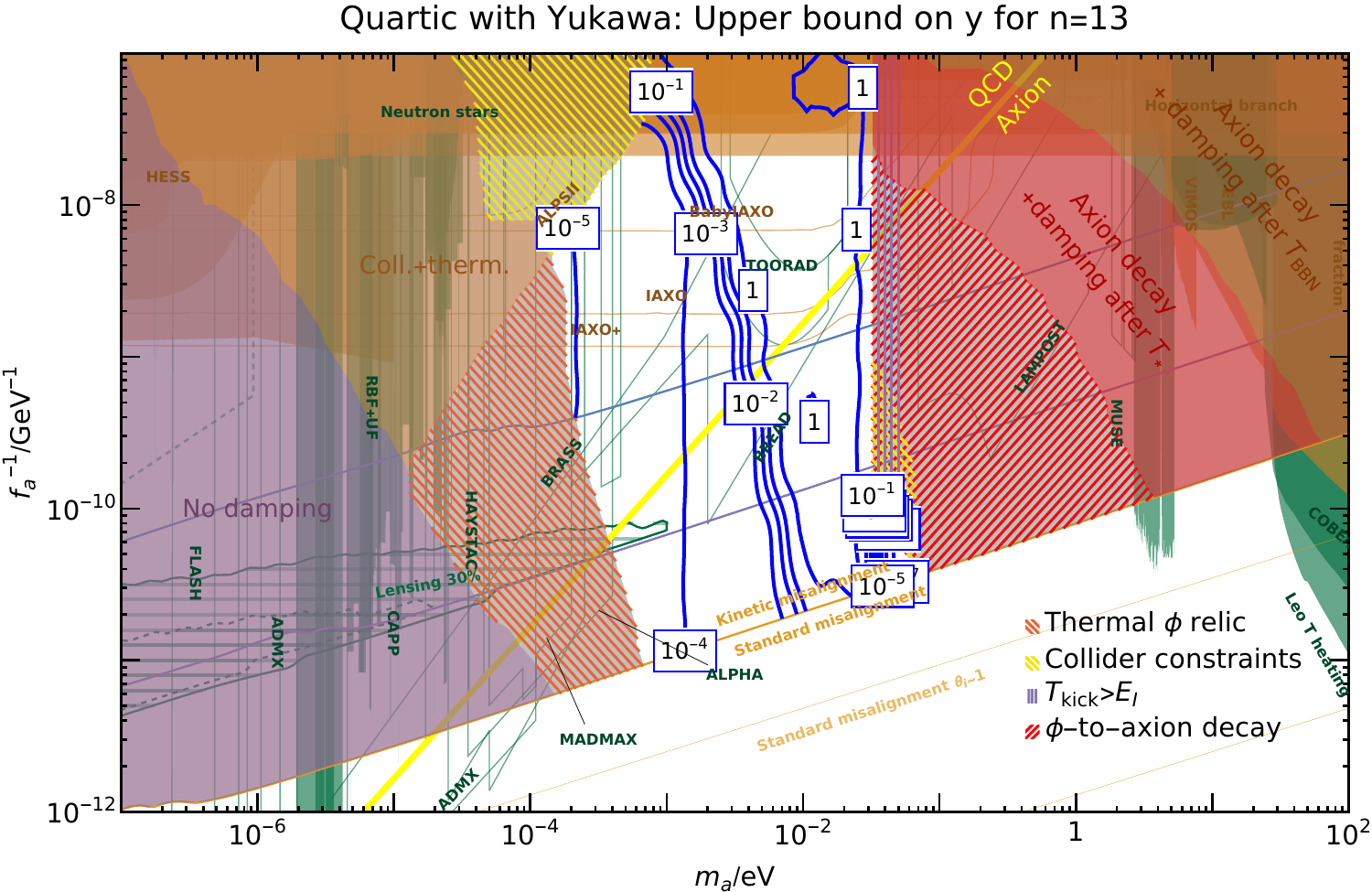}
	
	\vspace{0.5cm}
	\includegraphics[width=\standardwidth \textwidth]{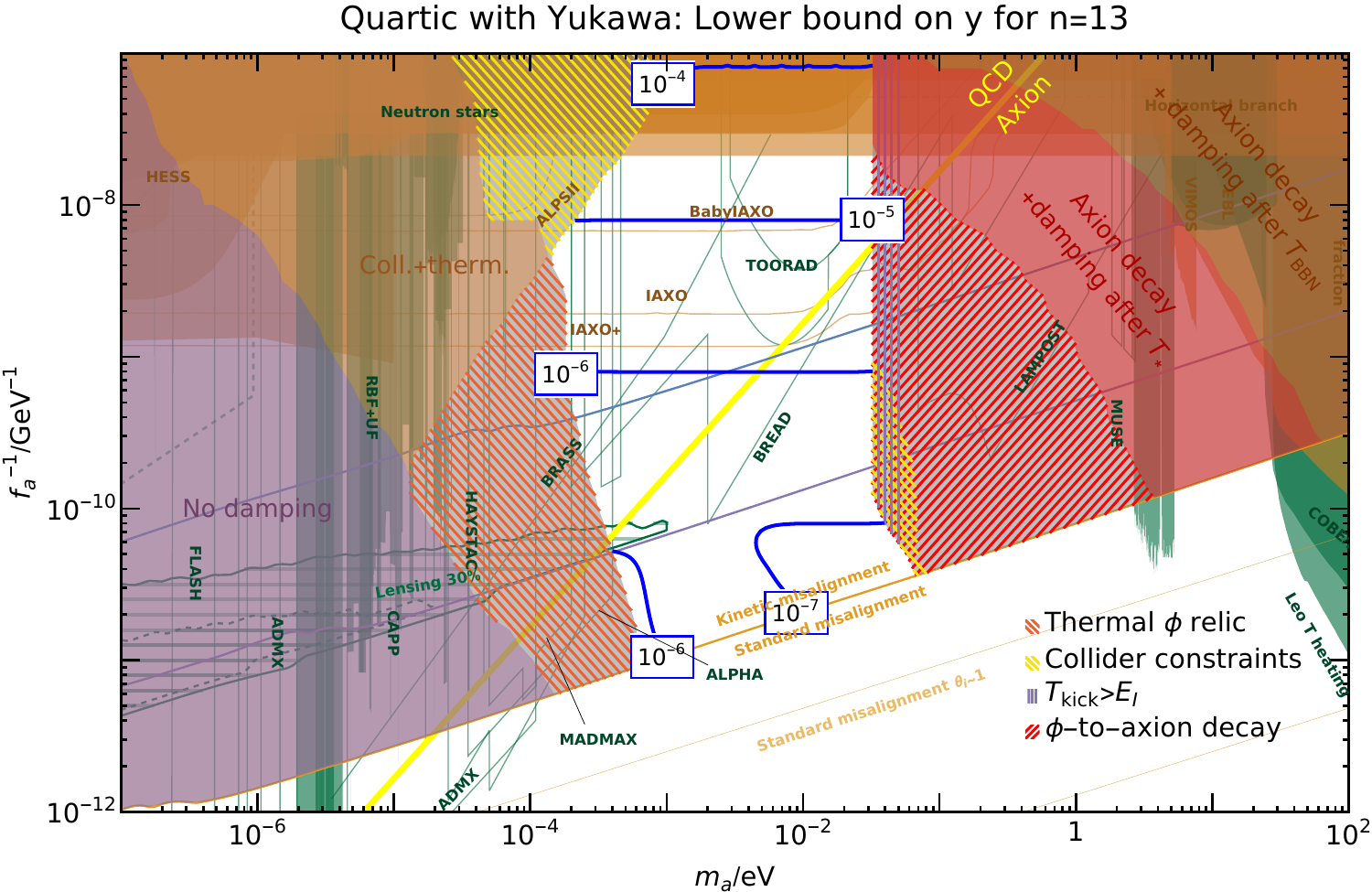}
	
	\caption{\small\it Upper and lower bounds on the Yukawa coupling $y$ in models with quartic potentials, higher dimensional operators with $n=13$, and damping implemented with a Yukawa interaction.}
	\label{fig:YukawaPlotQHigherDimYukawaMmpn13}
\end{figure}


\begin{figure}
	\centering
	\includegraphics[width=\standardwidth \textwidth]{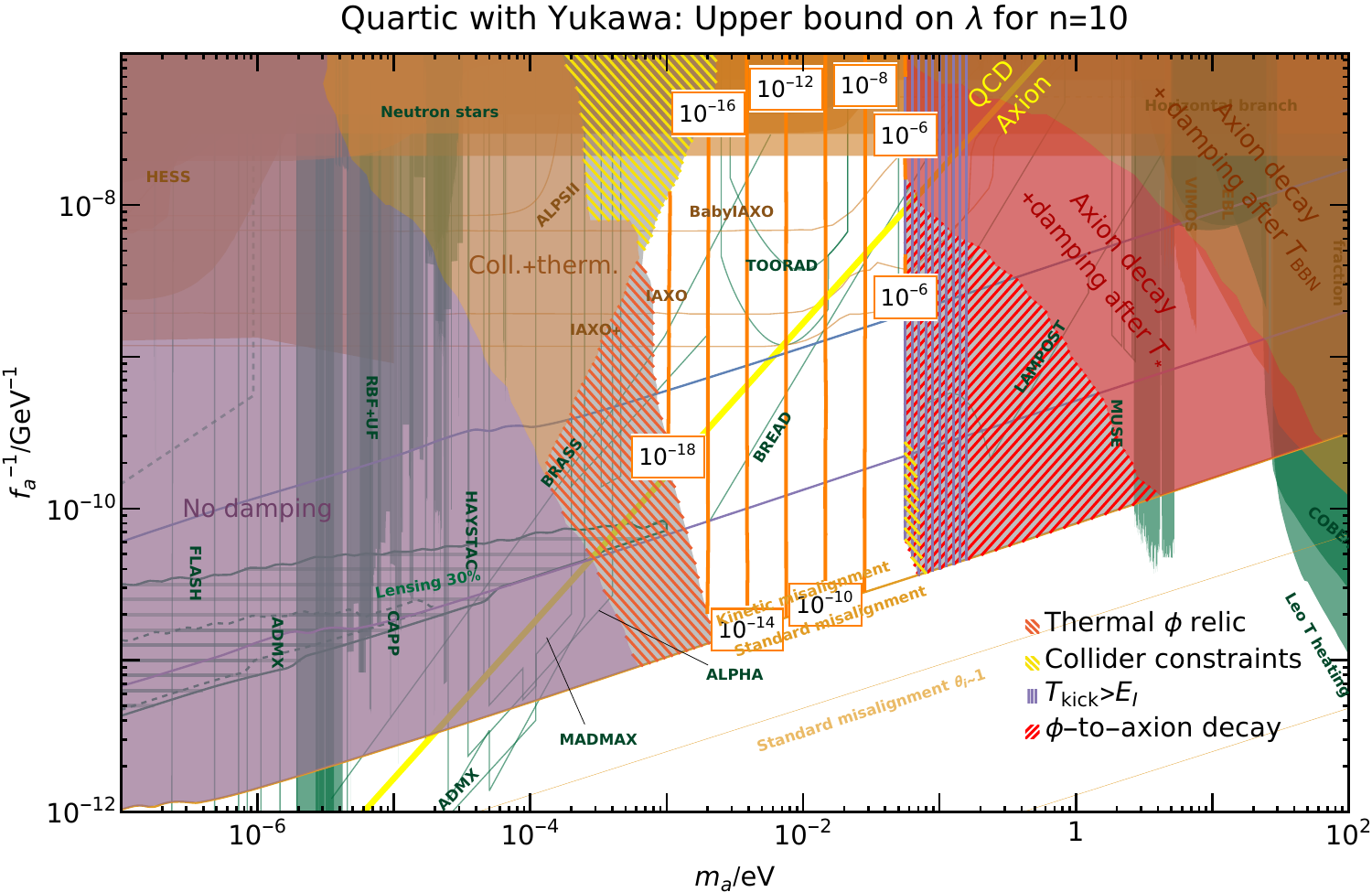}
	
	\vspace{0.5cm}
	\includegraphics[width=\standardwidth \textwidth]{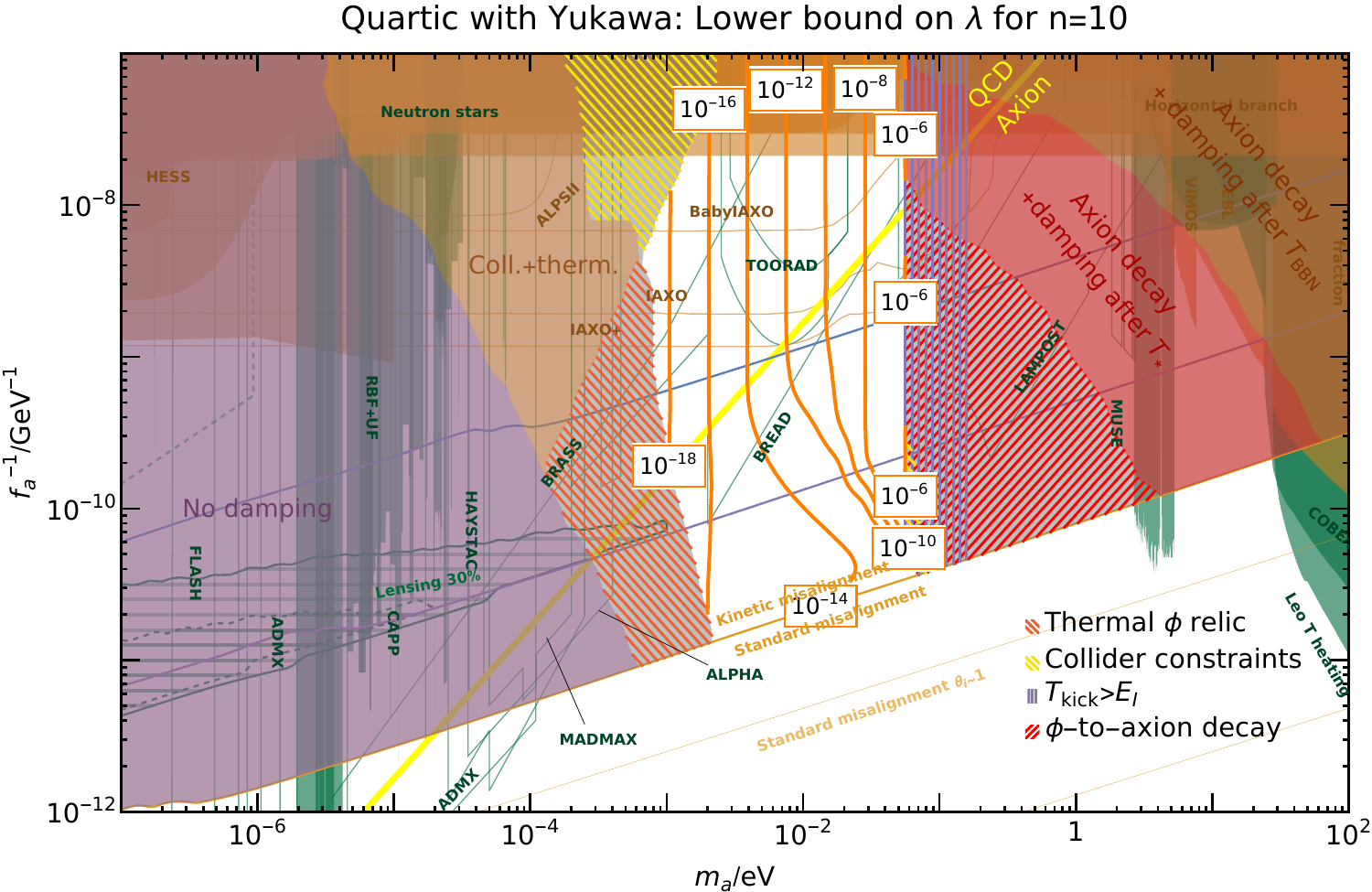}
	
	\caption{\small\it Upper and lower bounds on the quartic coupling $\lambda$ in models with quartic potentials, higher dimensional operators with $n=10$, and damping implemented with a Yukawa interaction.}
	\label{fig:QuarticPlotQHigherDimYukawaMmpn10}
\end{figure}

\begin{figure}
	\centering
	\includegraphics[width=\standardwidth \textwidth]{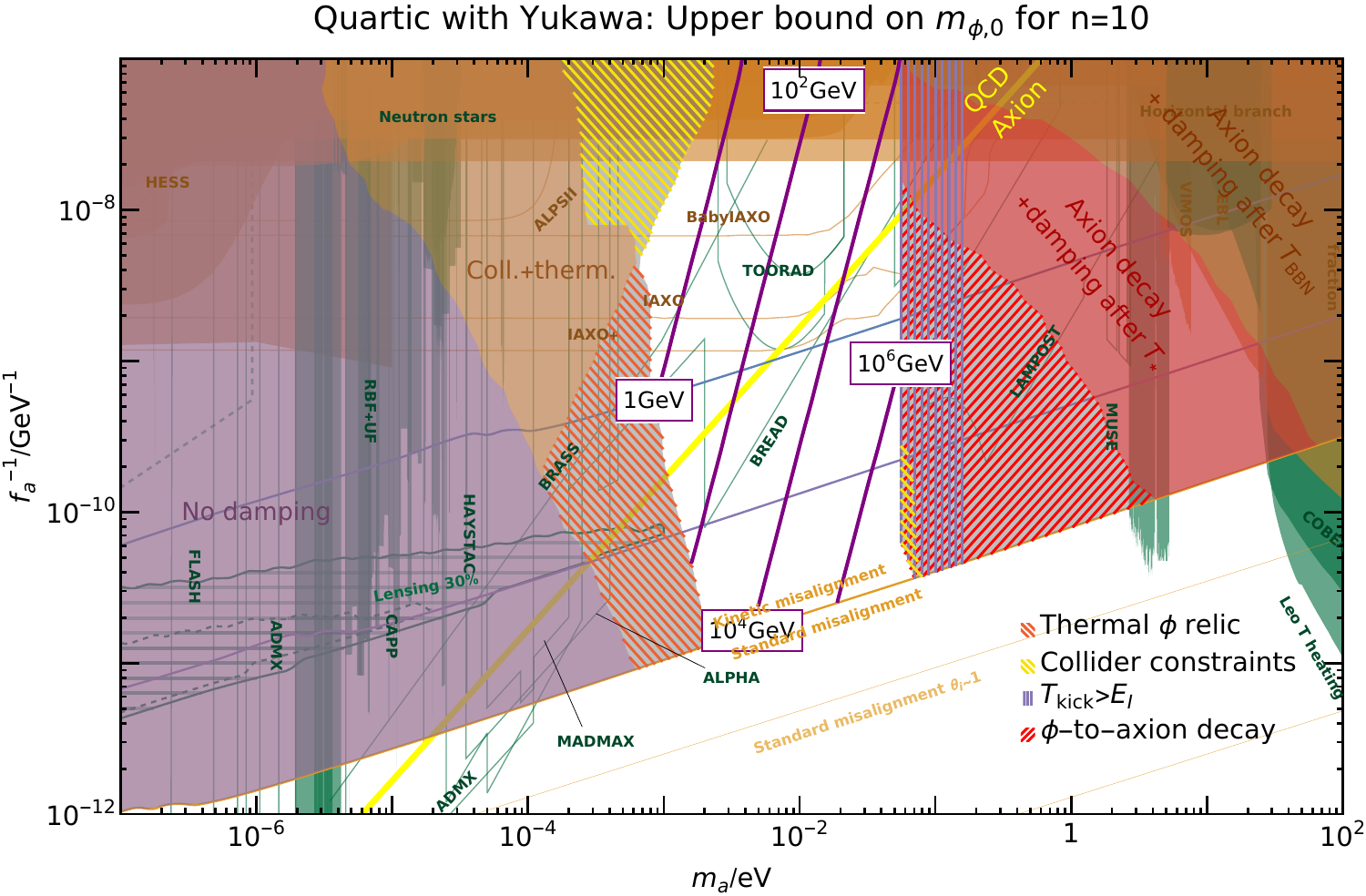}
	
	\vspace{0.5cm}
	\includegraphics[width=\standardwidth \textwidth]{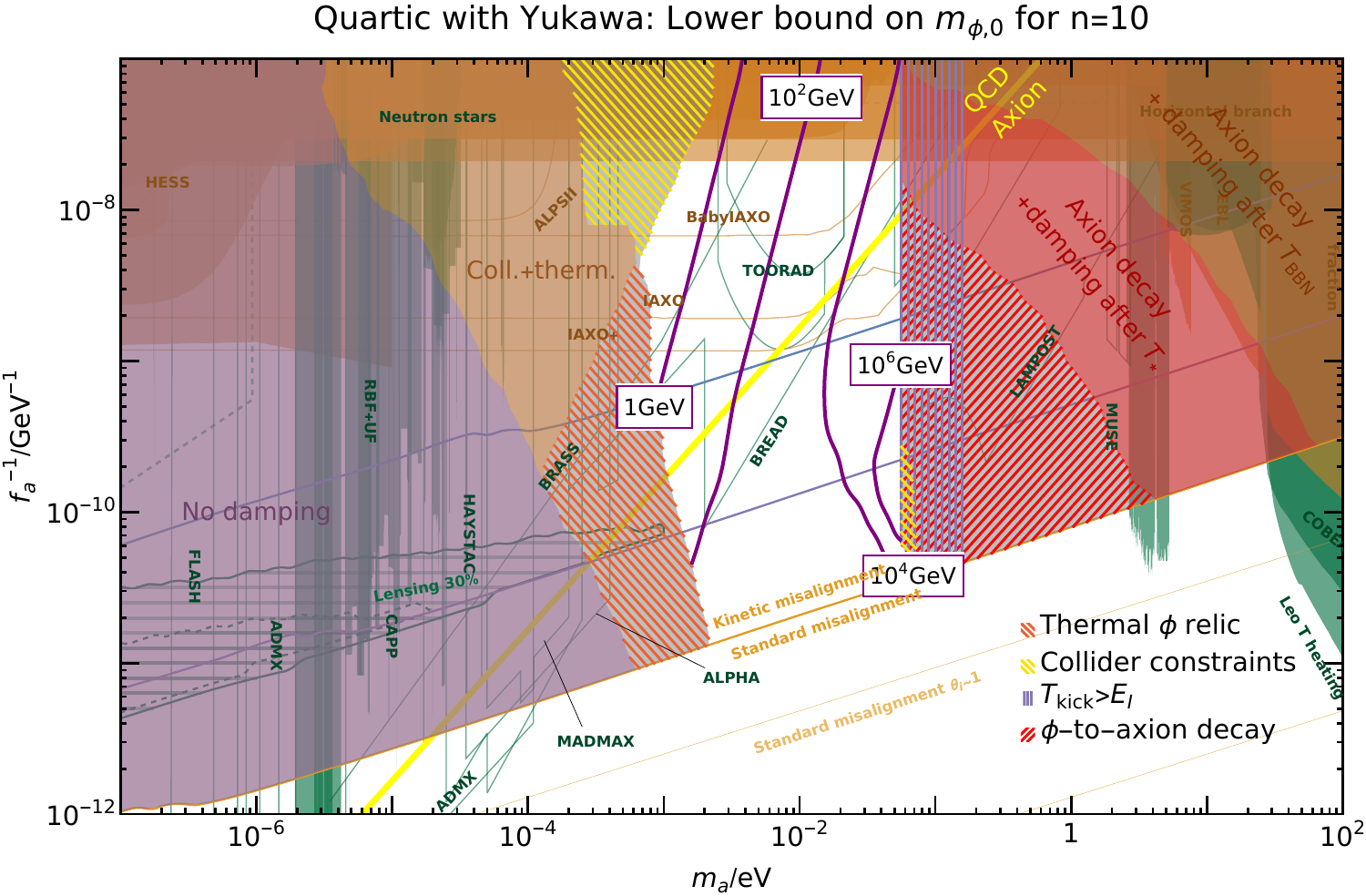}
	
	\caption{\small\it Upper and lower bounds on the radial mass $m_\phi$ in models with quartic potentials, higher dimensional operators with $n=10$, and damping implemented with a Yukawa interaction.}
	\label{fig:msPlotQHigherDimYukawaMmpn10}
\end{figure}

\begin{figure}
	\centering
	\includegraphics[width=\standardwidth \textwidth]{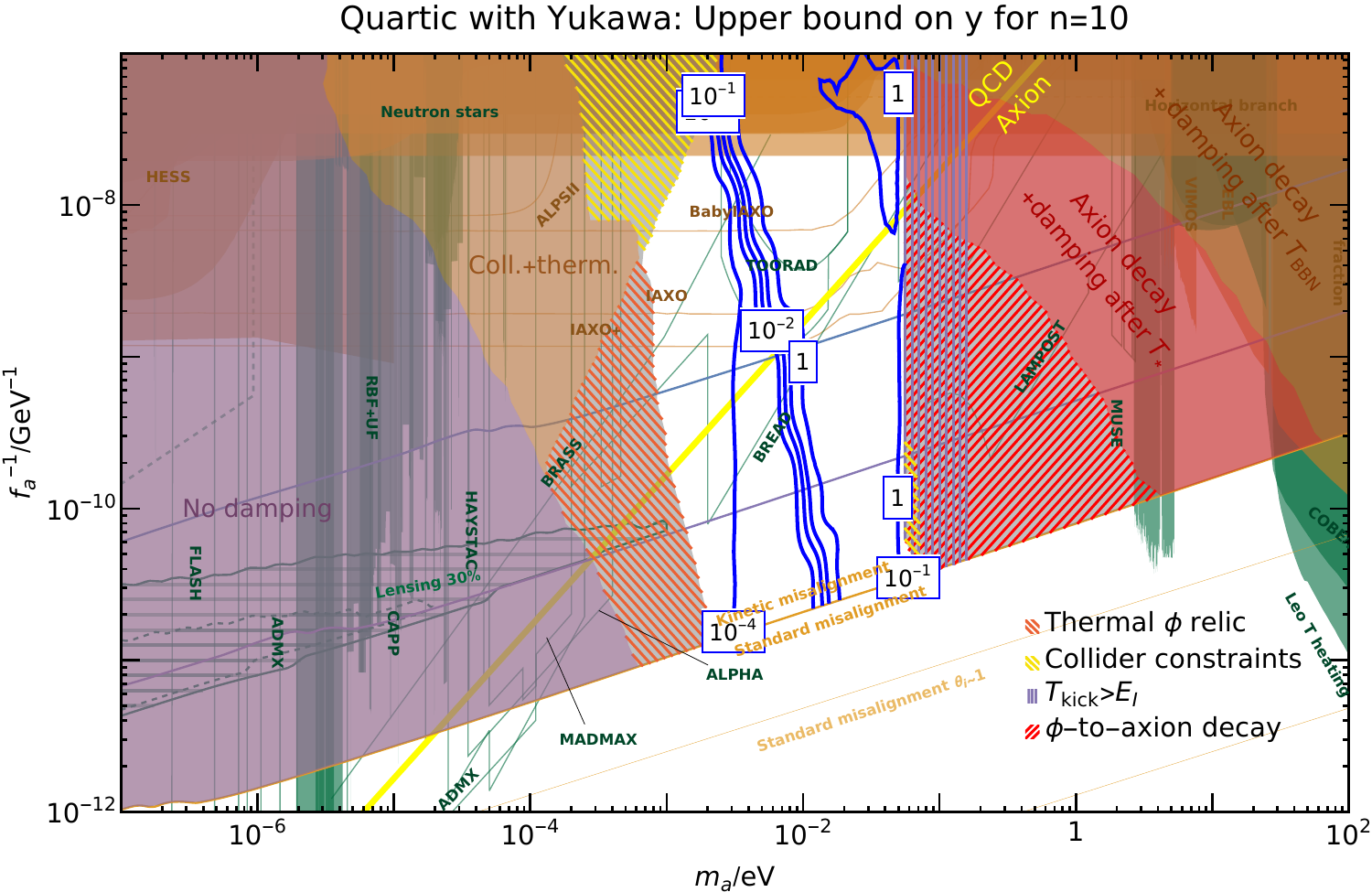}
	
	\vspace{0.5cm}
	\includegraphics[width=\standardwidth \textwidth]{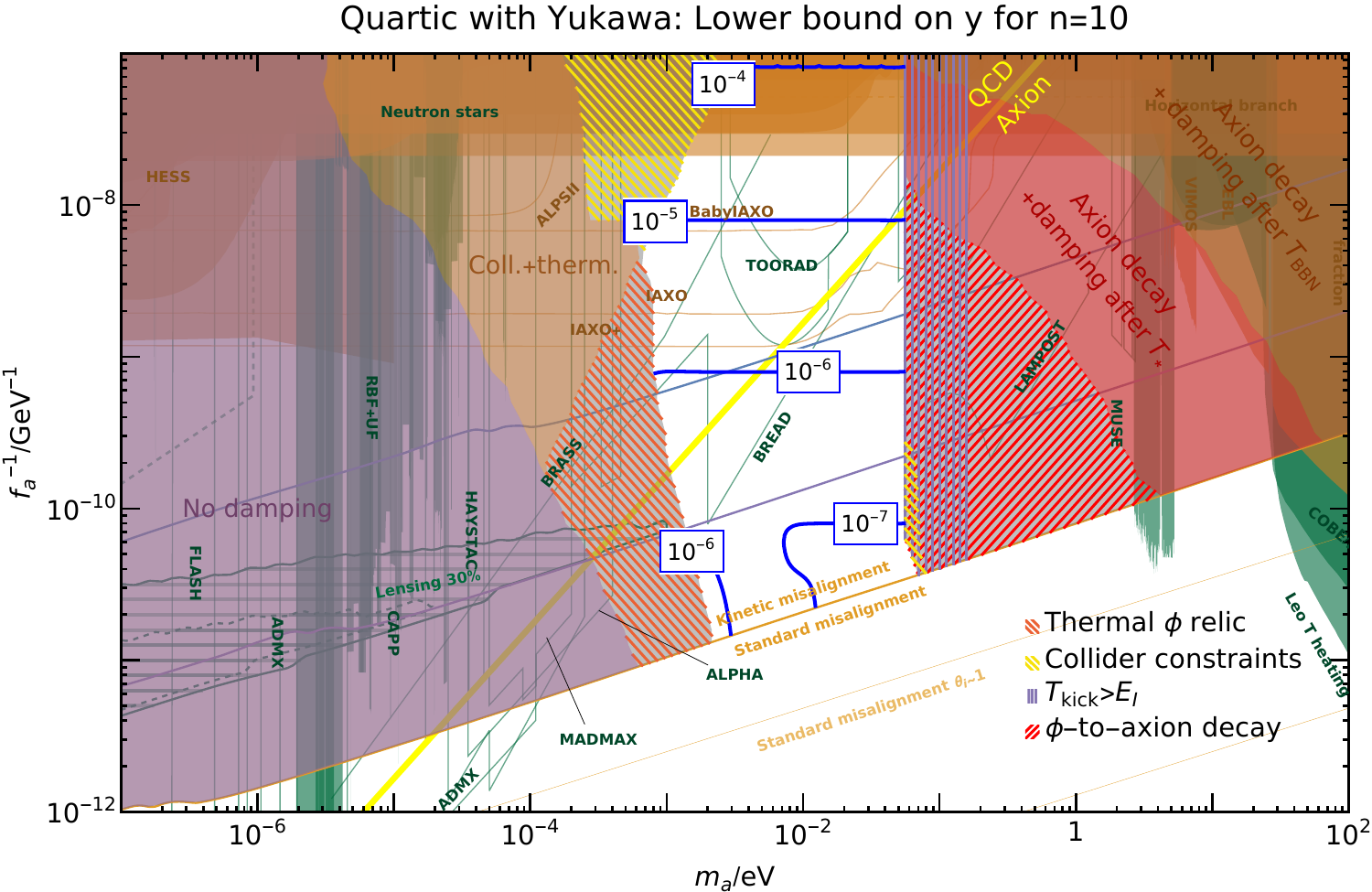}
	
	\caption{\small\it Upper and lower bounds on the Yukawa coupling $y$ in models with quartic potentials, higher dimensional operators with $n=10$, and damping implemented with a Yukawa interaction.}
	\label{fig:YukawaPlotQHigherDimYukawaMmpn10}
\end{figure}


\begin{figure}
	\centering
	\includegraphics[width=\standardwidth \textwidth]{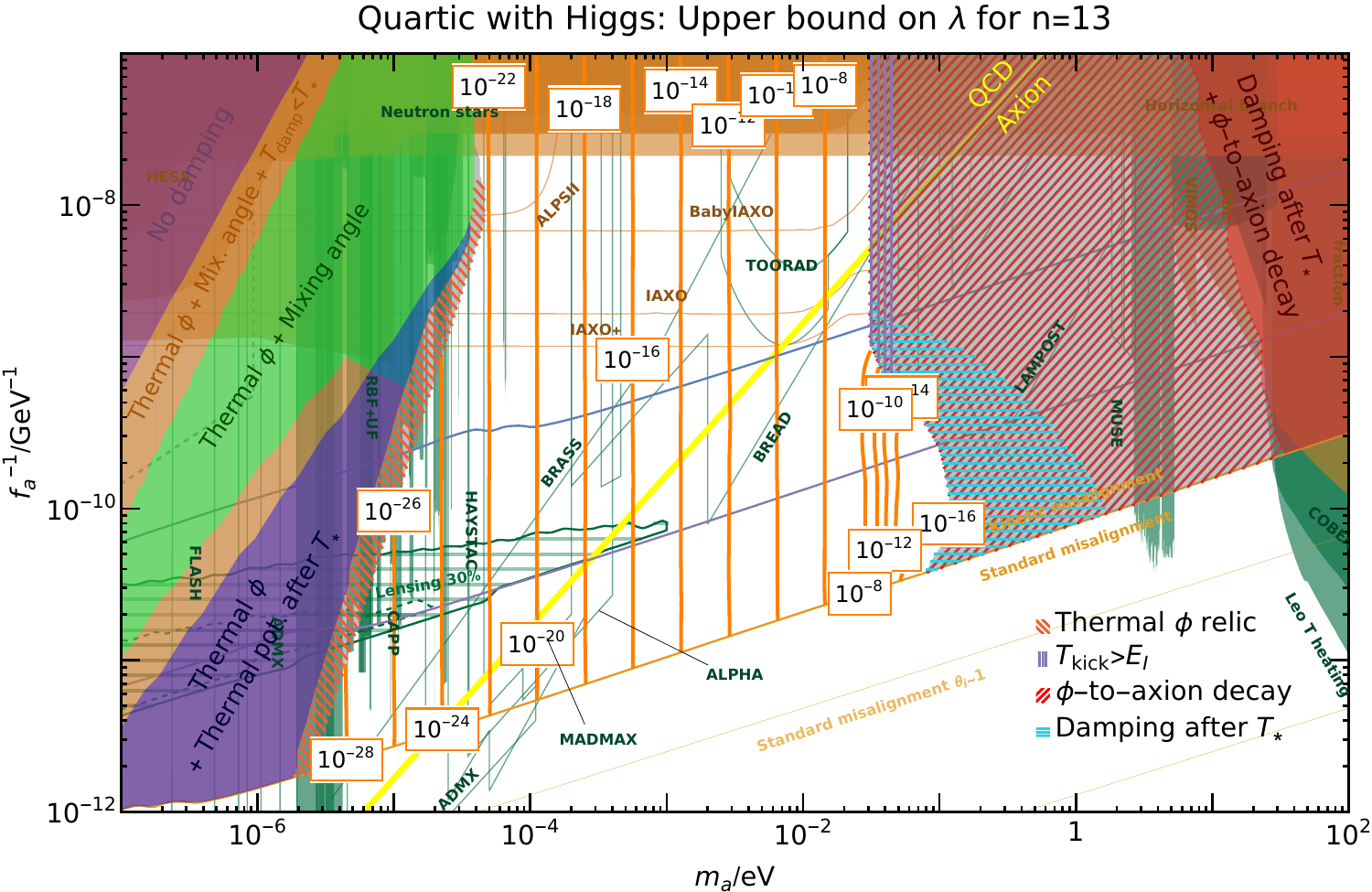}
	
	\vspace{0.5cm}
	\includegraphics[width=\standardwidth \textwidth]{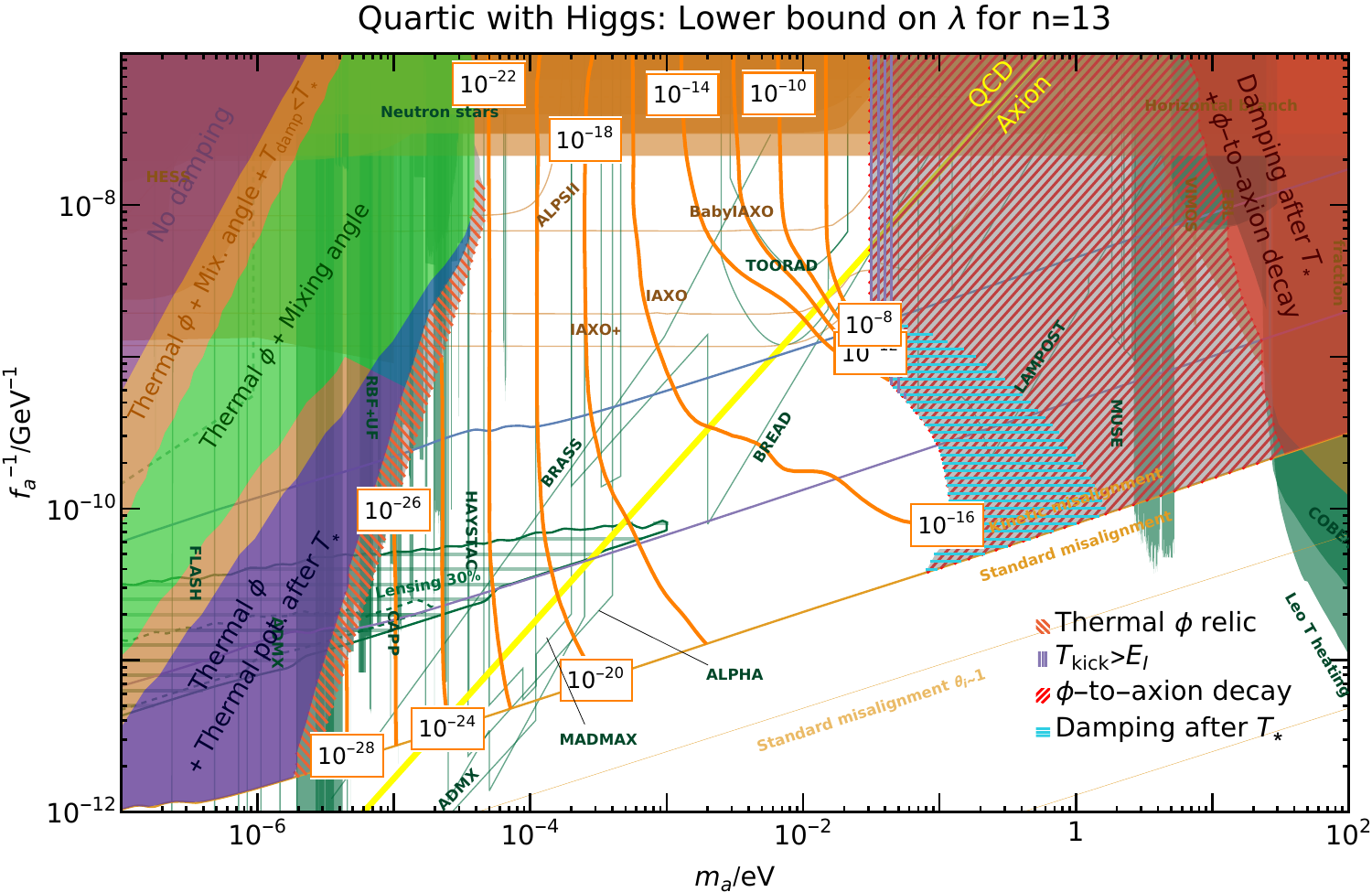}
	
	\caption{\small\it Upper and lower bounds on the quartic coupling $\lambda$ in models with quartic potentials, higher dimensional operators with $n=13$, and damping implemented with a Higgs interaction.}
	\label{fig:QuarticPlotQHigherDimHiggsMmpn13}
\end{figure}

\begin{figure}
	\centering
	\includegraphics[width=\standardwidth \textwidth]{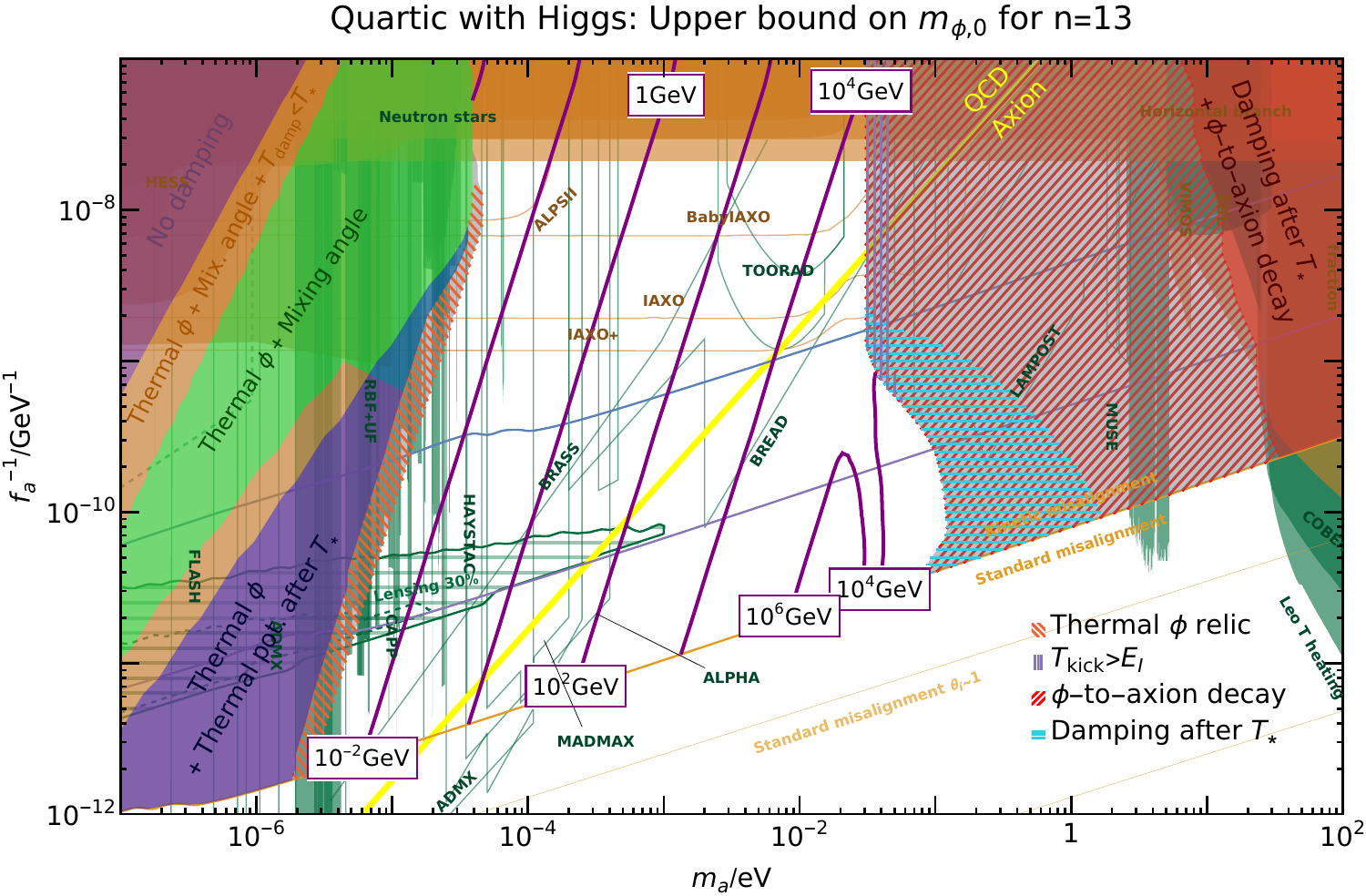}
	
	\vspace{0.5cm}
	\includegraphics[width=\standardwidth \textwidth]{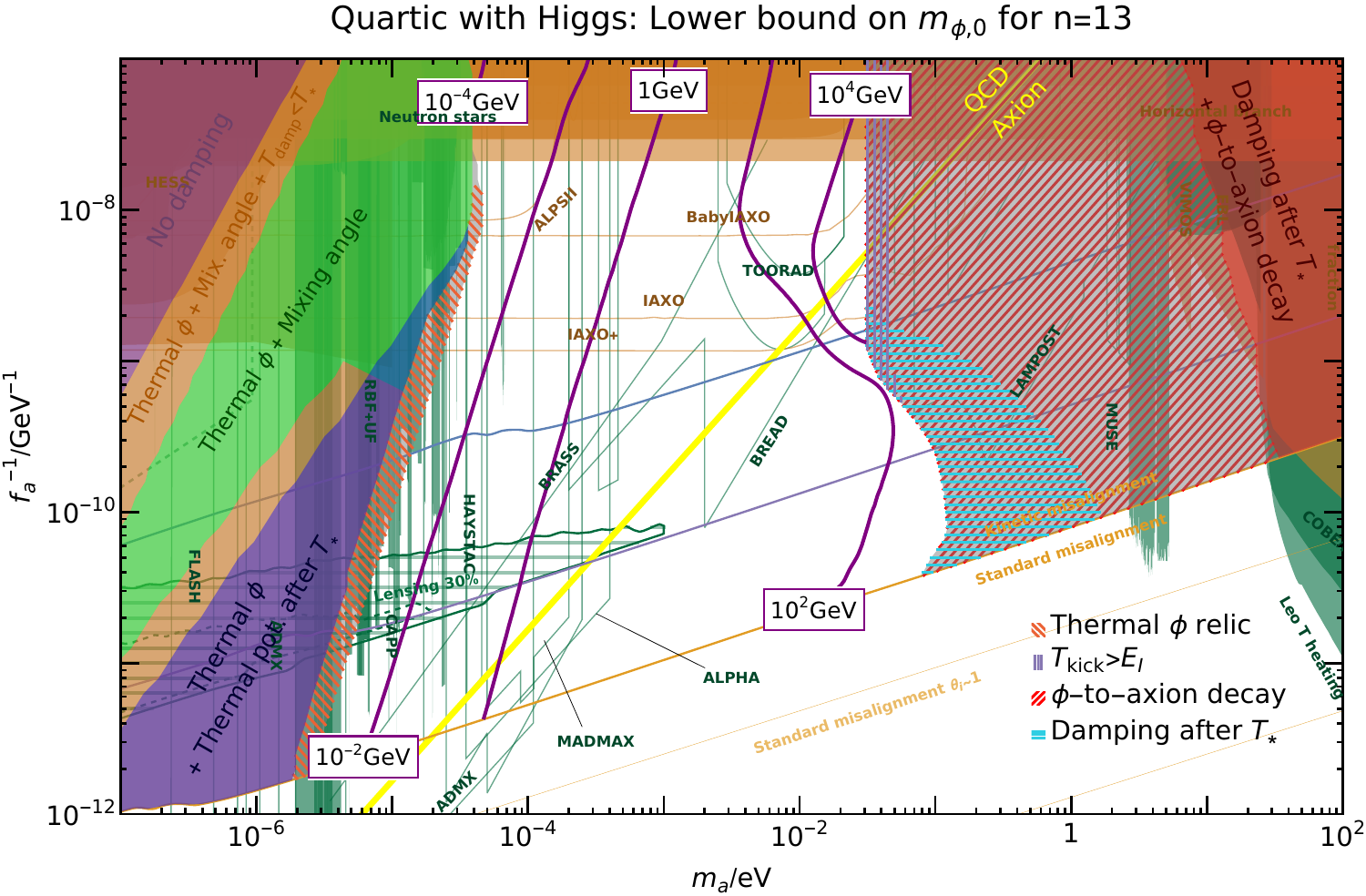}
	
	\caption{\small\it Upper and lower bounds on the radial mass $m_\phi$ in models with quartic potentials, higher dimensional operators with $n=13$, and damping implemented with a Higgs interaction.}
	\label{fig:msPlotQHigherDimHiggsMmpn13}
\end{figure}

\begin{figure}
	\centering
	\includegraphics[width=\standardwidth \textwidth]{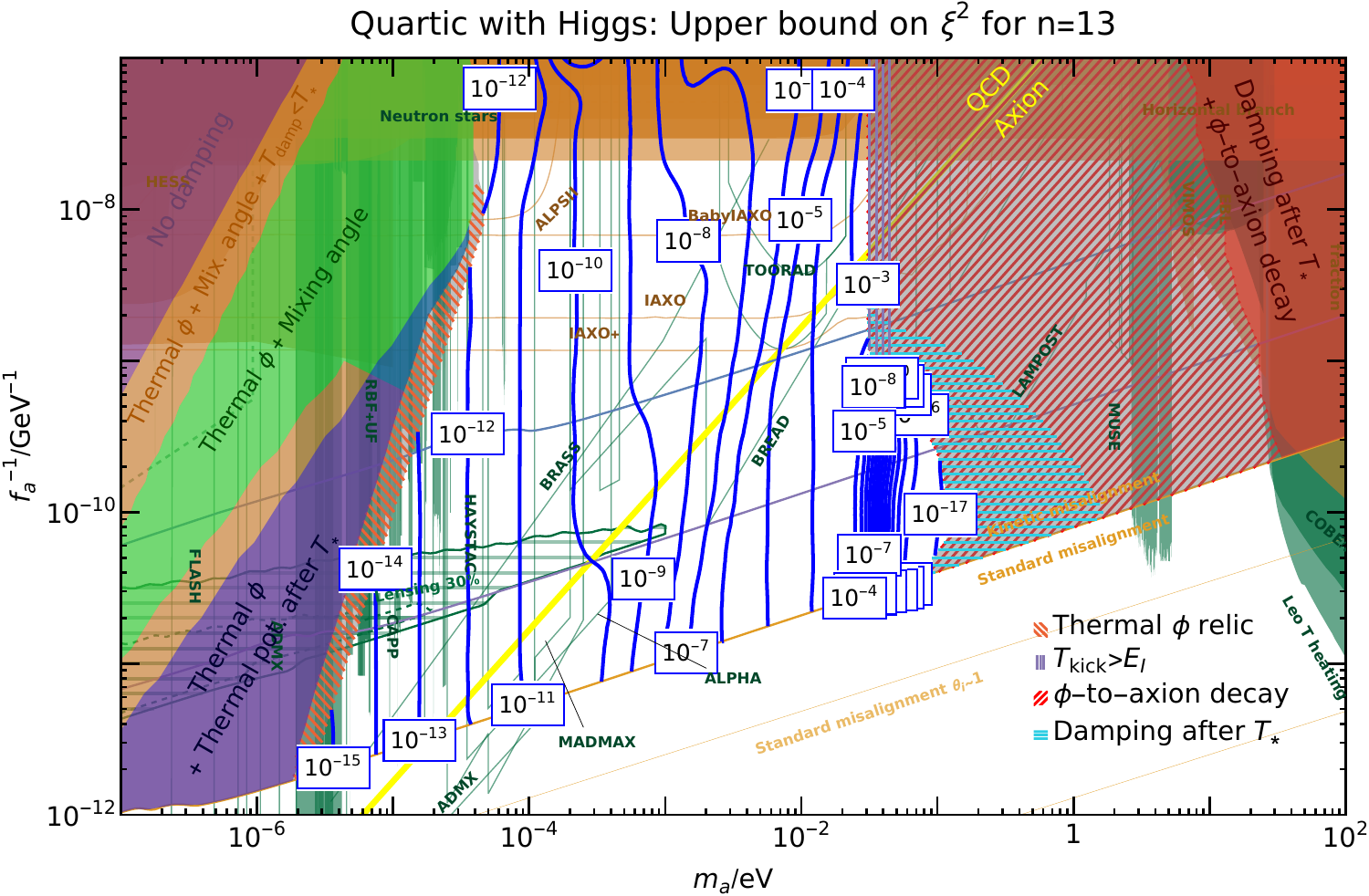}
	
	\vspace{0.5cm}
	\includegraphics[width=\standardwidth \textwidth]{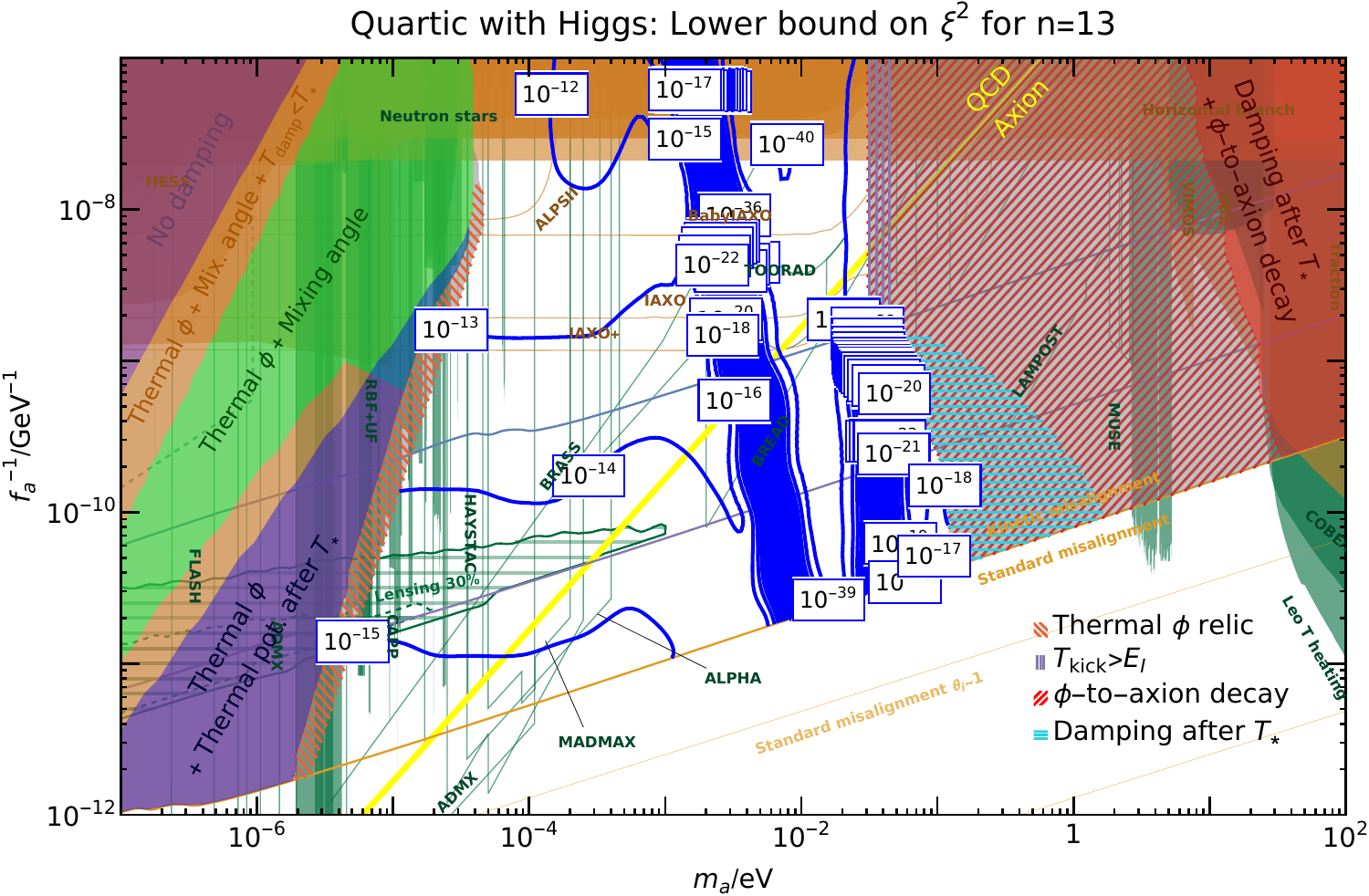}
	
	\caption{\small\it Upper and lower bounds on the Higgs coupling $\xi^2$ in models with quartic potentials, higher dimensional operators with $n=13$, and damping implemented with a Higgs interaction. The steep drop observed around $m_a\sim 10^{-2}$ eV is because $\phi$-to-axion decay can account for damping along in that regime, and the axion-Higgs coupling $\xi^{2}$ can therefore be set to zero. The small-but-finite value displayed in this regime is caused by the minimum numerical range of our scan, but it should be interpreted as zero. See Fig.~\ref{fig:quarticHubble13WithAxionDamping}.}
	\label{fig:HiggsPlotQHigherDimHiggsMmpn13}
\end{figure}


\begin{figure}
	\centering
	\includegraphics[width=\standardwidth \textwidth]{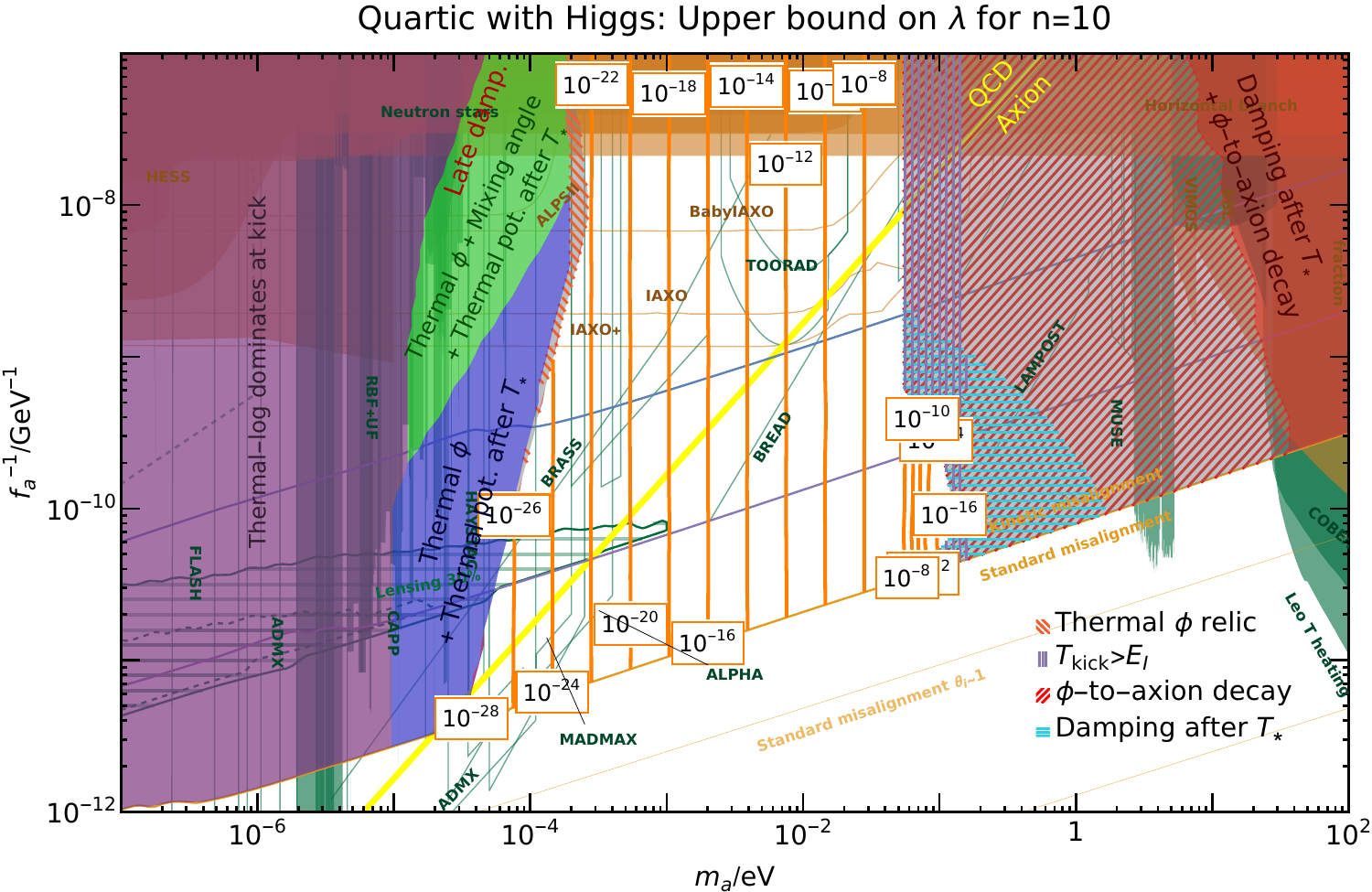}
	
	\vspace{0.5cm}
	\includegraphics[width=\standardwidth \textwidth]{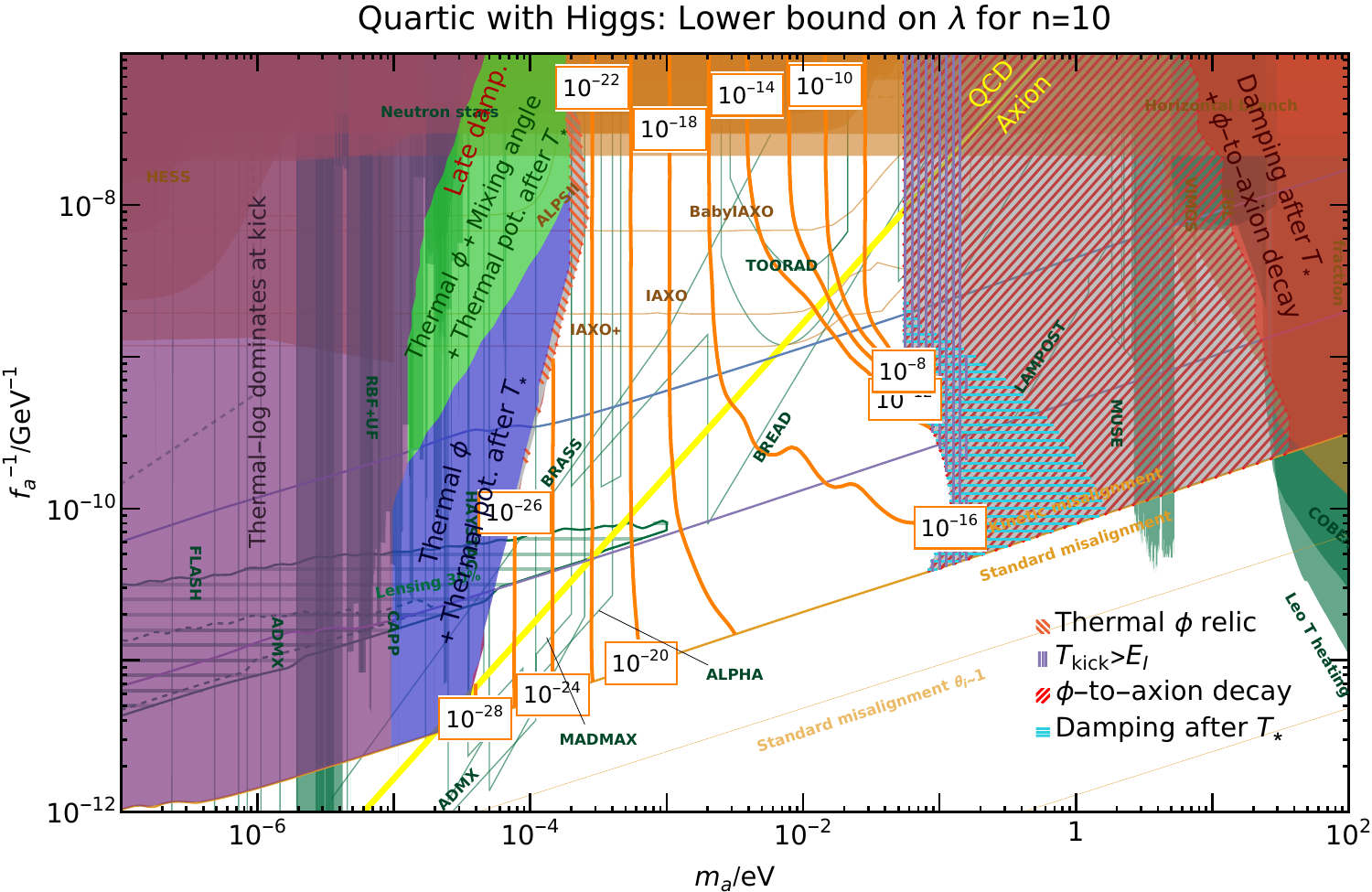}
	
	\caption{\small\it Upper and lower bounds on the quartic coupling $\lambda$ in models with quartic potentials, higher dimensional operators with with $n=10$, and damping implemented with a Higgs interaction.}
	\label{fig:QuarticPlotQHigherDimHiggsMmpn10}
\end{figure}

\begin{figure}
	\centering
	\includegraphics[width=\standardwidth \textwidth]{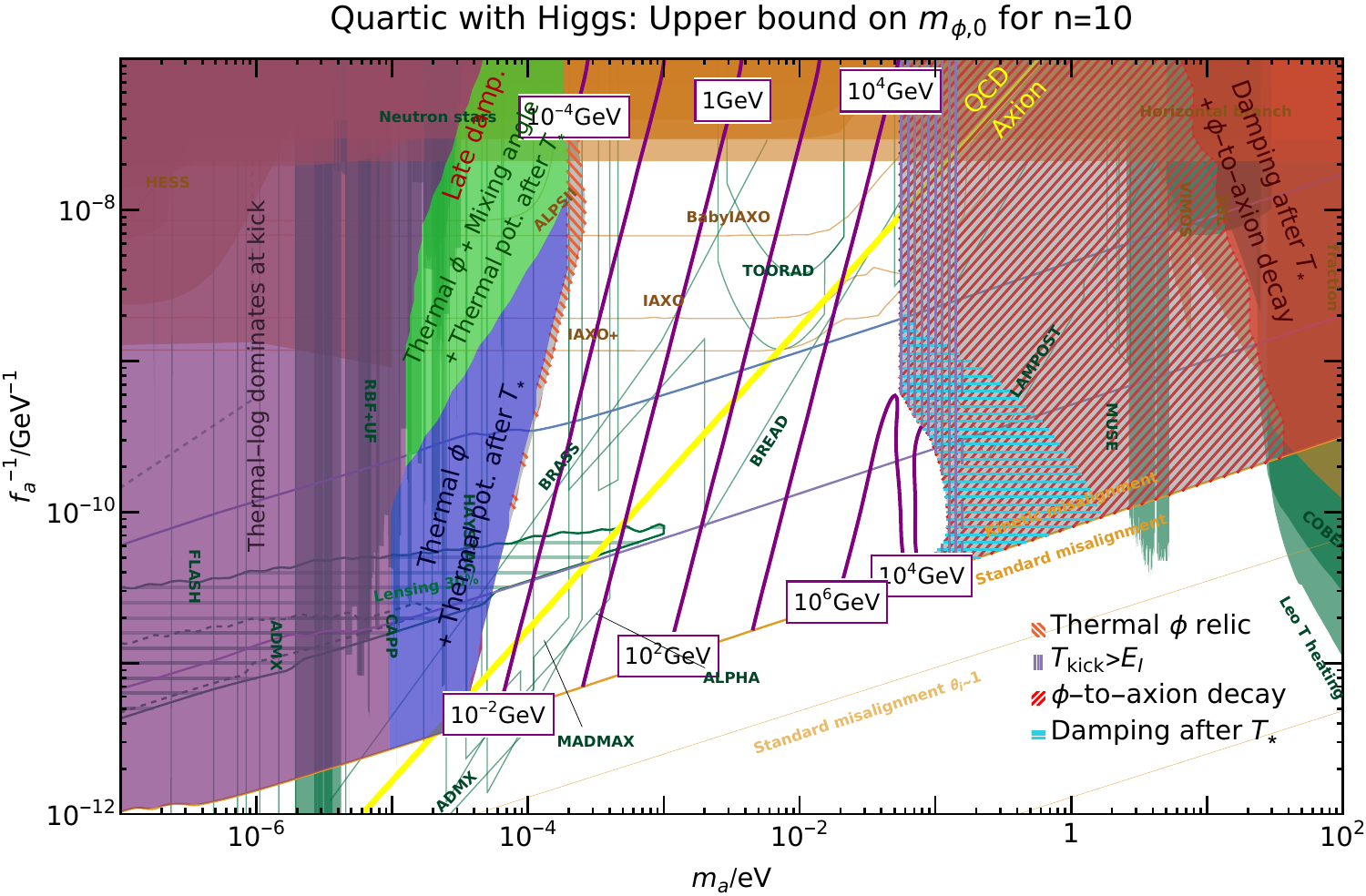}
	
	\vspace{0.5cm}
	\includegraphics[width=\standardwidth \textwidth]{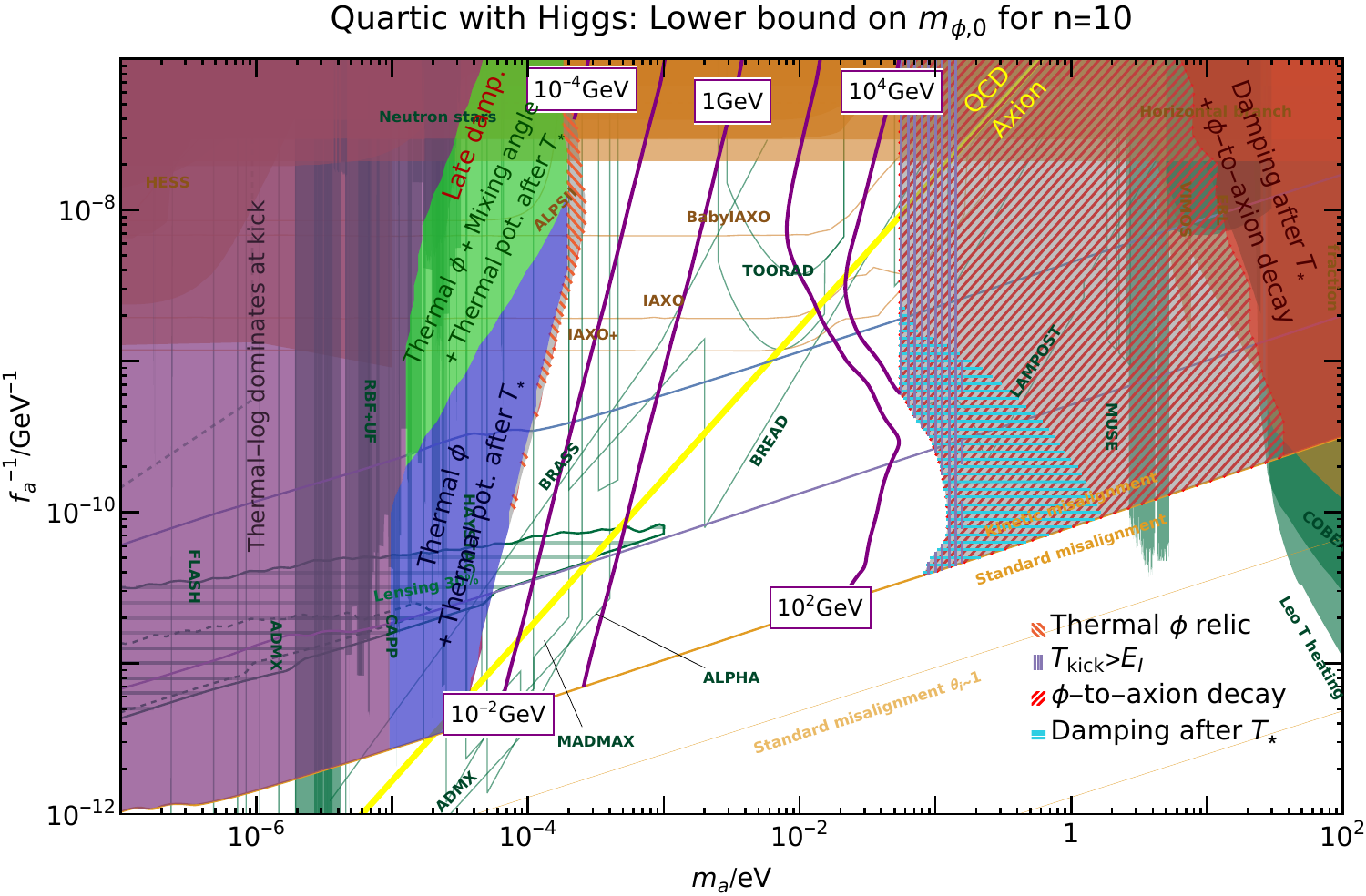}
	
	\caption{\small\it Upper and lower bounds on the radial mass $m_\phi$ in models with quartic potentials, higher dimensional operators with $n=10$, and damping implemented with a Higgs interaction.}
	\label{fig:msPlotQHigherDimHiggsMmpn10}
\end{figure}

\begin{figure}
	\centering
	\includegraphics[width=\standardwidth \textwidth]{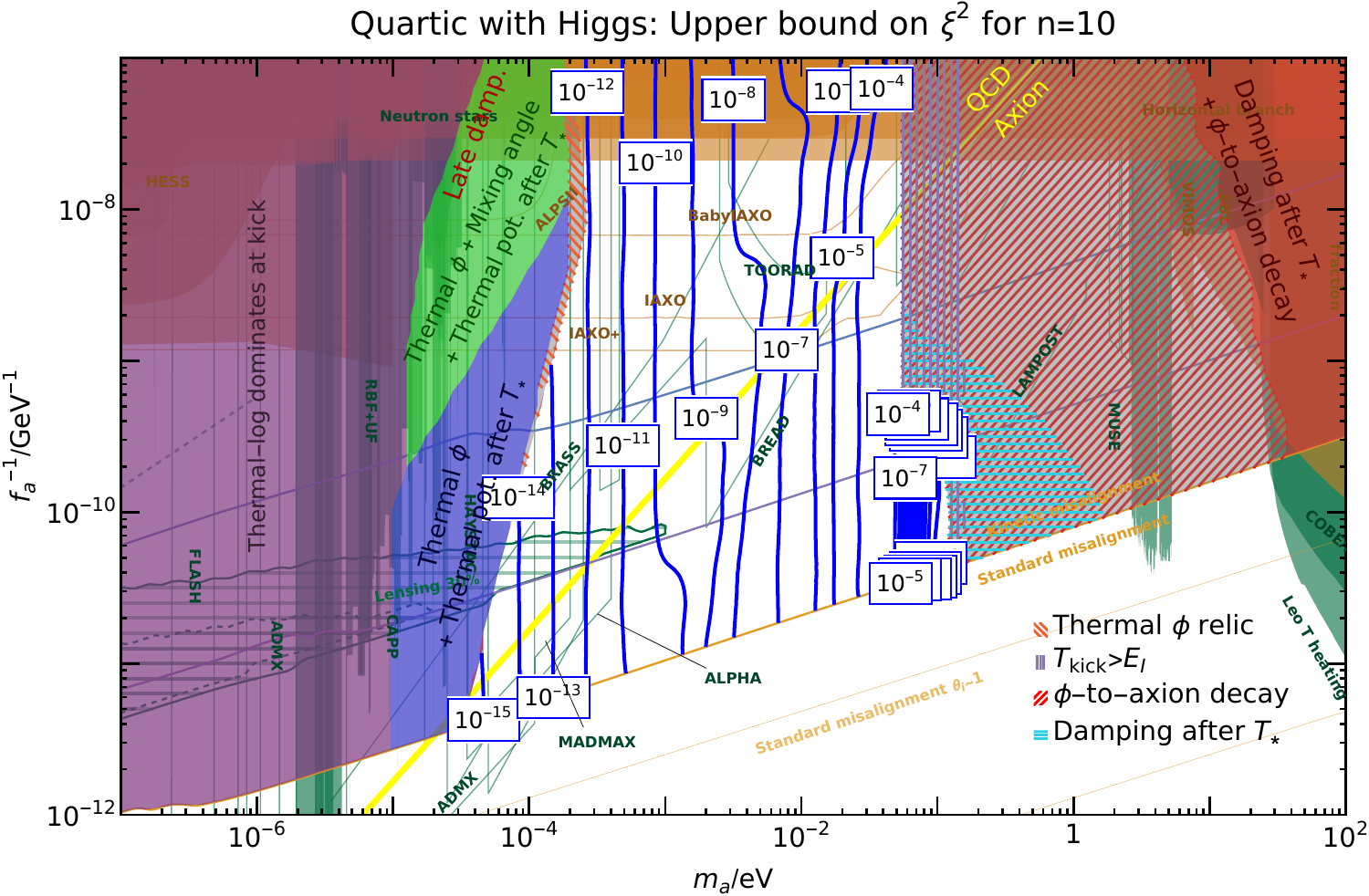}
	
	\vspace{0.5cm}
	\includegraphics[width=\standardwidth \textwidth]{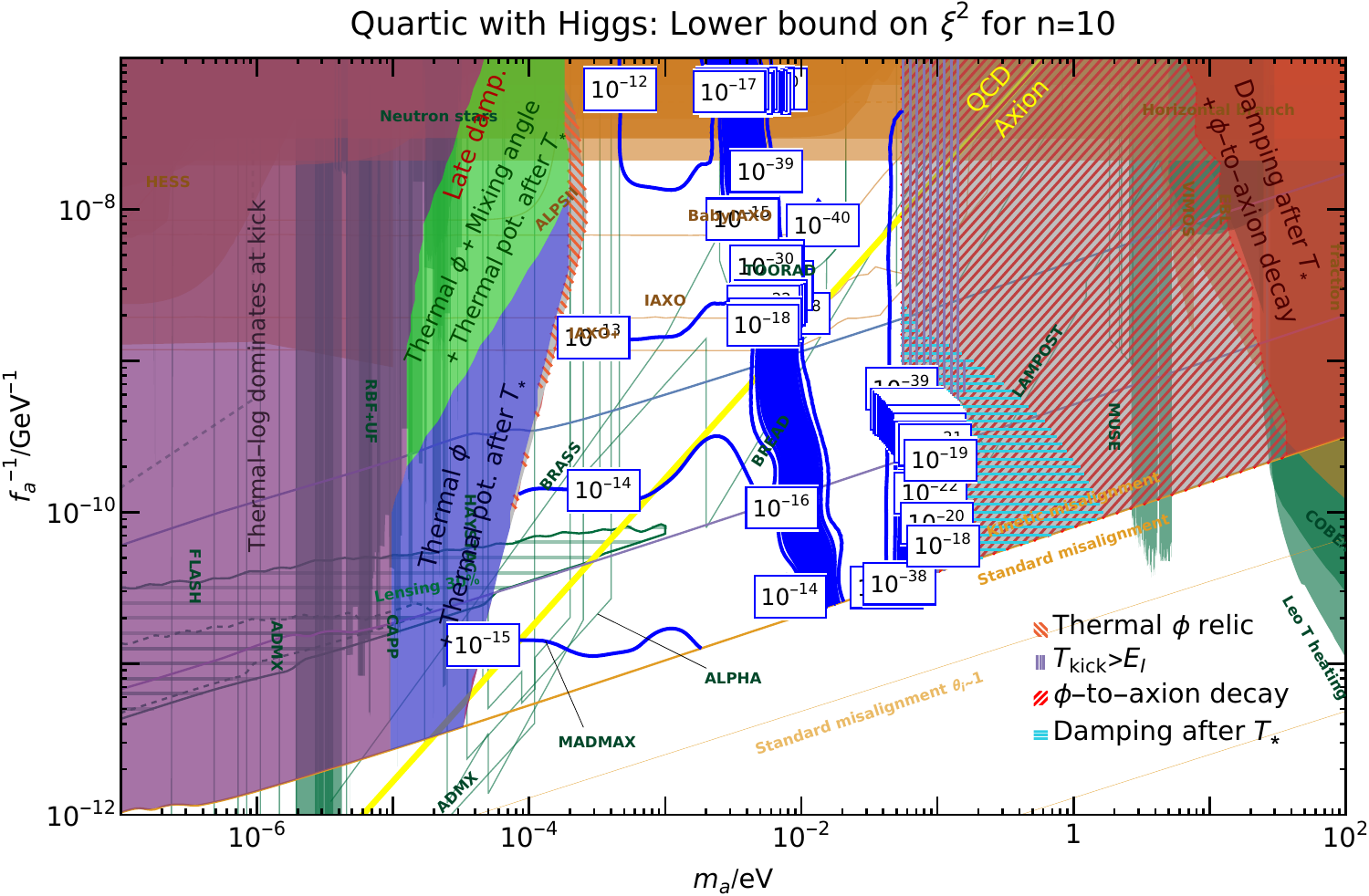}
	
	\caption{\small\it Upper and lower bounds on the Higgs coupling $\xi^2$ in models with quartic potentials, higher dimensional operators with $n=10$, and damping implemented with a Higgs interaction. Similar to the $n=13$ case the steep drop around $m_a \sim 10^{-2}$ eV is because $\phi$-to-axion decay can entirely account for damping in this regime. The small-but-finite value of $\xi^2$ in this regime should be interpreted as zero. }
	\label{fig:HiggsPlotQHigherDimHiggsMmpn10}
\end{figure}

\FloatBarrier
\bibliography{UVKMM}

\end{document}